\documentclass[reprint,amsmath,amssymb,showpacs,aps,pra,superscriptaddress,nobibnotes]{revtex4-2}
\usepackage{dcolumn}
\usepackage[utf8]{inputenc}
\usepackage{amsmath}
\usepackage{mathtools}
\usepackage{adjustbox}
\usepackage{lipsum}
\usepackage[font=small,labelfont=bf,
   justification=Justified,
   format=plain]{caption}
\usepackage{comment}
\usepackage{enumitem}
\usepackage[normalem]{ulem}
\usepackage[clock]{ifsym}
\usepackage{nicefrac}
\usepackage{eurosym}
\usepackage{graphicx}
\usepackage{subcaption}
\usepackage{appendix}
\usepackage{nomencl}
\usepackage{amsfonts}
\usepackage{tabto}
\usepackage{subcaption}
\numberwithin{equation}{section}
\usepackage{amssymb}
\usepackage{fancyhdr}
\usepackage{amscd}
\usepackage{quantikz}
\usepackage{braket}
\usepackage{booktabs}
\usepackage[english]{babel}
\usepackage[dvipsnames]{xcolor}
\usepackage[colorlinks=true,linkcolor=blue,citecolor=blue,urlcolor=blue]{hyperref}
\usepackage[capitalise]{cleveref}
\usepackage[extra]{tipa}
\usepackage{newfloat}
\DeclareFloatingEnvironment[fileext=loa,name=Algorithm,placement=tbp,within=none]{algorithm}
\usepackage{algpseudocode}
\algrenewcommand\algorithmicrequire{\textbf{Input:}}
\algrenewcommand\algorithmicensure{\textbf{Output:}}
\newcommand{\Phase}[1]{\Statex\vspace{\dimexpr 5pt-\baselineskip}\Statex\textbf{#1}\vspace{1pt}}

\usetikzlibrary{calc}
\usetikzlibrary{spy}
\usetikzlibrary{arrows.meta,bending}
\usetikzlibrary{shapes.geometric}

\usetikzlibrary{decorations.markings,calc}

\newcommand{\gdr}[1]{\textcolor{ForestGreen}{#1}}

\usepackage{listings}
\usepackage{epstopdf}

\tikzset{
    gradient border/.style={
        draw=none,
        fill=none,
        inner sep=6pt,
        append after command={
            \pgfextra
            \begin{scope}

                \clip[rounded corners=4pt] (\tikzlastnode.north west) rectangle (\tikzlastnode.south east);
                \fill[left color=red!60, right color=blue!60, middle color=purple!30] 
                    (\tikzlastnode.north west) rectangle (\tikzlastnode.south east);
            \end{scope}
            
            \begin{scope}
                \clip [rounded corners=4pt] (\tikzlastnode.north west) rectangle (\tikzlastnode.south east);
                
                \shade[left color=red, right color=blue, middle color=purple] 
                    (\tikzlastnode.north west) rectangle (\tikzlastnode.south east);
                
                \fill[left color=red!20, right color=blue!20, middle color=purple!20, rounded corners=3.2pt] 
                    ([xshift=0.8pt, yshift=-0.8pt]\tikzlastnode.north west) 
                    rectangle 
                    ([xshift=-0.8pt, yshift=0.8pt]\tikzlastnode.south east);
            \end{scope}
            \endpgfextra
        }
    }
}

\newcounter{parrow}
\tikzset{
  record path/.style={
    /utils/exec=\tikzset{parrow/.cd,#1},
    decorate,
    decoration={
      markings,
      mark=at position 0 with {
        \setcounter{parrow}{1}
        \path (0,\pgfkeysvalueof{/tikz/parrow/dist}/2)
              coordinate (parrowt-\pgfkeysvalueof{/tikz/parrow/name}-\number\value{parrow})
               (0,-\pgfkeysvalueof{/tikz/parrow/dist}/2)
              coordinate (parrowb-\pgfkeysvalueof{/tikz/parrow/name}-\number\value{parrow});
        \pgfmathsetmacro{\mystep}{(\pgfdecoratedpathlength-4pt)/int(1+(\pgfdecoratedpathlength-4pt)/2pt)}
        \xdef\mystep{\mystep}
      },
      mark=between positions 2pt and 1 step \mystep pt with {
        \stepcounter{parrow}
        \path (0,\pgfkeysvalueof{/tikz/parrow/dist}/2)
              coordinate (parrowt-\pgfkeysvalueof{/tikz/parrow/name}-\number\value{parrow})
              (0,-\pgfkeysvalueof{/tikz/parrow/dist}/2)
              coordinate (parrowb-\pgfkeysvalueof{/tikz/parrow/name}-\number\value{parrow})
              (0,0)
              coordinate (parrowm-\pgfkeysvalueof{/tikz/parrow/name}-\number\value{parrow});
      }
    }
  },
  reconstruct top/.style={
    insert path={
      plot[variable=\t,samples at={1,...,\number\value{parrow}},smooth]
        (parrowt-#1-\t)
    }
  },
  reconstruct bottom/.style={
    insert path={
      plot[variable=\t,samples at={\number\value{parrow},\the\numexpr\value{parrow}-1,...,1},smooth]
        (parrowb-#1-\t)
    }
  },
  parrow/.cd,
  dist/.initial=4pt,
  step/.initial=2pt,
  name/.initial={}
}

\newcommand{\circleopen}{\tikz[baseline=-0.5ex]{\draw[black, thick] (0,0) circle (0.5ex);}}

\newcommand{\circlefill}{\tikz[baseline=-0.5ex]{\fill[black, thick] (0,0) circle (0.5ex);}}

\newcommand{\redcircleopen}{\tikz[baseline=-0.5ex]{\draw[red, thick] (0,0) circle (0.5ex);}}

\newcommand{\bluecircleopen}{\tikz[baseline=-0.5ex]{\draw[blue, thick] (0,0) circle (0.5ex);}}

\newcommand{\redcirclefill}{\tikz[baseline=-0.5ex]{\fill[red,thick] (0,0) circle (0.5ex);}}

\newcommand{\bluecirclefill}{\tikz[baseline=-0.5ex]{\fill[blue,thick] (0,0) circle (0.5ex);}}

\renewcommand{\gdr}[1]{#1}

\begin{document}

\renewcommand{\thesection}{\Roman{section}}
\numberwithin{equation}{section}

\title{Technical analysis of  the Resource-efficient Quantum Walkers Quantum Random Access Memory }

\author{Giuseppe De Riso}
\affiliation{Scuola Normale Superiore, Piazza dei Cavalieri 7, I-56126, Pisa, Italy}

\author{Giuseppe Catalano}
\affiliation{NEST-CNR Scuola Normale Superiore, Piazza dei Cavalieri 7, I-56126, Pisa, Italy}

\author{Seth Lloyd}
\affiliation{Dept. of Mech. Eng., Massachusetts Institute of Technology, 77 Mass. Av., Cambridge, MA 02139, USA}

\author{Vittorio Giovannetti}
\affiliation{NEST-CNR Scuola Normale Superiore, Piazza dei Cavalieri 7, I-56126, Pisa, Italy}

\author{Dario De Santis}
\affiliation{Scuola Normale Superiore, Piazza dei Cavalieri 7, I-56126, Pisa, Italy}

\begin{abstract}
Quantum Random Access Memory (qRAM) is a critical component for achieving quantum advantage in algorithms ranging from database search to quantum machine learning. In a recently introduced model [\href{https://www.arxiv.org/abs/2508.02855}{arXiv:2508.02855}], we proposed a resource-efficient qRAM architecture based on discrete-time quantum walkers. This article serves as a comprehensive technical follow-up, providing the full mathematical derivations, detailed protocol specifications, and in-depth resource analysis. 
Moreover, we extend the original proposal with novel techniques for the purpose of making the qRAM implementation more realistic.
Our model resolves the primary drawbacks of leading qRAM proposals: it avoids the exponential number of active nodes {$\mathcal{O}(2^n)$} required by the ``Bucket Brigade" architecture \cite{BB_architecture} by employing a number of quantum walkers that scales linearly with the address and message sizes, $n$ and $m$, respectively. Simultaneously, it overcomes the spatial bottlenecks of previous quantum-walker schemes \cite{Asaka_breve} by eliminating the need for multiple parallel trees. 
We propose two algorithmic paradigms: the long- and the short-range approaches, which differ by the length of interaction of the main routing gates employed in the qRAM. We formalize the routing, message-copy, and walker retrieval phases for both variants and we show how, while the long-range scheme employs controlled gates having a multitude of target systems, the short-range approach decomposes these interactions into sequences of 2- and 3-body \textit{local} gates, which improves the architecture's feasibility for near-term experimental implementation. 
Finally, our comprehensive resource analysis confirms that the short-range approach achieves the optimal $\mathcal{O}(n+m)$ circuit depth.
Within these two paradigms, we explore the potential of different types of quantum walkers, \gdr{namely bosons, dual-rail qubits and four-level qudits}.

\end{abstract}
\maketitle

\section{Introduction}
\label{sec:intro}
A central challenge in quantum computing is the efficient coherent retrieval of information from large datasets. While classical computers rely on Random Access Memory (RAM) to  retrieve information from specific memory cells, quantum algorithms often require querying data in coherent superpositions: this capability is provided by a quantum Random Access Memory (qRAM)~\cite{nielsen_chuang}.
Formally, consider a database containing $N = 2^n$ memory cells. Each cell $M^{(\boldsymbol{a})}$ is identified by a unique $n$-bit binary address $\boldsymbol{a} \in \{0,1\}^n$ and stores an $m$-bit classical message $\boldsymbol{b}^{(\boldsymbol{a})}$.
When a specific address is encoded into the state $\ket{\boldsymbol{a}}_A$ of an address register $A$, the qRAM accesses the corresponding memory cell and retrieves the content $\boldsymbol{b}^{(\boldsymbol{a})}$ into a data register $D$. 
In case we need a specific superposition of the information contained into multiple memory cells, we can initialize our address register $A$ in a superposition of the corresponding addresses
$\sum_{\boldsymbol{a}}\alpha_{\boldsymbol{a}}\ket{\boldsymbol{a}}_A$. In this case, we require the qRAM to implement the unitary transformation:
\begin{equation}\label{generalqRAM}
\sum_{\boldsymbol{a}}\alpha_{\boldsymbol{a}}\ket{\boldsymbol{a}}_A\ket{0}_D\xrightarrow{\text{qRAM}} \sum_{\boldsymbol{a}}\alpha_{\boldsymbol{a}}\ket{\boldsymbol{a}}_A\ket{\boldsymbol{b}^{(\boldsymbol{a})}}_D,
\end{equation}
where $\ket{0}_D$ is a fiducial initial state of the data register.

The development of a scalable, resource-efficient, and noise-resilient qRAM architecture is a critical prerequisite for achieving quantum advantage in several scenarios. It is important to stress that such advantage is guaranteed only if the qRAM query time scales efficiently, typically as $\mathcal{O}(\text{poly} \,n)$; otherwise, an exponential overhead would negate any algorithmic speedup.
\gdr{A well-known example of an algorithm that can benefit from efficient qRAM is Grover’s database search~\cite{Grover_search}}. Additionally, it is fundamental for linear algebra subroutines~\cite{Harrow_Linear_algebra,Low_Linear_algebra, Clader_Linear_algebra, Giovannetti_Linear_algebra, Childs_Linear_algebra}, quantum chemistry simulations~\cite{Kassal_quantum_chemistry, Whitfield_Quantum_chemistry, Babbush_Quantum_chemistry, Cao_Quantum_chemistry}, quantum machine learning~\cite{Lloyd_machine_learning, Poggiali_machine_learning, Rebentrost_machine_learning,Lloyd_machine_learning_2, Zhao_Machine_learning, Kerenidis_Machine_learning}, and cryptography~\cite{Giovannetti_Cryptography,Kuperberg_cryptography}.

The first major architecture proposed is the ``Bucket Brigade" (BB) model by Giovannetti, Lloyd, and Maccone ~\cite{Bucket_Brigade}. 
This proposal is modeled as a binary tree, where the $N=2^n$ memory cells are its final leaves.
Its routing mechanism involves sequentially sending $n$ address qubits into the tree; each address qubit sets the state of a three-level system (qutrit), which in turn directs the subsequent address and data particles. The primary advantage of the BB model is its celebrated resilience to noise, as only $\mathcal{O}(n)$ components are activated along a single path during any classical query~\cite{Hann_noise_resilience}. However, its principal drawback is the exponential hardware complexity: the architecture requires $\mathcal{O}(2^n)$ active quantum elements, i.e. the qutrits at each node, whose coherence must be preserved throughout the entire protocol. Indeed, while for a classical query only $\mathcal{O}(n)$ nodes are activated along a single path, a superposition query engages the entire array of $\mathcal{O}(2^n)$ nodes coherently, representing a concrete technological barrier for large-scale implementations.
A different paradigm, pioneered by Asaka, Sakai, and Yahagi (ASY), employs discrete-time quantum walkers (QWs) as information carriers~\cite{Asaka_1,Asaka_2,Asaka_breve}. 
The principal improvement of this model is the employment of a linear number of quantum particles, namely $n$ address walkers and $m$ data walkers, instead of exponential, as in the case of the BB model.
However, this proposal suffers from two significant scalability challenges: it requires $2(n+m)$ distinct parallel binary trees, and it relies on long-range interactions to correlate the internal states of walkers that are physically separated and travel along these multitude of binary trees.

These limitations are further contextualized by Ref.~\cite{Jaques_QRAM_survey}, which provides a systematic critique of existing qRAM proposals. While the authors argue that a practically viable qRAM would ideally be ``passive'' or ``ballistic'', they acknowledge that no existing proposal satisfactorily meets this requirement. Within the landscape of active architectures, our proposal offers concrete advantages over both the BB and ASY models: it reduces the number of active quantum carriers from $\mathcal{O}(2^n)$ to $\mathcal{O}(n+m)$, operates on a constant number of binary trees, and achieves optimal $\mathcal{O}(n+m)$ circuit depth using exclusively short-range interactions.

In a recent work~\cite{ourPRL}, we introduced a novel architecture that bridges the gap between the BB and ASY models. In this paper, we present a comprehensive technical characterization of this architecture together with novel additions. We expand significantly upon the initial proposal by establishing a unified framework that adapts the routing logic to diverse physical platforms, embracing the specific constraints and advantages of \gdr{ bosonic and qudit-based information carriers.} 
Crucially, our proposal synthesizes the advantages of the leading paradigms while eliminating their primary drawbacks. 

It avoids the linear number of parallel binary trees required by previous walker-based schemes, operating instead on a minimal, constant number of binary trees. Unlike the BB model, the nodes of our routing tree act as static, passive gates rather than active quantum elements whose coherence must be maintained throughout the protocol. Most importantly, when compared against any existing qRAM architecture, our proposal always exhibits favorable resource scaling under at least one key metric: number of active quantum carriers, circuit depth, spatial extent, or interaction range.

Our contribution is structured around three main pillars. First, we establish the operational logic of our qRAM by introducing a {\it \gdr{basic} protocol} based on bosonic carriers. We use this model as a pedagogical baseline: while conceptually simple, it allows us to rigorously define the fundamental routing mechanisms and memory access operations without the overhead of particle-number conservation constraints. Second, addressing the critical requirement for experimental feasibility, we introduce and analyze the {\it backup variant}. We prove that by introducing a constant overhead in the number of carriers, each non-local interaction in the routing protocol can be decomposed into a sequence of purely local, short-range operations (up to range-2 interactions). Furthermore, through a suitable parallelization of the routing operations at different levels of the binary tree, this variant achieves an optimal $\mathcal{O}(n+m)$ circuit depth for the query. This linear scaling decisively outperforms ASY in superposition queries and matches the best theoretical depth of parallelized BB implementations, but without the challenge of maintaining the quantum correlation of an exponential number of quantum particles. Third, we provide a detailed architectural blueprint for realistic hardware implementations. \gdr{We step away from the abstract bosonic model to explicitly construct the protocol for dual-rail systems and four-level qudits, adapting the framework to platforms governed by strict particle conservation laws.}

This paper is organized as follows. After formalizing the definition of a qRAM in Section~\ref{sec:definition}, we review the existing BB and ASY architectures to establish the necessary benchmarks in Section~\ref{sec:notable}. Sections~\ref{sec:our_model} and~\ref{sec:tech_desc} define our \gdr{basic} protocol in the bosonic picture, where the former gives a high-level overview of the routing dynamics, while the latter details the mathematical definitions of the employed unitary gates and how the query of the memory cells is made possible. In Section~\ref{sec:backup}, we present the backup variant, proving how local operations can replace long-range gates through the use of auxiliary carriers. Section~\ref{sec:dualrail} is dedicated to introduce a dual-rail variant of our qRAM, while in Section~\ref{sec:qudit} we present a qudit-based encodings. A detailed resource analysis of our models is provided in Section~\ref{sec:scaling}, comparing circuit depth and qubit count against the state of the art. Finally, Section~\ref{sec:conclusion} concludes with a discussion 
of open questions and future directions.

\section{qRAM definition}
\label{sec:definition}

A quantum Random Access Memory (qRAM) is a physical device designed to efficiently access a database $M$ containing $N=2^n$ memory cells. Each memory cell contains an $m$-bit classical string $\boldsymbol{b}=b_1\dots b_m\in\{0,1\}^m$, uniquely identified by an $n$-bit binary address $\boldsymbol{a}=a_1\dots a_n\in\{0,1\}^n$. Let $\mathcal{A}=\{0,1\}^n$ denote the set of possible $n$-bit addresses. We denote by $M^{(\boldsymbol{a})}$ the memory cell associated with address $\boldsymbol{a}$, storing the classical message $\boldsymbol{b}^{(\boldsymbol{a})}$.

Three basic components needed for the functioning of a qRAM are the address, the data and the memory registers, which we define as follows: 
\begin{itemize}[label=--]
    \item 
    \textit{Address register}: an $n$-qubit register $A=\{A_1,\dots,A_n\}$ with Hilbert space $\mathcal{H}_A\cong(\mathbb{C}^2)^{\otimes n}$. The fundamental role of this register is to allow a correct routing towards the desired memory cells. Its initialization strictly depends on the particular memory cell(s) to be reached, possibly in superposition;
    
    \item 
    \textit{Data register}: an $m$-qubit register $D=\{D_1,\dots,D_m\}$ with Hilbert space $\mathcal{H}_{D}\cong(\mathbb{C}^2)^{\otimes m}$, initialized in the fiduciary state $\ket{0}^{\otimes m}$. This register has the role to store the information retrieved from the desired memory cells;
    
    \item \textit{Memory register}: a collection of $N$ registers $M=\{M^{(\boldsymbol{a})}\}_{\boldsymbol{a}\in\mathcal{A}}$, where each $M^{(\boldsymbol{a})}$ consists of $m$ qubits with Hilbert space $\mathcal{H}_{M^{(\boldsymbol{a})}}\cong(\mathbb{C}^2)^{\otimes m}$. Each register physically represents a memory cell indexed by $\boldsymbol{a}$. The overall memory Hilbert space is $\mathcal{H}_{M}=\bigotimes_{\boldsymbol{a}\in\mathcal{A}}\mathcal{H}_{M^{(\boldsymbol{a})}}$. 
\end{itemize} 
In the simplest scenario, let us consider the memory register to be in a classical state $\ket{\Psi_{\boldsymbol{b}}}_M = \bigotimes_{\boldsymbol{a}\in\mathcal{A}} \ket{\boldsymbol{b}^{(\boldsymbol{a})}}_{\boldsymbol{M}^{(\boldsymbol{a})}}$, representing the database content, where each $\ket{\boldsymbol{b}^{(\boldsymbol{a})}}_{M^{(\boldsymbol{a})}} = \bigotimes_{j=1}^{m}\ket{b_j^{(\boldsymbol{a})}}_{M_j^{(\boldsymbol{a})}}$ is a classical bit string stored in a product state across the $m$ qubits of each memory cell. If we initialize the address register in a classical state $\ket{\boldsymbol{a}}_A$ and the data register in the fiducial state $\ket{\boldsymbol{0}}_D$, the extraction of information via the qRAM can be modeled as the unitary map:
\begin{equation}            \hat{U}_{qRAM}\ket{\boldsymbol{a}}_A\ket{\boldsymbol{0}}_D\ket{\Psi_{\boldsymbol{b}}}_M = \ket{\boldsymbol{a}}_A\ket{\boldsymbol{b}^{(\boldsymbol{a})}}_D\ket{\Psi_{\boldsymbol{b}}}_M.
\end{equation}
Crucially, this action extends to superpositions of addresses, and this feature is the main difference between an ordinary RAM and a qRAM. If the address register is prepared in a generic state $\sum_{\boldsymbol{a}}\alpha_{\boldsymbol{a}}\ket{\boldsymbol{a}}_A$, the qRAM outputs an entangled state between the address and the data registers:

\begin{multline}\label{qRAMgeneral}
\hat{U}_{qRAM}\left[\sum_{\boldsymbol{a}\in\mathcal{A}}\alpha_{\boldsymbol{a}}\ket{\boldsymbol{a}}_A\ket{\boldsymbol{0}}_D\right]\ket{\Psi_{\boldsymbol{b}}}_M \\ = \left[\sum_{\boldsymbol{a}\in\mathcal{A}}\alpha_{\boldsymbol{a}}\ket{\boldsymbol{a}}_A\ket{\boldsymbol{b}^{(\boldsymbol{a})}}_D\right] \ket{\Psi_{\boldsymbol{b}}}_M.
\end{multline}

In other words, the qRAM unitary accepts a superposition of addresses as input and returns a coherent superposition of address states correlated with the corresponding classical information stored in the memory cells. From an operational perspective, $\hat{U}_{qRAM}$ acts as a dynamical bridge connecting the three registers. 

Specifically, depending on the initial state of $A$, it routes both the address $A$ and data $D$ registers to the target memory cells $M^{\bf(a)}$ in the memory register $M$. Hence, it writes in $D$ the information contained in these cells, preserving the overall coherence of the superposition between the address and the corresponding memory cell information. Finally, it routes $A$ and $D$ to the output port.

\section{Notable architectures}
\label{sec:notable}

Several architectures have been proposed to implement a qRAM. In this section, we briefly review two representative examples that serve as benchmarks for the schemes proposed in this work.

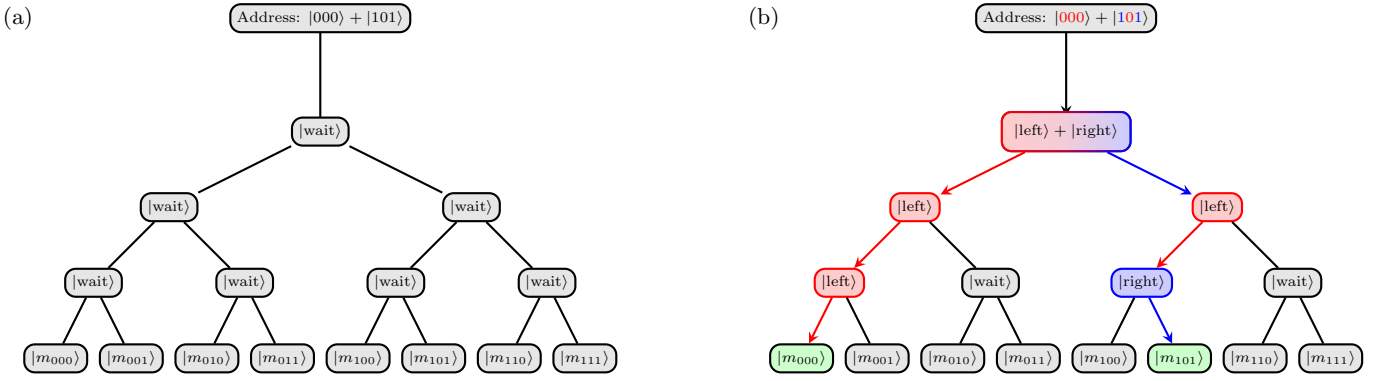
\begin{figure*}[htb]
    \begin{minipage}{0.45\textwidth}
    \begin{tikzpicture}[edge from parent/.style={draw},
  level distance=1cm,
  level 1/.style={sibling distance=4cm},
  level 2/.style={sibling distance=2cm},
  level 3/.style={sibling distance=1cm},
  every node/.style={draw=black, fill=gray!20, rectangle, rounded corners, align=center, scale=0.8}
  ]
  \draw[solid, thick] (0, 1.5) -- (0, 0.0);
  \node[thick] at (0, 1.5) {\scriptsize{Address: $\ket{000}+\ket{101}$}};
  \node[thick] {\scriptsize {$\ket{\text{wait}}$}}
    child[draw=black, thick]  {node {\scriptsize $\ket{\text{wait}}$}
      child[draw=black, thick]{node {\scriptsize $\ket{\text{wait}}$}
        child[draw=black, thick] {node {\scriptsize $\ket{m_{000}}$}}
        child[draw=black] {node {\scriptsize $\ket{m_{001}}$}}
      }
      child[draw=black] {node {\scriptsize $\ket{\text{wait}}$}
        child {node {\scriptsize $\ket{m_{010}}$}}
        child {node {\scriptsize $\ket{m_{011}}$}}
      }
    }
    child [draw=black, thick] {node {\scriptsize $\ket{\text{wait}}$}
      child [draw=black, thick] {node {\scriptsize $\ket{\text{wait}}$}
        child [draw=black] {node {\scriptsize $\ket{m_{100}}$}}
        child [draw=black, thick] {node {\scriptsize $\ket{m_{101}}$}}
      }
      child [draw=black] {node {\scriptsize $\ket{\text{wait}}$}
        child[draw=black] {node {\scriptsize $\ket{m_{110}}$}}
        child [draw=black] {node {\scriptsize $\ket{m_{111}}$}}
      }
    };

      \node[draw=none, fill=none, scale=1.2, transform shape] at(-4,1.5){(a)};
    \end{tikzpicture}
    \end{minipage}
    \hfill
    \begin{minipage}{0.45\textwidth}
    \begin{tikzpicture}[edge from parent/.style={draw},
  level distance=1cm,
  level 1/.style={sibling distance=4cm},
  level 2/.style={sibling distance=2cm},
  level 3/.style={sibling distance=1cm},
  every node/.style={draw=black, fill=gray!20, rectangle, rounded corners, align=center, scale=0.8}
  ]
  \draw[-stealth,solid, thick] (0, 1.5) -- (0, 0.2);
  
  \node[thick] at (0, 1.5) {\scriptsize{Address: $\ket{\textcolor{red}{000}}+\ket{\textcolor{blue}{1}\textcolor{red}{0}\textcolor{blue}{1}}$}};
  % MODIFICA QUI per lo step 2
  \node[gradient border]{\scriptsize {$\ket{\text{left}}+\ket{\text{right}}$}}
    child[-stealth,draw=red, thick]  {node[draw=red,fill=red!20] {\scriptsize $\ket{\text{left}}$}
      child[-stealth,draw=red, thick]{node[draw=red,fill=red!20] {\scriptsize $\ket{\text{left}}$}
        child[draw=red, thick] {node [fill=green!20] {\scriptsize $\ket{m_{000}}$}}
        child[-,draw=black] {node {\scriptsize $\ket{m_{001}}$}}
      }
      child[-,draw=black] {node {\scriptsize $\ket{\text{wait}}$}
        child {node {\scriptsize $\ket{m_{010}}$}}
        child {node {\scriptsize $\ket{m_{011}}$}}
      }
    }
    child [-stealth,draw=blue, thick] {node[draw=red,fill=red!20] {\scriptsize $\ket{\text{left}}$}
      child [draw=red, thick] {node[draw=blue,fill=blue!20] {\scriptsize $\ket{\text{right}}$}
        child [-,draw=black] {node {\scriptsize $\ket{m_{100}}$}}
        child [draw=blue, thick] {node [fill=green!20] {\scriptsize $\ket{m_{101}}$}}
      }
      child [-,draw=black] {node {\scriptsize $\ket{\text{wait}}$}
        child[draw=black] {node {\scriptsize $\ket{m_{110}}$}}
        child [draw=black] {node {\scriptsize $\ket{m_{111}}$}}
      }
    };

    \node[draw=none, fill=none, scale=1.2, transform shape] at(-4,1.5){(b)};
    \end{tikzpicture}
    \end{minipage}
    \caption{The Bucket Brigade scheme for a superposition of addresses. (a) All qutrits are initialized in the wait state $\ket{\text{wait}}$, leaving the routing paths to the memory cells closed. (b) The address register qubits are injected sequentially into the tree, changing the states of the qutrits (red for $\ket{\text{left}}$ and blue for $\ket{\text{right}}$), effectively opening a superposition of paths toward the target memory cells (green).}
    \label{BBarchitecture}
\end{figure*}

\paragraph*{Bucket Brigade (BB):} 
The Bucket Brigade architecture, introduced by Giovannetti, Lloyd, and Maccone~\cite{Bucket_Brigade}, is designed to bypass the need for an exponential number of active gates, typical of fan-out architectures \cite{BB_architecture}. The scheme relies on a perfect binary tree of depth $n$, where the $N=2^n$ final leaves coincide with the memory cells. Each internal node is equipped with a three-level quantum system, namely a qutrit, initialized in a passive state $\ket{\mathrm{wait}}$ (see Fig. ~\ref{BBarchitecture}a).
The routing protocol operates via a sequential cascade mechanism. The address qubits $A_1, \dots, A_n$ are injected one by one into the root of the tree. When the $k$-th address qubit $A_k$ enters the system, it traverses the first $k-1$ levels guided by the switches activated by the previous qubits. Once it reaches a node at level $k$ in the $\ket{\mathrm{wait}}$ state, a unitary interaction transfers the logical information from the qubit to the spatial degree of freedom of the switch. Specifically, the mapping is defined as:
\begin{align}
    \ket{\mathrm{wait}}_j \ket{0}_{A_k} &\mapsto \ket{\mathrm{left}}_j \ket{\varnothing}_{A_k}, \nonumber \\
    \ket{\mathrm{wait}}_j \ket{1}_{A_k} &\mapsto \ket{\mathrm{right}}_j \ket{\varnothing}_{A_k},
\end{align}
where $\ket{\emptyset}_{A_k}$ denotes that the address carrier has been absorbed or reflected, effectively transferring its information to the router node. This step extends the active path by one level, enabling the subsequent qubit $A_{k+1}$ to travel further down the tree. After $n$ such injections, a unique physical path is established from the root to the target memory cell. This process constructs a physical path connecting the root to the target memory cell (see Fig. ~\ref{BBarchitecture}b).
Once the path is established, the data register traverses the tree to access the memory cell. Finally, the path is dismantled by reversing the operations.

While the \gdr{basic} sequential protocol requires a query time of $\mathcal{O}(n^2)$, subsequent optimizations based on pipelining have demonstrated a possible reduction to $\mathcal{O}(n)$~\cite{Hann_noise_resilience}.
A key feature of the BB scheme is its noise resilience: for a classical address, only $\mathcal{O}(n)$ nodes are activated. 
While theoretical models suggest favorable error scaling~\cite{Hann_noise_resilience}, the experimental overhead of maintaining $N$ coherent, addressable three-level systems remains a significant technological barrier.

\paragraph*{Quantum Walker based scheme (ASY):}A more recent approach, introduced by Asaka, Sakai, and Yahagi, proposes a qRAM implementation using quantum walkers as information carriers. In this scheme, purpose-specific unitaries are introduced to route the walkers toward the memory cells. In particular, these gates eliminate the presence of active routing elements that must be entangled with the information carriers and whose coherence must be preserved throughout the protocol. The scheme, however, introduces other types of hardware overhead. For every walker, a dual rail encoding is required, which implies the need for $2(n+m)$ binary trees over which the information carriers move and are manipulated. More precisely, the protocol requires interaction between the carriers; thus, highly nonlocal operations between different binary trees are needed for the proper functioning of this scheme.

\section{General functioning of our qRAM}
\label{sec:our_model}

In this section, we provide a general overview of our qRAM architecture and introduce the fundamental principles guiding the coherent routing protocol. The step-by-step workflow of our scheme is summarized in Algorithm~\ref{High_level_pseudocode}.

Our proposal employs quantum walkers as dynamic information carriers propagating along a single binary
tree structure. Similarly to the BB and the ASY models, the tree has depth $n$ and $N=2^n$ leaves corresponding to the cells of the quantum memory. Each leaf stores an $m$-bit classical message that can be accessed coherently through the routing dynamics of the walkers.
We employ quantum walkers that consist in two-level quantum systems with an internal degree of freedom referred to as its color, which is used to encode logical information. Walkers are injected into the tree in a spatially ordered configuration, with a fixed separation $\Delta x$ between consecutive particles. This spatial structure is an integral part of the encoding and routing protocol. The logical information is encoded both in the internal states of the walkers and, when relevant, in their spatial arrangement, depending on the implementation details of the routing protocol, as discussed in the following sections. We anticipate that, later in this work, we also introduce an alternative qudit version of our qRAM which, instead of qubits, employs four-level systems.

Operationally, our qRAM scheme consists in the application of a particular combination of quantum gates to a precise set of quantum walkers traveling along the binary tree.
These walkers are divided into two distinct registers, called the address register $\boldsymbol{A}$ and the data register $\boldsymbol{D}$. Importantly, data walkers are initialized to always follow address walkers: this arrangement is fundamental both for the correct routing to the memory cells and to the output port. The physical routing is dictated by the address information encoded in $\boldsymbol{A}$, while the walkers in $\boldsymbol{D}$ are employed to perform operations on the memory cells and to coherently retrieve information from them.
The application of address-independent quantum gates on $A$ and $D$, allows the data register $D$ to reach the desired memory cells. At this stage,
the (classical) information in the memory cells is coherently copied into the data register particles. Finally, the registers $A$ and $D$ undergo an inverse-routing phase, where they are routed to the qRAM output port.

We design a routing strategy for the walkers train traveling towards the memory cells, ensuring that the state of the $d$-th address particle dictates the direction taken by all subsequent address and data walkers at the bifurcation at depth
$d$. \gdr{To achieve this coherent routing toward the target memory cells we apply a modular unitary block at each bifurcation of the binary tree. The block is composed of two operations arranged in sequence: an address-transmission gate $\hat{U}^{(d)}$ which acts on the walkers while they travel along the edge at
depth $d$ and a scattering gate $\hat{S}$, placed at the bifurcation.}

The first gate encountered by the walkers, $\hat{U}^{(d)}$, has the role to transfer the path information carried by the (active) address walker $A_d$ to all the subsequent walkers.
The explicit form of this gate depends on the implementation we refer to and therefore we discuss them in the dedicated sections.

Subsequently, the scattering gate $\hat{S}$ acts as a passive unitary that conditionally routes the walkers from level $d$ to level $d+1$. The routing choice is uniquely determined by the internal state of each walker, thereby associating the color information with a unique path choice. This gate remains identical for all the protocols described below, apart from a variation needed in the qudit case.

A unique feature of our protocol is the path taken by the address walkers during the routing phase and the subsequent output phase. At any given depth $d$, only the color information of the $d$-th address walker is relevant for the routing and must be propagated to the walkers following it, as described. Once $A_d$ has transmitted its information via $\hat{U}^{(d)}$, its routing function is complete. This sequential dependence implies that, in general, each address walker will take a different path from the others following it. Consequently, at the end of the routing protocol, only the data walkers $\boldsymbol{D}$
are ensured to reach the targeted memory cells. 
Nonetheless, during the inverse routing phase, the {\it dispersed} address particles are sequentially recollected, guaranteeing the correct routing towards the output port. Finally, the state of the address and data registers are transformed as in Eq.~(\ref{qRAMgeneral}), a superposition of the desired addresses and the information contained in the corresponding memory cells.

In the following sections, we present the different variants of our qRAM strategy, all sharing this common underlying logic to efficiently route particles along the binary tree.
We start by explaining the {\it \gdr{basic}} architecture, in Section~\ref{sec:tech_desc}, which is particularly pedagogical to introduce the reader to the main features shared by all the schemes. Nonetheless, this strategy relies on long-range interactions that can be replaced with short-range ones, as demonstrated by the {\it backup} variant in Section~\ref{sec:backup}. This improvement, along with a better scaling for the time of execution of our qRAM and other structural trade offs, motivates the development of the alternative qRAM models we introduce here. Specifically, in Section~\ref{sec:dualrail} we introduce the {\it dual-rail} protocol alongside its corresponding backup implementation. Finally, in Section~\ref{sec:qudit}, we 
present the {\it qudit}-based architecture and its purely 
short-range counterpart.

\begin{algorithm}[t]
\setlength{\abovecaptionskip}{0pt}
\setlength{\belowcaptionskip}{0pt}
\hrule height.8pt \vspace{2pt}
\caption{Quantum Walkers qRAM}
\label{High_level_pseudocode}
\vspace{1pt}\hrule \vspace{2pt}
\begin{algorithmic}[1]
\algrenewcommand\algorithmicrequire{\textbf{Input:}}
\algrenewcommand\algorithmicensure{\textbf{Output:}}
\Require Address register $\boldsymbol{A}$ encoding the target memory cell address $\ket{\boldsymbol{a}}_{\boldsymbol{A}}$, and data register $\boldsymbol{D}$ initialized in a reference state. A depth $n$ binary tree with $2^n$ memory cells, where the target cell $M^{(\boldsymbol{a})}$ stores the $m$ bit message $\boldsymbol{b}^{(\boldsymbol{a})}$.
\Ensure A state $\ket{\boldsymbol{b}^{(\boldsymbol{a})}}_{\boldsymbol{D}}\ket{\boldsymbol{a}}_{\boldsymbol{A}}$ where the data walkers encode the queried message.

\Phase{Initialization}
\State Encode information into the walker train
$\boldsymbol{W}$ composed by the registers $\boldsymbol{D}$ and $\boldsymbol{A}$,
according to the architecture logic.
\State Inject the train into the root of the routing tree.

\Phase{Routing}
\For{$d = 1,\dots,n$}
    \State Apply address-transmission gate $\hat{U}^{(d)}$ controlled by $A_d$ on all following walkers
    \State Propagate the train from level $d$ to level $d+1$ using the scattering gate $\hat{S}$
\EndFor

\Phase{Memory Query}
\State Trigger the copy operation $\hat{U}_{copy}$ at the target memory cell $M^{(\boldsymbol{a})}$
\State Transfer the stored message $\boldsymbol{b}^{(\boldsymbol{a})}$ into the internal state of the data walkers $\{D_j\}$

\Phase{Inverse Routing}
\For{$d = n+1,\dots,2$}
\State Propagate the walkers from level $d$ to level $d-1$ using the scattering gate $\hat{S}^{\dagger}$
\State Apply the address-transmission gate $\hat{U}^{(d-1)}$ controlled by $A_{d-1}$ on all following walkers
\EndFor
\State \Return Data register $\boldsymbol{D}$ in state $\ket{\boldsymbol{b}^{(\boldsymbol{a})}}_{\boldsymbol{D}}$
and address register $\boldsymbol{A}$ in state $\ket{\boldsymbol{a}}_{\boldsymbol{A}}$.
\end{algorithmic}
\vspace{3pt}\hrule height.8pt
\end{algorithm}

\begin{figure*}[t]
\begin{tikzpicture}[transform shape,
  grow=east,
  level 1/.style={level distance=1.5cm, sibling distance=3.2cm},
  level 2/.style={level distance=1.5cm, sibling distance=1.6cm},
  level 3/.style={level distance=1.5cm, sibling distance=0.8cm},
  edge from parent/.style={draw, thick},
 every node/.style={rectangle, draw, thick, minimum size=5mm, inner sep=0pt, fill=orange!20},
  leaf/.style={rectangle, draw, fill=gray!30, minimum height=8mm, minimum width=15mm, anchor=west}
]

\node[draw=none, fill=none] at (-2.5,4.6){\textbf{Depth}};

\draw[dashed] (0,5) -- (0,0);
\node[fill=none, draw=none] at (-.75,4.6){$d=0$};

\draw[dashed] (1.5,5) -- (1.5,-1.6);
\node[fill=none, draw=none] at (0.75,4.6){$d=1$};

\draw[dashed] (3,5) -- (3,-2.5) ;
\node[fill=none, draw=none] at (2.25,4.6){$d=2$};

%\draw[dashed] (4.5,5) -- (4.5,-2.5) ;
\node[fill=none, draw=none] at (3.75,4.6){$d=3$};

% PRIMO ALBERO (grow=east)
\node (root) {}
  child {node (L1_0) {}
    child {node (L2_0) {}
      child {node[leaf] {\Large$000$} edge from parent {}}
      child {node[leaf] {\Large$001$} edge from parent {}}
      edge from parent {}
    }
    child {node (L2_1) {}
      child {node[leaf] {\Large$010$} edge from parent {}}
      child {node[leaf] {\Large$011$} edge from parent {}}
      edge from parent {}
    }
    edge from parent {}
  }
  child {node (L1_1) {}
    child {node (L2_2) {}
      child {node[leaf] {\Large$100$} edge from parent {}}
      child {node[leaf] {\Large$101$} edge from parent {}}
      edge from parent {}
    }
    child {node (L2_3) {}
      child {node[leaf] {\Large$m_{110}$} edge from parent {}}
      child {node[leaf] {\Large$111$} edge from parent {}}
      edge from parent {}
    }
    edge from parent {}
  };

\draw[dashed] (10.5,5) -- (10.5,0);
\node[fill=none, draw=none] at (11.25,4.6){$d=0$};

\draw[dashed] (9,5) -- (9,-1.6);
\node[fill=none, draw=none] at (9.75,4.6){$d=1$};

\draw[dashed] (7.5,5) -- (7.5,-2.5) ;
\node[fill=none, draw=none] at (8.25,4.6){$d=2$};

\node[fill=none, draw=none] at (6.75,4.6){$d=3$};

% SECONDO ALBERO (grow=west)
\begin{scope}[grow=west,
  level 1/.style={level distance=1.5cm, sibling distance=3.2cm},
  level 2/.style={level distance=1.5cm, sibling distance=1.6cm},
  level 3/.style={level distance=1.5cm, sibling distance=0.8cm},
  edge from parent/.style={draw, thick},
  every node/.style={rectangle, draw, thick, minimum size=5mm, inner sep=0pt, fill=orange!20},
  leaf/.style={rectangle, draw, fill=gray!30, minimum height=8mm, minimum width=15mm, anchor=east}
]
\node (root2) at (10.525,0) {}
  child {node (R1_0) {}
    child {node (R2_0) {}
      child {node[leaf] {$\boldsymbol{M}^{(000)}$} edge from parent {}}
      child {node[leaf] {$\boldsymbol{M}^{(001)}$} edge from parent {}}
      edge from parent {}
    }
    child {node (R2_1) {}
      child {node[leaf] {$\boldsymbol{M}^{(010)}$} edge from parent {}}
      child {node[leaf] {$\boldsymbol{M}^{(011)}$} edge from parent {}}
      edge from parent {}
    }
    edge from parent {}
  }
  child {node (R1_1) {}
    child {node (R2_2) {}
      child {node[leaf] {$\boldsymbol{M}^{(100)}$} edge from parent {}}
      child {node[leaf] {$\boldsymbol{M}^{(101)}$} edge from parent {}}
      edge from parent {}
    }
    child {node (R2_3) {}
      child {node[leaf] {$\boldsymbol{M}^{(110)}$} edge from parent {}}
      child {node[leaf] {$\boldsymbol{M}^{(111)}$} edge from parent {}}
      edge from parent {}
    }
    edge from parent {}
  };
\end{scope}

\def\n{10}
\def\spacing{0.4} 
\draw[thick] (-3,0) -- (-0.25,0);
\foreach \i/\col in {0/red, 1/white, 2/red, 3/red, 4/red, 5/red} {
\edef\colorname{\col} 
\filldraw[black,thick, fill=\colorname] ( -\i*\spacing-0.75,0) circle (4pt);}

\draw[thick] (10.775,0) -- (13.5,0);

\def\n{10}
\def\spacing{0.4} 
\foreach \i/\col in {0/red, 1/white, 2/red, 3/white, 4/red, 5/red} {
\edef\colorname{\col} 
\filldraw[black,thick, fill=\colorname] ( -\i*\spacing-0.75+14,0) circle (4pt);}

\draw[decorate,decoration={brace,amplitude=5pt}, thick] (-0.55,-0.3) -- (-1.75,-0.3);

\draw[decorate,decoration={brace,amplitude=5pt}, thick] (-1.8,-0.3) -- (-2.95,-0.3);

\node[align=center, draw=none, text width=1.5cm, fill=none] at (-1.15, -0.7) {$\boldsymbol{A}$};

\node[align=center, draw=none, text width=1.5cm, fill=none] at (-2.4, -0.7) {$\boldsymbol{D}$};

\draw[decorate,decoration={brace,amplitude=5pt},very thick] (4.5,3.5) -- (6,3.5);

\node[align=center, draw=none, text width=2cm, fill=none] at (5.25, 4) {$N=2^3$};

\node[align=left, draw=none, fill=none] at (-3, 0.5) {
 $\ket{\psi_{in}}$
};

\draw[-stealth, thick](-2.5,0.5)--(-1,0.5)node[midway, draw=none, fill=none, above] {\text{input}};

\draw[-stealth, thick](11.5,0.5)--(13,0.5)node[midway, draw=none, fill=none, above] {\text{output}};

\end{tikzpicture}

\caption{Schematic depiction of our \gdr{basic} qRAM implementation on a binary tree of depth $d=3$ with $N=2^3$ memory cells. Red circles denote walkers in state $\ket{R}$ while white circles denote walkers in state $\ket{\varnothing}$. The address register $\boldsymbol{A}$ uses two walkers to encode the classical address $\boldsymbol{a}=101$, while the data register $\boldsymbol{D}$ uses two data walkers plus a flag walker for message retrieval. The overall initial state injected at the root node of the tree is $\ket{\psi_0} = \ket{R}_{\boldsymbol{D}}^{\otimes 3} \otimes \ket{R\varnothing R}_{\boldsymbol{A}}$. %\dds{\it [Magari metterei una scrittina con "input" e "output" come avevamo detto l'altra volta, e magari anche uno stato per l'output, dove i $D$ walker sono diversi da quelli dell'input.]}
}
\label{binarytree}
\end{figure*}
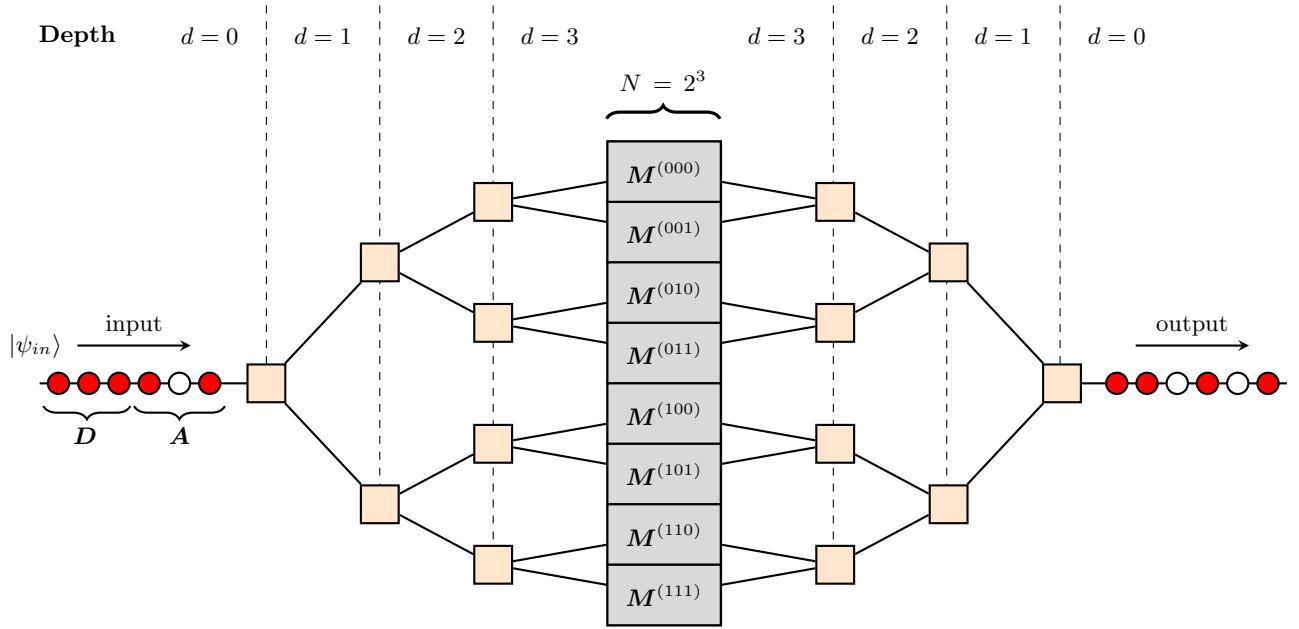

\section{Basic variant}
\label{sec:tech_desc}

In this section, we describe the main components and the functioning of our \gdr{basic} qRAM proposal. We define structure of the address and data walkers alongside with their initialization. We follow by describing the unitary operations required to route them toward the target memory cells and the operations to copy the contained information into the data register particles. Finally, we discuss the inverse routing toward the output port of the binary tree.

\subsection{Encoding and initialization}

In the \gdr{basic} qRAM protocol proposed here, a natural physical realization of our walkers is given by bosonic 
walkers, such as single photons. Due to the bosonic statistics, the state space for a single mode includes different numbers of particles. In the operational regime considered for our architecture, each walker $W_i$ either  occupies a single-particle state with a two-level internal degree of freedom, referred to as a color state, or the vacuum state. 

We denote with $d\in\{0,\dots,n\}$ the depth in the binary tree and $l\in\{1,\dots,2^{d}\}$ the branch index at that depth, see Fig.~\ref{binarytree}. The single-particle basis vector $\ket{w^{l,d}}_W$ represents a walker localized on branch $(l,d)$ with internal color ``red" or ``blue", namely $w\in\{R,B\}$. The vacuum $\ket{\varnothing^{d}}_W$ denotes the absence of a particle at depth $d$; crucially, because the vacuum carries no localized excitation, the branch index $l$ is redundant and we therefore omit it in the vacuum label. Physically, this reflects the fact that the absence of a particle is insensitive to the branching structure of the tree.
The Fock space of a {generic} walker $W_i$ is therefore:
\begin{multline}
\mathcal{F}_{W_i}=\mathrm{span}\{\ket{\varnothing^{d}}_{W_i}\}\oplus
\mathrm{span}\{\ket{w^{l,d}}_{W_i}: w\in\{R,B\},\\\,
d\in\{0,\dots,n\},\, l\in\{1,\dots,2^{d}\}\},
\end{multline}

The protocol employs two registers of quantum walkers: an address register  $\boldsymbol{A}$ and a data register  $\boldsymbol{D}$.
%We remind that $N=2^n$ is the number of memory cells in the qRAM and $m$ is the size of the information stored into each  memory cell.
The address register $\boldsymbol{A}=\{A_1,\dots,A_n\}$
%is composed of at most n quantum walkers, each encoding a single bit of the address of the target memory cells.
encodes an $n$-bit classical address $\boldsymbol{a}=a_n\dots a_1$ associated to a target memory cell.
The encoding of this address into the particles of the register works as follows: each bit $a_i$ is associated to a walker $A_i$, and the internal state of this quantum walker is initialized as:
\begin{equation}\label{addressencoding}
a_i=0\mapsto\ket{\varnothing}_{A_i},\quad a_i=1\mapsto\ket{R}_{A_i}.
\end{equation}
For instance, if we are required to access the memory cell with address $\boldsymbol{a}=101$, the address register is initialized in the state $\ket{\boldsymbol{a}}_A=\ket{R\varnothing R}_A$ (See Fig.~\ref{binarytree}).

Note that the exact number of address walkers required depends on the particular address selected. Indeed, given the encoding strategy~(\ref{addressencoding}), the number of bits with $a_i=1$ in $\boldsymbol{a}$ dictates how many address particles are needed. 
For example, the classical address $\boldsymbol{a}=00\dots0$ requires no walkers, while $\boldsymbol{a}=1\dots1$ requires all $n$ walkers. \gdr{As a consequence, a superposition of addresses involves states with different total particle numbers, a demanding requirement for a physical implementation and
one of the motivations for the particle-conserving variants of
Secs.~\ref{sec:dualrail} and~\ref{sec:qudit}.} %In the case of a superposition of addresses, at most $n$ address walkers are employed.

The data register $\boldsymbol{D}=\{D_0,\dots,D_m\}$ is used to perform operations on the memory cells and it is composed of $m+1$ quantum walkers. Each walker $D_i$ is initialized in the red state $\ket{R}_{D_i}$.

Notice that in our protocol the address walkers precede the data ones. Thus, we fix the spatial ordering of the walkers along the tree as
\begin{equation*}
    \{W_{n+m+1},\dots,W_{1}\}=\{D_{m},D_{m-1}\dots,D_0,A_n,\dots,A_1\} \ ,
\end{equation*}
where the rightmost walker, $A_1$, is the first entering the binary tree root node and $D_m$ is the last.
Hence, within the address register, the ordering is such that the first/rightmost walker $A_1$ carries the most significant bit of the address, namely $a_1$, while $A_n$ carries the least significant one, that is $a_n$.

In case we are required to query a superposition of qRAM memory cells, where this superposition is defined by a given set of amplitudes $\alpha_{\boldsymbol{a}}$, we prepare the walkers in the address register  in a superposition of states using the same amplitudes as weights for the corresponding states $\ket{\boldsymbol{a}}_A$, while $D$ is again initialized with red walkers. In this case, the initial state of our walkers is:
\begin{equation}
\begin{split}
  \ket{\psi_{\mathrm{in}}} 
&= \sum_{\boldsymbol{a}\in\mathcal{A}} \alpha_{\boldsymbol{a}}
\bigotimes_{j=0}^{m} \ket{R}_{D_j} 
\bigotimes_{i=1}^{n} \ket{a_i}_{A_i}
\\&= \sum_{\boldsymbol{a}\in\mathcal{A}} \alpha_{\boldsymbol{a}}
\ket{R}^{\otimes m+1}_D \ket{\boldsymbol{a}}_A.
\end{split}
\end{equation}

\begin{figure*}[hbt!]
\centering
\begin{subfigure}{0.6\textwidth}
        \resizebox{\linewidth}{!}{
     \begin{tikzpicture}[scale=1.2, transform shape]
         \def\w{8}
  \def\h{1.5}
  \def\dr{3pt}
  \def\dist{0.3}
  
%%%%%%%% PRIMO SCOPE %%%%%%%
\begin{scope}[shift={(1,0)}]

% Parametri per coerenza
    \def\dotRadius{0.5pt}
% rail aux
\draw[thick] (7,-0.3) -- (4.8,-0.3);
\draw[thick] (2,-0.3) -- (4.2,-0.3);

    \foreach \x in {4.3, 4.5, 4.7} {
        \fill ( \x, -0.3) circle (\dotRadius);
    }

%branch S up
\draw[thick] (7,-0.3) to[out=0, in=180] (8,0.5);

%branch S down
\draw[thick] (7,-0.3) to[out=0, in=180] (8,-1.1);

\node[rectangle, draw, minimum size=6mm,fill=cyan!30] at (7,-0.3){$\hat{S}$};

%distance
\draw[thick, |-|] (5.5,-0.6) -- (6,-0.6) 
        node[midway, below=2pt] {{\scalebox{0.7}{$\Delta x$}}};
% 2l-1,d+1
\draw[thick] (8,0.5) -- (12,0.5);
% 2l,d+1
\draw[thick] (8,-1.1) -- (12,-1.1);

\draw[thick,fill=red] (6,-0.3) circle (\dr);
\draw[thick,fill=red] (5.5,-0.3) circle (\dr);
\draw[thick,fill=white] (5,-0.3) circle (\dr);
\draw[thick,fill=red] (4,-0.3) circle (\dr);
\draw[thick,fill=red] (3.5,-0.3) circle (\dr);

 \node at (3.5,0) {{\scalebox{0.7}{$D_m$}}};
 \node at (4,0) {{\scalebox{0.7}{$D_{m-1}$}}};

 \node at (5,0) {{\scalebox{0.7}{$A_{d+2}$}}};
\node at (5.5,0) {{\scalebox{0.7}{$A_{d+1}$}}};
\node at (6,0) {{\scalebox{0.7}{$A_d$}}};

\node at (2.4,0) {{\scalebox{0.8}{$(l,d)$}}};
\node at (12,0.8) {{\scalebox{0.8}{$(2l-1,d+1)$}}};
\node at (12,-0.8) {{\scalebox{0.8}{$(2l,d+1)$}}};

\draw[decorate, decoration={brace, amplitude=5pt}, thick] (3.3,0.2) -- (5.7,0.2)node[midway, above=6pt, font=\scriptsize] {$T_d$};

\end{scope}

%%%%%%%% SECONDO SCOPE %%%%%%%
\begin{scope}[shift={(1,-3)}]

% Parametri per coerenza
    \def\dotRadius{0.5pt}

\node[rectangle, draw, minimum height=10mm,minimum width=30mm,fill=blue!20,rounded corners] at (4.75,-0.3){};
% rail aux
\draw[thick] (7,-0.3) -- (4.8,-0.3);
\draw[thick] (2,-0.3) -- (4.2,-0.3);

    \foreach \x in {4.3, 4.5, 4.7} {
        \fill ( \x, -0.3) circle (\dotRadius);
    }

%branch S up
\draw[thick] (7,-0.3) to[out=0, in=180] (8,0.5);

%branch S down
\draw[thick] (7,-0.3) to[out=0, in=180] (8,-1.1);

\node[rectangle, draw, minimum size=6mm,fill=cyan!30] at (7,-0.3){$\hat{S}$};

% 2l-1,d+1
\draw[thick] (8,0.5) -- (12,0.5);
% 2l,d+1
\draw[thick] (8,-1.1) -- (12,-1.1);

\draw[thick,fill=red] (6,-0.3) circle (\dr);
\draw[thick,fill=blue] (5.5,-0.3) circle (\dr);
\draw[thick,fill=white] (5,-0.3) circle (\dr);
\draw[thick,fill=blue] (4,-0.3) circle (\dr);
\draw[thick,fill=blue] (3.5,-0.3) circle (\dr);

 \node at (3.5,0) {{\scalebox{0.7}{$D_m$}}};
 \node at (4,0) {{\scalebox{0.7}{$D_{m-1}$}}};

 \node at (5,0) {{\scalebox{0.7}{$A_{d+2}$}}};
\node at (5.5,0) {{\scalebox{0.7}{$A_{d+1}$}}};
\node at (6,0) {{\scalebox{0.7}{$A_d$}}};

\node at (4.5,-0.6) {{\scalebox{0.7}{$\hat{U}^{(d)}$}}};

\node at (2.4,0) {{\scalebox{0.8}{$(l,d)$}}};
\node at (12,0.8) {{\scalebox{0.8}{$(2l-1,d+1)$}}};
\node at (12,-0.8) {{\scalebox{0.8}{$(2l,d+1)$}}};

\end{scope}

%%%%%%%% terzo SCOPE %%%%%%%
\begin{scope}[shift={(1,-6)}]

% Parametri per coerenza
    \def\dotRadius{0.5pt}
    
% rail aux
\draw[thick] (7,-0.3) -- (2,-0.3);

  \foreach \x in {9.3, 9.5, 9.7} {
        \fill ( \x, -1.1) circle (\dotRadius);
    }

  \foreach \x in {9.3, 9.5, 9.7} {
        \fill ( \x, 0.5) circle (\dotRadius);
    }

%branch S up
\draw[thick] (7,-0.3) to[out=0, in=180] (8,0.5);

%branch S down
\draw[thick] (7,-0.3) to[out=0, in=180] (8,-1.1);

\node[rectangle, draw, minimum size=6mm,fill=cyan!30] at (7,-0.3){$\hat{S}$};

% 2l-1,d+1
\draw[thick] (8,0.5) -- (9.2,0.5);
\draw[thick] (9.8,0.5) -- (12,0.5);
% 2l,d+1
\draw[thick] (8,-1.1) -- (9.2,-1.1);
\draw[thick] (9.8,-1.1) -- (12,-1.1);

\draw[dashed] (11,0.5)--(11,-1.1);
\draw[dashed] (10.5,0.5)--(10.5,-1.1);
\draw[thick, |-|] (10.5,-0.3) -- (11,-0.3) 
        node[midway, below=2pt] {{\scalebox{0.7}{$\Delta x$}}};

\draw[thick,fill=red] (11,0.5) circle (\dr);
\draw[thick,fill=red] (10.5,-1.1) circle (\dr);
\draw[thick,fill=white] (10,0.5) circle (\dr);
\draw[thick,fill=white] (10,-1.1) circle (\dr);
\draw[thick,fill=red] (9,-1.1) circle (\dr);
\draw[thick,fill=red] (8.5,-1.1) circle (\dr);

\node at (8.5,-1.4) {{\scalebox{0.7}{$D_m$}}};
\node at (9,-1.4) {{\scalebox{0.7}{$D_{m-1}$}}};

\node at (10,0.8) {{\scalebox{0.7}{$A_{d+2}$}}};
\node at (10,-1.4) {{\scalebox{0.7}{$A_{d+2}$}}};
\node at (10.5,-1.4) {{\scalebox{0.7}{$A_{d+1}$}}};
\node at (11,0.8) {{\scalebox{0.7}{$A_d$}}};

\node at (2.4,0) {{\scalebox{0.8}{$(l,d)$}}};
\node at (12,0.8) {{\scalebox{0.8}{$(2l-1,d+1)$}}};
\node at (12,-0.8) {{\scalebox{0.8}{$(2l,d+1)$}}};

\end{scope}

\node at (3,1){(a)};
     \end{tikzpicture}
     }
\end{subfigure}\hfill
\vrule width 1pt
\hfill
     \begin{subfigure}{0.35\textwidth}
\begin{tikzpicture}[gateX/.style={rectangle, draw, fill=green!20, minimum size=5mm}, rotate = 90, scale=0.9,transform shape]
\node [draw, rounded corners, rectangle, fill=blue!20, inner sep=5pt, minimum width=15mm, minimum height=50mm] 
        at (0,0) {};
\def\n{5}
\def\spacing{1} 
\draw[thick] (0,-3.5) -- (0,{(\spacing*(\n-1))-0.5 });
\foreach \i/\col in {0/white, 1/black, 2/black, 3/black, 4/black,5/black,6/white} {
\edef\colorname{\col} 
\filldraw[black, fill=\colorname] (0, \i*\spacing-3) circle (2pt);}

\draw[very thick] (0,-2)--(0,2);

\node[midway, rotate = -90, xshift=-2cm,yshift=0cm,gateX] {$\hat{X}$};

\node[midway, rotate = -90, xshift=-1cm,yshift=0cm,gateX] {$\hat{X}$};

\node[midway,rotate = -90,xshift=1cm,yshift=0cm,gateX] {$\hat{X}$};

\fill (0,-2) circle (3pt);

\node[fill=blue!20, rectangle, minimum size=7mm] at (0,0) {};

 \foreach \x in {-0.2, 0, 0.2} {
        \fill (0, \x) circle (1pt);
    }

\draw[decorate,rotate=-90, decoration={brace, amplitude=5pt}, thick, transform shape] (-2.5,1) -- (1.5,1)node[midway, above=6pt, align=center] {$n+m-(d-1)$\\ targets};    

\node[rotate=-90] at (0.5,-2){$\hat{U}^{(d)}$};

\node[rotate=-90] at (2.1,2.8){(b)};

\end{tikzpicture}

\vspace{0.5cm}
\noindent\tikz\draw[dashed, thick] (-2,0) -- (4,0);
\vspace{0.5cm}
     
\begin{tikzpicture}[scale=0.8,rotate=90, transform shape, line join=round]
     
  % Parametri
  \def\stemthick{5pt}
  \def\branchthick{6pt}
  \def\branchlen{1.4}
  \def\bendangle{20}
  \def\dr{3pt}

  % Coordinate principali
  \coordinate (O) at (0,0);
  \coordinate (B) at (0,-0.8); % punto di biforcazione
  \coordinate (L) at ($ (B) + (-\branchlen,-\branchlen) $);
  \coordinate (R) at ($ (B) + (\branchlen,-\branchlen) $);

  % Gambo centrale
  \draw[red,line width=\stemthick] (O) -- (B);

  % Ramo sinistro: rosso sottile con freccia
  \draw[red,line width=\branchthick,-{Latex[length=5pt, width=7pt]}]
    (B) to[bend left=\bendangle] (L);

  % Ramo destro: sfumatura blu→rosso
  \path[record path={name=ramo,dist=6pt}] (B) to[bend right=\bendangle] (R);

    % Freccia sopra la sfumatura: con stessa head del ramo sinistro
  \draw[opacity=0, line width=0pt] (B) to[bend right=\bendangle] (R); % aiuta compilatori "pigri"
  \draw[line width=\branchthick, -{Latex[length=8pt,width=8pt]}, draw=none] (B) to[bend right=\bendangle] (R);
  \path[draw=red, postaction={draw=red, -{Latex[length=5pt,width=7pt]}, line width=\branchthick}]
    (B) to[bend right=\bendangle] (R);

  % Ricostruzione sfumata — nessun bordo visibile
  %\addtocounter{parrow}{-2}
  %\draw let \p1=($(R)-(B)$),\n1={atan2(\y1,\x1)} in
    %[left color=blue, right color=red, shading angle=\n1+90, draw=none]
    %[reconstruct top=ramo] -- (parrowm-ramo-\the\numexpr\value{parrow}+2)
    %-- (parrowb-ramo-\number\value{parrow})
    %[reconstruct bottom=ramo] %-- (parrowm-ramo-3)
    %-- (parrowt-ramo-1);

\node[rectangle, draw=none, fill=blue, inner sep=0pt, outer sep=0pt, minimum width=5pt, minimum height=12mm
] at (-0.08,-0.45){}; 

\node[rectangle, draw=none, fill=red, inner sep=0pt, outer sep=0pt, minimum width=5pt, minimum height=12mm
] at (0.08,-0.45){}; 

\node[rectangle, fill=cyan!30, inner sep=0pt, outer sep=0pt, minimum width=20pt, minimum height=20pt, rotate=-90
] at (0,-1){$\hat{S}$};

\draw[blue, line width=1.5pt, -{Latex[length=3.5pt, width=4.5pt]}, shorten <=5pt, shorten >=5pt]
    ($ (B) + (-0.5, 1) $) to[bend left=30] ($ (L) + (0, 0.6) $);

\draw[red, line width=1.5pt, -{Latex[length=3.5pt, width=4.5pt]}, shorten <=5pt, shorten >=5pt]
    ($ (B) + (0.5, 1) $) to[bend right=30] ($ (R) + (0, 0.6) $);

\draw[thick,fill=blue] (-0.5,0.5) circle (\dr);

\draw[thick,fill=red] (-1.5,-1.7) circle (\dr);

\draw[thick,fill=red] (0.5,0.5) circle (\dr);

\draw[thick,fill=red] (1.5,-1.7) circle (\dr);

\node[rotate=-90, transform shape] at (0,1){\large$(l,d)$};

\node[rotate=-90, transform shape] at (2,-3){\large$(2l-1,d+1)$};

\node[rotate=-90, transform shape] at (-2,-3){\large$(2l,d+1)$};

\node[rotate=-90, scale=1.2, transform shape] at (2.5,2){(c)};

\end{tikzpicture} 
     \end{subfigure}
    \caption{Routing elements for our qRAM architecture. (a) Joint action of the operations within the unitary block at a generic level $d$. The target walkers $T_d$ first undergo the action of the address-transmission gate $\hat{U}^{(d)}$, conditioned on the color of $A_d$, and are then routed to the next tree level by the scattering gate $\hat{S}$. (b) Address-transmission gate $\hat{U}^{(d)}$: it is activated by a walker in the red color state and has $n+m-(d-1)$ targets at level $d$. (c) Scattering gate $\hat{S}$: particles are routed left or right according to their color degree of freedom, but always exit the gate in the red state.}
    \label{ForwardRouting}
\end{figure*}
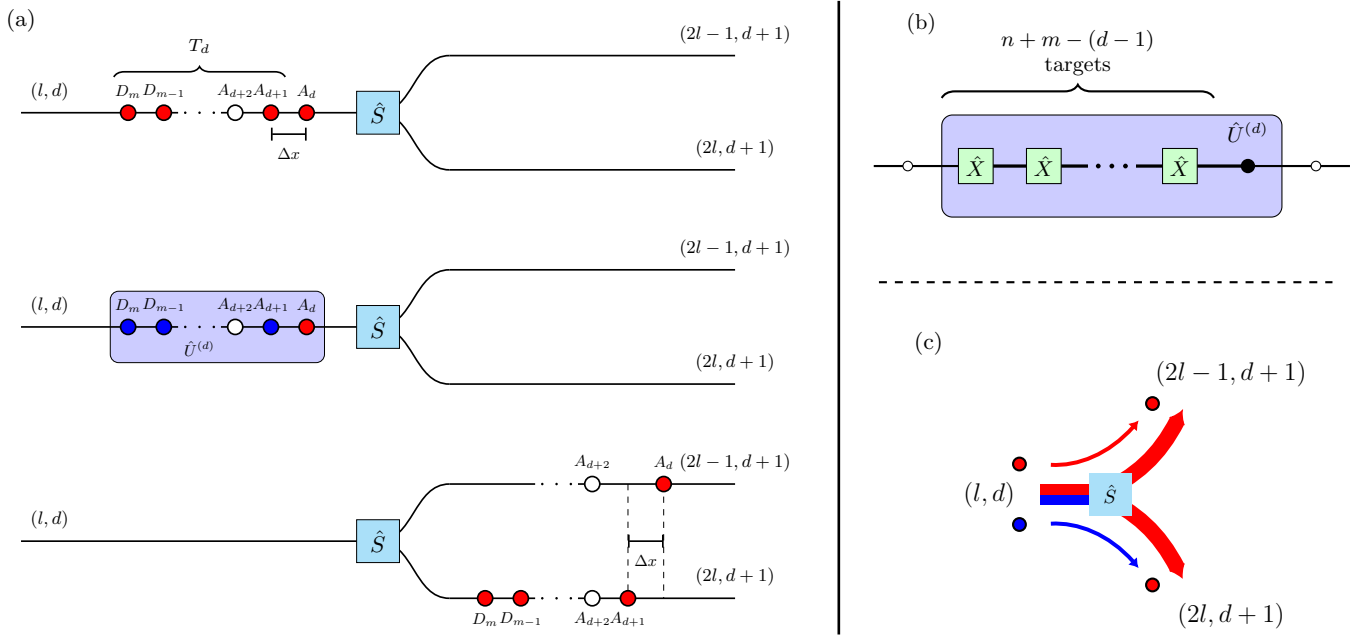

\subsection{Routing operations}

Our routing strategy is designed such that the $d$-th address particle $A_d$ dictates whether all the following walkers have to turn left or right at the $d$-th bifurcation of the tree.
To realize such a mechanism, we define a modular unitary block that is applied at each bifurcation of the tree and consists of two fundamentally different operations, as shown in  Fig.~\ref{ForwardRouting}a.
First, we apply $\hat U^{(d)}$, which transfers the address information of $A_d$ to the following walkers. Immediately after, depending on this information received, the walkers are physically routed left or right at the $d$-th bifurcation by the scattering gate $\hat S$. 
This strategy is repeated at all depths $d=0,\dots,n-1$, where the path at the $d$-th bifurcation is dictated by the address walker $A_d$.
In what follows, we describe the structure of the operations comprised in the block and how their sequential application allows for a coherent routing of the walkers.

\subsubsection{Address-transmission gate}\label{sec:address_transmission}

The address-transmission gate $\hat{U}^{(d)}$ is used to propagate the path information carried by the walker $A_d$ to all the following walkers, namely $A_{d+1}, A_{d+2},\dots, A_n$ and the data register particles, for all $d=0,\dots,n-1$. A straightforward definition for the operations is achieved by considering a global unitary $\hat{U}^{(d)}$, which takes as control the state of walker $A_d$, and acts on all the states of the walkers following it depending on its color. We remind that the state of $A_d$ encodes, through Eq.~(\ref{addressencoding}), the information regarding the $d$-th address bit $a_d$. Equivalently, $\hat{U}^{(d)}$ is a C-NOT-...-NOT gate controlled by the 
presence of walker $A_d$ in the red state, acting on all following particles in the train (see Fig.~\ref{ForwardRouting}b).

We define:
\begin{equation}\label{WalkerSets}
    P_d=\{A_i:i<d\},\quad
    T_d =\boldsymbol{D}\cup \{A_i:i>d\},
\end{equation}
which are, respectively, the sets of walkers preceding and following $A_d$.
The action of $\hat{U}^{(d)}$ can be split according to the state $a_d$ of the control walker $A_d$, where its (generic) internal state is $\ket{a_d}_{A_d}$. 

Let $\{\hat{\Pi}^{(A_d)}_{a_d}:=\ket{a_d}_{A_d}\bra{a_d},a_d\in\{\varnothing,R,B\}\}$ be the set of projectors onto the local subspaces of the walker $A_d$. Then the address-transmission gate can be expressed as
\begin{equation}\label{infoTransm}
    \hat{U}^{(d)}(T_d|A_d):=\sum_{a_d}\hat{V}^{(T_d)}_{a_d}\otimes\hat{\Pi}^{(A_d)}_{a_d}
\end{equation}
where the sum is over $a_d=\{\varnothing,R,B\}$ and
$\hat V^{(T_d)}_{a_d}$ is a unitary acting only on the target walkers $T_d$, namely those following $A_d$. We define these operators as:
\begin{equation}\label{targetOp}
    \hat{V}^{(T_d)}_{a_d=R}\equiv \bigotimes_{j\in T_d}\hat{X}_j,\quad \hat{V}^{(T_d)}_{a_d=B}=\hat{V}^{(T_d)}_{a_d=\varnothing}\equiv \bigotimes_{j\in T_d}\hat{I}_j
\end{equation}
Hence, when the control walker is in state $\ket{R}_{A_d}$, the gate flips the internal color state of all the target walkers by applying a Pauli $\hat{X}$; here $\hat{X}\ket{R}=\ket{B}, \hat{X}\ket{B}=\ket{R}, \hat{X}\ket{\varnothing}=\ket{\varnothing}$. Otherwise, if $\ket{a_d}\neq\ket{R}$, its action has no effect.
%\begin{equation*}
%\begin{split}
    %\hat{U}^{(d)}=\sum_{a_d}\left(\bigotimes_{j\in T_d}\hat{X}_j^{2-\delta_{R,a_d}}\right)\otimes \ket{a_d}_{A_d}\bra{a_d}\otimes \left(\bigotimes_{i=1}^{d-1}\hat{I}_{A_{i}}\right)
    %\end{split}
%\end{equation*}

Thus, if we consider a generic state of the form $\ket{d_m}_{D_m}\otimes\dots\otimes \ket{R}_{A_d}\otimes\dots\otimes\ket{a_1}_{A_1}\equiv\ket{\varphi}_{T_d}\otimes\ket{R}_{A_d}\otimes \ket{\psi}_{P_d}$ the effect of $\hat{U}^{(d)}$ reads

\begin{multline*} \hat{U}^{(d)}\left(\ket{\varphi}_{T_d}\otimes\ket{R}_{A_d}\otimes \ket{\psi}_{P_d}\right)=\\    \left(\bigotimes_{j\in T_d}\hat{X}_j\ket{\varphi}_{T_d}\right)\otimes\ket{R}_{A_d}\otimes\ket{\psi}_{P_d}\quad
\end{multline*}
Here $\ket{\varphi}_{T_d}$ is the joint state of the targets $T_d$, $\ket{R}_{A_d}$ is the state of the control and $\ket{\psi}_{P_d}$ the joint state of the walkers preceding $A_d$ in the train. Notice that the walkers $P_d$ are not affected by the action of $\hat U^{(d)}$, no matter the state of $A_d$. 

Consider a generic depth $d=0,\dots,n-1$.
The gate $\hat U^{(d)}$ is the first operation applied in the routing unitary block placed at the bifurcation $l=1,\dots,2^{d}$ at this level. For instance, as can be seen in Fig.~\ref{binarytree}, at depth $d=3$ we have 4 bifurcations: we place a $\hat U^{(3)}$ gate right before each one of these bifurcations.

Notice that the number of targets of $\hat U^{(d)}$ depends on the depth $d$ at which it is activated. More precisely, at level $d$, we have $\lvert T_d\rvert=(n-d) +m+1$. Thus, as we go deeper into the tree, this number shortens, eventually reaching $\lvert T_d\rvert=\lvert D\rvert=m+1$ at depth $n$.

As we show in detail in the following, the same operation is used during the inverse routing of the quantum walkers from the memory cells to the output node.

\gdr{It is worth stressing that, unlike the scattering gate discussed below,
$\hat{U}^{(d)}$ is a conditional operation acting on different information
carriers. In an optical realization the gate must therefore rely on highly
nonlinear transformations, requiring an interaction between two photons, which is
unavailable in linear optics and makes such operations costly to implement.}

\subsubsection{Scattering gate $\hat S$.}\label{sec:S}

The scattering gate implements locally at node $(l,d)$ the routing of a walker from a node at depth $d$ to one of its two children at depth $d+1$ according to the walker's color, while also acting on the internal color degree of freedom. Specifically, as illustrated in Fig.~\ref{ForwardRouting}c, a red walker 
is routed to the left branch while retaining its red color, while 
a blue walker is routed to the right branch together with a color change to red.

\gdr{Using the branch index convention above, the left and right children of node $(l,d)$ are $(2l-1,d+1)$ and $(2l,d+1)$, respectively. 
Since the walker traverses the node in a definite direction, we label the
scattering gate by the direction of propagation and denote by
$\hat S^{(l,d)}_{\rightarrow}$ the gate acting at node $(l,d)$ on a walker
propagating from the root towards the leaves.
We define the action of $\hat S_{\rightarrow}^{(l,d)}$ on the single-particle subspace by
\begin{align*}
\hat S^{(l,d)}_{\rightarrow}\bigl(\ket{R^{l,d}}_W\bigr) &= \ket{R^{\,2l-1,d+1}}_W,\\
\hat S^{(l,d)}_{\rightarrow}\bigl(\ket{B^{\,l,d}}_W\bigr) &= \ket{R^{\,2l,d+1}}_W.
\end{align*}
In a more compact form, we can write 
\begin{equation}\label{bosonicS}
\hat S_{\rightarrow}^{(l,d)}\ket{w^{l,d}}_W = \ket{R^{2l-\delta_{w,R},\,d+1}}_W,\qquad w\in\{R,B\},
\end{equation}
where $\delta_{w,R}=1$ if $\ket{w}_W=\ket{R}_W$ and $0$ otherwise.
On the vacuum subspace the scattering gate preserves the absence of excitations,
\begin{equation*}
\hat S^{(l,d)}_{\rightarrow}\ket{\varnothing^d}_W = \ket{\varnothing^{d+1}}_W ,
\end{equation*}
where the branch label of the vacuum state can be omitted.
Unlike a walker, the vacuum carries no color
and therefore no routing information: there is nothing for the gate to direct, and
accordingly no branching takes place. By linearity, this fixes the action of
$\hat S^{(l,d)}_{\rightarrow}$ on arbitrary superpositions of vacuum and
single-particle states.}

%Formally, the unitary operator $\hat{S}$ is defined as 
%\begin{equation}\label{bosonicSdef}
%\begin{split}
%\hat{S}&:=
%\ket{R^{2l-1,d+1}}\bra{R^{l,d}}+ 
%\ket{R^{2l,d+1}}\bra{B^{l,d}}+
%\ket{\varnothing^{d+1}}\bra{\varnothing^{d}} \\
%&+\ket{B^{2l,d+1}}\bra{R^{2l-1,d+1}} 
%+ \ket{B^{2l-1,d+1}}\bra{R^{2l,d+1}} 
%\\& + \ket{B^{l,d}}\bra{B^{2l-1,d+1}}
%+ \ket{R^{l,d}}\bra{B^{2l,d+1}}
%+ \ket{\varnothing^{d}}\bra{\varnothing^{d+1}} 
%\end{split}
%\end{equation}
%where the first three terms are those allowing the routing scheme proposed by our qRAM scheme, while the others ensure the unitarity of the operator. In other words, this operator acts as a permutation $\sigma_{(l,d)}$ of the six single-particle modes
%$\{R^{l,d},B^{l,d},R^{2l-1,d+1},B^{2l-1,d+1},R^{2l,d+1},B^{2l,d+1}\}$
%according to the cycles $(R^{l,d}\rightarrow R^{2l-1,d+1}\rightarrow B^{2l,d+1})$ and $(B^{l,d}\rightarrow R^{2l,d+1}\rightarrow B^{2l-1,d+1})$.

\gdr{The action of $\hat S_{\rightarrow}^{(l,d)}$ is analogous to that of a polarizing
beam splitter (PBS) on photons, with the two colors playing the role of two orthogonal
polarizations. Indeed, a PBS transmits one polarization into one output port and reflects
the orthogonal one into the other, exactly as $\hat S_{\rightarrow}^{(l,d)}$ sends red and blue walkers
into the left and right branch, respectively. The color reset that follows after
the routing of a blue walker can in turn be seen as a polarization rotation operation, i.e. a
half-wave plate placed at the output of the right
branch, so that the outgoing color is the same on both branches. Both elements are optically passive: the branch taken by the walker is
decided by the color carried by the walker itself, so that the scattering gate
requires no external activation, no classical control signal and no
synchronization with the arrival of the walker at the node. Being passive and
lossless, such an element is unitary by construction.}

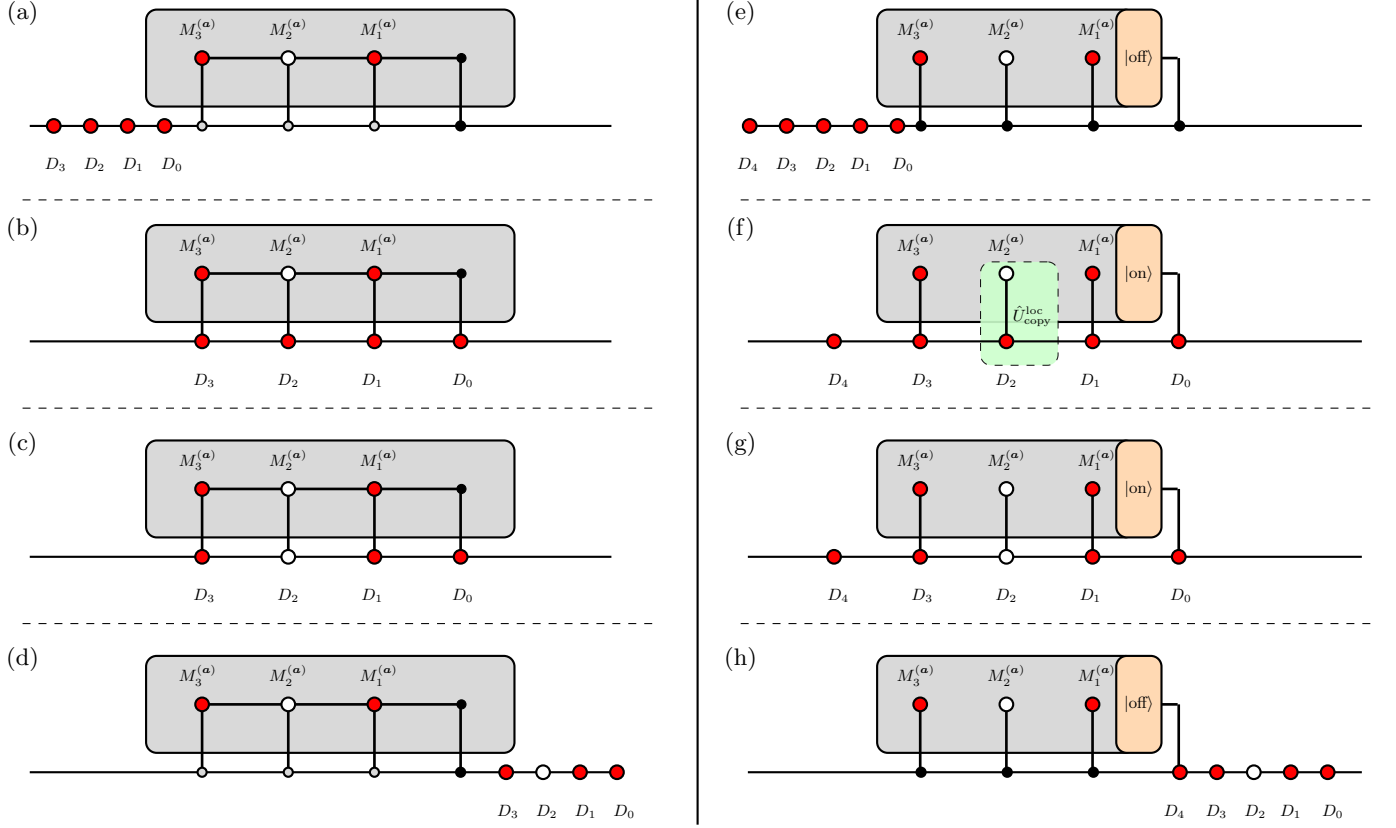
\begin{figure*}[t]
\centering
\begin{tikzpicture}[x=1cm,y=1cm, scale=0.95]
  \def\w{8}
  \def\h{1.5}
  \def\dr{3pt}
  \def\dist{0.3}

  % --- Primo disegno originale ---
  \begin{scope}[shift={(-2,2)}, scale=0.9, transform shape]
    % Rettangolo
    \draw[rounded corners, thick, fill=gray!30] (0.8,0) rectangle (\w-1.5,\h);

    % Linee verticali dai pallini sopra a quelli sotto
    \foreach \i in {1,...,4} {
      \pgfmathsetmacro{\x}{(\i+0.25)/6*\w}
      \draw[line width=1pt] (\x, \h/2) -- (\x, {-\dist});
    }

       % Linee spesse pallini superiori
      \draw[line width=1pt] (1.7, {\h/2}) -- (5.68, {\h/2});
      \filldraw[fill=black, draw=black] (5.68,{\h/2}) circle (\dr-1);
    
    % Pallini sopra
    \foreach[count=\k from 1] \col in {red, white, red} {
        \pgfmathsetmacro{\x}{(\k+0.25)/6*\w}  % \n non definito, quindi uso 6 fisso coerente con sopra
        \filldraw[fill=\col, draw=black, thick] (\x, {\h/2}) circle [radius=\dr];
      }

    % Linea sotto
    \draw[thick] (-1,-\dist) -- (\w,-\dist);

   % Pallini sotto
    \foreach[count=\k from 1] \col in {red,red,red,red} {
        \pgfmathsetmacro{\x}{(\k-2.1)/7*(\w/2)}
        \pgfmathsetmacro{\y}{-\dist}
        \filldraw[fill=\col, draw=black, thick] (\x, \y) circle [radius=\dr];
      }

  % Pallini neri (spazi vuoti)
    \foreach[count=\k from 1] \col in {gray!30,gray!30,gray!30,black} {
        \pgfmathsetmacro{\x}{(\k+0.25)/6*(\w)}
        \pgfmathsetmacro{\y}{-\dist}
        \filldraw[fill=\col, draw=black, thick] (\x, \y) circle [radius=\dr-1];
      }

     % Label pallini sopra
    \node at (1.6,1.2) {{\scalebox{0.8}{$M_3^{(\boldsymbol{a})}$}}};
    
    \node at (3,1.2) {{\scalebox{0.8}{$M_2^{(\boldsymbol{a})}$}}};
    
    \node at (4.4,1.2) {{\scalebox{0.8}{$M_1^{(\boldsymbol{a})}$}}};

    % Label pallini inferiori

    \node at (-0.6,-0.9) {{\scalebox{0.8}{$D_{3}$}}};
      
    \node at (0,-0.9) {{\scalebox{0.8}{$D_2$}}};
    
    \node at (0.6,-0.9) {{\scalebox{0.8}{$D_1$}}};
    
    \node at (1.2,-0.9) {{\scalebox{0.8}{$D_{0}$}}};
  \end{scope}

\draw[dashed] (-3,{\h-0.8})--({\w-2.2},{\h-0.8});

% Secondo disegno prima colonna
  \begin{scope}[shift={(-2,-1)}, scale=0.9, transform shape]
    % Rettangolo
    \draw[rounded corners, thick, fill=gray!30] (0.8,0) rectangle (\w-1.5,\h);

    % Linee verticali dai pallini sopra a quelli sotto
    \foreach \i in {1,...,4} {
      \pgfmathsetmacro{\x}{(\i+0.25)/6*\w}
      \draw[line width=1pt] (\x, \h/2) -- (\x, {-\dist});
    }

       % Linee spesse pallini superiori
      \draw[line width=1pt] (1.7, {\h/2}) -- (5.68, {\h/2});
      \filldraw[fill=black, draw=black] (5.68,{\h/2}) circle (\dr-1);
    
    % Pallini sopra
    \foreach[count=\k from 1] \col in {red, white, red} {
        \pgfmathsetmacro{\x}{(\k+0.25)/6*\w}  % \n non definito, quindi uso 6 fisso coerente con sopra
        \filldraw[fill=\col, draw=black, thick] (\x, {\h/2}) circle [radius=\dr];
      }

    % Linea sotto
    \draw[thick] (-1,-\dist) -- (\w,-\dist);

   % Pallini sotto
    \foreach[count=\k from 1] \col in { red, red,red,red} {
        \pgfmathsetmacro{\x}{(\k+0.25)/6*(\w)}
        \pgfmathsetmacro{\y}{-\dist}
        \filldraw[fill=\col, draw=black, thick] (\x, \y) circle [radius=\dr];
      }

  % Label pallini sopra
    \node at (1.6,1.2) {{\scalebox{0.8}{$M_3^{(\boldsymbol{a})}$}}};
    
    \node at (3,1.2) {{\scalebox{0.8}{$M_2^{(\boldsymbol{a})}$}}};
    
    \node at (4.4,1.2) {{\scalebox{0.8}{$M_1^{(\boldsymbol{a})}$}}};

    % Label pallini inferiori
    \node at (1.7,-0.9) {{\scalebox{0.8}{$D_3$}}};

    \node at (3,-0.9) {{\scalebox{0.8}{$D_2$}}};
 
    \node at (4.3,-0.9) {{\scalebox{0.8}{$D_1$}}};
    
    \node at (5.7,-0.9) {{\scalebox{0.8}{$D_{0}$}}};

  \end{scope}

\draw[dashed] (-3,{-(2*\h)+0.8})--({\w-2.2},{-(2*\h)+0.8});

%terzo disegno prima colonna
     \begin{scope}[shift={(-2,-4)}, scale=0.9, transform shape]
     % Rettangolo
    \draw[rounded corners, thick, fill=gray!30] (0.8,0) rectangle (\w-1.5,\h);

    % Linee verticali dai pallini sopra a quelli sotto
    \foreach \i in {1,...,4} {
      \pgfmathsetmacro{\x}{(\i+0.25)/6*\w}
      \draw[line width=1pt] (\x, \h/2) -- (\x, {-\dist});
    }

       % Linee spesse pallini superiori
      \draw[line width=1pt] (1.7, {\h/2}) -- (5.68, {\h/2});
      \filldraw[fill=black, draw=black] (5.68,{\h/2}) circle (\dr-1);
    
    % Pallini sopra
    \foreach[count=\k from 1] \col in {red, white, red} {
        \pgfmathsetmacro{\x}{(\k+0.25)/6*\w}  % \n non definito, quindi uso 6 fisso coerente con sopra
        \filldraw[fill=\col, draw=black, thick] (\x, {\h/2}) circle [radius=\dr];
      }

    % Linea sotto
    \draw[thick] (-1,-\dist) -- (\w,-\dist);

   % Pallini sotto
    \foreach[count=\k from 1] \col in { red, white,red,red} {
        \pgfmathsetmacro{\x}{(\k+0.25)/6*(\w)}
        \pgfmathsetmacro{\y}{-\dist}
        \filldraw[fill=\col, draw=black, thick] (\x, \y) circle [radius=\dr];
      }

    % Label pallini sopra
    \node at (1.6,1.2) {{\scalebox{0.8}{$M_3^{(\boldsymbol{a})}$}}};
    
    \node at (3,1.2) {{\scalebox{0.8}{$M_2^{(\boldsymbol{a})}$}}};
    
    \node at (4.4,1.2) {{\scalebox{0.8}{$M_1^{(\boldsymbol{a})}$}}};

 % Label pallini inferiori
    
    \node at (1.7,-0.9) {{\scalebox{0.8}{$D_3$}}};

    \node at (3,-0.9) {{\scalebox{0.8}{$D_2$}}};
 
    \node at (4.3,-0.9) {{\scalebox{0.8}{$D_1$}}};
    
    \node at (5.7,-0.9) {{\scalebox{0.8}{$D_{0}$}}};

  \end{scope}

   \draw[dashed] (-3,{-(3*\h)-0.7})--({\w-2.2},{-(3*\h)-0.7});

 \begin{scope}[shift={(-2,-7)}, scale=0.9, transform shape]
    % Rettangolo
    \draw[rounded corners, thick, fill=gray!30] (0.8,0) rectangle (\w-1.5,\h);

    % Linee verticali dai pallini sopra a quelli sotto
    \foreach \i in {1,...,4} {
      \pgfmathsetmacro{\x}{(\i+0.25)/6*\w}
      \draw[line width=1pt] (\x, \h/2) -- (\x, {-\dist});
    }

       % Linee spesse pallini superiori
      \draw[line width=1pt] (1.7, {\h/2}) -- (5.68, {\h/2});
      \filldraw[fill=black, draw=black] (5.68,{\h/2}) circle (\dr-1);
    
    % Pallini sopra
    \foreach[count=\k from 1] \col in {red, white, red} {
        \pgfmathsetmacro{\x}{(\k+0.25)/6*\w}  % \n non definito, quindi uso 6 fisso coerente con sopra
        \filldraw[fill=\col, draw=black, thick] (\x, {\h/2}) circle [radius=\dr];
      }

    % Linea sotto
    \draw[thick] (-1,-\dist) -- (\w,-\dist);

   % Pallini sotto
    \foreach[count=\k from 1] \col in {red,white,red,red} {
        \pgfmathsetmacro{\x}{(\k-2.1)/7*(\w/2)}
        \pgfmathsetmacro{\y}{-\dist}
        \filldraw[fill=\col, draw=black, thick] (\x+7, \y) circle [radius=\dr];
      }

  % Pallini neri (spazi vuoti)
    \foreach[count=\k from 1] \col in {gray!30,gray!30,gray!30,black} {
        \pgfmathsetmacro{\x}{(\k+0.25)/6*(\w)}
        \pgfmathsetmacro{\y}{-\dist}
        \filldraw[fill=\col, draw=black, thick] (\x, \y) circle [radius=\dr-1];
      }

     % Label pallini sopra
    \node at (1.6,1.2) {{\scalebox{0.8}{$M_3^{(\boldsymbol{a})}$}}};
    
    \node at (3,1.2) {{\scalebox{0.8}{$M_2^{(\boldsymbol{a})}$}}};
    
    \node at (4.4,1.2) {{\scalebox{0.8}{$M_1^{(\boldsymbol{a})}$}}};

    % Label pallini inferiori

    \node at (6.4,-0.9) {{\scalebox{0.8}{$D_{3}$}}};
      
    \node at (7,-0.9) {{\scalebox{0.8}{$D_2$}}};
    
    \node at (7.6,-0.9) {{\scalebox{0.8}{$D_1$}}};
    
    \node at (8.2,-0.9) {{\scalebox{0.8}{$D_{0}$}}};
  \end{scope}

% Linea separazione verticale
\draw[thick](6.4,4)--(6.4,-8);

%primo disegno seconda colonna
 \begin{scope}[shift={(8,2)}, scale=0.9, transform shape]
    % Rettangolo
    \draw[rounded corners, thick, fill=gray!30] (1,0) rectangle ({\w-3},\h);

       % Linee verticali dai pallini sopra a quelli sotto
    \foreach \i in {1,...,4} {
      \pgfmathsetmacro{\x}{(\i+0.25)/6*\w}
      \draw[line width=1pt] (\x, \h/2) -- (\x, {-\dist});
    }
  
    %linea switch
    \draw[line width=1pt] ({\w-2.6},{\h/2})--({\w-2.35},{\h/2});

% Switch on/off
    \draw[rounded corners, thick, fill=orange!30] ({\w-3.3},0) rectangle ({\w-2.6},\h) node[midway,anchor=center] {{\scalebox{0.8}{$\ket{\text{off}}$}}};

 % Pallini sopra
    \foreach[count=\k from 1] \col in {red, white, red} {
        \pgfmathsetmacro{\x}{(\k+0.25)/6*\w}  % \n non definito, quindi uso 6 fisso coerente con sopra
        \filldraw[fill=\col, draw=black, thick] (\x, {\h/2}) circle [radius=\dr];
        }

    % Linea sotto
    \draw[thick] (-1,-\dist) -- ({\w+0.5},-\dist);

     % Pallini sotto
    \foreach[count=\k from 1] \col in {red,red,red,red,red} {
        \pgfmathsetmacro{\x}{(\k-2.7)/7*(\w/2)}
        \pgfmathsetmacro{\y}{-\dist}
        \filldraw[fill=\col, draw=black, thick] (\x, \y) circle [radius=\dr];
      }

  % Pallini neri (spazi vuoti)
    \foreach[count=\k from 1] \col in {black,black,black,black} {
        \pgfmathsetmacro{\x}{(\k+0.26)/6*(\w)}
        \pgfmathsetmacro{\y}{-\dist}
        \filldraw[fill=\col, draw=black, thick] (\x, \y) circle [radius=\dr-1];
      }

     % Label pallini sopra
    \node at (1.6,1.2) {{\scalebox{0.8}{$M_3^{(\boldsymbol{a})}$}}};
    
    \node at (3,1.2) {{\scalebox{0.8}{$M_2^{(\boldsymbol{a})}$}}};
    
    \node at (4.4,1.2) {{\scalebox{0.8}{$M_1^{(\boldsymbol{a})}$}}};

     % Label pallini inferiori
    \node at (-1,-0.9) {{\scalebox{0.8}{$D_{4}$}}};

    \node at (-0.4,-0.9) {{\scalebox{0.8}{$D_{3}$}}};
      
    \node at (0.2,-0.9) {{\scalebox{0.8}{$D_2$}}};
    
    \node at (0.75,-0.9) {{\scalebox{0.8}{$D_1$}}};
    
    \node at (1.4,-0.9) {{\scalebox{0.8}{$D_{0}$}}};
  \end{scope}

  \draw[dashed] (7,{\h-0.8})--({\w+7.8},{\h-0.8});

% Secondo disegno seconda colonna
  \begin{scope}[shift={(8,-1)}, scale=0.9, transform shape]
    % Rettangolo
    \draw[rounded corners, thick, fill=gray!30] (1,0) rectangle ({\w-3},\h);

     % Rettangolo gate U_copy^loc
   \node[draw, dashed, rounded corners, fill=green!20, minimum width=12mm, minimum height=16mm, opacity=0.9] at (3.2,0.13) {};

       % Linee verticali dai pallini sopra a quelli sotto
    \foreach \i in {1,...,4} {
      \pgfmathsetmacro{\x}{(\i+0.25)/6*\w}
      \draw[line width=1pt] (\x, \h/2) -- (\x, {-\dist});
    }
  
    %linea switch
    \draw[line width=1pt] ({\w-2.6},{\h/2})--({\w-2.35},{\h/2});

% Switch on/off
    \draw[rounded corners, thick, fill=orange!30] ({\w-3.3},0) rectangle ({\w-2.6},\h) node[midway,anchor=center] {{\scalebox{0.8}{$\ket{\text{on}}$}}};

 % Pallini sopra
    \foreach[count=\k from 1] \col in {red, white, red} {
        \pgfmathsetmacro{\x}{(\k+0.25)/6*\w}  % \n non definito, quindi uso 6 fisso coerente con sopra
        \filldraw[fill=\col, draw=black, thick] (\x, {\h/2}) circle [radius=\dr];
        }

    % Linea sotto
    \draw[thick] (-1,-\dist) -- (\w+0.5,-\dist);

   % Pallini sotto
    \foreach[count=\k from 1] \col in { red, red,red,red,red} {
        \pgfmathsetmacro{\x}{(\k-0.75)/6*(\w)}
        \pgfmathsetmacro{\y}{-\dist}
        \filldraw[fill=\col, draw=black, thick] (\x, \y) circle [radius=\dr];
      }

  % Label pallini sopra
    \node at (1.6,1.2) {{\scalebox{0.8}{$M_3^{(\boldsymbol{a})}$}}};
    
    \node at (3,1.2) {{\scalebox{0.8}{$M_2^{(\boldsymbol{a})}$}}};
    
    \node at (4.4,1.2) {{\scalebox{0.8}{$M_1^{(\boldsymbol{a})}$}}};

      % Label pallini inferiori
    \node at (0.4,-0.9) {{\scalebox{0.8}{$D_{4}$}}};
       
    \node at (1.7,-0.9) {{\scalebox{0.8}{$D_3$}}};

    \node at (3,-0.9) {{\scalebox{0.8}{$D_2$}}};
 
    \node at (4.3,-0.9) {{\scalebox{0.8}{$D_1$}}};
    
    \node at (5.7,-0.9) {{\scalebox{0.8}{$D_{0}$}}};

% Label gates
    \node at (3.4,0.1) {{\scalebox{0.8}{$\hat{U}_{\text{copy}}^{\text{loc}}$}}};
 \end{scope}

\draw[dashed] (7,{-(2*\h)+0.8})--({\w+7.8},{-(2*\h)+0.8});

% Terza figura seconda colonna
   \begin{scope}[shift={(8,-4)}, scale=0.9, transform shape]
        % Rettangolo
    \draw[rounded corners, thick, fill=gray!30] (1,0) rectangle ({\w-3},\h);

       % Linee verticali dai pallini sopra a quelli sotto
    \foreach \i in {1,...,4} {
      \pgfmathsetmacro{\x}{(\i+0.25)/6*\w}
      \draw[line width=1pt] (\x, \h/2) -- (\x, {-\dist});
    }
  
    %linea switch
    \draw[line width=1pt] ({\w-2.6},{\h/2})--({\w-2.35},{\h/2});

% Switch on/off
    \draw[rounded corners, thick, fill=orange!30] ({\w-3.3},0) rectangle ({\w-2.6},\h) node[midway,anchor=center] {{\scalebox{0.8}{$\ket{\text{on}}$}}};

 % Pallini sopra
    \foreach[count=\k from 1] \col in {red, white, red} {
        \pgfmathsetmacro{\x}{(\k+0.25)/6*\w}  % \n non definito, quindi uso 6 fisso coerente con sopra
        \filldraw[fill=\col, draw=black, thick] (\x, {\h/2}) circle [radius=\dr];
        }

    % Linea sotto
    \draw[thick] (-1,-\dist) -- (\w+0.5,-\dist);

   % Pallini sotto
    \foreach[count=\k from 1] \col in { red, red,white,red,red} {
        \pgfmathsetmacro{\x}{(\k-0.75)/6*(\w)}
        \pgfmathsetmacro{\y}{-\dist}
        \filldraw[fill=\col, draw=black, thick] (\x, \y) circle [radius=\dr];
      }

  % Label pallini sopra
    \node at (1.6,1.2) {{\scalebox{0.8}{$M_3^{(\boldsymbol{a})}$}}};
    
    \node at (3,1.2) {{\scalebox{0.8}{$M_2^{(\boldsymbol{a})}$}}};
    
    \node at (4.4,1.2) {{\scalebox{0.8}{$M_1^{(\boldsymbol{a})}$}}};

      % Label pallini inferiori
    \node at (0.4,-0.9) {{\scalebox{0.8}{$D_{4}$}}};
       
    \node at (1.7,-0.9) {{\scalebox{0.8}{$D_3$}}};

    \node at (3,-0.9) {{\scalebox{0.8}{$D_2$}}};
 
    \node at (4.3,-0.9) {{\scalebox{0.8}{$D_1$}}};
    
    \node at (5.7,-0.9) {{\scalebox{0.8}{$D_{0}$}}};

 \end{scope}

 \draw[dashed] (7,{-(3*\h)-0.7})--({\w+7.8},{-(3*\h)-0.7});

 % Quarta figura seconda colonna
   \begin{scope}[shift={(8,-7)}, scale=0.9, transform shape]
   % Rettangolo
    \draw[rounded corners, thick, fill=gray!30] (1,0) rectangle ({\w-3},\h);

       % Linee verticali dai pallini sopra a quelli sotto
    \foreach \i in {1,...,4} {
      \pgfmathsetmacro{\x}{(\i+0.25)/6*\w}
      \draw[line width=1pt] (\x, \h/2) -- (\x, {-\dist});
    }
  
    %linea switch
    \draw[line width=1pt] ({\w-2.6},{\h/2})--({\w-2.35},{\h/2});

% Switch on/off
    \draw[rounded corners, thick, fill=orange!30] ({\w-3.3},0) rectangle ({\w-2.6},\h) node[midway,anchor=center] {{\scalebox{0.8}{$\ket{\text{off}}$}}};

 % Pallini sopra
    \foreach[count=\k from 1] \col in {red, white, red} {
        \pgfmathsetmacro{\x}{(\k+0.25)/6*\w}  % \n non definito, quindi uso 6 fisso coerente con sopra
        \filldraw[fill=\col, draw=black, thick] (\x, {\h/2}) circle [radius=\dr];
        }

    % Linea sotto
    \draw[thick] (-1,-\dist) -- ({\w+0.5},-\dist);

  % Pallini neri (spazi vuoti)
    \foreach[count=\k from 1] \col in {black,black,black,black} {
        \pgfmathsetmacro{\x}{(\k+0.26)/6*(\w)}
        \pgfmathsetmacro{\y}{-\dist}
        \filldraw[fill=\col, draw=black, thick] (\x, \y) circle [radius=\dr-1];
      }

     % Pallini sotto
    \foreach[count=\k from 1] \col in {red,red,white,red,red} {
        \pgfmathsetmacro{\x}{(\k+8.95)/7*(\w/2)}
        \pgfmathsetmacro{\y}{-\dist}
        \filldraw[fill=\col, draw=black, thick] (\x, \y) circle [radius=\dr];
      }

     % Label pallini sopra
    \node at (1.6,1.2) {{\scalebox{0.8}{$M_3^{(\boldsymbol{a})}$}}};
    
    \node at (3,1.2) {{\scalebox{0.8}{$M_2^{(\boldsymbol{a})}$}}};
    
    \node at (4.4,1.2) {{\scalebox{0.8}{$M_1^{(\boldsymbol{a})}$}}};

      % Label pallini inferiori
    \node at (5.6,-0.9) {{\scalebox{0.8}{$D_{4}$}}};

    \node at (6.25,-0.9) {{\scalebox{0.8}{$D_{3}$}}};
      
    \node at (6.85,-0.9) {{\scalebox{0.8}{$D_2$}}};
    
    \node at (7.4,-0.9) {{\scalebox{0.8}{$D_1$}}};
    
    \node at (8.05,-0.9) {{\scalebox{0.8}{$D_{0}$}}};
 \end{scope}

\node at (-3,3.3){(a)};
\node at (-3,0.3){(b)};
\node at (-3,-2.7){(c)};
\node at (-3,-5.7){(d)};

\node at (7,3.3){(e)};
\node at (7,0.3){(f)};
\node at (7,-2.7){(g)};
\node at (7,-5.7){(h)};

\end{tikzpicture}

\caption{Schematic representation of the control-copy schemes introduced in Section~\ref{memoryOpStandard}, for a message of length $m=3$. The left column illustrates the scheme based on a global unitary acting on all walkers: (a) after traversing the binary tree, the data walkers reach the targeted memory cell; (b) the flag walker $D_0$ acts as the control for the global unitary operation; (c) the message is copied to the data walkers according to Eq.~(\ref{GlobCopy}); (d) the particles then exit the cell.
The right column shows the main steps of the cell state switch variant: (e) after the routing process, the particles arrive at the memory cell, which is initially in the state $\ket{\text{off}}_{F^{(\boldsymbol{a})}}$; (f) the flag walker $D_0$ switches the cell state to $\ket{\text{on}}_{F^{(\boldsymbol{a})}}$; (g) the cell activates $m=3$ local copy operations between the memory and data walkers, copying the message according to Eq.~(\ref{locCopy}); (h) finally, the particles leave the cell, and the last data walker, $D_{4}$, acts as a flag to reset the switch state back to $\ket{\text{off}}_{F^{(\boldsymbol{a})}}$.  }
\label{copy_standard}
\end{figure*}

\subsection{Memory Operations}\label{memoryOpStandard}

The memory cells at the leaves of the binary tree contain an $m$ bit classical message $\boldsymbol{b}^{(\boldsymbol{a})}$ encoded in the state of an $m$ qubit register $\boldsymbol{M}^{\boldsymbol{(a)}}=\{M^{(\boldsymbol{a})}_1,\dots,M^{(\boldsymbol{a})}_m\}$, associated with leaf address $\boldsymbol{a}$. 
Our goal is to design memory operations that act only on the data walkers and the message walkers contained in a target memory cell. In practice, one must ensure that these memory operations act exclusively on the message register associated with the addressed leaf, while remaining idle on all other cells traversed along the routing path.

To this purpose, we employ the ``flag" data walker 
$D_{0}$, which carries no information and whose role is to trigger memory access exclusively at the desired memory cell location.
Each logical COPY operation corresponds, at the hardware level, to a collection of $\mathcal{O}(m)$ two-body primitives between the walkers in the data register $D$ and the memory qubits $M_j^{(\boldsymbol{a})}$ of the target cell. 
Since each memory cell stores classical information, each primitive COPY step can be implemented as a permutation of basis states of the joint spaces of the registers $D$ and $M^{(\boldsymbol{a})}$.
Denoting by $\ket{m_j}_{M_j^{(\boldsymbol{a})}}$ the local state of the $j$-th memory qubit, with $m_j\in\{\varnothing,R,B\}$, the elementwise action we require for a single data–memory pair reads

\begin{equation}\label{copyMapping}
    \begin{split}\ket{R}_{D_j}\ket{m_j=R}_{M_j^{(\boldsymbol{a})}}\xmapsto{\text{copy}} \ket{R}_{D_j}\ket{m_j=R}_{M_j^{(\boldsymbol{a})}},\\
    \ket{R}_{D_j}\ket{m_j=\varnothing}_{M_j^{(\boldsymbol{a})}}\xmapsto{\text{copy}} \ket{\varnothing}_{D_j}\ket{m_j=\varnothing}_{M_j^{(\boldsymbol{a})}}.
    \end{split}
\end{equation}
We propose two alternative mechanisms for selective activation and manipulation of the message information: (i) a global controlled COPY triggered by the flag walker, and (ii) a local cell-activation mechanism based on an internal switch degree of freedom of each memory cell.

\subsubsection{Global Controlled COPY operation}

The simplest approach to executing read/write operations on the selected memory cells is to define a global unitary operation acting jointly on the data register $D$ and on the target memory register $M^{(\boldsymbol{a})}$. In this setting, the flag walker $D_0$ acts as a control to trigger a COPY operation acting on the joint Hilbert space of the particles in data and message registers. An example of the procedure is shown in Fig.~\ref{copy_standard}(a-d).

We define a unitary $\hat{U}^{(\boldsymbol{a})}_{copy}$ that applies a local primitive copy operation $\hat{U}^{loc}_{copy}(D_j\vert M_j^{(\boldsymbol{a})})$ between the $j$-th data walker $D_j$ and the $j$-th message qubit $M_j^{(\boldsymbol{a})}$, conditioned on the state of $D_0$. The operation is defined as 
\begin{equation}\label{GlobCopy}  
\begin{split}
\hat{U}^{(\boldsymbol{a})}_{copy}&:=\ket{R}_{D_0}\bra{R}\otimes \left(\bigotimes_{j=1}^{m}\hat{U}_{copy}^{loc}(D_j,M_j^{(\boldsymbol{a})})\right)\\&+\left(\ket{\varnothing}_{D_0}\bra{\varnothing}+\ket{B}_{D_0}\bra{B}\right)\otimes \left(\bigotimes_{j=1}^{m}\hat{I}_{M^{(\boldsymbol{a})}_j,D_j}\right)
\end{split}
\end{equation}
where now $\hat{U}^{loc}_{copy}(D_j\vert  M^{(\boldsymbol{a})}_j)$ is a two-body operation generally defined as 
\begin{equation}\label{locCopy}
\begin{split}
    \hat{U}_{copy}^{loc}(D_i\vert M_i):=(\ket{B}_{M_j^{(\boldsymbol{a})}}\bra{B}+\ket{R}_{M_j^{(\boldsymbol{a})}}\bra{R})\otimes \hat{I}_{D_j}+\\\ket{\varnothing}_{M_j^{(\boldsymbol{a})}}\bra{\varnothing}\otimes (\ket{\varnothing}_{D_j}\bra{R}+\ket{R}_{D_j}\bra{\varnothing}+\ket{B}_{D_j}\bra{B}).
\end{split}
\end{equation}
In other words, the COPY operation activates if the flag walker is in the state $\ket{R}_{D_0}$. The local copy operation transforms the data walkers only if the corresponding message walker is in the empty state $\ket{\varnothing}_{M_j^{(\boldsymbol{a})}}$. The flag walker $D_0$, therefore, ensures that the COPY operation is performed only at the addressed memory cell.

\subsubsection{COPY operation through memory cell activation}

An alternative implementation uses memory cells equipped with a switch. More precisely, each memory cell has a switch $F^{(\boldsymbol{a})}$ that carries a binary internal degree of freedom $\ket{s}_{F^{(\boldsymbol{a})}}\in\{\ket{\mathrm{on}}_{F^{(\boldsymbol{a})}},\ket{\mathrm{off}}_{F^{(\boldsymbol{a})}}\}$ that serves as an enable/disable flag for local memory operations. Thus, in this protocol, the data train is extended to $m+2$ data walkers $\boldsymbol{D}=\{D_{m+1}, D_m, \dots, D_1, D_0\}$. %\dds{(Introdurrei prima il ruolo di $D_0$ e poi quello di $D_{m+1}$)}
{The leading flag walker $D_0$ is used to switch the memory cell on before the COPY step, while the trailing flag $D_{m+1}$ is used to switch it off after the COPY operation.} Since all walkers involved in the memory manipulation are now located at the same leaf node, we no longer attach $(l,d)$ position labels to their states.

An example of the protocol is shown in Fig.~\ref{copy_standard}(e-h). The required steps are as follows:
the switch $F^{(\boldsymbol{a})}$ of the target memory cell is initially prepared in the state $\ket{\mathrm{off}}_{F^{(\boldsymbol{a})}}$. When the data walkers train $\boldsymbol{D}$ reaches the cell, $D_0$ is used to trigger a controlled unitary transformation $\hat{U}^{(\boldsymbol{a})}_{\mathrm{on}}(F^{(\boldsymbol{a})}\vert D_0)$, where $D_0$ is the control and $F^{(\boldsymbol{a})}$ the target. Explicitly we have
    \begin{equation*}
\hat{U}^{(\boldsymbol{a})}_{\mathrm{on}}(F^{(\boldsymbol{a})}\vert D_0)\left(\ket{\mathrm{off}}_{F^{(\boldsymbol{a})}}\otimes\ket{R}_{D_0}\right)=\ket{\mathrm{on}}_{F^{(\boldsymbol{a})}}\otimes\ket{R}_{D_0}
    \end{equation*}
which changes the state of the cell from $\ket{\mathrm{off}}_{F^{(\boldsymbol{a})}}$ to $\ket{\mathrm{on}}_{F^{(\boldsymbol{a})}}$ if {and only if} the flag data walker $D_0$ is in the red state. The form of the activation operator $\hat{U}^{(\boldsymbol{a})}_{\mathrm{on}}$ is
%\begin{equation*}
\begin{multline}\label{activeCellCopy}
    \hat{U}^{(\boldsymbol{a})}_{\mathrm{on}}(F^{(\boldsymbol{a})}\vert D_0):=\ket{R}_{D_0}\bra{R}\otimes \left(\ket{\mathrm{on}}_{F^{(\boldsymbol{a})}}\bra{\mathrm{off}}\right.\\\left.+\ket{\mathrm{off}}_{F^{(\boldsymbol{a})}}\bra{\mathrm{on}}\right)+\left(\ket{\varnothing}_{D_0}\bra{\varnothing}+\ket{B}_{D_0}\bra{B}\right)\otimes\hat{I}_{F^{(\boldsymbol{a})}}.
\end{multline}
%\end{equation*}

In case the memory cell is switched on, a collection of local copy operations take place, as described by the unitary:
\begin{equation*}    \begin{split}\hat{U}^{(\boldsymbol{a})}_{copy}&:=\ket{\mathrm{on}}_{F^{(\boldsymbol{a})}}\bra{\mathrm{on}}\otimes \left[\bigotimes_{j=1}^m\hat{U}_{copy}^{loc}(D_j\vert M^{(\boldsymbol{a})}_j)\right]\\&+\ket{\mathrm{off}}_{F^{(\boldsymbol{a})}}\bra{\mathrm{off}}\otimes\left[\bigotimes_{i=1}^m\hat{I}_{M_i^{(\boldsymbol{a})},D_i}\right],
    \end{split}
\end{equation*}
where $\hat{U}_{copy}^{loc}(D_j\vert M_j^{(\boldsymbol{a})})$ is defined in Eq.~(\ref{locCopy}). 

Finally, once the message is copied, the walker $D_{m+1}$ triggers another unitary transformation to switch off the cell state:
\begin{multline*}  \hat{U}^{(\boldsymbol{a})}_{\mathrm{off}}(F^{(\boldsymbol{a})}\vert D_{m+1})\left(\ket{R}_{D_{m+1}}\otimes\ket{\mathrm{on}}_{F^{(\boldsymbol{a})}}\right)=\\\ket{R}_{D_{m+1}}\otimes\ket{\mathrm{off}}_{F^{(\boldsymbol{a})}}.
\end{multline*}

The switch-off unitary $\hat{U}_{\mathrm{off}}^{(\boldsymbol{a})}$ is defined analogously to Eq.~(\ref{activeCellCopy}) as
\begin{multline*}
\hat{U}^{(\boldsymbol{a})}_{\mathrm{off}}(F^{(\boldsymbol{a})}\vert D_{m+1}):=\ket{R}_{D_{m+1}}\bra{R}\otimes \left(\ket{\mathrm{on}}_{F^{(\boldsymbol{a})}}\bra{\mathrm{off}}\right.\\\left.+\ket{\mathrm{off}}_{F^{(\boldsymbol{a})}}\bra{\mathrm{on}}\right)+\left(\ket{\varnothing}_{D_{m+1}}\bra{\varnothing}+\ket{B}_{D_{m+1}}\bra{B}\right)\otimes\hat{I}_{F^{(\boldsymbol{a})}}.
\end{multline*}
With this strategy, we ensure that the copy operation is effectively executed only on the memory cells reached by the data walkers by activating only the memory cells targeted by the input addresses. 
\gdr{As for the address-transmission gate, the copy operation is a conditional gate acting on different carriers, and therefore requires the same kind of nonlinearity in a
photonic implementation, as discussed in Sec.~\ref{sec:address_transmission}.}

\subsection{Inverse routing}

Once the message stored in the memory cells is copied into the data register $\boldsymbol{D}$, all walkers, from both registers, must be recollected at the output node of the binary tree. As we show in this section, the inverse routing is obtained using the same operation defined in the routing procedure, namely the address-transmission gate $\hat{U}^{(d)}$ introduced above and the inverse of the scattering gate $\hat{S}^{\dagger}$. Finally, we also explain how the spatial separation between walkers, together with their synchronous dynamics, enables their progressive recollection as they approach the output node.

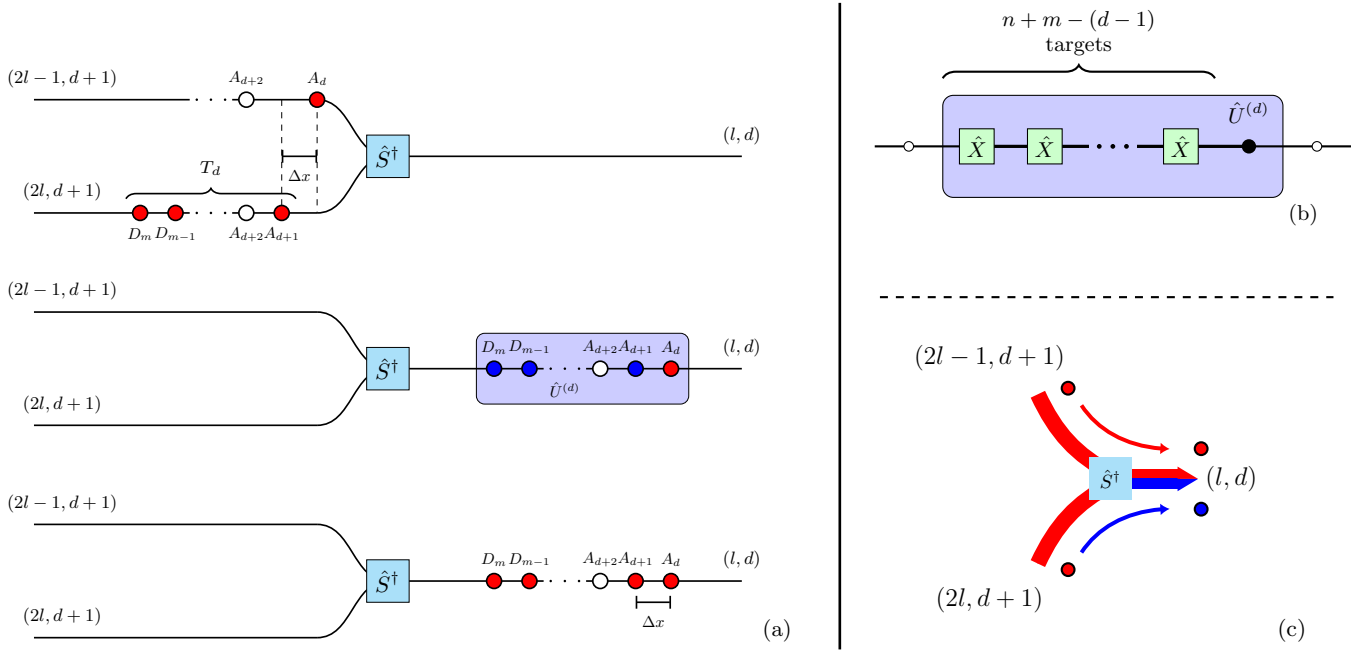
\begin{figure*}[hbt!]
\centering
\begin{subfigure}{0.6\textwidth}
        \resizebox{\linewidth}{!}{
     \begin{tikzpicture}[scale=1.2, transform shape]
         \def\w{8}
  \def\h{1.5}
  \def\dr{3pt}
  \def\dist{0.3}
  
%%%%%%%% PRIMO SCOPE %%%%%%%
\begin{scope}[shift={(1,0)}]

% Parametri per coerenza
    \def\dotRadius{0.5pt}
    
% rail aux
\draw[thick] (7,-0.3) -- (12,-0.3);

  \foreach \x in {4.3, 4.5, 4.7} {
        \fill ( \x, -1.1) circle (\dotRadius);
    }

  \foreach \x in {4.3, 4.5, 4.7} {
        \fill ( \x, 0.5) circle (\dotRadius);
    }

%branch S up
\draw[thick] (7,-0.3) to[out=180, in=0] (6,0.5);

%branch S down
\draw[thick] (7,-0.3) to[out=180, in=0] (6,-1.1);

\node[rectangle, draw, minimum size=6mm,fill=cyan!30] at (7,-0.3){$\hat{S}^{\dagger}$};

% 2l-1,d+1
\draw[thick] (6,0.5) -- (4.8,0.5);
\draw[thick] (4.2,0.5) -- (2,0.5);
% 2l,d+1
\draw[thick] (6,-1.1) -- (4.8,-1.1);
\draw[thick] (4.2,-1.1) -- (2,-1.1);

\draw[dashed] (6,0.5)--(6,-1.1);
\draw[dashed] (5.5,0.5)--(5.5,-1.1);
\draw[thick, |-|] (5.5,-0.3) -- (6,-0.3) 
        node[midway, below=2pt] {{\scalebox{0.7}{$\Delta x$}}};

\draw[thick,fill=red] (6,0.5) circle (\dr);
\draw[thick,fill=red] (5.5,-1.1) circle (\dr);
\draw[thick,fill=white] (5,0.5) circle (\dr);
\draw[thick,fill=white] (5,-1.1) circle (\dr);
\draw[thick,fill=red] (4,-1.1) circle (\dr);
\draw[thick,fill=red] (3.5,-1.1) circle (\dr);

\node at (3.5,-1.4) {{\scalebox{0.7}{$D_m$}}};
\node at (4,-1.4) {{\scalebox{0.7}{$D_{m-1}$}}};

\node at (5,0.8) {{\scalebox{0.7}{$A_{d+2}$}}};
\node at (5,-1.4) {{\scalebox{0.7}{$A_{d+2}$}}};
\node at (5.5,-1.4) {{\scalebox{0.7}{$A_{d+1}$}}};
\node at (6,0.8) {{\scalebox{0.7}{$A_d$}}};

\node at (12,0) {{\scalebox{0.8}{$(l,d)$}}};
\node at (2.4,0.8) {{\scalebox{0.8}{$(2l-1,d+1)$}}};
\node at (2.4,-0.8) {{\scalebox{0.8}{$(2l,d+1)$}}};

\draw[decorate, decoration={brace, amplitude=5pt}, thick] (3.3,-0.9) -- (5.7,-0.9)node[midway, above=6pt, font=\scriptsize] {$T_d$};

\end{scope}

%%%%%%%% SECONDO SCOPE %%%%%%%
\begin{scope}[shift={(1,-3)}]

% Parametri per coerenza
    \def\dotRadius{0.5pt}

\node[rectangle, draw, minimum height=10mm,minimum width=30mm,fill=blue!20,rounded corners] at (9.75,-0.3){};
% rail aux
\draw[thick] (7,-0.3) -- (9.2,-0.3);
\draw[thick] (9.8,-0.3) -- (12,-0.3);

    \foreach \x in {9.3, 9.5, 9.7} {
        \fill ( \x, -0.3) circle (\dotRadius);
    }

%branch S up
\draw[thick] (7,-0.3) to[out=180, in=0] (6,0.5);

%branch S down
\draw[thick] (7,-0.3) to[out=180, in=0] (6,-1.1);

\node[rectangle, draw, minimum size=6mm,fill=cyan!30] at (7,-0.3){$\hat{S}^{\dagger}$};

% 2l-1,d+1
\draw[thick] (6,0.5) -- (2,0.5);
% 2l,d+1
\draw[thick] (6,-1.1) -- (2,-1.1);

\draw[thick,fill=red] (11,-0.3) circle (\dr);
\draw[thick,fill=blue] (10.5,-0.3) circle (\dr);
\draw[thick,fill=white] (10,-0.3) circle (\dr);
\draw[thick,fill=blue] (9,-0.3) circle (\dr);
\draw[thick,fill=blue] (8.5,-0.3) circle (\dr);

 \node at (8.5,0) {{\scalebox{0.7}{$D_m$}}};
 \node at (9,0) {{\scalebox{0.7}{$D_{m-1}$}}};

 \node at (10,0) {{\scalebox{0.7}{$A_{d+2}$}}};
\node at (10.5,0) {{\scalebox{0.7}{$A_{d+1}$}}};
\node at (11,0) {{\scalebox{0.7}{$A_d$}}};

\node at (9.5,-0.6) {{\scalebox{0.7}{$\hat{U}^{(d)}$}}};

\node at (12,0) {{\scalebox{0.8}{$(l,d)$}}};
\node at (2.4,0.8) {{\scalebox{0.8}{$(2l-1,d+1)$}}};
\node at (2.4,-0.8) {{\scalebox{0.8}{$(2l,d+1)$}}};

\end{scope}

%%%%%%%% terzo SCOPE %%%%%%%
\begin{scope}[shift={(1,-6)}]

% Parametri per coerenza
    \def\dotRadius{0.5pt}
% rail aux
\draw[thick] (7,-0.3) -- (9.2,-0.3);
\draw[thick] (9.8,-0.3) -- (12,-0.3);

    \foreach \x in {9.3, 9.5, 9.7} {
        \fill ( \x, -0.3) circle (\dotRadius);
    }

%branch S up
\draw[thick] (7,-0.3) to[out=180, in=0] (6,0.5);

%branch S down
\draw[thick] (7,-0.3) to[out=180, in=0] (6,-1.1);

\node[rectangle, draw, minimum size=6mm,fill=cyan!30] at (7,-0.3){$\hat{S}^{\dagger}$};

%distance
\draw[thick, |-|] (10.5,-0.6) -- (11,-0.6) 
        node[midway, below=2pt] {{\scalebox{0.7}{$\Delta x$}}};
% 2l-1,d+1
\draw[thick] (6,0.5) -- (2,0.5);
% 2l,d+1
\draw[thick] (6,-1.1) -- (2,-1.1);

\draw[thick,fill=red] (8.5,-0.3) circle (\dr);
\draw[thick,fill=red] (9,-0.3) circle (\dr);
\draw[thick,fill=white] (10,-0.3) circle (\dr);
\draw[thick,fill=red] (10.5,-0.3) circle (\dr);
\draw[thick,fill=red] (11,-0.3) circle (\dr);

 \node at (8.5,0) {{\scalebox{0.7}{$D_m$}}};
 \node at (9,0) {{\scalebox{0.7}{$D_{m-1}$}}};

 \node at (10,0) {{\scalebox{0.7}{$A_{d+2}$}}};
\node at (10.5,0) {{\scalebox{0.7}{$A_{d+1}$}}};
\node at (11,0) {{\scalebox{0.7}{$A_d$}}};

\node at (12,0) {{\scalebox{0.8}{$(l,d)$}}};
\node at (2.4,0.8) {{\scalebox{0.8}{$(2l-1,d+1)$}}};
\node at (2.4,-0.8) {{\scalebox{0.8}{$(2l,d+1)$}}};

\end{scope}

\node at (13.5,-7){(a)};
     \end{tikzpicture}
     }
\end{subfigure}\hfill
\vrule width 1pt
\hfill
     \begin{subfigure}{0.35\textwidth}
\begin{tikzpicture}[gateX/.style={rectangle, draw, fill=green!20, minimum size=5mm}, rotate = 90, scale=0.9,transform shape]
\node [draw, rounded corners, rectangle, fill=blue!20, inner sep=5pt, minimum width=15mm, minimum height=50mm] 
        at (0,0) {};
\def\n{5}
\def\spacing{1} 
\draw[thick] (0,-3.5) -- (0,{(\spacing*(\n-1))-0.5 });
\foreach \i/\col in {0/white, 1/black, 2/black, 3/black, 4/black,5/black,6/white} {
\edef\colorname{\col} 
\filldraw[black, fill=\colorname] (0, \i*\spacing-3) circle (2pt);}

\draw[very thick] (0,-2)--(0,2);

\node[midway, rotate = -90, xshift=-2cm,yshift=0cm,gateX] {$\hat{X}$};

\node[midway, rotate = -90, xshift=-1cm,yshift=0cm,gateX] {$\hat{X}$};

\node[midway,rotate = -90,xshift=1cm,yshift=0cm,gateX] {$\hat{X}$};

\fill (0,-2) circle (3pt);

\node[fill=blue!20, rectangle, minimum size=7mm] at (0,0) {};

 \foreach \x in {-0.2, 0, 0.2} {
        \fill (0, \x) circle (1pt);
    }

\draw[decorate,rotate=-90, decoration={brace, amplitude=5pt}, thick, transform shape] (-2.5,1) -- (1.5,1)node[midway, above=6pt, align=center] {$n+m-(d-1)$\\ targets};    

\node[rotate=-90] at (0.5,-2){$\hat{U}^{(d)}$};

\node[rotate=-90] at (-1,-2.8){(b)};

\end{tikzpicture}

\vspace{0.5cm}
\noindent\tikz\draw[dashed, thick] (-2,0) -- (4,0);
\vspace{0.5cm}
     
\begin{tikzpicture}[scale=0.8,rotate=-90, transform shape, line join=round]
     
   % Parametri
  \def\stemthick{5pt}
  \def\branchthick{6pt}
  \def\branchlen{1.4}
  \def\bendangle{20}
  \def\dr{3pt}

  % Coordinate principali
  \coordinate (O) at (0,0);
  \coordinate (B) at (0,-0.8); % punto di biforcazione
  \coordinate (L) at ($ (B) + (-\branchlen,-\branchlen) $);
  \coordinate (R) at ($ (B) + (\branchlen,-\branchlen) $);

  % Gambo centrale
  \draw[red,line width=\stemthick] (O) -- (B);

  % Ramo sinistro: rosso sottile con freccia
  \draw[red,line width=\branchthick,{Latex[length=5pt, width=7pt]}-]
    (B) to[bend left=\bendangle] (L);

    % Freccia sopra la sfumatura: con stessa head del ramo sinistro
  \draw[opacity=0, line width=0pt] (B) to[bend right=\bendangle] (R); % aiuta compilatori "pigri"
  \draw[line width=\branchthick,{Latex[length=8pt,width=8pt]}-, draw=none] (B) to[bend right=\bendangle] (R);
  \path[draw=red, postaction={draw=red,{Latex[length=5pt,width=7pt]}-, line width=\branchthick}]
    (B) to[bend right=\bendangle] (R);

\node[rectangle, draw=none, fill=red, inner sep=0pt, outer sep=0pt, minimum width=5pt, minimum height=12mm
] at (-0.08,-0.45){}; 

\node[rectangle, draw=none, fill=blue, inner sep=0pt, outer sep=0pt, minimum width=5pt, minimum height=12mm
] at (0.08,-0.45){}; 

\node[rectangle, fill=cyan!30, inner sep=0pt, outer sep=0pt, minimum width=20pt, minimum height=20pt, rotate=-90,
xscale=-1, yscale=-1] at (0,-1){$\hat{S}^{\dagger}$};

\begin{scope}[shift={(0,0.45)}, scale=1.7]
  % triangolo sinistro rosso
  \fill[red]  (-0.12,-0.2) -- (0,0) -- (0,-0.2) -- cycle;
  % triangolo destro blu
  \fill[blue] (0.12,-0.2) -- (0,0) -- (0,-0.2) -- cycle;
\end{scope}

\draw[red, line width=1.5pt, {Latex[length=3.5pt, width=4.5pt]}-, shorten <=5pt, shorten >=5pt]
    ($ (B) + (-0.5, 1) $) to[bend left=30] ($ (L) + (0, 0.6) $);

\draw[blue, line width=1.5pt, {Latex[length=3.5pt, width=4.5pt]}-, shorten <=5pt, shorten >=5pt]
    ($ (B) + (0.5, 1) $) to[bend right=30] ($ (R) + (0, 0.6) $);

\draw[thick,fill=red] (-0.5,0.5) circle (\dr);

\draw[thick,fill=red] (-1.5,-1.7) circle (\dr);

\draw[thick,fill=blue] (0.5,0.5) circle (\dr);

\draw[thick,fill=red] (1.5,-1.7) circle (\dr);

\node[rotate=90, transform shape] at (0,1){\large$(l,d)$};

\node[rotate=90, transform shape] at (2,-3){\large$(2l,d+1)$};

\node[rotate=90, transform shape] at (-2,-3){\large$(2l-1,d+1)$};

\node[rotate=90, scale=1.2, transform shape] at (2.5,2){(c)};

\end{tikzpicture} 
     \end{subfigure}
       \caption{Inverse routing elements for our qRAM architecture. (a) Joint action of the operations within the unitary block at a generic level $d$. The target walkers $T_d$ move along one branch entering the node, while the address walker $A_d$ move along the other. First, the particles are directed onto the same parent branch by the scattering gate $\hat S^{\dagger}$, while maintaining their spatial ordering. Later, they undergo the action of the address transmission gate $\hat{U}^{(d)}$, conditioned on the color of $A_d$. (b) Address-transmission gate $\hat{U}^{(d)}$. As in the forward routing phase, the gate is activated by a walker in the red color state and targets $n+m-(d-1)$ particles at level $d$. %\gc{Check: nel testo dici $n + m − (d − 1)$ targets.}
       (c) Inverse scattering gate $\hat{S}^{\dagger}$. All particles enter the gate in the red state and are routed onto the same parent branch. However, their color is changed by the action of the gate from red to blue depending on their specific entry branch.}  \label{InversedRouting}
\end{figure*}

{\subsubsection{Inverse routing operations}

As for the routing protocol, we place a unitary block on every node of the tree. Here, while the sequence of gates is reversed compared to the routing phase, the internal structure of each gate is preserved, as explained below.

\gdr{The first operation of the block is the inverse of the scattering gate. Consistently with the notation of Sec.~\ref{sec:S}, we denote it by
\begin{equation*}
\hat S^{(l,d)}_{\leftarrow} := \bigl(\hat S^{(l,d)}_{\rightarrow}\bigr)^{\dagger},
\end{equation*}
the subscript indicating that the walker now propagates from the leaves towards
the root. It is placed at the same node $(l,d)$ of $\hat{S}_{\rightarrow}^{(l,d)}$ in the output tree and moves the particles from the child branches $(2l-1,d)$ and $(2l,d)$ to the parent branch $(l,d-1)$, as shown in Fig.~\ref{InversedRouting}c. By design of the protocol, in the output phase, every walker, if present, arrives at every node with color $c=R$. Hence, in analogy with the action of $\hat{S}_{\rightarrow}^{(l,d)}$ during the routing phase, $\hat{S}_{\leftarrow}^{(l,d)}$ must also change the internal state of the walkers according to the entrance branch. The action of $\hat{S}_{\leftarrow}^{(l,d)}$ on a single particle state is defined as 
\begin{equation*}
\begin{split}
&\hat S^{(l,d)}_{\leftarrow}\ket{R^{2l-1,d+1}}_W=\ket{R^{l,d}}_W\\
&\hat S^{(l,d)}_{\leftarrow}\ket{R^{2l,d+1}}_W=\ket{B^{l,d}}_W
\end{split}
\end{equation*}
i.e., a walker coming from the left branch is directed toward the previous level, retaining its color, while one coming from the right branch is directed toward the previous level, and its color is changed to blue. We can write this in a more compact form as
\begin{equation}
\hat S^{(l,d)}_{\leftarrow}\ket{R^{j,d+1}}_W
=\hat X^{\,j-2l+1}\ket{R^{l,d}}_W,\qquad j\in\{2l-1,2l\},
\end{equation}
where $\hat{X}^0=\hat{I}$ and  $\hat{X}\ket{R}=\ket{B},\hat{X}\ket{B}=\ket{R},\hat{X}\ket{\varnothing}=\ket{\varnothing}$.}

\gdr{As for the forward gate, the vacuum carries no color and hence no routing
information, and is simply propagated to the previous level,
\begin{equation*}
\hat S^{(l,d)}_{\leftarrow}\ket{\varnothing^{d+1}}_W=\ket{\varnothing^d}_W,
\end{equation*}
thus we can omit the branch index.
%The explicit form of $\hat{S}^{\dagger}$ is defined as 
%\begin{equation}
%\begin{split}
%\hat{S}^{\dagger}&=\ket{B^{l,d-1}} \bra{R^{2l,d}} +\ket{R^{l,d-1}}\bra{R^{2l-1,d}}+\ket{\varnothing^{d-1}}\bra{\varnothing^d}\\&+\ket{R^{2l-1,d}}\bra{B^{2l,d}}+\ket{R^{2l,d}}\bra{B^{2l-1,d}}\\&+\ket{B^{2l,d}}\bra{R^{l,d-1}} + \ket{B^{2l-1,d}
%} \bra{B^{l,d-1}} + \ket{\varnothing^d} \bra{\varnothing^{d-1}
%} 
%\end{split}
%\end{equation}
%where the first line consists of the three terms that allow for the inverse routing scheme, while the others ensure unitarity.
In optical terms, $\hat S^{(l,d)}_{\leftarrow}$ is the very same PBS and half-wave
plate of Sec.~\ref{sec:S}, traversed in the opposite direction: a walker coming
from the left branch is transmitted into the parent branch unchanged, while one
coming from the right branch first crosses the half-wave plate, which now turns
its color from red to blue, and is then reflected into the parent branch. No
additional element is thus required, and the inverse routing is passive exactly as
the direct one.}

The second operation of the block is the address transmission gate $\hat{U}^{(d)}$, placed on 
each branch at depth $d\in\{0,\dots,n-1\}$. The action of this gate is the same as the one in the forward routing,
as defined in Eq.~(\ref{infoTransm}). In the output phase $\hat{U}^{(d)}$ is placed symmetrically with respect to the memory cells, yet maintains the same internal structure: the rightmost walker of the train remains the control for the gate (see Fig.~\ref{InversedRouting}b), dictating the path to take from level $d$ to $d-1$, with the preceding walker set $P_d$ and following target set $T_d$ defined in Eq.~(\ref{WalkerSets}). Crucially, as explained in the next section, our architecture recollects address particles level by level during the output phase. Thus, the train of recollected particles grows as we move toward the output port, making the set of target walkers $T_d$ progressively larger as the depth decreases.

\subsubsection{Recollection of walkers}
A distinctive feature of our qRAM architecture is that, during the first routing of the data particles towards the memory cells, the action of the routing unitary blocks induces a progressive separation of the address walkers across different branches, once the address information has been transmitted. 

This effect originates from the action of the address-transmission gate $\hat{U}^{(d)}$, which acts on all walkers in the target set $T_d$ following the address walker $A_d$. As a consequence, the color of the walkers in $T_d$ is modified, while the state of $A_d$ remains unchanged. The subsequent application of the scattering gate $\hat{S}$, which routes walkers according to their color, cause this separation: the address walker $A_d$ propagates along one branch of the tree, while the $T_d$ walkers propagate along the other. 

At first sight, this dispersion of address walkers may appear problematic, as the same particles are required during the inverse routing to coherently route the data register $\boldsymbol{D}$ back to the output node of the tree.
We now explain how  our protocol ensures the correct recollection of the address walkers during the output phase. 

The recollection mechanism is the result of three main ingredients: (i) the synchronous propagation of the walkers on the binary tree, (ii) the fixed ordering and spatial separation $\Delta x$ between each walker in the train, and (iii) the specific placement and functioning of the unitary blocks in the output tree.
Indeed, because the walkers are arranged in a precise order which is preserved throughout the protocol, and because their mutual spatial separation $\Delta x$ is maintained by the synchronous evolution of the train, any address walker $A_d$ that is diverted from the path followed by the corresponding target set $T_d$ remains displaced by exactly $\Delta x$ with respect to them. 
Once it has taken a different path, the address $A_d$ no longer interacts with other address-transmission gates before reaching the memory cells; hence, its color remains fixed at $c=R$. Consequently, for the remaining $k=n-d$ levels, it deterministically propagates from level $(l,d)$ to level $(2l-1,d+1)$. By construction, the output tree is the mirror image of the input tree with respect to the locations of the memory cells. As a result, during the inverse routing, the address walker traverses the first $n-d$ levels of the output tree along the branch that is symmetric to the one followed in the last $n-d$ levels of the routing phase. Since this mechanism applies to every walker in the address register $\boldsymbol{A}$, it follows that the address walker $A_d$ meets again its corresponding target set $T_d$ precisely at level $d$ of the output tree (see Fig.~\ref{InversedRouting}a). Moreover, since the spatial separation $\Delta x$ and the ordering of the train have been preserved, all walkers reappear at that level in exactly the same ordering as in the routing phase. This ensures the correct activation of the unitary block placed at that level, thereby routing the entire train coherently toward the next level.

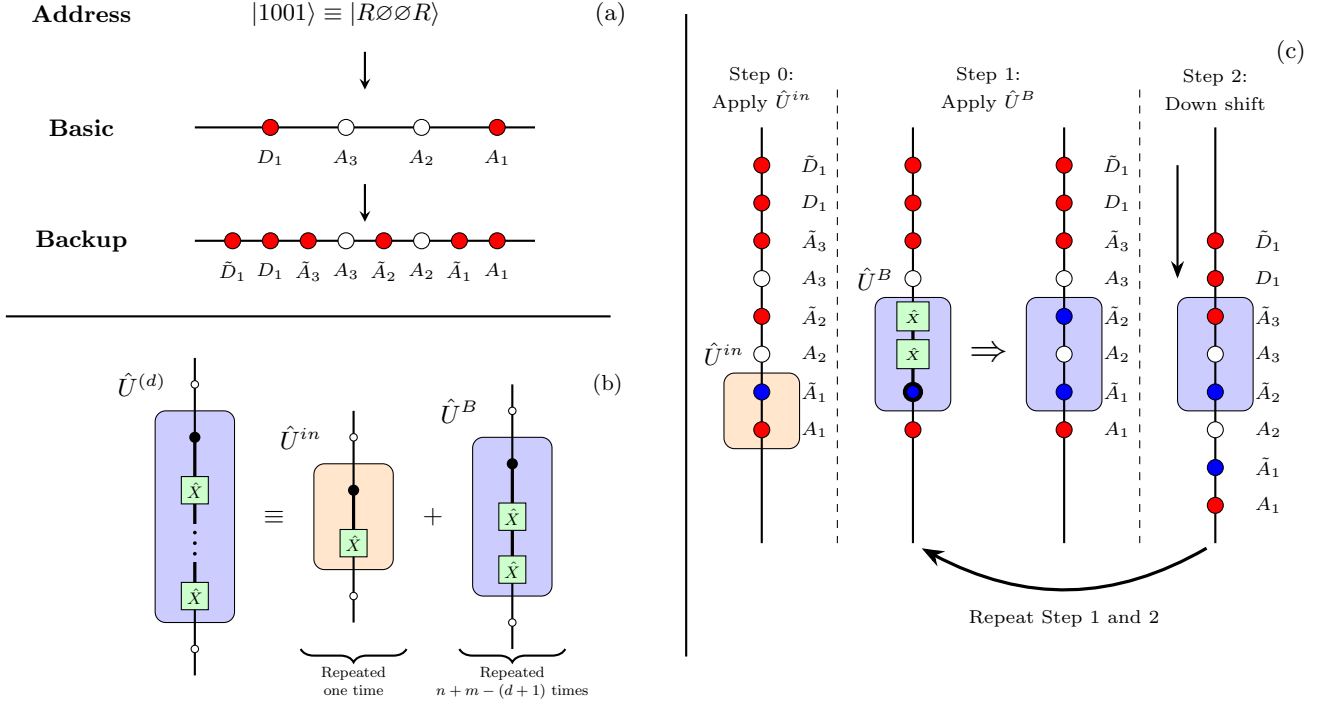
\begin{figure*}[htb!]
\centering

\begin{tikzpicture}
\begin{scope}[xshift=-3cm]

\begin{scope}[yshift=-2cm]
    \def\n{8}
\def\spacing{0.5}

\node at (-0.5,5){$\ket{1001}\equiv\ket{R\varnothing\varnothing R}$};

\draw[-stealth, thick] (-0.25,4.5)--(-0.25,4);

\node at (-4,5){\textbf{Address}};

\draw[thick] (2,{(\spacing*(\n-1))-1.5+1.5 }) -- (-2.5,3.5);

\def\none{none}
\foreach \i/\col in {0/none, 1/red, 2/none, 3/white, 4/none,5/white,6/none,7/red} {
\edef\colorname{\col} 
\ifx\colorname\none
\else
  \filldraw[black, fill=\colorname] ( \i*\spacing-2,3.5) circle (3pt);
\fi
}

\draw[-stealth, thick] (-0.25,2.75)--(-0.25,2.25);

\node at (-4,3.5){\textbf{\gdr{Basic}}};

\node at (1.5,3.1){\scriptsize$A_1$};

\node at (0.5,3.1){\scriptsize$A_2$};

\node at (-0.5,3.1){\scriptsize$A_3$};

\node at (-1.5,3.1){\scriptsize$D_1$};

\draw[thick] (2,{(\spacing*(\n-1))-1.5 }) -- (-2.5,2);

\foreach \i/\col in {0/red, 1/red, 2/red, 3/white, 4/red,5/white,6/red,7/red} {
\edef\colorname{\col} 
\filldraw[black, fill=\colorname] ( \i*\spacing-2,2) circle (3pt);}

\node at (-4,2){\textbf{Backup}};

\node at (1.5,1.6){\scriptsize$A_1$};

\node at (1,1.6){\scriptsize$\tilde A_1$};

\node at (0.5,1.6){\scriptsize$A_2$};

\node at (0,1.6){\scriptsize$   \tilde A_2$};

\node at (-0.5,1.6){\scriptsize$A_3$};

\node at (-1,1.6){\scriptsize$\tilde A_3$};

\node at (-1.5,1.6){\scriptsize$D_1$};

\node at (-2,1.6){\scriptsize$\tilde D_1$};

\node at (3,5){$\text{(a)}$};

\end{scope}

 \draw[thick] (-5, -1) -- (3, -1);

\begin{scope}[yshift=-4cm, xshift=-2.5cm, gateX/.style={rectangle, draw, fill=green!20, minimum size=5mm}, rotate = 180, scale=0.7,transform shape]
\node [draw, rounded corners, rectangle, fill=blue!20, inner sep=5pt, minimum width=15mm, minimum height=40mm] 
        at (0,-0.5) {};
\def\n{4}
\def\spacing{1} 
\draw[thick] (0,-3.5) -- (0,{(\spacing*(\n-1))-0.5 });
\foreach \i/\col in {0/white, 1/black, 2/black, 3/black, 4/black,5/white} {
\edef\colorname{\col} 
\filldraw[black, fill=\colorname] (0, \i*\spacing-3) circle (2pt);}

\draw[very thick] (0,-2)--(0,1);

\node[midway,rotate = 180,xshift=0cm,yshift=1cm,gateX] {$\hat{X}$};

\node[midway, rotate = 180, xshift=0cm,yshift=-1cm,gateX] {$\hat{X}$};

\fill (0,-2) circle (3pt);

\node[fill=blue!20, rectangle, minimum size=7mm] at (0,0) {};

 \foreach \x in {-0.2, 0, 0.2} {
        \fill (0, \x) circle (1pt);
    }

\node[rotate=180, scale = 1.5, transform shape] at (1,-3){$\hat U^{(d)}$};

\node at (-1.5,-0.5){\Large$\equiv$};

\node [draw, rounded corners, rectangle, fill=orange!20, inner sep=5pt, minimum width=15mm, minimum height=20mm] 
        at (-3,-0.5) {};
\def\n{3}
\def\spacing{1} 
\draw[thick] (-3,-2.5) -- (-3,{(\spacing*(\n-1))-0.5 });
\foreach \i/\col in {0/white, 1/black, 2/black, 3/white} {
\edef\colorname{\col} 
\filldraw[black, fill=\colorname] (-3, \i*\spacing-2) circle (2pt);}

\draw[very thick] (-3,-1)--(-3,0);

\node[midway, rotate = 180, xshift=3cm,yshift=0cm,gateX] {$\hat{X}$};

\fill (-3,-1) circle (3pt);

\node[fill=blue!20, rectangle, minimum size=7mm] at (0,0) {};

 \foreach \x in {-0.2, 0, 0.2} {
        \fill (0, \x) circle (1pt);
    }

\node[rotate=180, scale = 1.5, transform shape] at (-2,-2){$\hat U^{in}$};

\draw[decorate, rotate=180,align=center, decoration={brace, amplitude=5pt}, thick, transform shape] (4,-2) -- (2,-2)node[midway, below=6pt, font=\footnotesize] {Repeated \\ one time};

\node at (-4.5,-0.5){\Large$+$};

\node [draw, rounded corners, rectangle, fill=blue!20, inner sep=5pt, minimum width=15mm, minimum height=30mm] 
        at (-6,-0.5) {};
\def\n{5}
\def\spacing{1} 
\draw[thick] (-6,-3) -- (-6,{(\spacing*(\n-1))-2 });
\foreach \i/\col in {0/white, 1/black, 2/black, 3/black, 4/white} {
\edef\colorname{\col} 
\filldraw[black, fill=\colorname] (-6, \i*\spacing-2.5) circle (2pt);}

\draw[very thick] (-6,-1.5)--(-6,0.5);

\node[midway,rotate = 180,xshift=6cm,yshift=0.5cm,gateX] {$\hat{X}$};

\node[midway, rotate = 180, xshift=6cm,yshift=-0.5cm,gateX] {$\hat{X}$};

\fill (-6,-1.5) circle (3pt);

\node[rotate=180, scale = 1.5, transform shape] at (-5,-2.5){$\hat U^{B}$};

\draw[decorate, rotate=180, align=center,decoration={brace, amplitude=5pt}, thick, transform shape] (7,-2) -- (5,-2)node[midway, below=6pt, font=\footnotesize] {Repeated \\ $n+m-(d+1)$ times};

\node[scale=1.3, rotate=180, transform shape] at (-7.8,-3){$\text{(b)}$};

\end{scope}

\end{scope}

 \draw[ thick] (1, -5.5) -- (1, 3);

\begin{scope}[xshift=0cm, yshift=-0.5cm, scale=1, gateX/.style={rectangle, draw, fill=green!20, minimum size=2mm}, transform shape]

%copy1

\node [draw, rounded corners, fill=orange!20, inner sep=5pt, minimum width=10mm, minimum height=10mm] 
        at (2,-1.75) {};

\def\n{8}
\def\spacing{0.5} 
\draw[thick] (2,-3.5) -- (2,{(\spacing*(\n-1))-1.5 });

%\draw[very thick] (2,-1.5)--(2,-0.5);

\foreach \i/\col in {0/red, 1/blue, 2/white, 3/red, 4/white,5/red,6/red,7/red} {
\edef\colorname{\col} 
\filldraw[black, fill=\colorname] (2, \i*\spacing-2) circle (3pt);}

%\draw[ultra thick] (2,-1.5) circle (3pt);

\node at (1.5,-1){$\hat{U}^{in}$};

\node at (2.7,1.5){\scriptsize$\tilde{D}_1$};

\node at (2.7,1){\scriptsize$D_1$};

\node at (2.7,0.5){\scriptsize$\tilde{A}_3$};

\node at (2.7,0){\scriptsize$A_3$};

\node at (2.7,-0.5){\scriptsize$\tilde{A}_2$};

\node at (2.7,-1){\scriptsize$A_2$};

\node at (2.7,-1.5){\scriptsize$\tilde{A}_1$};

\node at (2.7,-2){\scriptsize$A_1$};

%copy2

\node [draw, rounded corners, fill=blue!20, inner sep=5pt, minimum width=10mm, minimum height=15mm] 
        at (4,-1) {};

\def\n{8}
\def\spacing{0.5} 
\draw[thick] (4,-3.5) -- (4,{(\spacing*(\n-1))-1.5 });

\draw[very thick] (4,-1.5)--(4,-0.5);

\foreach \i/\col in {0/red, 1/blue, 2/white, 3/blue, 4/white,5/red,6/red,7/red} {
\edef\colorname{\col} 
\filldraw[black, fill=\colorname] (4, \i*\spacing-2) circle (3pt);}

\node[midway,yshift=-0.5cm,xshift=4cm,gateX] {\tiny$\hat{X}$};

\node[midway,yshift=-1cm,xshift=4cm,gateX] {\tiny$\hat{X}$};

\draw[ultra thick] (4,-1.5) circle (3pt);

%\node at (4.7,1.5){\scriptsize$\tilde{m}_1$};

%\node at (4.7,1){\scriptsize$m_1$};

%\node at (4.7,0.5){\scriptsize$\tilde{a}_3$};

%\node at (4.7,0){\scriptsize$a_3$};

%\node at (4.7,-0.5){\scriptsize$\tilde{a}_2$};

%\node at (4.7,-1){\scriptsize$a_2$};

%\node at (4.7,-1.5){\scriptsize$\tilde{a}_1$};

%\node at (4.7,-2){\scriptsize$a_1$};

\node at (5,-1){\Large $\Rightarrow$};

\node at (3.5,0){$\hat{U}^{B}$};

%copy3

\node [draw, rounded corners, fill=blue!20, inner sep=5pt, minimum width=10mm, minimum height=15mm] 
        at (6,-1) {};

\def\n{8}
\def\spacing{0.5} 
\draw[thick] (6,-3.5) -- (6,{(\spacing*(\n-1))-1.5 });

%\draw[very thick] (6,-1.5)--(6,-0.5);

%\draw[very thick] (6,-1.2) circle (3pt);

\foreach \i/\col in {0/red, 1/blue, 2/white, 3/blue, 4/white,5/red,6/red,7/red} {
\edef\colorname{\col} 
\filldraw[black, fill=\colorname] (6, \i*\spacing-2) circle (3pt);}

\node at (6.7,1.5){\scriptsize$\tilde{D}_1$};

\node at (6.7,1){\scriptsize$D_1$};

\node at (6.7,0.5){\scriptsize$\tilde{A}_3$};

\node at (6.7,0){\scriptsize$A_3$};

\node at (6.7,-0.5){\scriptsize$\tilde{A}_2$};

\node at (6.7,-1){\scriptsize$A_2$};

\node at (6.7,-1.5){\scriptsize$\tilde{A}_1$};

\node at (6.7,-2){\scriptsize$A_1$};

%copy4

\node [draw, rounded corners, fill=blue!20, inner sep=5pt, minimum width=10mm, minimum height=15mm] 
        at (8,-1) {};

\def\n{8}
\def\spacing{0.5} 
\draw[thick] (8,-3.5) -- (8,{(\spacing*(\n-1))-1.5 });

%\draw[very thick] (8,-1.5)--(8,0);

%\draw[very thick] (8,-1.2) circle (3pt);

\foreach \i/\col in {0/white, 1/blue, 2/white, 3/red, 4/red,5/red} {
\edef\colorname{\col} 
\filldraw[black, fill=\colorname] (8, \i*\spacing-2) circle (3pt);}

\filldraw[color=black, fill=blue](8,-2.5) circle (3pt);

\filldraw[color=black, fill=red](8,-3) circle (3pt);

\node at (8.7,0.5){\scriptsize$\tilde{D}_1$};

\node at (8.7,0){\scriptsize$D_1$};

\node at (8.7,-0.5){\scriptsize$\tilde{A}_3$};

\node at (8.7,-1){\scriptsize$A_3$};

\node at (8.7,-1.5){\scriptsize$\tilde{A}_2$};

\node at (8.7,-2){\scriptsize$A_2$};

\node at (8.7,-2.5){\scriptsize$\tilde{A}_1$};

\node at (8.7,-3){\scriptsize$A_1$};

\draw[->, -Stealth,thick] (7.5,1.5)--(7.5,0);

\draw[dashed] (3,2.5)--(3,-3.5);

\draw[dashed] (7,2.5)--(7,-3.5);

\node(0) at (2,2.5){};

\node(1a) at (4,2.5){};

\node(1b) at (6,2.5){};

\node(2) at (8,2.5){};

\node(in) at (4,-3.5){};

\node(out) at (8,-3.5){};

\node[align=center] at (2,2.5){\scriptsize Step 0: \\ \scriptsize Apply $\hat{U}^{in}$};

%\draw[->, >={Stealth[bend]}, very thick, out=30, in=150] (0) to (1a);

\node[align=center] at (5,2.5){\scriptsize Step 1: \\ \scriptsize Apply $\hat{U}^B$ };

%\draw[->, >={Stealth[bend]}, very thick, out=30, in=150] (1b) to (2);

\node[align=center] at (8,2.5){\scriptsize Step 2: \\ \scriptsize Down shift};

\draw[->, >={Stealth[bend]}, very thick, out=-150, in=-30] (out) to (in);

\node at (6,-4.5){\scriptsize Repeat Step 1 and 2 };

\node at (9,3){$\text{(c)}$};

\end{scope}

\end{tikzpicture}
\caption{Example of the  backup variant implementation of the gate
$\hat{U}^{(d)}$ for $d=1$ and target address $(a_1,a_2,a_3)=(1,0,1)$.
(a) State encoding comparison: in the backup variant the walkers are implemented in pairs, with backup particles always in the red state. (b) Decomposition of the address transmission gate $\hat{U}^{(d)}$ through two gates: the initialization gate $\hat{U}^{in}$ (C-NOT between $d$-th address and its backup) and the block gate $\hat{U}^B$ (C-NOT-NOT among walker triples). Through repeated applications of $\hat{U}^B$, the path information is propagated  along the particle train.
(c) Steps required to transmit the address information in the backup variant.
Step 0: the first address bit transfers its path information to its corresponding backup walker via $\hat{U}^{in}$.
Step 1: the backup, the next address walker, and its associated backup undergo the action of the gate $\hat{U}^B$. Step 2: all three particles shift forward by two positions. As a result, the second backup now holds the same path information originally carried by the first. Steps 1 and 2 are repeated until the last data walker receive the action of $\hat U^B$, namely $n+m-(d+1)$ times.}
 
\label{backupscheme}
\end{figure*}

{\subsection{Resource scaling}}
\label{StandardScaling}
We now describe the resources needed to implement the \gdr{basic} version of the protocol. In particular, we account for spatial resources, the number of active elements per query, and the overall depth of the algorithm.
The total number of carriers consists of at most $n$ walkers for the address register $\boldsymbol{A}$ and $m+1$ walkers for the data register $\boldsymbol{D}$. 
Thus, the total number of walkers is at most $n+m+1\sim\mathcal{O}(n+m)$. 

The protocols utilize a single binary tree of depth $n$, where each bifurcation of the tree is associated with a unitary block composed of two operations, resulting in a total of $2(2^n-1)\sim\mathcal{O}(2^n)$ required gates. While the scattering gate $\hat{S}$ in the unitary block is local, the address-transmission gate $\hat{U}^{(d)}$ acts on $n+m-d$ walkers. Consequently, the branch capacity of each edge, i.e., the number of sites required to host all walkers, at level $d$, scales linearly with the number of walkers it hosts, $n_{\text{sites}}^d=n+m-d\sim\mathcal{O}(n+m)$. 

While the physical longitudinal extension of the entire architecture inherently scales as $\mathcal{O}(2^n)$ to physically reach all memory cells, the relevant metric for the carrier dynamics is the extent of a single root-to-leaf path. This path extent is obtained by summing these contributions over all levels
\begin{equation}\label{StandardSitesScaling}
\begin{split}   n_{\text{sites}}^{tot}&=\sum_{d=1}^{n}n_{\text{sites}}^d=\sum_{d=1}^{n}(n+m-d)%=n(n+m)-\frac{n(n+1)}{2}
\\&= {\frac{n^2+n(2m-1)}2}\sim\mathcal{O}(n^2+nm).
\end{split}
\end{equation}

Regarding the active elements per query, the scaling depends on the input state. For classical single address query or a sparse superposition, the number of required unitary blocks scales as $\mathcal{O}(n)$. However, in the worst-case scenario in which we have a fully delocalized superposition of all possible addresses, the scaling becomes $\mathcal{O}(2^n)$, a cost inherent to any qRAM architecture. 

Finally, we analyze the circuit depth. In the serial scheme, the global unitary operation at depth $d+1$
can only be applied once all walkers have reached the corresponding branch, introducing a strict sequential dependence between tree levels.
Each level requires $\mathcal{O}(n+m)$ propagation steps.
Summing over all the $n$ levels, the total execution time scales as $T_{\text{serial}} \sim \mathcal{O}(n(n+m))$, exhibiting a quadratic scaling in $n$.

These results provide a fundamental benchmark for the resource overhead of our \gdr{basic} protocol. While the quadratic scaling in time and the spatial extent are derived here for bosonic carriers in a binary tree, the structural logic of the architecture remains a versatile blueprint. In the following sections, we show how this framework can be generalized to different carriers, \gdr{such as dual-rail qubits and four-level qudits,} discussing how these changes allow for both reduced circuit depth and simpler gate implementations via short-range interactions, alongside the trade-offs required to achieve these more favorable properties.

\section{Backup variant}
\label{sec:backup}

The address-transmission gate of the \gdr{basic} walker-based qRAM, introduced in Eq.~(\ref{infoTransm}), can be alternatively implemented using only short-range coupling between information carriers. This modification leads to a simpler structure for the gate as well as to a modular workflow for the overall protocol. In particular, the full operation can be decomposed into repeated applications of a basic local block acting on a limited number of walkers at a time, which is identical at all depths. We remind that, in the qRAM scheme presented above, the gates $\hat U^{(d)}$ had a different number of targets depending on the particular depth $d$ considered.

This new scheme admits a parallelization pipeline that drastically reduces the circuit depth from quadratic to linear, namely $\mathcal{O}(n+m)$. Moreover, it requires fewer spatial resources to implement the binary tree, which in this case scales linearly as $\mathcal{O}(n+m)$, and allows memory operations to be performed through strictly short-range interactions, without the need for global control. 

In this section, we show how to achieve these results through an overall modification of the previous scheme by using a constant overhead in the number of walkers needed.

\subsection{Encoding and initialization}
In this variant, we double the total number of information carriers, considering now a train of $2(n+m)$ walkers composed of the original $n$ address walkers and $m$ data walkers, each followed by an auxiliary particle, called ``backup" walker. Thus, the address and data registers in this variant are defined as $\boldsymbol{A}=\{\tilde A_n,A_n,\tilde A_{n-1}\dots,A_2,\tilde A_1,A_1\}$ and $\boldsymbol{D}=\{\tilde D_m,D_m,\tilde D_{m-1},\dots,D_2,\tilde D_1, D_1\}$. The last data backup $\tilde D_{m}$ is not required for the execution of the protocol, since, as explained in the next section, backup walkers serve as a bridge to communicate the routing information among walkers. Nonetheless, we leave it in the register definition for clarity.
Each one of these walkers, regardless of the encoded address or whether it belongs to the address or data register,} is initialized in the reference red state $\ket{R}$. Hence, the global initialization of our walkers is
\begin{align*}
\ket{\psi_{\mathrm{in}}}&=\ket{\tilde d_m d_m \dots \tilde d_1 d_1 \tilde a_n a_n \dots \tilde a_1 a_1}_{DA}\\
&=\bigotimes_{j=1}^m\left(\ket{R}_{\tilde D_j}\otimes \ket{R}_{D_j}\right)\otimes\bigotimes_{i=1}^n\left(\ket{R}_{\tilde A_i}\otimes\ket{a_i}_{A_i}\right)
\end{align*}
so that each address and data walker, respectively, $A_i$ and $D_j$ for $i=1,\dots,n$ and $j=1,\dots,m$, is followed by its red backup. We use the notation $\tilde A_i$ to indicate the backup walker corresponding to the $i$-th address walker and, similarly, $\tilde D_j$ is the backup of $D_j$. 

The chosen ordering reflects the physical sequence of modes in the walker train (see Fig.~\ref{backupscheme}a). We will employ this redundancy in information carriers to redefine the action of the address-transmission gate $\hat{U}^{(d)}$, which was the most costly element of our long-range protocol due to the requirement of highly nonlocal interactions between carriers.

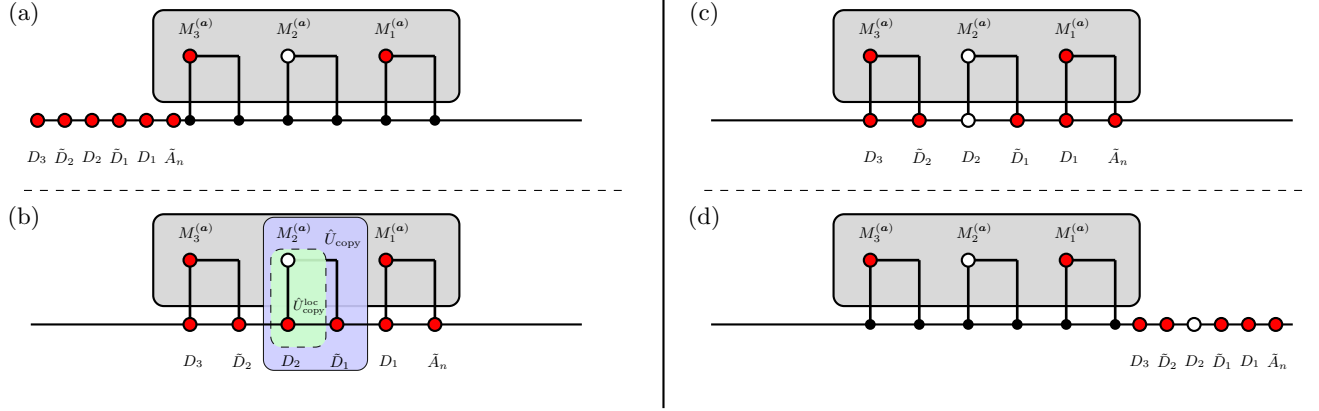
\begin{figure*}[ht]
    \centering
    \begin{tikzpicture}[x=1cm,y=1cm, scale=0.9]
  \def\w{8}
  \def\h{1.5}
  \def\dr{3pt}
  \def\dist{0.3}

  % --- Primo disegno originale ---
  \begin{scope}[shift={(-2,2)}, scale=0.9, transform shape]
    % Rettangolo
    \draw[rounded corners, thick, fill=gray!30] (1,0) rectangle ({\w-2},\h);

    % Linee verticali dai pallini sopra a quelli sotto
    \foreach \i in {1,...,3} {
      \pgfmathsetmacro{\x}{\i/5*\w}
      \draw[line width=1pt] (\x, \h/2) -- (\x, -\dist);
    }

      % Linee verticali backup 
    \foreach \i in {1,...,3} {
      \pgfmathsetmacro{\x}{\i/5*\w}
      \draw[line width=1pt] ({\x+0.8}, \h/2) -- ({\x+0.8}, -\dist);
    }

       % Linee spesse pallini superiori
    \foreach \j in {1,3,5} {
      \pgfmathsetmacro{\xStart}{(\j+1)/5*(\w/2)}
      \pgfmathsetmacro{\xEnd}{(\j+2)/5*(\w/2)}
      \draw[line width=1pt] (\xStart, {\h/2}) -- (\xEnd, {\h/2});
    }
    
    % Pallini sopra
    \foreach[count=\k from 1] \col in {red, white, red} {
        \pgfmathsetmacro{\x}{\k/5*\w}  % \n non definito, quindi uso 6 fisso coerente con sopra
        \filldraw[fill=\col, draw=black, thick] (\x, {\h/2}) circle [radius=\dr];
      }

    % Linea sotto
    \draw[thick] (-1,-\dist) -- (\w,-\dist);

   % Pallini sotto
    \foreach[count=\k from 1] \col in {red,red,red,red,red,red} {
        \pgfmathsetmacro{\x}{(\k-3)/9*(\w/2)}
        \pgfmathsetmacro{\y}{-\dist}
        \filldraw[fill=\col, draw=black, thick] (\x, \y) circle [radius=\dr];
      }

  % Pallini neri (spazi vuoti)
    \foreach[count=\k from 1] \col in {black,black,black,black,black,black} {
        \pgfmathsetmacro{\x}{(\k+1)/5*(\w/2)}
        \pgfmathsetmacro{\y}{-\dist}
        \filldraw[fill=\col, draw=black, thick] (\x, \y) circle [radius=\dr-1];
      }

     % Label pallini sopra
    \node at (1.7,1.2) {{\scalebox{0.8}{$M_{3}^{(\boldsymbol{a})}$}}};

    \node at (3.3,1.2) {{\scalebox{0.8}{$M_{2}^{(\boldsymbol{a})}$}}};
    
    \node at (4.9,1.2) {{\scalebox{0.8}{$M_{1}^{(\boldsymbol{a})}$}}};

    % Label pallini inferiori
      \node at (-0.9,-0.9) {{\scalebox{0.8}{$D_3$}}};

     \node at (-0.45,-0.9) {{\scalebox{0.8}{$\tilde{D}_2$}}};

    \node at (0,-0.9) {{\scalebox{0.8}{$D_2$}}};

    \node at (0.45,-0.9) {{\scalebox{0.8}{$\tilde{D}_1$}}};

    \node at (0.9,-0.9) {{\scalebox{0.8}{$D_1$}}};

    \node at (1.35,-0.9) {{\scalebox{0.8}{$\tilde{A}_n$}}};
    
  \end{scope}

\draw[dashed] (-3,{\h-0.8})--({\w-2.2},{\h-0.8});

% Secondo disegno prima colonna
  \begin{scope}[shift={(-2,-1)}, scale=0.9, transform shape]
    % Rettangolo
    \draw[rounded corners, thick, fill=gray!30] (1,0) rectangle ({\w-2},\h);

% Rettangolo gate U_copy
   \node[draw, rounded corners, fill=blue!20, minimum width=17mm, minimum height=25mm, opacity=0.9] at (3.65,0.2) {};

      % Linee verticali backup 
    \foreach \i in {1,...,3} {
      \pgfmathsetmacro{\x}{\i/5*\w}
      \draw[line width=1pt] ({\x+0.8}, \h/2) -- ({\x+0.8}, -\dist);
    }

       % Linee spesse pallini superiori
    \foreach \j in {1,3,5} {
      \pgfmathsetmacro{\xStart}{(\j+1)/5*(\w/2)}
      \pgfmathsetmacro{\xEnd}{(\j+2)/5*(\w/2)}
      \draw[line width=1pt] (\xStart, {\h/2}) -- (\xEnd, {\h/2});
    }

% Rettangolo gate U_copy^loc
   \node[draw, dashed, rounded corners, fill=green!20, minimum width=9mm, minimum height=16mm, opacity=0.9] at (3.37,0.13) {};

 % Linee verticali dai pallini sopra a quelli sotto
    \foreach \i in {1,...,3} {
      \pgfmathsetmacro{\x}{\i/5*\w}
      \draw[line width=1pt] (\x, \h/2) -- (\x, -\dist);
    }

    % Pallini sopra
    \foreach[count=\k from 1] \col in {red, white, red} {
        \pgfmathsetmacro{\x}{\k/5*\w}  % \n non definito, quindi uso 6 fisso coerente con sopra
        \filldraw[fill=\col, draw=black, thick] (\x, {\h/2}) circle [radius=\dr];
      }

    % Linea sotto
    \draw[thick] (-1,-\dist) -- (\w,-\dist);

   % Pallini sotto
    \foreach[count=\k from 1] \col in { red, red,red,red,red, red} {
        \pgfmathsetmacro{\x}{(\k+1)/5*(\w/2)}
        \pgfmathsetmacro{\y}{-\dist}
        \filldraw[fill=\col, draw=black, thick] (\x, \y) circle [radius=\dr];
      }

   % Label pallini sopra
    \node at (1.7,1.2) {{\scalebox{0.8}{$M_{3}^{(\boldsymbol{a})}$}}};

    \node at (3.3,1.2) {{\scalebox{0.8}{$M_{2}^{(\boldsymbol{a})}$}}};
    
    \node at (4.9,1.2) {{\scalebox{0.8}{$M_{1}^{(\boldsymbol{a})}$}}};

    % Label pallini inferiori

    \node at (1.65,-0.9) {{\scalebox{0.8}{$D_3$}}};

    \node at (2.45,-0.9) {{\scalebox{0.8}{$\tilde{D}_2$}}};

    \node at (3.25,-0.9) {{\scalebox{0.8}{$D_2$}}};

    \node at (4.05,-0.9) {{\scalebox{0.8}{$\tilde{D}_1$}}};

    \node at (4.85,-0.9) {{\scalebox{0.8}{$D_1$}}};

    \node at (5.65,-0.9) {{\scalebox{0.8}{$\tilde{A}_n$}}};
    
%label gate

\node at (4.1,1.1) {{\scalebox{0.8}{$\hat{U}_{\text{copy}}$}}};

\node at (3.55,0) {{\scalebox{0.7}{$\hat{U}_{\text{copy}}^{\text{loc}}$}}};

  \end{scope}

% Linea separazione
\draw[thick](6.4,4)--(6.4,-2.5);

%terzo disegno prima colonna
   \begin{scope}[shift={(8,2)}, scale=0.9, transform shape]
       % Rettangolo
    \draw[rounded corners, thick, fill=gray!30] (1,0) rectangle ({\w-2},\h);

      % Linee verticali backup 
    \foreach \i in {1,...,3} {
      \pgfmathsetmacro{\x}{\i/5*\w}
      \draw[line width=1pt] ({\x+0.8}, \h/2) -- ({\x+0.8}, -\dist);
    }

       % Linee spesse pallini superiori
    \foreach \j in {1,3,5} {
      \pgfmathsetmacro{\xStart}{(\j+1)/5*(\w/2)}
      \pgfmathsetmacro{\xEnd}{(\j+2)/5*(\w/2)}
      \draw[line width=1pt] (\xStart, {\h/2}) -- (\xEnd, {\h/2});
    }

 % Linee verticali dai pallini sopra a quelli sotto
    \foreach \i in {1,...,3} {
      \pgfmathsetmacro{\x}{\i/5*\w}
      \draw[line width=1pt] (\x, \h/2) -- (\x, -\dist);
    }

    % Pallini sopra
    \foreach[count=\k from 1] \col in {red, white, red} {
        \pgfmathsetmacro{\x}{\k/5*\w}  % \n non definito, quindi uso 6 fisso coerente con sopra
        \filldraw[fill=\col, draw=black, thick] (\x, {\h/2}) circle [radius=\dr];
      }

    % Linea sotto
    \draw[thick] (-1,-\dist) -- (\w+0.5,-\dist);

   % Pallini sotto
    \foreach[count=\k from 1] \col in { red, red,white,red,red, red} {
        \pgfmathsetmacro{\x}{(\k+1)/5*(\w/2)}
        \pgfmathsetmacro{\y}{-\dist}
        \filldraw[fill=\col, draw=black, thick] (\x, \y) circle [radius=\dr];
      }

   % Label pallini sopra
    \node at (1.7,1.2) {{\scalebox{0.8}{$M_{3}^{(\boldsymbol{a})}$}}};

    \node at (3.3,1.2) {{\scalebox{0.8}{$M_{2}^{(\boldsymbol{a})}$}}};
    
    \node at (4.9,1.2) {{\scalebox{0.8}{$M_{1}^{(\boldsymbol{a})}$}}};

    % Label pallini inferiori

    \node at (1.65,-0.9) {{\scalebox{0.8}{$D_3$}}};

    \node at (2.45,-0.9) {{\scalebox{0.8}{$\tilde{D}_2$}}};

    \node at (3.25,-0.9) {{\scalebox{0.8}{$D_2$}}};

    \node at (4.05,-0.9) {{\scalebox{0.8}{$\tilde{D}_1$}}};

    \node at (4.85,-0.9) {{\scalebox{0.8}{$D_1$}}};

    \node at (5.65,-0.9) {{\scalebox{0.8}{$\tilde{A}_n$}}};
     
\end{scope}

%secondo disegno seconda colonna
\begin{scope}[shift={(8,-1)}, scale=0.9, transform shape]
    % Rettangolo
    \draw[rounded corners, thick, fill=gray!30] (1,0) rectangle ({\w-2},\h);

    % Linee verticali dai pallini sopra a quelli sotto
    \foreach \i in {1,...,3} {
      \pgfmathsetmacro{\x}{\i/5*\w}
      \draw[line width=1pt] (\x, \h/2) -- (\x, -\dist);
    }

      % Linee verticali backup 
    \foreach \i in {1,...,3} {
      \pgfmathsetmacro{\x}{\i/5*\w}
      \draw[line width=1pt] ({\x+0.8}, \h/2) -- ({\x+0.8}, -\dist);
    }

       % Linee spesse pallini superiori
    \foreach \j in {1,3,5} {
      \pgfmathsetmacro{\xStart}{(\j+1)/5*(\w/2)}
      \pgfmathsetmacro{\xEnd}{(\j+2)/5*(\w/2)}
      \draw[line width=1pt] (\xStart, {\h/2}) -- (\xEnd, {\h/2});
    }
    
    % Pallini sopra
    \foreach[count=\k from 1] \col in {red, white, red} {
        \pgfmathsetmacro{\x}{\k/5*\w}  % \n non definito, quindi uso 6 fisso coerente con sopra
        \filldraw[fill=\col, draw=black, thick] (\x, {\h/2}) circle [radius=\dr];
      }

    % Linea sotto
    \draw[thick] (-1,-\dist) -- (\w+0.5,-\dist);

   % Pallini sotto
    \foreach[count=\k from 1] \col in {red,red,white,red,red,red} {
        \pgfmathsetmacro{\x}{(\k+12.5)/9*(\w/2)}
        \pgfmathsetmacro{\y}{-\dist}
        \filldraw[fill=\col, draw=black, thick] (\x, \y) circle [radius=\dr];
      }

  % Pallini neri (spazi vuoti)
    \foreach[count=\k from 1] \col in {black,black,black,black,black,black} {
        \pgfmathsetmacro{\x}{(\k+1)/5*(\w/2)}
        \pgfmathsetmacro{\y}{-\dist}
        \filldraw[fill=\col, draw=black, thick] (\x, \y) circle [radius=\dr-1];
      }

     % Label pallini sopra
    \node at (1.7,1.2) {{\scalebox{0.8}{$M_{3}^{(\boldsymbol{a})}$}}};

    \node at (3.3,1.2) {{\scalebox{0.8}{$M_{2}^{(\boldsymbol{a})}$}}};
    
    \node at (4.9,1.2) {{\scalebox{0.8}{$M_{1}^{(\boldsymbol{a})}$}}};
   
    % Label pallini inferiori
    \node at (6,-0.9) {{\scalebox{0.8}{$D_3$}}};

    \node at (6.45,-0.9) {{\scalebox{0.8}{$\tilde{D}_2$}}};

    \node at (6.9,-0.9) {{\scalebox{0.8}{$D_2$}}};

    \node at (7.35,-0.9) {{\scalebox{0.8}{$\tilde{D}_1$}}};

    \node at (7.8,-0.9) {{\scalebox{0.8}{$D_1$}}};

    \node at (8.25,-0.9) {{\scalebox{0.8}{$\tilde{A}_n$}}};
  \end{scope}

\draw[dashed] (7,{\h-0.8})--({\w+7.8},{\h-0.8});

\node at (-3,3.3){(a)};
\node at (-3,0.3){(b)};
\node at (7,3.3){(c)};
\node at (7,0.3){(d)};

\end{tikzpicture}
    \caption{Steps involved in the message copy process in the backup variant of the qRAM protocol. (a) The data particles, their corresponding backup particles, and the final address backup arrive at the memory cell; (b) since the backup particles $\tilde{D}_{i-1}$ are always red (the control particle necessary to write the message on $D_1$ is $\tilde A_n$), they serve as control particles for activating the local unitary gate defined in Eq.~(\ref{backup_copy}); (c) the message is copied into the data walkers; (d) finally the particles exit the memory cell and enter the inverse routing phase to the final output port.}
    \label{copy_backup}
\end{figure*}

\subsection{Address-transmission gate decomposition} 
The constant multiplicative overhead given by the doubling of information carriers allows us to replace the nonlocal multi-controlled unitary $\hat U^{(d)}$ with the combined action of two simpler short-range operations. More precisely, we replace the action of the gate $\hat{U}^{(d)}$ applying (i)~an initialization operation $\hat U_{\mathrm{in}}$ acting on the $d$-th address walker $A_d$ and its backup $\tilde A_d$ a single time, followed by (ii)~the sequential application of the unitary gate $\hat U_B$ on the walkers following $A_d$, used to propagate the stored path information stepwise along the train up to $\tilde D_m$ (see Fig.~\ref{backupscheme}b).
{These gates can be understood as controlled operations acting on two and three walkers, respectively, conditioned on the color of the control walker. More precisely, $\hat U_{in}$ corresponds to a $\hat {CX}$ gate conditioned on the control walker being in its red internal state, while $\hat{U}_{B}$ to a $\hat {CXX}$ gate conditioned on the control walker being in its blue internal state.

Concretely, in analogy with the definition of $\hat{U}^{(d)}$, we introduce the set of projectors $\{\hat \Pi_{a_d}^{(A_d)}\}$ onto local subspaces for the address walker $A_d$ and write the initialization gate as 
\begin{equation}
    \hat{U}_{in}(\tilde A_d|A_d):=\sum_{a_d}\hat{V}_{a_d}^{(\tilde A_d)}\otimes \hat{\Pi}_{a_d}^{(A_d)}.
\end{equation}
Here, $\hat{V}_{a_d}^{(\tilde A_d)}$ is a unitary acting on the companion backup walker $\tilde A_d$ following the address walker $A_d$. The form of this operator is given by 
\begin{equation*}
    \hat{V}_{a_d=R}^{(\tilde A_d)}\equiv \hat{X}_{\tilde A_d},\quad \hat{V}_{a_d=B}=\hat{V}_{a_d=\varnothing}\equiv \hat{I}_{\tilde A_d}
\end{equation*}
that is, it has the same form as in Eq.~(\ref{targetOp}) with $T_d\equiv\{\tilde A_d\}$.
Denoting with $P_d=\{\tilde A_{d-1},A_{d-1},\dots,\tilde A_1,A_1\}$ and with $F_d=\boldsymbol{D}\cup\{\tilde A_n,A_n,\dots,\tilde A_{d+1},A_{d+1}\}$ the sets of walkers respectively preceding the $d$-th address walker and following the $d$-th backup walker, we can consider a generic state of the form
\begin{multline*}
    \ket{\tilde d_m}_{\tilde D_m}\otimes\dots\otimes\ket{R}_{\tilde A_d}\otimes\ket{a_d}_{A_d}\otimes\dots\otimes \ket{a_1}_{A_1}\\\equiv\ket{\varphi}_{F_d}\otimes \ket{R}_{\tilde A_d}\otimes \ket{a_d}_{A_d}\otimes \ket{\psi}_{P_d}
\end{multline*}
The action of $\hat{U}_{in}$ can then be expressed as 
\begin{equation*}
\begin{split}
    \hat{U}_{in}(\tilde A_d|A_d)&\left(\ket{\varphi}_{F_d}\otimes \ket{R}_{\tilde A_d}\otimes \ket{a_d}_{A_d}\otimes \ket{\psi}_{P_d}\right)\\ &=
    \ket{\varphi}_{F_d}\otimes \hat{U}_{in}\left(\ket{R}_{\tilde A_d}\otimes \ket{a_d}_{A_d}\right)\otimes \ket{\psi}_{P_d}\\ &=
    \ket{\varphi}_{F_d}\otimes \left(\hat{X}_{\tilde A_d}\ket{R}_{\tilde A_d}\right)\otimes \ket{a_d}_{A_d}\otimes \ket{\psi}_{P_d}.
\end{split}
\end{equation*}
i.e., a $\hat{CX}$ gate, having as control $A_d$ and target $\tilde A_d$, which flips the backup state only when the control is in $\ket{R}$, thereby copying the relevant routing information into the local backup register. Otherwise, it acts as the identity on $\tilde A_d$.

Once the path information is transferred to the backup $\tilde A_d$, the recursive transmission begins: the propagation block $\hat U_B$ acts on the triplet formed by the current backup $\tilde A_d$, used as control, and the next carrier–backup pair $(A_{d+1},\tilde A_{d+1})$, which are the targets. In general, this gate acts on each triplet of backup, subsequent information carrier, and corresponding backup $(\tilde W_{i+1},W_{i+1},\tilde W_i), i\in\{1,\dots,n+m\}$ present in the walker train. 

Introducing the set of projectors $\{\hat \Pi_{\tilde w_i}^{(\tilde W_i)}\}$ onto the subspaces of the $i$-th backup walker $\tilde W_i$, we can write the block operator for a generic triplet in the usual form as
%\begin{equation*}
%\hat U_B=\sum_{w\in\{R,B,\varnothing\}}\bigl(\hat X_{W_{i+1}}\otimes\hat X_{\tilde W_{i+1}}\bigr)^{\,2-\delta_{B;w}}\otimes\ket{w}_{\tilde W_i}\!\bra{w},
%\end{equation*}
\begin{equation}
    \hat{U}_B(\tilde W_{i+1},W_{i+1}|\tilde W_i):=\sum_{\tilde w_i}\hat{V}_{\tilde w_i}^{(\tilde W_{i+1},W_{i+1})}\otimes \hat{\Pi}_{\tilde w_i}^{(\tilde W_i)}
\end{equation}
where now the unitary $\hat{V}_{\tilde w_i}^{(\tilde W_{i+1},W_{i+1})}$ acting on the targets $(\tilde W_{i+1},W_{i+1})$ has the form 
\begin{flalign*}
    &\hat{V}_{\tilde w_i=B}^{(\tilde W_{i+1},W_{i+1})}\equiv \hat{X}_{\tilde W_{i+1}}\otimes \hat{X}_{W_{i+1}},\\&\hat{V}_{\tilde w_i=R}^{(\tilde W_{i+1},W_{i+1})} = \hat{V}_{\tilde w_i=\varnothing}^{(\tilde W_{i+1},W_{i+1})} \equiv \hat{I}_{\tilde W_{i+1}}\otimes \hat{I}_{W_{i+1}}.
\end{flalign*}

In other words, whenever the control backup is in the state $\ket{B}_{\tilde W_i}$ the gate implements a Pauli $\hat{X}$ on both $W_{i+1}$ and $\tilde W_{i+1}$, effectively mapping $\ket{R}\mapsto\ket{B}$. Instead, a walker in the vacuum state $\ket{\varnothing}$ is left unchanged by this operation. After applying $\hat U_B$ once, the information stored in $\tilde W_i$ is effectively passed to $W_{i+1}$ and $\tilde W_{i+1}$, and the required logical action on the $(i+1)$-th carrier is made effective. By moving the entire train forward by two positions, the newly updated backup $\tilde W_{i+1}$ becomes the control for the subsequent block application. Repeating this elementary step iteratively on triplets of walkers $(\tilde W_{d+k+1},W_{d+k+1},\tilde W_{d+k})$ for $k=0,1,\dots,(n-d)+m-1$, we propagate the original path information from the initialized backup $\tilde A_d$ to all the following walkers. After $t_B=n+m-d$ iterations, the net effect on the non-backup walkers is identical to that of (the nonlocal) gate $\hat U^{(d)}=\hat U^{(d)}(T_d|A_d)$} in the \gdr{basic} qRAM variant, yet every primitive operation involves at most three carriers and only short-range interactions. A complete example of how these two gates works together is provided in Fig.~\ref{backupscheme}(c).

The backup variant trades a constant multiplicative overhead in the number of walkers for a dramatic simplification of the control structure: the global 
multi-target transformation is replaced by a sequence of local unitaries having only two target walkers. This modular construction reduces the interaction range and the many-body control requirements, making the protocol particularly suitable for architectures with nearest-neighbor couplings or with limited multi-qubit gate availability. Moreover, as shown in Sec.~\ref{DepthReduction}, this variant allows for a reduction in circuit depth through a suitable pipeline scheduling of the operations $\hat{U}_{in}$ and $\hat{U}_{B}$, achieving an overall $\mathcal{O}(n+m)$ scaling, which corresponds to a quadratic speedup over the \gdr{basic} protocol.

\subsection{Memory operations}\label{SecMemoStandard}

The backup variant can also be employed to reduce the complexity requirements for the operations needed at the memory cells. Indeed, the presence of a backup walker next each data walker can be exploited to execute the message copy operations using only local operations between a data walker $D_i$, the backup of the preceding data walker $\tilde D_{i-1}$, and the corresponding memory cell walker $M_i^{(\boldsymbol a)}$ (see Fig.~\ref{copy_backup}). 
The backup walker $\tilde D_{i-1}$ can in fact be used as a control walker for a local unitary operation $\hat{U}_{copy}(D_i,M_i^{(\boldsymbol{a})}|\tilde D_{i-1})$ having as targets the $i$-th data walker $D_i$ and the message $M_i^{(\boldsymbol{a})}$ containing the classical information bit $b_i^{(\boldsymbol{a})}$ 
Thus, the message is copied only if the walker $\tilde D_{i-1}$ is present. In this way, the message contained in the cell is copied only if the data register reaches such cell, avoiding unwanted activation of cells not addressed by the routed walkers. 

Formally, the unitary operation takes the form 
\begin{multline}\label{backup_copy}
    \hat{U}_{copy}(D_i,M_i^{(\boldsymbol{a})}|\tilde D_{i-1}):=\\\ket{R}_{\tilde D_{i-1}}\bra{R}\otimes \hat{U}^{loc}_{copy}(D_i,M_i^{(\boldsymbol{a})})+\\\left(\ket{\varnothing}_{\tilde D_{i-1}}\bra{\varnothing}+\ket{B}_{\tilde D_{i-1}}\bra{B}\right)\otimes\left( \hat{I}_{D_i}\otimes \hat{I}_{M_i^{\boldsymbol{(a)}}}\right)
\end{multline}
where the operation $\hat{U}_{copy}^{loc}(D_i,M_i^{(\boldsymbol{a})})$ is the one defined in Eq.~(\ref{locCopy}).
{Notice that by design the routing procedure always ensures that the last backup walker $\tilde A_n$ reaches the addressed memory cells. Thus, it can be used to control the copy operation on the first data walker $D_1$.
This guarantees that all data walkers are manipulated by strictly local operations applied at the targeted memory cells.}

Overall, the backup-assisted COPY implements the same logical transformation as the global and activation-based schemes discussed in Sec.~\ref{memoryOpStandard}, but replaces their multi-body control structure with a purely local, nearest-neighbor mechanism. This preserves the correctness of the memory-access protocol while significantly easing its physical implementation.

\begin{figure*}[ht]
    \centering
\begin{tikzpicture}[scale=0.75, transform shape,
  level 1/.style={level distance=2cm, sibling distance=4cm},
  level 2/.style={level distance=1cm, sibling distance=1cm},
  grow=east, % Cresce verso destra
  edge from parent/.style={draw, thick}, % Stile dei rami
  every node/.style={rectangle, draw, minimum size=6mm,fill=cyan!30}, % Nodi interni
  leaf/.style={rectangle, draw, fill=gray!30, minimum size=6mm}, % Foglie grigie
  block/.style={rectangle, draw=none,dashed,fill=none, minimum width=1.8cm, minimum height=5mm, rotate around={0:(0,0)}}, % Blocco blu chiaro
  gateX/.style={rectangle, draw, fill=green!20, minimum size=5mm, rotate around={0:(0,0)}} % Blocco quadrato rosso
]

\begin{scope}[shift={(0,0)}]

% Nodo radice
\node (root1) {$\hat{S}$}
  child {
    % Primo nodo intermedio (verso l'alto)
    node {$\hat{S}$}
    child { 
      %node[leaf] {$\boldsymbol{x}_3$} % Foglia sinistra-sinistra (grigia)
      edge from parent {}
    }
    child { 
      %node[leaf] {$\boldsymbol{x}_2$} % Foglia sinistra-destra (grigia)
      edge from parent{
%node[midway,xshift=-1cm,gateX, yshift=-0.5cm,rotate=27] {$\hat{X}$} % Blocco rosso ruotato
      }
    }
    edge from parent {
      node[midway,xshift=0cm, yshift=-0cm,  block, rotate=-45] {} % Blocco blu ruotato
    }
  }
  child {
    % Secondo nodo intermedio (verso l'alto)
    node {$\hat{S}$}
    child { 
      %node[leaf] {$\boldsymbol{x}_1$} % Foglia destra-sinistra (grigia)
      %edge from parent {}
    }
    child { 
      %node[leaf] {$\boldsymbol{x}_0$} % Foglia destra-destra (grigia)
      edge from parent{ 
      %node[midway,xshift=-1cm, yshift=-0.5cm,gateX, rotate=27] {$\hat{X}$} % Blocco rosso ruotato
      }
    }
    edge from parent {
      %node[midway, xshift=-1.3cm,yshift=-1.3cm,gateX, rotate=45] {$\hat{X}$} % Blocco rosso ruotato
      node[midway, xshift=0cm,yshift=0cm, block, rotate=45] {} % Blocco blu ruotato
    }
  };

% Ramo in ingresso al nodo radice
\draw (-7, 0) -- (root1) node(Ugateroot)[xshift=-1.37cm, block, rotate=0] {};
% Linea di ingresso al nodo radice
%\draw[thick] (root2) -- (20,0) node[midway, block] {$\hat{U}^{(1)}$};

%\node[block, rotate = 27, draw = none, minimum width = 0.5cm](b1) at (3,2.5){};
%\draw[thick] (b1.south west) -- (b1.south east); % lato basso
%\draw[thick] (b1.south west) -- (b1.north west);         % lato destro
%\draw[thick] (b1.north east) -- (b1.north west); % lato alto

%\node[block, rotate = -27, draw = none, minimum width = 0.5cm](b2) at (3,1.5){};
%\draw[thick] (b2.south west) -- (b2.south east); % lato basso
%\draw[thick] (b2.south west) -- (b2.north west);         % lato destro
%\draw[thick] (b2.north east) -- (b2.north west); % lato alto

%\node[block, rotate = 27, draw = none, minimum width = 0.5cm](b3) at (3,-1.5){};
%\draw[thick] (b3.south west) -- (b3.south east); % lato basso
%\draw[thick] (b3.south west) -- (b3.north west);         % lato destro
%\draw[thick] (b3.north east) -- (b3.north west); % lato alto

%\node[block, rotate = -27, draw = none, minimum width = 0.5cm](b4) at (3,-2.5){};
%\draw[thick] (b4.south west) -- (b4.south east); % lato basso
%\draw[thick] (b4.south west) -- (b4.north west);         % lato destro
%\draw[thick] (b4.north east) -- (b4.north west); % lato alto

%nodes before root node

\draw[fill=black] (-0.7,0) circle (2pt)node[above=3pt, fill = none, draw=none]{\scriptsize$4$};
\draw[fill=black] (-1.15,0) circle (2pt)node[above=3pt, fill = none, draw=none]{\scriptsize$3$};
\draw[fill=black] (-1.6,0) circle (2pt)node[above=3pt, fill = none, draw=none]{\scriptsize$2$};
\draw[fill=black] (-2.05,0) circle (2pt)node[above=3pt, fill = none, draw=none]{\scriptsize$1$};
\node[dashed, fill=none, minimum width=50pt, minimum height=20pt]at (-1.375,0.2){};

\draw[thick, fill=red] (-2.5,0) circle (3pt) node[below=3pt, fill = none, draw=none]{\scriptsize$A_1$};
\draw[thick, fill=red] (-2.95,0) circle (3pt) node[below=3pt, fill = none, draw=none]{\scriptsize$\tilde{A}_1$};
\draw[thick, fill=red] (-3.4,0) circle (3pt)node[below=3pt, fill = none, draw=none]{\scriptsize$A_{2}$};
\draw[thick, fill=red] (-3.85,0) circle (3pt)node[below=3pt, fill = none, draw=none]{\scriptsize$\tilde{A}_{2}$};
\draw[thick, fill=white] (-4.3,0) circle (3pt)node[below=3pt, fill = none, draw=none]{\scriptsize$A_{3}$};
\draw[thick, fill=red] (-4.75,0) circle (3pt)node[below=3pt, fill = none, draw=none]{\scriptsize$\tilde{A}_{3}$};
\draw[thick, fill=red] (-5.2,0) circle (3pt)node[below=3pt, fill = none, draw=none]{\scriptsize$A_{4}$};
\draw[thick, fill=red] (-5.65,0) circle (3pt)node[below=3pt, fill = none, draw=none]{\scriptsize$\tilde{A}_{4}$};
\draw[thick, fill=red] (-6.1,0) circle (3pt)node[below=3pt, fill = none, draw=none]{\scriptsize$A_{5}$};
\draw[thick, fill=red] (-6.55,0) circle (3pt)node[below=3pt, fill = none, draw=none]{\scriptsize$\tilde{A}_{5}$};

%nodes first level upper branch

\draw[fill=black] (1.45,1.45) circle (2pt);
\draw[fill=black] (1.15,1.15) circle (2pt);
\draw[fill=black] (0.85,0.85) circle (2pt);
\draw[fill=black] (0.55,0.55) circle (2pt);
%\node[dashed, fill=none, minimum width=50pt, minimum height=20pt, rotate=45, transform shape]at (1,1){};

%nodes first level lower branch

\draw[ fill=black] (1.45,-1.45) circle (2pt);
\draw[fill=black] (1.15,-1.15) circle (2pt);
\draw[fill=black] (0.85,-0.85) circle (2pt);
\draw[fill=black] (0.55,-0.55) circle (2pt);

\node[draw=none, fill=none] at (-1.1,-0.8){\scalebox{1.2}{$(1,1)$}};

\node[draw=none, fill=none, rotate=-45] at (0.5,-1.5){\scalebox{1.2}{$(1,2)$}};

\node[draw=none, fill=none, rotate=45] at (0.5,1.5){\scalebox{1.2}{$(2,2)$}};

%in-out arrows

\draw[->, >={Stealth[length=6pt]}, thick] (-4.5,0.5) -- (-3,0.5)node[fill=none,draw=none,midway,above] {};

\node[draw=none, fill=none, align=left] at (-4.5,-2)
{\large Step 1: initialization of the walkers\\ \large at the root node };

\node[draw=none, fill=none] at (-7,2){\Large $(a)$};

\end{scope}

%%%%%% SECONDO DISEGNO%%%%%%%%%%%
\begin{scope}[shift={(0,-6)}]

% Nodo radice
\node (root1) {$\hat{S}$}
  child {
    % Primo nodo intermedio (verso l'alto)
    node {$\hat{S}$}
    child { 
      %node[leaf] {$\boldsymbol{x}_3$} % Foglia sinistra-sinistra (grigia)
      edge from parent {}
    }
    child { 
      %node[leaf] {$\boldsymbol{x}_2$} % Foglia sinistra-destra (grigia)
      edge from parent{
%node[midway,xshift=-1cm,gateX, yshift=-0.5cm,rotate=27] {$\hat{X}$} % Blocco rosso ruotato
      }
    }
    edge from parent {
      node[midway,xshift=0cm, yshift=-0cm,  block, rotate=-45] {} % Blocco blu ruotato
    }
  }
  child {
    % Secondo nodo intermedio (verso l'alto)
    node {$\hat{S}$}
    child { 
      %node[leaf] {$\boldsymbol{x}_1$} % Foglia destra-sinistra (grigia)
      %edge from parent {}
    }
    child { 
      %node[leaf] {$\boldsymbol{x}_0$} % Foglia destra-destra (grigia)
      edge from parent{ 
      %node[midway,xshift=-1cm, yshift=-0.5cm,gateX, rotate=27] {$\hat{X}$} % Blocco rosso ruotato
      }
    }
    edge from parent {
      %node[midway, xshift=-1.3cm,yshift=-1.3cm,gateX, rotate=45] {$\hat{X}$} % Blocco rosso ruotato
      node[midway, xshift=0cm,yshift=0cm, block, rotate=45] {} % Blocco blu ruotato
    }
  };

% Ramo in ingresso al nodo radice
\draw (-7, 0) -- (root1) node(Ugateroot)[xshift=-1.37cm, block, rotate=0] {};
% Linea di ingresso al nodo radice
%\draw[thick] (root2) -- (20,0) node[midway, block] {$\hat{U}^{(1)}$};

%\node[block, rotate = 27, draw = none, minimum width = 0.5cm](b1) at (3,2.5){};
%\draw[thick] (b1.south west) -- (b1.south east); % lato basso
%\draw[thick] (b1.south west) -- (b1.north west);         % lato destro
%\draw[thick] (b1.north east) -- (b1.north west); % lato alto

%\node[block, rotate = -27, draw = none, minimum width = 0.5cm](b2) at (3,1.5){};
%\draw[thick] (b2.south west) -- (b2.south east); % lato basso
%\draw[thick] (b2.south west) -- (b2.north west);         % lato destro
%\draw[thick] (b2.north east) -- (b2.north west); % lato alto

%\node[block, rotate = 27, draw = none, minimum width = 0.5cm](b3) at (3,-1.5){};
%\draw[thick] (b3.south west) -- (b3.south east); % lato basso
%\draw[thick] (b3.south west) -- (b3.north west);         % lato destro
%\draw[thick] (b3.north east) -- (b3.north west); % lato alto

%\node[block, rotate = -27, draw = none, minimum width = 0.5cm](b4) at (3,-2.5){};
%\draw[thick] (b4.south west) -- (b4.south east); % lato basso
%\draw[thick] (b4.south west) -- (b4.north west);         % lato destro
%\draw[thick] (b4.north east) -- (b4.north west); % lato alto

%gates

\node[block,draw, thick,solid, minimum width = 0.8cm, fill=orange!20,label={[yshift=1mm]above:$\hat{U}_{in}$}] at (-1.37,0){};

%nodes before root node

\draw[fill=black] (-0.7,0) circle (2pt);
\draw[thick, fill=red] (-1.15,0) circle (3pt);
\draw[fill=blue,thick] (-1.6,0) circle (3pt);
\draw[fill=red,thick] (-2.05,0) circle (3pt);
\draw[thick, fill=red] (-2.5,0) circle (3pt);
\draw[thick, fill=white] (-2.95,0) circle (3pt);
\draw[thick, fill=red] (-3.4,0) circle (3pt);
\draw[thick, fill=red] (-3.85,0) circle (3pt);
\draw[thick, fill=red] (-4.3,0) circle (3pt);
\draw[thick, fill=red] (-4.75,0) circle (3pt);
\draw[thick, fill=red] (-5.2,0) circle (3pt);

%nodes first level upper branch

\draw[fill=black] (1.45,1.45) circle (2pt);
\draw[fill=black] (1.15,1.15) circle (2pt);
\draw[fill=black] (0.85,0.85) circle (2pt);
\draw[fill=black] (0.55,0.55) circle (2pt);

%nodes first level lower branch

\draw[ fill=black] (1.45,-1.45) circle (2pt);
\draw[fill=black] (1.15,-1.15) circle (2pt);
\draw[fill=black] (0.85,-0.85) circle (2pt);
\draw[fill=black] (0.55,-0.55) circle (2pt);

\node[draw=none, fill=none, align=left] at (-4.5,-2)
{\large Step 2: shift of two positions and \\ \large application of $\hat{U}_{in}$};

\node[draw=none, fill=none] at (-7,2){\Large $(b)$};
   
\end{scope}

%%%%%%%%%%% TERZO DISEGNO %%%%%%%%%

\begin{scope}[shift={(0,-12)}]

% Nodo radice
\node (root1) {$\hat{S}$}
  child {
    % Primo nodo intermedio (verso l'alto)
    node {$\hat{S}$}
    child { 
      %node[leaf] {$\boldsymbol{x}_3$} % Foglia sinistra-sinistra (grigia)
      edge from parent {}
    }
    child { 
      %node[leaf] {$\boldsymbol{x}_2$} % Foglia sinistra-destra (grigia)
      edge from parent{
%node[midway,xshift=-1cm,gateX, yshift=-0.5cm,rotate=27] {$\hat{X}$} % Blocco rosso ruotato
      }
    }
    edge from parent {
      node[midway,xshift=0cm, yshift=-0cm,  block, rotate=-45] {} % Blocco blu ruotato
    }
  }
  child {
    % Secondo nodo intermedio (verso l'alto)
    node {$\hat{S}$}
    child { 
      %node[leaf] {$\boldsymbol{x}_1$} % Foglia destra-sinistra (grigia)
      %edge from parent {}
    }
    child { 
      %node[leaf] {$\boldsymbol{x}_0$} % Foglia destra-destra (grigia)
      edge from parent{ 
      %node[midway,xshift=-1cm, yshift=-0.5cm,gateX, rotate=27] {$\hat{X}$} % Blocco rosso ruotato
      }
    }
    edge from parent {
      %node[midway, xshift=-1.3cm,yshift=-1.3cm,gateX, rotate=45] {$\hat{X}$} % Blocco rosso ruotato
      node[midway, xshift=0cm,yshift=0cm, block, rotate=45] {} % Blocco blu ruotato
    }
  };

% Ramo in ingresso al nodo radice
\draw (-7, 0) -- (root1) node(Ugateroot)[xshift=-1.37cm, block, rotate=0] {};
% Linea di ingresso al nodo radice
%\draw[thick] (root2) -- (20,0) node[midway, block] {$\hat{U}^{(1)}$};

%\node[block, rotate = 27, draw = none, minimum width = 0.5cm](b1) at (3,2.5){};
%\draw[thick] (b1.south west) -- (b1.south east); % lato basso
%\draw[thick] (b1.south west) -- (b1.north west);         % lato destro
%\draw[thick] (b1.north east) -- (b1.north west); % lato alto

%\node[block, rotate = -27, draw = none, minimum width = 0.5cm](b2) at (3,1.5){};
%\draw[thick] (b2.south west) -- (b2.south east); % lato basso
%\draw[thick] (b2.south west) -- (b2.north west);         % lato destro
%\draw[thick] (b2.north east) -- (b2.north west); % lato alto

%\node[block, rotate = 27, draw = none, minimum width = 0.5cm](b3) at (3,-1.5){};
%\draw[thick] (b3.south west) -- (b3.south east); % lato basso
%\draw[thick] (b3.south west) -- (b3.north west);         % lato destro
%\draw[thick] (b3.north east) -- (b3.north west); % lato alto

%\node[block, rotate = -27, draw = none, minimum width = 0.5cm](b4) at (3,-2.5){};
%\draw[thick] (b4.south west) -- (b4.south east); % lato basso
%\draw[thick] (b4.south west) -- (b4.north west);         % lato destro
%\draw[thick] (b4.north east) -- (b4.north west); % lato alto

%gates

\node[block,draw,thick,minimum width = 1.3cm, solid, fill=blue!20,label={[yshift=1mm]above:$\hat{U}_B$}] at (-1.15,0){};

%nodes before root node

\draw[fill=blue, thick] (-0.7,0) circle (3pt);
\draw[fill=blue,thick] (-1.15,0) circle (3pt);
\draw[fill=blue,thick] (-1.6,0) circle (3pt);
\draw[fill=white,thick] (-2.05,0) circle (3pt);
\draw[thick, fill=red] (-2.5,0) circle (3pt);
\draw[thick, fill=red] (-2.95,0) circle (3pt);
\draw[thick, fill=red] (-3.4,0) circle (3pt);
\draw[thick, fill=red] (-3.85,0) circle (3pt);
\draw[thick, fill=red] (-4.3,0) circle (3pt);

%nodes first level upper branch

\draw[fill=black] (1.45,1.45) circle (2pt);
\draw[fill=black] (1.15,1.15) circle (2pt);
\draw[fill=black] (0.85,0.85) circle (2pt);
\draw[fill=black] (0.55,0.55) circle (2pt);

%nodes first level lower branch

\draw[ fill=black] (1.45,-1.45) circle (2pt);
\draw[fill=black] (1.15,-1.15) circle (2pt);
\draw[fill=black] (0.85,-0.85) circle (2pt);
\draw[fill=red,thick] (0.55,-0.55) circle (3pt);

\node[draw=none, fill=none, align=left] at (-4.5,-2)
{\large Step 3: shift of two positions and \\ \large application of $\hat{U}_{B}$};

\node[draw=none, fill=none] at (-7,2){\Large $(c)$};

\end{scope}

%%%%%%%%% QUARTO DISEGNO %%%%%

\begin{scope}[shift={(12,0)}]

% Nodo radice
\node (root1) {$\hat{S}$}
  child {
    % Primo nodo intermedio (verso l'alto)
    node {$\hat{S}$}
    child { 
      %node[leaf] {$\boldsymbol{x}_3$} % Foglia sinistra-sinistra (grigia)
      edge from parent {}
    }
    child { 
      %node[leaf] {$\boldsymbol{x}_2$} % Foglia sinistra-destra (grigia)
      edge from parent{
%node[midway,xshift=-1cm,gateX, yshift=-0.5cm,rotate=27] {$\hat{X}$} % Blocco rosso ruotato
      }
    }
    edge from parent {
      node[midway,xshift=0cm, yshift=-0cm,  block, rotate=-45] {} % Blocco blu ruotato
    }
  }
  child {
    % Secondo nodo intermedio (verso l'alto)
    node {$\hat{S}$}
    child { 
      %node[leaf] {$\boldsymbol{x}_1$} % Foglia destra-sinistra (grigia)
      %edge from parent {}
    }
    child { 
      %node[leaf] {$\boldsymbol{x}_0$} % Foglia destra-destra (grigia)
      edge from parent{ 
      %node[midway,xshift=-1cm, yshift=-0.5cm,gateX, rotate=27] {$\hat{X}$} % Blocco rosso ruotato
      }
    }
    edge from parent {
      %node[midway, xshift=-1.3cm,yshift=-1.3cm,gateX, rotate=45] {$\hat{X}$} % Blocco rosso ruotato
      node[midway, xshift=0cm,yshift=0cm, block, rotate=45] {} % Blocco blu ruotato
    }
  };

% Ramo in ingresso al nodo radice
\draw (-7, 0) -- (root1) node(Ugateroot)[xshift=-1.37cm, block, rotate=0] {};

%\node[block, rotate = 27, draw = none, minimum width = 0.5cm](b1) at (3,2.5){};
%\draw[thick] (b1.south west) -- (b1.south east); % lato basso
%\draw[thick] (b1.south west) -- (b1.north west);         % lato destro
%\draw[thick] (b1.north east) -- (b1.north west); % lato alto

%\node[block, rotate = -27, draw = none, minimum width = 0.5cm](b2) at (3,1.5){};
%\draw[thick] (b2.south west) -- (b2.south east); % lato basso
%\draw[thick] (b2.south west) -- (b2.north west);         % lato destro
%\draw[thick] (b2.north east) -- (b2.north west); % lato alto

%\node[block, rotate = 27, draw = none, minimum width = 0.5cm](b3) at (3,-1.5){};
%\draw[thick] (b3.south west) -- (b3.south east); % lato basso
%\draw[thick] (b3.south west) -- (b3.north west);         % lato destro
%\draw[thick] (b3.north east) -- (b3.north west); % lato alto

%\node[block, rotate = -27, draw = none, minimum width = 0.5cm](b4) at (3,-2.5){};
%\draw[thick] (b4.south west) -- (b4.south east); % lato basso
%\draw[thick] (b4.south west) -- (b4.north west);         % lato destro
%\draw[thick] (b4.north east) -- (b4.north west); % lato alto

%gates

\node[block,draw,thick,minimum width = 1.3cm, solid, fill=blue!20,label={[yshift=1mm]above:$\hat{U}_B$}] at (-1.15,0){};

%nodes before root node

\draw[fill=blue,thick] (-0.7,0) circle (3pt);
\draw[fill=white,thick] (-1.15,0) circle (3pt);
\draw[fill=blue,thick] (-1.6,0) circle (3pt);
\draw[fill=red,thick] (-2.05,0) circle (3pt);
\draw[thick, fill=red] (-2.5,0) circle (3pt);
\draw[thick, fill=red] (-2.95,0) circle (3pt);
\draw[thick, fill=red] (-3.4,0) circle (3pt);

%nodes first level upper branch

\draw[fill=black] (1.45,1.45) circle (2pt);
\draw[fill=black] (1.15,1.15) circle (2pt);
\draw[fill=red,thick] (0.85,0.85) circle (3pt);
\draw[fill=red,thick] (0.55,0.55) circle (3pt);

\draw[ fill=black] (1.45,-1.45) circle (2pt);
\draw[fill=red,thick] (1.15,-1.15) circle (3pt);
\draw[fill=black] (0.85,-0.85) circle (2pt);
\draw[fill=black] (0.55,-0.55) circle (2pt);

\node[draw=none, fill=none, align=left] at (-4.5,-2)
{\large Step 4: shift of two positions and \\ \large application of $\hat{U}_{B}$};

\node[draw=none, fill=none] at (-7,2){\Large $(d)$};

\end{scope}

%%%%%%%%%% QUINTO DISEGNO %%%%%%%%

\begin{scope}[shift={(12,-6)}]

% Nodo radice
\node (root1) {$\hat{S}$}
  child {
    % Primo nodo intermedio (verso l'alto)
    node {$\hat{S}$}
    child { 
      %node[leaf] {$\boldsymbol{x}_3$} % Foglia sinistra-sinistra (grigia)
      edge from parent {}
    }
    child { 
      %node[leaf] {$\boldsymbol{x}_2$} % Foglia sinistra-destra (grigia)
      edge from parent{
%node[midway,xshift=-1cm,gateX, yshift=-0.5cm,rotate=27] {$\hat{X}$} % Blocco rosso ruotato
      }
    }
    edge from parent {
      node[midway,xshift=0cm, yshift=-0cm,  block, rotate=-45] {} % Blocco blu ruotato
    }
  }
  child {
    % Secondo nodo intermedio (verso l'alto)
    node {$\hat{S}$}
    child { 
      %node[leaf] {$\boldsymbol{x}_1$} % Foglia destra-sinistra (grigia)
      %edge from parent {}
    }
    child { 
      %node[leaf] {$\boldsymbol{x}_0$} % Foglia destra-destra (grigia)
      edge from parent{ 
      %node[midway,xshift=-1cm, yshift=-0.5cm,gateX, rotate=27] {$\hat{X}$} % Blocco rosso ruotato
      }
    }
    edge from parent {
      %node[midway, xshift=-1.3cm,yshift=-1.3cm,gateX, rotate=45] {$\hat{X}$} % Blocco rosso ruotato
      node[midway, xshift=0cm,yshift=0cm, block, rotate=45] {} % Blocco blu ruotato
    }
  };

% Ramo in ingresso al nodo radice
\draw (-7, 0) -- (root1) node(Ugateroot)[xshift=-1.37cm, block, rotate=0] {};
% Linea di ingresso al nodo radice
%\draw[thick] (root2) -- (20,0) node[midway, block] {$\hat{U}^{(1)}$};

%\node[block, rotate = 27, draw = none, minimum width = 0.5cm](b1) at (3,2.5){};
%\draw[thick] (b1.south west) -- (b1.south east); % lato basso
%\draw[thick] (b1.south west) -- (b1.north west);         % lato destro
%\draw[thick] (b1.north east) -- (b1.north west); % lato alto

%\node[block, rotate = -27, draw = none, minimum width = 0.5cm](b2) at (3,1.5){};
%\draw[thick] (b2.south west) -- (b2.south east); % lato basso
%\draw[thick] (b2.south west) -- (b2.north west);         % lato destro
%\draw[thick] (b2.north east) -- (b2.north west); % lato alto

%\node[block, rotate = 27, draw = none, minimum width = 0.5cm](b3) at (3,-1.5){};
%\draw[thick] (b3.south west) -- (b3.south east); % lato basso
%\draw[thick] (b3.south west) -- (b3.north west);         % lato destro
%\draw[thick] (b3.north east) -- (b3.north west); % lato alto

%\node[block, rotate = -27, draw = none, minimum width = 0.5cm](b4) at (3,-2.5){};
%\draw[thick] (b4.south west) -- (b4.south east); % lato basso
%\draw[thick] (b4.south west) -- (b4.north west);         % lato destro
%\draw[thick] (b4.north east) -- (b4.north west); % lato alto

%gates

%\node[block,draw, thick,solid, minimum width = 0.8cm, fill=orange!20] at (-1.37,0){};

\node[block,draw,thick,minimum width = 1.3cm, solid, fill=blue!20,label={[yshift=1mm]above:$\hat{U}_B$}] at (-1.15,0){};

\node[block,draw, thick,solid, minimum width = 0.8cm, fill=orange!20, rotate = 45,label={[yshift=1mm]above:$\hat{U}_{in}$}] at (1,1){};

%\node[block,draw,thick,minimum width = 1.3cm, solid, fill=blue!20, rotate = 45,label={[yshift=1mm]above:$\hat{U}_B$}] at (1.15,1.15){};

%nodes before root node

%\draw[thick] (-4.3,0) circle (3pt);
%\draw[thick] (-4,0) circle (3pt);
%\draw[thick] (-3.7,0) circle (3pt);

\draw[fill=blue, thick] (-0.7,0) circle (3pt);
\draw[fill=blue, thick] (-1.15,0) circle (3pt);
\draw[fill=blue,thick] (-1.6,0) circle (3pt);
\draw[fill=red,thick] (-2.05,0) circle (3pt);
\draw[thick, fill=red] (-2.5,0) circle (3pt);
%\draw[thick, fill=red] (-2.95,0) circle (3pt);
%\draw[thick, fill=red] (-3.4,0) circle (3pt);
%\draw[thick, fill=red] (-3.85,0) circle (3pt);
%\draw[thick, fill=white] (-4.3,0) circle (3pt);
%\draw[thick, fill=red] (-4.75,0) circle (3pt);
%\draw[thick, fill=red] (-5.2,0) circle (3pt);
%\draw[thick, fill=red] (-5.65,0) circle (3pt);
%\draw[thick, fill=red] (-6.1,0) circle (3pt);
%\draw[thick, fill=red] (-6.55,0) circle (3pt);

%\draw[thick, fill=blue] (-1.4,0) circle (3pt);
%\halfcoloredcircle{-1.4}{0}{blue}{blue}
%\draw[thick,fill=white] (-1.1,0) circle (3pt);
%\halfcoloredcircle{-1.1}{0}{white}{white}
%\draw[thick, fill=red] (-0.8,0) circle (3pt);
%\halfcoloredcircle{-0.8}{0}{red}{red}

%nodes first level upper branch

\draw[fill=red,thick] (1.45,1.45) circle (3pt);
\draw[fill=red,thick] (1.15,1.15) circle (3pt);
\draw[fill=blue,thick] (0.85,0.85) circle (3pt);
\draw[fill=white,thick] (0.55,0.55) circle (3pt);

%\draw[thick,fill=white] (3.4,3.4) circle (3pt);
%\draw[thick,fill=white] (3.2,3.2) circle (3pt);
%\draw[thick,fill=red] (3,3) circle (3pt);

%nodes first level lower branch

\draw[ fill=black] (1.45,-1.45) circle (2pt);
\draw[fill=black] (1.15,-1.15) circle (2pt);
\draw[fill=black] (0.85,-0.85) circle (2pt);
\draw[fill=white,thick] (0.55,-0.55) circle (3pt);

%nodes second level first branch

%\draw[thick,fill=red] (6,5) circle (3pt);
%\draw[thick,fill=white] (6.25,5.125) circle (3pt);
%\draw[thick,fill=white] (6.5,5.25) circle (3pt);

%nodes second level second branch

%\draw[thick, fill=red] (6,3) circle (3pt);
%\draw[thick, fill=white] (6.25,2.875) circle (3pt);
%\draw[thick,fill=white] (6.5,2.745) circle (3pt);

%nodes second level third branch

%\draw[thick] (6,-3) circle (3pt);
%\draw[thick] (6.25,-2.875) circle (3pt);
%\draw[thick] (6.5,-2.745) circle (3pt);

%nodes second level fourth branch

%\draw[thick,fill=white] (6,-5) circle (3pt);
%\draw[thick,fill=white] (6.25,-5.125) circle (3pt);
%\draw[thick, fill=red] (6.5,-5.25) circle (3pt);

%nodes third level first branch

%\draw[thick,fill=red] (9.5,5.25) circle (3pt);
%\draw[thick,fill=white] (9.75,5.125) circle (3pt);
%\draw[thick,fill=white] (10,5) circle (3pt);

%\node[draw=none, fill=none] at (-1.1,-0.8){\scalebox{1.2}{$(d,l)$}};

%\node[draw=none, fill=none, rotate=-45] at (0.5,-1.5){\scalebox{1.2}{$(d+1,2l)$}};

%\node[draw=none, fill=none, rotate=45] at (0.5,1.5){\scalebox{1.2}{$(d+1,2l+1)$}};

%in-out arrows

%\draw[->, >={Stealth[length=6pt]}, thick] (-4.5,0.5) -- (-3,0.5)node[fill=none,draw=none,midway,above] {};
%\draw[->, >={Stealth[length=6pt]}, very thick] (19,0.5) -- (20,0.5)node[fill=none,draw=none,midway,above] {\Large$\ket{\psi_{out}}$};

\node[draw=none, fill=none, align=left] at (-4.5,-2)
{\large Step 5: shift of two positions and \\ \large parallel application of $\hat{U}_{in}$ and $\hat{U}_B$\\\large  at different levels};

\node[draw=none, fill=none] at (-7,2){\Large $(e)$};

\end{scope}

%%%%%%%%%% SESTO DISEGNO %%%%%%%%

\begin{scope}[shift={(12,-12)}]

% Nodo radice
\node (root1) {$\hat{S}$}
  child {
    % Primo nodo intermedio (verso l'alto)
    node {$\hat{S}$}
    child { 
      %node[leaf] {$\boldsymbol{x}_3$} % Foglia sinistra-sinistra (grigia)
      edge from parent {}
    }
    child { 
      %node[leaf] {$\boldsymbol{x}_2$} % Foglia sinistra-destra (grigia)
      edge from parent{
%node[midway,xshift=-1cm,gateX, yshift=-0.5cm,rotate=27] {$\hat{X}$} % Blocco rosso ruotato
      }
    }
    edge from parent {
      node[midway,xshift=0cm, yshift=-0cm,  block, rotate=-45] {} % Blocco blu ruotato
    }
  }
  child {
    % Secondo nodo intermedio (verso l'alto)
    node {$\hat{S}$}
    child { 
      %node[leaf] {$\boldsymbol{x}_1$} % Foglia destra-sinistra (grigia)
      %edge from parent {}
    }
    child { 
      %node[leaf] {$\boldsymbol{x}_0$} % Foglia destra-destra (grigia)
      edge from parent{ 
      %node[midway,xshift=-1cm, yshift=-0.5cm,gateX, rotate=27] {$\hat{X}$} % Blocco rosso ruotato
      }
    }
    edge from parent {
      %node[midway, xshift=-1.3cm,yshift=-1.3cm,gateX, rotate=45] {$\hat{X}$} % Blocco rosso ruotato
      node[midway, xshift=0cm,yshift=0cm, block, rotate=45] {} % Blocco blu ruotato
    }
  };

% Ramo in ingresso al nodo radice
\draw (-7, 0) -- (root1) node(Ugateroot)[xshift=-1.37cm, block, rotate=0] {};
% Linea di ingresso al nodo radice
%\draw[thick] (root2) -- (20,0) node[midway, block] {$\hat{U}^{(1)}$};

%\node[block, rotate = 27, draw = none, minimum width = 0.5cm](b1) at (3,2.5){};
%\draw[thick] (b1.south west) -- (b1.south east); % lato basso
%\draw[thick] (b1.south west) -- (b1.north west);         % lato destro
%\draw[thick] (b1.north east) -- (b1.north west); % lato alto

%\node[block, rotate = -27, draw = none, minimum width = 0.5cm](b2) at (3,1.5){};
%\draw[thick] (b2.south west) -- (b2.south east); % lato basso
%\draw[thick] (b2.south west) -- (b2.north west);         % lato destro
%\draw[thick] (b2.north east) -- (b2.north west); % lato alto

%\node[block, rotate = 27, draw = none, minimum width = 0.5cm](b3) at (3,-1.5){};
%\draw[thick] (b3.south west) -- (b3.south east); % lato basso
%\draw[thick] (b3.south west) -- (b3.north west);         % lato destro
%\draw[thick] (b3.north east) -- (b3.north west); % lato alto

%\node[block, rotate = -27, draw = none, minimum width = 0.5cm](b4) at (3,-2.5){};
%\draw[thick] (b4.south west) -- (b4.south east); % lato basso
%\draw[thick] (b4.south west) -- (b4.north west);         % lato destro
%\draw[thick] (b4.north east) -- (b4.north west); % lato alto

%gates

\node[block,draw,thick,minimum width = 1.3cm, solid, fill=blue!20,label={[yshift=1mm]above:$\hat{U}_B$}] at (-1.15,0){};

\node[block,draw,thick,minimum width = 1.3cm, solid, fill=blue!20, rotate = 45,label={[yshift=1mm]above:$\hat{U}_B$}] at (1.15,1.15){};

%nodes before root node

\draw[fill=blue,thick] (-0.7,0) circle (3pt);
\draw[fill=blue,thick] (-1.15,0) circle (3pt);
\draw[fill=blue,thick] (-1.6,0) circle (3pt);
\draw[fill=black] (-2.05,0) circle (2pt);

%nodes first level upper branch

\draw[fill=blue,thick] (1.45,1.45) circle (3pt);
\draw[fill=white,thick] (1.15,1.15) circle (3pt);
\draw[fill=blue,thick] (0.85,0.85) circle (3pt);
\draw[fill=red,thick] (0.55,0.55) circle (3pt);

%\draw[thick,fill=white] (3.4,3.4) circle (3pt);
%\draw[thick,fill=white] (3.2,3.2) circle (3pt);
%\draw[thick,fill=red] (3,3) circle (3pt);

%nodes first level lower branch

\draw[ fill=black] (1.45,-1.45) circle (2pt);
\draw[fill=white,thick] (1.15,-1.15) circle (3pt);
\draw[fill=black] (0.85,-0.85) circle (2pt);
\draw[fill=black] (0.55,-0.55) circle (2pt);

\node[draw=none, fill=none, align=left] at (-4.5,-2)
{\large Step 6: shift of two positions and \\ \large parallel application of $\hat{U}_{B}$ \\ \large at different levels};

\node[draw=none, fill=none] at (-7,2){\Large $(f)$};

\end{scope}

\draw[dashed] (4,3)--(4,-15);

\draw[dashed] (-8,-3)--(16,-3);

\draw[dashed] (-8,-9)--(16,-9);

\end{tikzpicture}
\caption{%\dds{\it[Confonde leggermente il fatto che nel resto del paper, quando si hanno due colonne, l'ordine crescente degli indici è dall'alto in basso sulla stessa colonna, per poi passare alla colonna successiva. Mentre qui si fa destra - sinistra.]}
Example of parallelization schedule, in the backup variant setting, between the first two levels of the binary tree: (a) the walkers are initialized at level $d=1$ of the tree; (b) after a two-site shift, the initialization gate $\hat{U}_{in}$ is applied to the pair $(\tilde A_1,A_1)$; (c) following another two-site shift, the block gate $\hat{U}_B$ acts on the triplet $(\tilde A_{2},A_{2},\tilde A_1)$; (d) after the next shift $\hat{U}_B$ is applied to the $(\tilde A_{3},A_{3},\tilde A_{2})$; (e) subsequent shifts allow the simultaneous application $\hat{U}_{B}$ at level $d=1$ and $\hat{U}_{in}$ at level $d=2$; (f) from this point onward the applications of gates $\hat{U}_B$ at levels $d=1$ and $d=2$ can be in parallel.}
\label{parallelization}
\end{figure*}
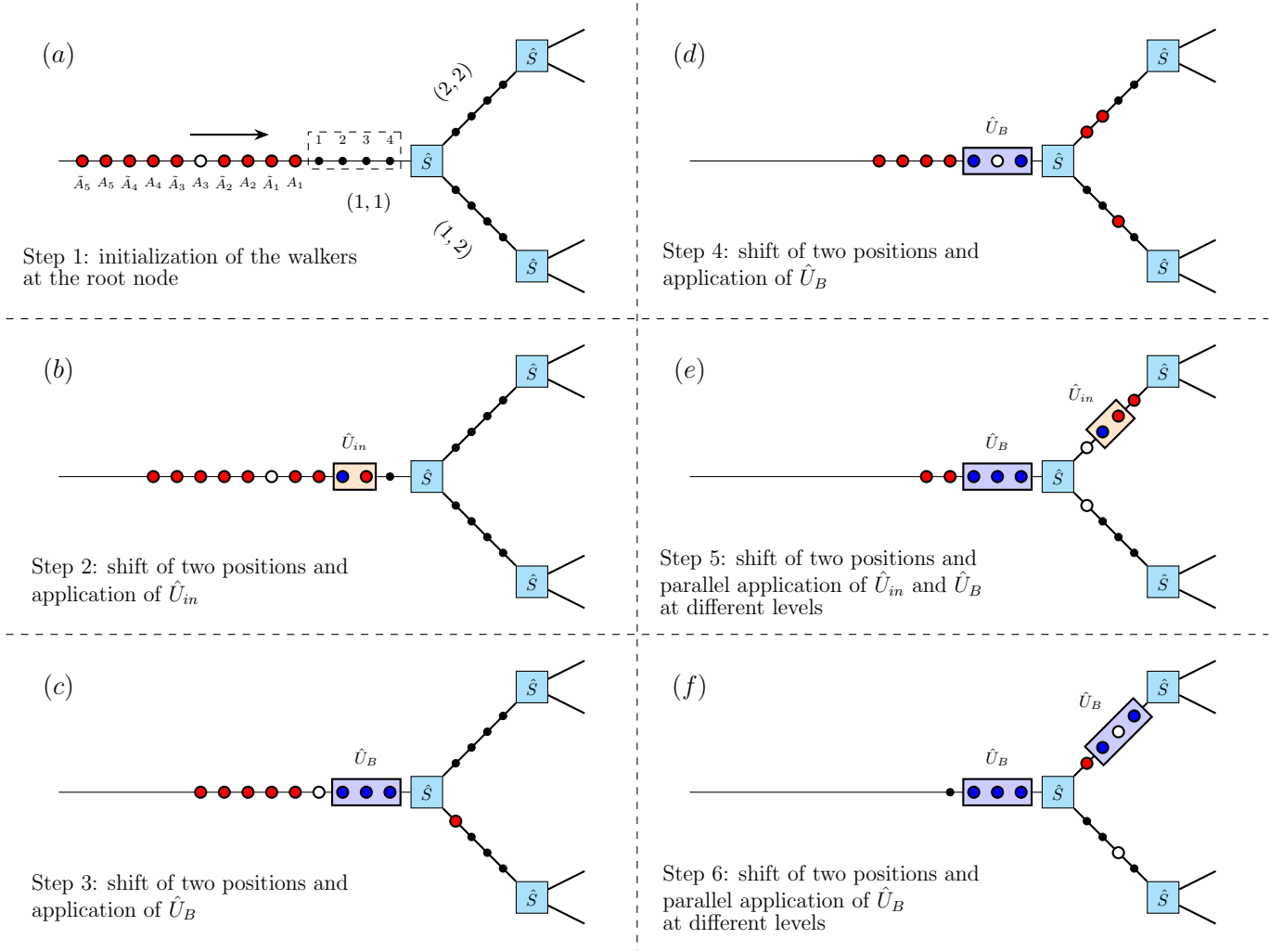

\subsection{Depth reduction}\label{DepthReduction}

{As established in Sec.~\ref{StandardScaling}, the execution time of the \gdr{basic} long-range protocol scaled as  $T_{\text{serial}} \sim \mathcal{O}(n^2+nm)$. In that sequential routine, the routing at depth $d$ consisted of two steps: shifting the entire train of $n+m-d$ walkers along the corresponding edge, and subsequently applying the global unitary $\hat{U}^{(d)}$. 
Because the physical extent of the edges must accommodate all walkers in transit, leading to the spatial scaling $n_{sites}^{tot}$ detailed in Eq.~(\ref{StandardSitesScaling}), waiting for the entire register to clear a level before initiating the operation at level $d+1$ imposed a strict sequential dependence. The quadratic scaling is thus a direct consequence of the edge lengths growing proportionally to the number of propagating walkers.}

In the backup variant, a substantial depth reduction becomes possible. Each elementary gate $\hat{U}_B$ acts on at most three walkers, and every application of $\hat{U}_B$ requires shifting the particle train by a constant number of sites (specifically, two). 
{Furthermore, the protocol exhibits a fixed synchronization pattern: once $\hat{U}_{in}$ is applied at level $d$, the walkers undergo a specific sequence of exactly two $\hat{U}_B$ operations before being processed by the corresponding $\hat{U}_{in}$ at level $d+1$.} (see Fig.\ref{parallelization}). This is true at all levels $d$. Since these constraints define a constant interaction window that does not depend on the depth $d$, it is possible to schedule all required operations in parallel without generating conflicts among overlapping groups of interacting walkers. 

Now we demonstrate how, in the backup variant, the depth of our qRAM scales linearly with $n$ and $m$.
We consider all edges of the tree to have a constant number of sites, equal to $n_{sites}=4$, with sites enumerated
in ascending order from level $d$ to level $d+1$, hence having site $1$ immediately after the $d$-th bifurcation and site $4$ immediately before the $d+1$-th bifurcation (see Fig.~\ref{parallelization}a).
Combined with a proper initialization and shift schedule, this choice for the number of sites allows backup walkers to act sequentially as targets and controls for successive applications of the block gate $\hat{U}_B$ on a given edge. Consequently, this layout ensures the modular propagation of address information along the train while enabling parallel gate activation across different levels.

Beyond this geometric layout, the two main ingredients that allows the design of an
efficient parallelization scheme are (i) the fact that each gate in the backup scheme involves at most
three walkers and that (ii) gates acting on disjoint triplets commute. The steps of this scheme are as follows:
\begin{itemize}[label=$\bullet$]
    \item At level $d$, the walkers $A_d$, $\tilde A_d$, and $A_{d+1}$ occupy sites $3$, $2$, and $1$ of the edge, respectively.
    
    \item The incoming operation $\hat{U}_{in}$ is applied to the pair $(\tilde A_d, A_d)$ (Fig.~\ref{parallelization}b).
    
    \item All walkers are shifted forward by two sites. After this shift, $A_d$ reaches site $1$ at level $d+1$
    and does not participate further in the routing, while $\tilde A_d$ moves to site $4$ at level $d$.
    The next walkers in the train, $A_{d+1}$, $\tilde A_{d+1}$, and $A_{d+2}$, now lie on sites $3$, $2$, and $1$.
    
    \item The operation $\hat{U}_B$ is applied to the triple $(\tilde A_{d+1}, A_{d+1}, \tilde A_d)$ (Fig.~\ref{parallelization}c).
    
    \item After another two-site shift, $\tilde A_d$ moves to site $2$ at level $d+1$, followed by $A_{d+1}$ at site $1$.
    Simultaneously, $\tilde A_{d+1}$ reaches site $4$ at level $d$, followed by $A_{d+2}$, $\tilde A_{d+2}$, and $A_{d+3}$.
    
    \item The next backup operation $\hat{U}_B$ is applied to the triple $(\tilde A_{d+2}, A_{d+2}, \tilde A_{d+1})$ (Fig.~\ref{parallelization}d).
    
    \item Another two-site shift now produces the ordering
    $(A_{d+2}, \tilde A_{d+1}, A_{d+1}, \tilde A_d)$ across the boundary between levels $d$ and $d+1$,
    while at level $d$ the ordering becomes $(A_{d+4}, \tilde A_{d+3}, A_{d+3}, \tilde A_{d+2})$. Importantly, the relative positioning of walkers at level $d+1$ mirrors that at level $d$: for example, $A_{d+1}$ and its backup occupy sites $3$ and $2$, just as $A_d$ and its backup did at the previous level.
    
    \item At this point, the next groups of walkers involved in gate operations occupy disjoint positions.
    Thus, $\hat{U}_B$ can be applied to the triple $(\tilde A_{d+3}, A_{d+3}, \tilde A_{d+2})$ at level $d$,
    simultaneously with the application of $\hat{U}_{in}$ to the pair $(\tilde A_{d+1}, A_{d+1})$ at level $d+1$ (Fig.~\ref{parallelization}e).
    
    \item From this point onward, the same structure repeats: two-site shifts preserve the alignment of
    walkers on each edge, and the required gates can be applied in parallel until the end of the train (Fig.~\ref{parallelization}f).
\end{itemize}

{This strategy enables a parallelization that transforms the routing depth from quadratic to linear. To provide an intuitive explanation of this exact linear scaling, we evaluate the total execution time as the interval between the injection of the first address walker into the tree and the exit of the last data walker. 
Since the length of each edge is fixed to $n_{sites}=4$, the number of time steps needed for the first walker to move from one level $d$ to the corresponding spot at the next level $d+1$ is constant and equal to $4$, regardless of the tree depth $n$ {and the starting depth $d$. As detailed in the steps above, the operations alternate the parallel action of the gates $\hat{U}_B$ and/or $\hat{U}_{in}$ with the synchronous forward shift of the particles by two sites. Therefore, the total time needed for the first address walker to reach the memory cells is $4n$. 

Once the head of the train reaches the memory cells, the time taken to travel through them depends on the specific query protocol. Assuming a sequential strategy through the memory register, the walker has to travel $2m$ sites, which is covered in $2m$ time steps. Following this logic and considering the symmetry of the inverse routing, the total time needed for the first walker to traverse the tree completely and finally exit it is $8n + 2m$. 
At the exact moment the first walker exits the tree, the tail of the train, i.e., the last data walker, is still located $2(n + m)$ sites away from the output port. Because the spatial propagation is fully parallelized and strictly pipelined, these remaining sites are covered in $2(n+m)$ additional time steps. Summing these contributions, the final execution time is given by:
\begin{equation}\label{Eq_BackupTime}
    T_{\text{backup}} = 10n + 4m + t_Q \sim \mathcal{O}(n + m),
\end{equation}
where $t_Q$ is the extra time needed to encode the information onto the data walkers during the query phase. Depending on the memory interaction scheme, this overhead can be minimized; for instance, if the information is written on the data particles in parallel with the execution of the routing gates during the inverse-routing phase, we obtain $t_Q = 0$.}

\subsection{Resource scalings}
We now quantify the resources required to implement the backup variant of the protocol, focusing on walker count, gate complexity, spatial extent, and circuit depth.
In contrast to the \gdr{basic} method, the routing requires doubling the total number of walkers, thus employing a $2n$-walker address register $\boldsymbol{A}$ and a $2m$-walker data register $\boldsymbol{D}$. While this represents a constant factor increase, it does not change the overall scaling of the total walker count, which remains linear: $2(n+m) \sim\mathcal{O}(n+m)$. 

\gdr{The architecture still utilizes a single binary tree of depth $n$, where a unitary block is associated with each bifurcation. However, we now have a three-operation unitary block:} the scattering gate $\hat{S}$ along with the gates $\hat{U}^{in}$ and $\hat{U}^B$ used to decompose the address-transmission operation. The total gate count is $3(2^n-1)\sim\mathcal{O}(2^n)$, preserving the same scaling as in the \gdr{basic} protocol. 

The main advantages of this backup variant in terms of scalability are the physical extent of the architecture and the circuit depth. As detailed in Sec.~\ref{DepthReduction}, the more local nature of the operations allows for a parallelization pipeline. This allows us to show that the number of sites required on each edge to host walkers can be taken as constant $n_{\text{sites}}^{d}=4=\mathcal{O}(1)$, regardless of the tree depth. Summing over all levels, the path extent changes from quadratic to linear, $n_{\text{sites}}^{\text{tot}}=4n\sim\mathcal{O}(n)$. 

Similarly, the circuit depth is reduced dramatically. Due to the parallelization routine, which allows for constant edge length, the execution time for the routing phase alone scales as $T_{\text{parallel}}=6n+2m\sim \mathcal{O}(n+m)$. 

Finally, the number of active elements per query remains consistent with the \gdr{basic} protocol, being $\mathcal{O}(n)$ for a classical or sparse query and $\mathcal{O}(2^n)$ for a dense superposition of addresses.

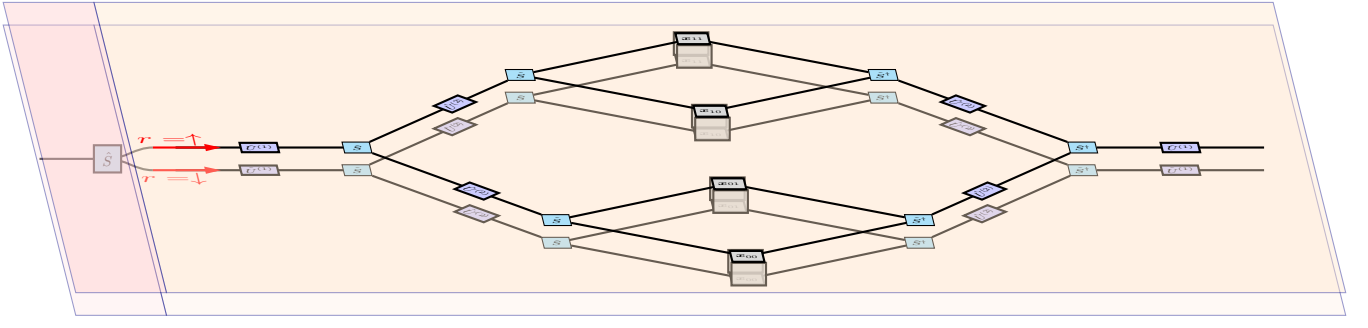
\begin{figure*}[htb]
\centering
\begin{tikzpicture}[scale=0.6, transform shape]

% Parametri
\def\dist{0.5} % distanza verticale tra i piani paralleli

\coordinate (S_center) at (-5.5, 0.25);

\draw[thick] (S_center) to[out=0, in=180] (-4.5,\dist);
\draw[thick] (S_center) to[out=0, in=180] (-4.5,0);

\draw[thick] (S_center)--(-7,0.25);

\node[rectangle, draw, minimum size=6mm, fill=cyan!30, thick] (S_gate) at (S_center) {$\hat{S}$};

% Piano inferiore con grafo
\begin{scope}[cm={1,0,-0.1,0.4,(0,0)},
  level 1/.style={level distance=4cm, sibling distance=8cm},
  level 2/.style={level distance=4cm, sibling distance=4cm},
  grow=east,
  edge from parent/.style={draw, thick},
   every node/.style={rectangle, draw, minimum size=6mm,fill=cyan!30},
  leaf/.style={rectangle, draw, fill=gray!30, minimum size=6mm},
  block/.style={rectangle, draw, fill=blue!20, minimum width=0.7cm, minimum height=5mm},
  gateX/.style={rectangle, draw, fill=green!20, minimum size=5mm}]

\fill[orange!10, draw=blue!50!black, opacity=0.4] (-5,-8) rectangle (21,8);

\fill[red!10, draw=blue!50!black, opacity=0.4] (-5,-8) rectangle (-7,8);

% Nodo radice
\node (root1) {$\hat{S}$}
  child {
    node {$\hat{S}$}
    child { 
      node[leaf] {$\boldsymbol{x}_3$} 
      edge from parent {}
    }
    child { 
      node[leaf] {$\boldsymbol{x}_2$} 
      edge from parent{
        %node[midway,xshift=-1cm,gateX, yshift=-0.5cm, rotate=27] {$\hat{X}$} 
      }
    }
    edge from parent {
      node[midway,xshift=0.4cm, yshift=-0.4cm,  block, rotate=-45] {$\hat{U}^{(2)}$} 
    }
  }
  child {
    node {$\hat{S}$}
    child { 
      node[leaf] {$\boldsymbol{x}_1$} 
      edge from parent {}
    }
    child { 
      node[leaf] {$\boldsymbol{x}_0$} 
      edge from parent{ 
        %node[midway,xshift=-1cm, yshift=-0.5cm,gateX, rotate=27] {$\hat{X}$} 
      }
    }
    edge from parent {
      %node[midway, xshift=-1.3cm,yshift=-1.3cm,gateX, rotate=45] {$\hat{X}$} 
      node[midway, xshift=0.4cm,yshift=0.4cm, block, rotate=45] {$\hat{U}^{(2)}$} 
    }
  };

\node (root2) at (16,0) {$\hat{S}^{\dagger}$}
[grow=left]
  child {
    node {$\hat{S}^{\dagger}$}
    child { 
      node[leaf] {$\boldsymbol{x}_{11}$} 
      edge from parent {
        %node[midway, xshift=1cm, gateX, yshift=-0.5cm, rotate=-27] {$\hat{X}$} 
      }
    }
    child { 
      node[leaf] {$\boldsymbol{x}_{10}$} 
      edge from parent {}
    }
    edge from parent {
      %node[midway, xshift=1.3cm, yshift=-1.3cm, gateX, rotate=-45] {$\hat{X}$} 
      node(Ugate)[midway,xshift=-0.4cm,yshift=0.4cm, block, rotate=-45] {$\hat{U}^{(2)}$} 
    }
  }
  child {
    node {$\hat{S}^{\dagger}$}
    child { 
      node[leaf] {$\boldsymbol{x}_{01}$} 
      edge from parent {
       %node[midway, xshift=1cm, gateX, yshift=-0.5cm, rotate=-27] {$\hat{X}$} 
      }
    }
    child { 
      node[leaf] {$\boldsymbol{x}_{00}$} 
      edge from parent {}
    }
    edge from parent {
      node[midway, xshift=-0.4cm, yshift=-0.4cm, block, rotate=45] {$\hat{U}^{(2)}$} 
    }
  };

\draw[thick] (-4, 0) -- (root1) node(Ugateroot)[midway, block, rotate=0] {$\hat{U}^{(1)}$};
\draw[thick] (root2) -- (20,0) node[midway, block] {$\hat{U}^{(1)}$};

%\draw[->,thick, red] (-4.5,0)coordinate (A_down)--(-3,0)node [below,pos=0.3, draw=none, fill=none] {\huge$r=\downarrow$};

\draw[thick, red] (-4.5,0) coordinate (A_down) -- (-3.1,0)node [below,pos=0.3, draw=none, fill=none] {\huge$r=\downarrow$};

\fill[red] (-3,0) -- (-3.35, 0.2) -- (-3.35, -0.2) -- cycle;

%back rectangles between planes
\node[thick,transform shape=false, fill=gray!30, rectangle, minimum height =8.5pt, minimum width=13pt, opacity=1] at (8.065,6.98) {}; 

\node[thick,transform shape=false, fill=gray!30, rectangle, minimum height =8.5pt, minimum width=13pt, opacity=1] at (8.065,2.98) {}; 

\node[thick,transform shape=false, fill=gray!30, rectangle, minimum height =8.5pt, minimum width=13pt, opacity=1] at (8.065,-1.02) {}; 

\node[thick, draw=black,transform shape=false, fill=gray!30, rectangle, minimum height =8.5pt, minimum width=13pt, opacity=1] at (8.065,-5.02) {}; 

%front rectangled between planes
\node[thick,transform shape=false, fill=gray!30, rectangle, minimum height =8.5pt, minimum width=13pt, fill opacity=0.8] at (8.065,6.27) {}; 

\node[thick,transform shape=false, fill=gray!30, rectangle, minimum height =8.5pt, minimum width=13pt, fill opacity=0.8] at (8.065,2.27) {}; 

\node[thick,transform shape=false, fill=gray!30, rectangle, minimum height =8.5pt, minimum width=13pt, fill opacity=0.8] at (8.065,-1.73) {}; 

\node[thick,transform shape=false, fill=gray!30, rectangle, minimum height =8.5pt, minimum width=13pt, fill opacity=0.8] at (8.065,-5.73) {};

\end{scope}

% Piano superiore con grafo trasparente
\begin{scope}[cm={1,0,-0.1,0.4,(0,\dist)},
  level 1/.style={level distance=4cm, sibling distance=8cm},
  level 2/.style={level distance=4cm, sibling distance=4cm},
  grow=east,
  edge from parent/.style={draw, thick},
every node/.style={rectangle, draw, minimum size=6mm,fill=cyan!30},
  leaf/.style={rectangle, draw, fill=gray!30, minimum size=6mm},
  block/.style={rectangle, draw, fill=blue!20, minimum width=0.7cm, minimum height=5mm},
  gateX/.style={rectangle, draw, fill=green!20, minimum size=5mm}]

\fill[orange!20, draw=blue!50!black, opacity=0.4] (-5,-8) rectangle (21,8);

\fill[red!20, draw=blue!50!black, opacity=0.4] (-5,-8) rectangle (-7,8);

% Nodo radice
\node (root1t) {$\hat{S}$}
  child {
    node {$\hat{S}$}
    child { 
      node[leaf] {$\boldsymbol{x}_3$} 
      edge from parent {}
    }
    child { 
      node[leaf] {$\boldsymbol{x}_2$} 
      edge from parent{
        %node[midway,xshift=-1cm,gateX, yshift=-0.5cm, rotate=27] {$\hat{X}$} 
      }
    }
    edge from parent {
      node[midway,xshift=0.4cm, yshift=-0.4cm,  block, rotate=-45] {$\hat{U}^{(2)}$} 
    }
  }
  child {
    node {$\hat{S}$}
    child { 
      node[leaf] {$\boldsymbol{x}_1$} 
      edge from parent {}
    }
    child { 
      node[leaf] {$\boldsymbol{x}_0$} 
      edge from parent{ 
        %node[midway,xshift=-1cm, yshift=-0.5cm,gateX, rotate=27] {$\hat{X}$} 
      }
    }
    edge from parent {
      %node[midway, xshift=-1.3cm,yshift=-1.3cm,gateX, rotate=45] {$\hat{X}$} 
      node[midway, xshift=0.4cm,yshift=0.4cm, block, rotate=45] {$\hat{U}^{(2)}$} 
    }
  };

\node (root2t) at (16,0) {$\hat{S}^{\dagger}$}
[grow=left]
  child {
    node {$\hat{S}^{\dagger}$}
    child { 
      node[leaf] {$\boldsymbol{x}_{11}$} 
      edge from parent {
        %node[midway, xshift=1cm, gateX, yshift=-0.5cm, rotate=-27] {$\hat{X}$} 
      }
    }
    child { 
      node[leaf] {$\boldsymbol{x}_{10}$} 
      edge from parent {}
    }
    edge from parent {
      %node[midway, xshift=1.3cm, yshift=-1.3cm, gateX, rotate=-45] {$\hat{X}$} 
      node(Ugate)[midway,xshift=-0.4cm,yshift=0.4cm, block, rotate=-45] {$\hat{U}^{(2)}$} 
    }
  }
  child {
    node {$\hat{S}^{\dagger}$}
    child { 
      node[leaf] {$\boldsymbol{x}_{01}$} 
      edge from parent {
       %node[midway, xshift=1cm, gateX, yshift=-0.5cm, rotate=-27] {$\hat{X}$} 
      }
    }
    child { 
      node[leaf] {$\boldsymbol{x}_{00}$} 
      edge from parent {}
    }
    edge from parent {
      node[midway, xshift=-0.4cm, yshift=-0.4cm, block, rotate=45] {$\hat{U}^{(2)}$} 
    }
  };

\draw[thick] (-4, 0) -- (root1t) node[midway, block, rotate=0] {$\hat{U}^{(1)}$};
\draw[thick] (root2t) -- (20,0) node[midway, block] {$\hat{U}^{(1)}$};

%\draw[->,thick, red] (-4.5,0) coordinate (A_up)--(-3,0)node [above,pos=0.3, draw=none, fill=none] {\huge$r=\uparrow$};

\draw[thick, red] (-4.5,0) coordinate (A_up) -- (-3.1,0)node [above,pos=0.3, draw=none, fill=none] at (-3.75,0) {\huge$r=\uparrow$};

\fill[red] (-3,0) -- (-3.35, 0.2) -- (-3.35, -0.2) -- cycle;

\end{scope}

\end{tikzpicture}
\caption{Schematic representation of the qRAM protocol in the dual-rail encoding. Each rail evolves independently throughout the protocol, with no direct interaction between them, except during the control-COPY step. In this phase, the position of the data walker on the rails is coherently updated based on the classical information stored in the memory cells.}
\label{dualrailQRAM}
\end{figure*}

%\dds{\it [Ho fatto una sez. per dual rail e un'altra per qudits. Prima erano sottosez.]}

\section{Dual-rail implementations}
\label{sec:dualrail}

\gdr{The introduced protocols can be easily generalized to a construction requiring a
constant number of information carriers. In what follows, we restate the schemes
introduced above in a dual-rail setting, highlighting the main differences in the
operations and in the resources required to implement these variants efficiently.
First, in Section~\ref{subsec:dual_rail}, we introduce a dual-rail implementation that adopts a strategy analogous to the long-range basic variant of Section~\ref{sec:tech_desc}. Later, in Section~\ref{subsec:backup_dual_rail}, we present the corresponding dual-rail backup variant.}

\subsection{Basic variant for the dual-rail implementation}
\label{subsec:dual_rail}

A natural way to extend the walker-based qRAM to an implementation with a constant number of information carriers is to give each particle two independent binary degrees of freedom: a rail index, which encodes the logical address bit, and a color label, which encodes the local routing choice. Concretely, this dual-rail encoding is realized by two parallel binary trees: a walker’s rail determines which tree it occupies, while its color stores the path information used by the routing primitives (Fig.~\ref{dualrailQRAM}).

\gdr{The routing operations of the protocol are placed on both rails, each gate acting
locally on the rail on which it is placed and being activated by the color of the
walkers travelling on that same rail. Importantly, no operation couples the two
rails, which therefore evolve independently of each other.
The primitives employed are the same as in the basic protocol, and so are their
physical requirements: the scattering gates are passive optical elements, whereas
the address-transmission and copy gates are conditional operations acting on
different carriers, and therefore require a strong nonlinearity. The dual-rail
encoding does not remove this requirement, but it removes another one: since every
walker occupies exactly one of the two rails, all the operations preserve the
particle number, and a superposition of addresses never involves states
with different particle numbers, differently from the basic protocol, where the
logical value $0$ is encoded in the absence of a walker.}

Below, we detail the {realization of the} dual-rail
protocol, describing encoding and initialization, the rail-local implementation of the routing primitives and the adaptations required for memory access.

\subsubsection{Encoding and initialization}\label{encodindDualStandard}

Formally, the Hilbert space of a single walker in this setting is $\mathcal{H}_W = \mathcal{H}_R\otimes\mathcal{H}_P \otimes \mathcal{H}_C$, where $\mathcal{H}_R$ denotes the rail degree of freedom, $\mathcal{H}_P$ the position along the binary tree, and $\mathcal{H}_C$ the color one. Hence, the total Hilbert space for a single walker $W$ is defined as
%\dds{(stessa cosa detta prima dei range $\in vs \leq \dots\leq$, convertirei tutto in $\in$)}
\begin{multline*} \mathcal{H}_{W}=\left\{\mathrm{span}\{\ket{r,c^{l,d}}_W : r\in\{\uparrow,\downarrow\},\right.\\\left.c\in\{R,B\},d\in\{0,\dots,n\}, l\in\{1,\dots,2^{d}\}\right\}
\end{multline*}
where $\ket{r,c^{l,d}}_W\equiv\ket{r}_R\otimes\ket{c}_C\otimes\ket{l,d}_P$ is a generic walker ket state vector: $r$ indicates whether the walkers is in the upper $r=\uparrow$ or lower $r=\downarrow$ rail, $c$ corresponds to its color, either red or blue, $d$ denotes the depth in the binary tree where the walker is located and $l$ the corresponding branch. 

\gdr{The overall Hilbert space of the walker can be decomposed into two logical
subspaces, representing logical $0$ and $1$, each associated with one of the
rails. More precisely, let $r=\uparrow$ denote a logical $0$ and $r=\downarrow$
denote a logical $1$, while the colors are used as routing degrees of freedom on
the single rails. The presence of the two rails then induces the decomposition
\begin{equation*}
\mathcal{H}_W=\mathcal{H}_W^{(0)}\oplus\mathcal{H}_W^{(1)}
\equiv\mathcal{H}_W^{(\uparrow)}\oplus\mathcal{H}_W^{(\downarrow)},
\end{equation*}
where
\begin{multline*}
\mathcal{H}_W^{(r)}=\mathrm{span}\{\ket{r,c^{l,d}}_W : c\in\{R,B\},\,\\
d\in\{0,\dots,n\},\, l\in\{1,\dots,2^{d}\}\}
\end{multline*}
is the subspace associated with rail $r$, so that $\mathcal{H}_W^{(\uparrow)}$
encodes a logical $0$ and $\mathcal{H}_W^{(\downarrow)}$ a logical $1$.}

%The overall Hilbert space of the walker can be decomposed into two logical subspaces, representing logical $0$ and $1$, each associated with one of the rails. More precisely, let $r=\uparrow$ denote a logical $0$ and $r=\downarrow$ denote a logical $1$, while the colors are used as routing degrees of freedom on the single rails. Then the internal Hilbert space for a single walker is decomposed as
%\begin{equation*}
%\mathcal{H}_{RC}=\mathcal{H}_{RC}^{(0)}\oplus\mathcal{H}_{RC}^{(1)}\equiv\mathcal{H}_{RC}^{(\uparrow)}\oplus\mathcal{H}_{RC}^{(\downarrow)}
%\end{equation*}
%where $\mathcal{H}_{RC}^{(\uparrow)}=\text{span}\{\ket{\uparrow,R}_{RC},\ket{\uparrow,B}_{RC}\}$ is the subspace encoding a logical $0$ and $\mathcal{H}_{RC}^{(\downarrow)}=\text{span}\{\ket{\downarrow,R}_{RC},\ket{\downarrow,B}_{RC}\}$ is the subspace encoding a logical $1$. This decomposition naturally induces a partition for the overall Hilbert space of the walker
%\begin{equation*}   \mathcal{H}_W=\left(\mathcal{H}_{RC}^{(\uparrow)}\otimes \mathcal{H}_P\right)\oplus\left(\mathcal{H}_{RC}^{(\downarrow)}\otimes \mathcal{H}_P\right).
%\end{equation*}

As for the \gdr{basic} protocol, we employ an $n$-walker address register $\boldsymbol{A}$ and an $(m+1)$-walker data register $\boldsymbol{D}$.
Address and data information are encoded in the walker's rail degree of freedom. We denote the two parallel trees (rails) by the states $\ket{r=\uparrow}$ and $\ket{r=\downarrow}$.
The logical address bits are naturally mapped into these rails as
\begin{equation}\label{dualrailencoding}
    a_i=0\mapsto\ket{r=\uparrow}_{A_i},\quad a_i=1\mapsto\ket{r=\downarrow}_{A_i}
\end{equation}
so that each address walker $A_i$ is placed on the rail corresponding to its bit value.
By convention, all data walkers $D_i$ instead occupy the rail state $\ket{r=\downarrow}_{D_i}$.%, as they must propagate along the tree where memory cells are located.
This rail assignment is fixed throughout the entire routing stage and is modified only during the memory-access operations.
\gdr{Notice that, differently from the schemes introduced before in this work (see Eq.~(\ref{addressencoding})), the vacuum state $\ket{\varnothing}_W$ no longer
carries any logical information, which is now entirely encoded in the rail degree of freedom. The vacuum is nonetheless still present: since each walker travels on a single rail, the other rail is empty in the corresponding time slot, so that
each rail individually carries a sequence of occupied and empty positions.} We anticipate that the same is true for the information carried by the data walkers, where the bits of information extracted from the memory cells are dual-rail encoded, similarly to Eq.~(\ref{dualrailencoding}).

The color degree of freedom, used exclusively for routing, is initially uniform: all $n+m$ walkers are initialized in the red color state $\ket{c=R}_W$. 
A generic initial state for the dual rail protocol is then given by 
\begin{equation}
\begin{split}
\ket{\psi_{in}} &=\sum_{\boldsymbol{a}\in\mathcal{A}}\alpha_{\boldsymbol{a}} \bigotimes_{j=0}^m\ket{\downarrow,R}_{D_i}\otimes\bigotimes_{i=1}^n\ket{a_i,R}_{A_i}\\ &=\sum_{\boldsymbol{a}\in\mathcal{A}}\alpha_{\boldsymbol{a}}\ket{\downarrow,R}_{\boldsymbol{D}}^{\otimes(m+1)}\ket{\boldsymbol{a},R}_{\boldsymbol{A}}.
\end{split}
\end{equation}
where the address walker states $\ket{\boldsymbol{a},R}_{\boldsymbol{A}}$ follows the encoding given in Eq.~(\ref{dualrailencoding}).

\gdr{It can be inconvenient to prepare the walker train by placing each particle directly on the rail associated with the carried information. An operational alternative is to encode the address and the data walkers' initial state in their color degree of freedom and convert that color encoding into a rail assignment afterwards.
We thus begin with all walkers on a single input rail, denoted by
$r=\mathrm{in}$ and located between the two protocol rails, with the logical address $\boldsymbol{a}$ encoded in the color of the address walker register $\boldsymbol{A}$, according to the following mapping:
\begin{equation}\label{inputrailencoding}
    a_i=0 \mapsto \ket{R}_{A_i},\quad a_i=1 \mapsto \ket{B}_{A_i}.
\end{equation}
Once the state is prepared, we transfer the information from the color degree of freedom to the rail degree of freedom, and simultaneously reset the color to the reference state $\ket{R}$. This conversion is performed by applying an additional scattering gate $\hat{S}^{(in)}$ placed between the input rail and the two protocol rails, as shown in Fig.~\ref{dual_rail_init}. This gate is the very same element introduced in Sec.~\ref{sec:S}: the input rail plays the role of the parent branch and the two protocol
rails that of its children. Concretely, the gate action is:
\begin{equation}\label{ScatterInputMapping}
\begin{split}
      &
\hat{S}^{(in)}\left(\ket{r=in, c=R}_{W_i}\right)=\ket{r=\uparrow, c=R}_{W_i},  \\ & \hat{S}^{(in)}\left(\ket{r=in, c=B}_{W_i}\right)=\ket{r=\downarrow, c=R}_{W_i}.
\end{split}   
\end{equation}
In other words, red walkers are deterministically sent to the upper rail $r=\uparrow$ (upper line of Eq.~(\ref{ScatterInputMapping})) and blue walkers to the lower one $r=\downarrow$ (lower line of Eq.~(\ref{ScatterInputMapping})), while resetting their color to $\ket{R}_{W_i}$.}

After this initialization procedure, the rail index remains fixed for the whole protocol, except during memory operations. The rail encodes the static logical information carried by a walker, and the color is again available as the dynamic routing degree of freedom for the subsequent routing operations. While this sequential initialization procedure greatly simplifies the experimental state preparation, it introduces a linear temporal overhead of $n+m+1$ steps, as the walkers must pass one-by-one through the input scattering gate.

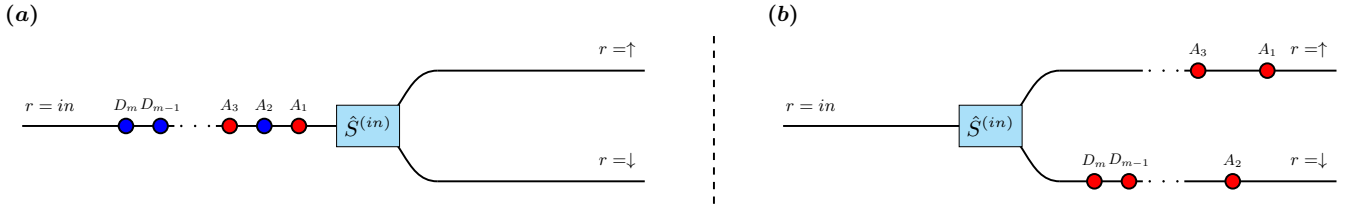
\begin{figure*}[htb]
\centering
        \resizebox{\linewidth}{!}{
     \begin{tikzpicture}
         \def\w{8}
  \def\h{1.5}
  \def\dr{3pt}
  \def\dist{0.3}
  
%%%%%%%% PRIMO SCOPE %%%%%%%
\begin{scope}[shift={(1,0)}]

% Parametri per coerenza
    \def\dotRadius{0.5pt}
% rail aux
\draw[thick] (7,-0.3) -- (4.8,-0.3);
\draw[thick] (2,-0.3) -- (4.2,-0.3);

    \foreach \x in {4.3, 4.5, 4.7} {
        \fill ( \x, -0.3) circle (\dotRadius);
    }

%branch S up
\draw[thick] (7,-0.3) to[out=0, in=180] (8,0.5);

%branch S down
\draw[thick] (7,-0.3) to[out=0, in=180] (8,-1.1);

\node[rectangle, draw, minimum size=6mm,fill=cyan!30] at (7,-0.3){$\hat{S}^{(in)}$};

% rail up
\draw[thick] (8,0.5) -- (11,0.5);
% new rail down
\draw[thick] (8,-1.1) -- (11,-1.1);

\draw[thick,fill=red] (6,-0.3) circle (\dr);
\draw[thick,fill=blue] (5.5,-0.3) circle (\dr);
\draw[thick,fill=red] (5,-0.3) circle (\dr);
\draw[thick,fill=blue] (4,-0.3) circle (\dr);
\draw[thick,fill=blue] (3.5,-0.3) circle (\dr);

 \node at (3.5,0) {{\scalebox{0.7}{$D_m$}}};
 \node at (4,0) {{\scalebox{0.7}{$D_{m-1}$}}};

 \node at (5,0) {{\scalebox{0.7}{$A_3$}}};
\node at (5.5,0) {{\scalebox{0.7}{$A_2$}}};
\node at (6,0) {{\scalebox{0.7}{$A_1$}}};

\node at (2.4,0) {{\scalebox{0.8}{$r=in$}}};
\node at (10.6,0.8) {{\scalebox{0.8}{$r=\uparrow$}}};
\node at (10.6,-0.8) {{\scalebox{0.8}{$r=\downarrow$}}};

\node at (2,1.3) {{\scalebox{1}{$\boldsymbol{(a)}$}}};

\end{scope}

\draw[thick, dashed](13,1)--(13,-1.5);

%%%%%%%% SECONDO SCOPE %%%%%%%
\begin{scope}[shift={(10,0)}]

% rail aux
\draw[thick] (7,-0.3) -- (4,-0.3);

%branch S up
\draw[thick] (7,-0.3) to[out=0, in=180] (8,0.5);

%branch S down
\draw[thick] (7,-0.3) to[out=0, in=180] (8,-1.1);

\node[rectangle, draw, minimum size=6mm,fill=cyan!30] at (7,-0.3){$\hat{S}^{(in)}$};

\def\dotRadius{0.5pt}
% rail up
\draw[thick] (8,0.5) -- (9.2,0.5);
\draw[thick] (9.8,0.5) -- (12,0.5);
  \foreach \x in {9.3, 9.5, 9.7} {
        \fill ( \x, 0.5) circle (\dotRadius);
    }
% new rail down
\draw[thick] (8,-1.1) -- (9.2,-1.1);
\draw[thick] (9.8,-1.1) -- (12,-1.1);
\foreach \x in {9.3, 9.5, 9.7} {
        \fill ( \x, -1.1) circle (\dotRadius);
    }

\draw[thick,fill=red] (11,0.5) circle (\dr);
\draw[thick,fill=red] (10.5,-1.1) circle (\dr);
\draw[thick,fill=red] (10,0.5) circle (\dr);
\draw[thick,fill=red] (9,-1.1) circle (\dr);
\draw[thick,fill=red] (8.5,-1.1) circle (\dr);

\node at (8.5,-0.8) {{\scalebox{0.7}{$D_m$}}};
\node at (9,-0.8) {{\scalebox{0.7}{$D_{m-1}$}}};

\node at (10,0.8) {{\scalebox{0.7}{$A_3$}}};
\node at (10.5,-0.8) {{\scalebox{0.7}{$A_2$}}};
\node at (11,0.8) {{\scalebox{0.7}{$A_1$}}};

\node at (4.4,0) {{\scalebox{0.8}{$r=in$}}};
\node at (11.6,0.8) {{\scalebox{0.8}{$r=\uparrow$}}};
\node at (11.6,-0.8) {{\scalebox{0.8}{$r=\downarrow$}}};

\node at (4,1.3) {{\scalebox{1}{$\boldsymbol{(b)}$}}};
\end{scope}
     \end{tikzpicture}
     }
     \caption{
     Initialization step for the \gdr{basic} dual-rail protocol. (a) Particles are initially placed on the input rail $r=in$, with the routing information encoded in their internal color states. (b) A scattering gate $\hat{S}^{(in)}$ is placed transversely to the routing trees and is used to sort the particles from the input rail onto the correct  $r=\uparrow$ or $r=\downarrow$ rails. Following this sorting step, the protocol proceeds according to the full architecture presented in Fig.~\ref{dualrailQRAM}. }
    \label{dual_rail_init}
\end{figure*}

\subsubsection{Routing operations}\label{DualRailRout}

The operational structure of the protocol remains the same as in the {single tree case with long range interactions,} 
relying on the two fundamental gates $\hat{S}$ and $\hat{U}^{(d)}$. {Crucially, however, these operations must now exhibit a behavior that is conditioned on the logical subspace occupied by the walker undergoing the operation.}

The address-transmission gate $\hat{U}^{(d)}$ is a multi-target unitary that acts on all walkers following the $d$-th address walker $A_d$, conditioned on its joint rail and color state $\ket{r,c}_{A_d}$. In the dual-rail architecture, this global operation naturally decomposes into two physically distinct gates, denoted by $\hat{U}^{(d, r)}$, applied independently to each rail $r\in\{\uparrow,\downarrow\}$. During the evolution, only the specific gate placed on the rail actually occupied by the control walker $A_d$ is activated, acting exclusively on the subsequent walkers traversing that same rail. As happens for the scattering gate $\hat{S}$, this spatial separation is perfectly equivalent to defining two independent operations acting directly on the two orthogonal logical subspaces $\mathcal{H}^{(\uparrow)}_W$ and $\mathcal{H}^{(\downarrow)}_W$, corresponding to the logical $0$ and $1$, respectively. 

More formally, we define the total gate acting on a given level $d$ in analogy with Eq.~(\ref{infoTransm}), that is as
\begin{equation}\label{InfoTransmDual}
\hat{U}^{(d)}=\sum_{r,c}\hat I_c^{(P_{d,r})}\otimes\hat{\Pi}_{r,c}^{(A_d)}\otimes \hat{V}^{\left(T_{d,r}\right)}_{c}.
\end{equation}
\gdr{Here, $P_{d,r}:=\{W_i:r_i=r,i<d\}$ represents the set of walkers on rail $r$ preceding $A_d$ while $T_{d,r}:=\{W_i:r_i=r,i>d\}$ denotes the corresponding set of target walkers.} Furthermore $$\{\hat{\Pi}_{r,c}^{(A_d)}:=\ket{r,c}_{A_d}\bra{r,c},r\in\{\uparrow,\downarrow\},c\in\{R,B\}\}$$ 
is the set of projectors onto the possible basis states of walker $A_d$, and $\hat{V}^{\left(T_{d,r}\right)}_c$ 
is a unitary operator applied to the target walkers according to the color of $A_d$. The form of this unitary is chosen as
\begin{equation*}\hat{V}^{\left(T_{d,r}\right)}_{c=R}=\bigotimes_{j\in T_{d,r}}\hat{X}_j,\quad \hat{V}^{\left(T_{d,r}\right)}_{c=B}=\bigotimes_{j\in T_{d,r}}\hat{I}_j,
\end{equation*}
Hence, analogously to the \gdr{basic} model presented above, when the address walker on rail $r$ is in state $\ket{r,R}_{A_d}$, the gate applies a color flip gate $\hat{X}$ to each following walker on that rail.
\gdr{The remaining terms are
required for unitarity but are never realized during the protocol: every address
walker reaches $\hat U^{(d)}$ in the reference color $R$, so that the control is
never blue, while on the rail not occupied by $A_d$ no control walker is present
at all and the gate acts trivially.}
%while if the address is $\ket{r,B}_{A_d}$ it does nothing.

In practice, this global gate can be split into the application of two distinct gates $\hat{U}^{(d, \uparrow)}$ and $\hat{U}^{(d, \downarrow)}$ applied in parallel on their respective rails. The gate $\hat{U}^{(d, \uparrow)}$ is placed on rail $r=\uparrow$ and acts only on particles whose state is in $\mathcal{H}_W^{(\uparrow)}$; on the other hand, the gate $\hat{U}^{(d, \downarrow)}$ placed on rail $r=\downarrow$ acts only on walkers whose state is in $\mathcal{H}_W^{(\downarrow)}$. The total operation can then be written as the direct sum of the two subspace-dependent operations:
\begin{equation*}
\hat{U}^{(d)}=\hat{U}^{(d, \uparrow)}\oplus\hat{U}^{(d, \downarrow)}
\end{equation*}
The operator $\hat{U}^{(d, \uparrow)}$ is then defined as
\begin{equation}
\hat{U}^{(d, \uparrow)}=\sum_{c\in\{R,B\}}\hat I_c^{(P_{d,\uparrow})}\otimes\hat{\Pi}_{\uparrow,c}^{(A_d)}\otimes\hat{V}_c^{\left(T_{d,\uparrow}\right)}
\end{equation}
where $\hat{V}_c^{\left(T_{d, \uparrow}\right)}$ is a multi-target unitary that acts on walkers in $T_{d, \uparrow}$, defined as
\begin{equation*}
\hat{V}_{c=R}^{\left(T_{d,\uparrow}\right)}=\bigotimes_{j\in T_{d, \uparrow}}\hat{X}_j,\quad \hat{V}_{c=B}^{\left(T_{d,\uparrow}\right)}=\bigotimes_{j\in T_{d, \uparrow}}\hat{I}_j.
\end{equation*}
In the exact same way, we define the operation $\hat{U}^{(d, \downarrow)}$ as
\begin{equation}
\hat{U}^{(d, \downarrow)}=\sum_{c\in\{R,B\}}\hat I_c^{(P_{d,\downarrow})}\otimes \hat{\Pi}_{\downarrow,c}^{(A_d)}\otimes\hat{V}_c^{\left(T_{d, \downarrow}\right)}
\end{equation}
with
\begin{equation*}
\hat{V}_{c=R}^{\left(T_{d, \downarrow}\right)}=\bigotimes_{j\in T_{d, \downarrow}}\hat{X}_j,\quad \hat{V}_{c=B}^{\left(T_{d,\downarrow}\right)}=\bigotimes_{j\in T_{d, \downarrow}}\hat{I}_j.
\end{equation*}

{In the \gdr{basic} setting, the} scattering gate $\hat{S}$ acts locally on single walkers, routing them conditionally according to their color state $\ket{c}_{W_i}$: if $c=R$, the walker is routed to the left and retains the color red; if $c=B$, the walker moves to the right and its color is reset to $c=R$ for the next step. 

In the dual-rail variant, scattering gates are placed on both rails, with their routing behavior depending on the logical subspace over which they act. In particular, a scattering gate placed on the lower rail $\hat S^{(\downarrow)}$ acts on the logical subspace $\mathcal{H}_W^{(\downarrow)}$, encoding a logical $1$, \gdr{in an analogous way to the action of the gate
$\hat S\equiv\hat S^{(l,d)}_{\rightarrow}$ introduced in Eq.~(\ref{bosonicS}),
whose node and direction labels we omit throughout this section.}
%in an analogous way to the action of the standard gate $\hat{S}$ introduced in Eq.~(\ref{bosonicS}). 
Conversely, a scattering gate placed on the upper rail $\hat{S}^{(\uparrow)}$ acts on the logical subspace $\mathcal{H}_W^{(\uparrow)}$, encoding a logical $0$, and routes the walkers in the exact {\it opposite} spatial direction relative to the \gdr{basic} protocol.
More formally, let $\ket{r,c^{l,d}}_{W_i}$ be the state of a walker $W_i$ moving on the rail $r\in\{\uparrow,\downarrow\}$ at position $(l,d)$. Each scattering
gate acts only on the states of the rail on which it is placed, and leaves the
rail label unchanged. \gdr{For the lower rail
\begin{equation}\label{ScattRail1}
    \begin{split}
    \hat{S}^{(\downarrow)} \ket{\downarrow,c^{l,d}}_{W_i}&=\ket{\downarrow}_{W_i}\otimes \left(\hat{S}\ket{c^{l,d}}_{W_i}\right)\\&=\ket{\downarrow}_{W_i}\otimes\ket{R^{2l-\delta_{c,R},d+1}}_{W_i},
    \end{split}
\end{equation}
where in the second equality we have used Eq.~(\ref{bosonicS}).
Analogously, on the upper rail
\begin{equation}\label{ScattRail0}
   \begin{split} \hat{S}^{(\uparrow)}\ket{\uparrow,c^{l,d}}_{W_i}&=\ket{\uparrow}_{W_i}\otimes \left(\hat{S}\hat{X}\ket{c^{l,d}}_{W_i}\right)\\&=\ket{\uparrow}_{W_i}\otimes\ket{R^{2l-\delta_{c,B},d+1}}_{W_i},
   \end{split}
\end{equation}
where $\hat{X}\ket{R}=\ket{B},\hat{X} \ket{B}=\ket{R}$.
Notice that $\hat{S}^{(\downarrow)}$ and $\hat{S}^{(\uparrow)}$ act
as $\hat S$ and $\hat S\hat X$ on their respective rails, thus they differ only by a local color operation.}
These scattering gates are strictly subspace preserving, thus we can rewrite the operator as a direct sum of operators acting on the two different logical subspaces
\begin{equation*}
\hat{S}^{(tot)}=\hat{S}^{(\uparrow)}\oplus \hat{S}^{(\downarrow)}
\end{equation*}
where $\hat{S}^{(\uparrow)}$ only acts on states in $\mathcal{H}_{W}^{(\uparrow)}$ and $\hat{S}^{(\downarrow)}$ only on states in $\mathcal{H}_{W}^{(\downarrow)}$, each of them vanishing on the complementary subspace.
%The explicit form of the operator can be taken as
%\begin{equation}
%\begin{split}  &\hat{S}^{(\uparrow)}=\ket{\uparrow,R^{2l,d+1}}\bra{\uparrow,R^{l,d}}+\ket{\uparrow,R^{2l-1,d+1}}\bra{\uparrow,B^{l,d}},\\& \hat{S}^{(\downarrow)}=\ket{\downarrow,R^{2l-1,d+1}}\bra{\downarrow,R^{l,d}}+\ket{\downarrow,R^{2l,d+1}}\bra{\downarrow,B^{l,d}}.   
%\end{split}
%\end{equation}
This behavior is essential to guarantee that data walkers are first routed to the correct memory cells and, subsequently, to the output node.

The direct-sum structure of the operations highlights a feature of the dual-rail architecture that is fundamental for the data particles routing: the two logical subspaces $\mathcal{H}^{(\uparrow)}_W$ and $\mathcal{H}^{(\downarrow)}_W$ evolve completely independently. All walkers of the data register $\boldsymbol{D}$ are assigned to rail $r=\downarrow$, which is fixed throughout the routing procedure, and initialized to color $c=R$. The routing of the data walkers is then determined entirely by whether the controlling address walker for level $d$ is on rail $r=\uparrow$ or $r=\downarrow$.

If the address walker $A_d$ at depth $d$ occupies the rail $a_d=r=\uparrow$, it will activate only the gate $\hat{U}^{(d, \uparrow)}$, which will not affect the color state of the data walkers: each particle of the data register remains in the color state $\ket{R}_{D_i}$ and, after the subsequent application of the scattering gate $\hat{S}^{(\downarrow)}$, they are routed from position $(l,d)$ to position $(2l-1,d+1)$. On the other hand, if the address walker is on rail $a_d=r=\downarrow$, the gate $\hat{U}^{(d, \downarrow)}$ is activated, flipping the color of each subsequent walker on rail $r=\downarrow$ from $\ket{R}$ to $\ket{B}$. After this color switch, the data walkers are routed from level $(l,d)$ to level $(2l,d+1)$ by the subsequent scattering gate $\hat{S}^{(\downarrow)}$.
%An address walker on rail $r=\uparrow$ leaves the data red and therefore sends them left, while an address on rail $r=\downarrow$ turns the data blue and therefore sends them right. 
Address walkers placed on rail $r=\uparrow$ therefore play a passive, structural role regarding the routing of data register, since they preserve train ordering but do not trigger data routing, whereas address walkers on rail $r=\downarrow$ perform the active control function.

While the upper rail might appear merely as an ancillary component to ensure particle-number conservation, this dual-rail encoding can actually be understood as the system running two parallel, independent \gdr{basic} routing protocols. As described in Section~\ref{sec:tech_desc}, the protocols are carried out using the usual degrees of freedom, i.e., internal color and empty state, and can be seen as each being the ``photographic negative" of the other. For a given logical address string $\boldsymbol{a}=a_n\dots a_1\in\{0,1\}^n$, e.g., $\boldsymbol{a}=101$, the lower rail physically processes the sequence corresponding to the $1$s, i.e., in our example $\ket{R\varnothing R}$, while the upper rail processes the exact bit-wise NOT of that string, i.e., the sequence physically corresponding to the $0$s, $\ket{\varnothing R\varnothing}$. 

To maintain the consistency of the protocol on both rails and ensure that the address walkers on the upper rail correctly separate from the main train at the right depths $d$, exactly as their counterparts do on the lower rail, the operations on $\mathcal{H}_W^{(\uparrow)}$ must be the mirrored version of those on $\mathcal{H}_W^{(\downarrow)}$. This is precisely why the scattering gate $\hat{S}^{(\uparrow)}$ acts with inverted routing rules compared to $\hat{S}^{(\downarrow)}$. 
Crucially, maintaining this structural consistency is essential for the output phase of the protocol. Since the data walkers, as we show below, move to both rails after interacting with the memory cells, the address walkers on $r=\uparrow$ must have been perfectly diverted at the correct nodes during the input phase to be re-collected in the exact reverse order. By doing so, they can actively guide the data walkers on the upper rail back through the output tree. The two parallel protocols operate independently of one another, processing complementary physical bit strings, and only interact at the very end of the circuit, where the dual-rail paths merge at the target memory cells to recover the full logical information.

\begin{figure*}[htb]
\begin{subfigure}[c]{0.48\linewidth}
        \centering
     \resizebox{0.9\linewidth}{!}{
    \begin{tikzpicture}[x=1cm,y=1cm, scale=0.9]
  \def\w{8}
  \def\h{1.5}
  \def\dr{3pt}
  \def\dist{0.3}
    % Rettangolo
    \draw[rounded corners, thick, fill=gray!30] (1,-2) rectangle ({\w+4.5},\h+1.5);

% Rettangolo gate U_copy
   \node[draw, rounded corners, fill=blue!20, minimum width=45mm, minimum height=20mm, opacity=0.9] at (3.6,-0.75) {};
   
    % linee pallini memoria
    \draw[thick] (1.5,-\h/2-0.5) -- (5.75,-\h/2-0.5);

 % Linee verticali dai pallini sopra a quelli sotto
    \foreach \i in {1,...,4} {
      \pgfmathsetmacro{\x}{\i/5*(\w*0.9)}
      \draw[line width=1pt] (\x, -\h/2-0.5) -- (\x, -\dist);
    }

% Pallini memoria
\foreach[count=\k from 1] \col in {red, blue, red,} {
        \pgfmathsetmacro{\x}{(\k)/5*(\w*0.9)}  % \n non definito, quindi uso 6 fisso coerente con sopra
        \filldraw[fill=\col, draw=black, thick] (\x, {-\h/2-0.5}) circle [radius=\dr];
      }

    % rail down
    \draw[thick] (0,-\dist) -- (\w-0.7,-\dist);

 % rail up
    \draw[thick] (0,2.1) -- (8.5,2.1);

   % Pallini rail down
    \foreach[count=\k from 1] \col in { red, red,red,red} {
        \pgfmathsetmacro{\x}{(\k)/5*(\w*0.9)}
        \pgfmathsetmacro{\y}{-\dist}
        \filldraw[fill=\col, draw=black, thick] (\x, \y) circle [radius=\dr];
      }

%%% CIRCUITO DI OUTPUT %%%%

% Rettangolo output circuit
   \node[draw, dashed, rounded corners, fill=pink!20, minimum width=50mm, minimum height=40mm, opacity=0.9] at (9,0.6) {};

% rail up in
\draw[thick] (6,2.1) -- (8.5,2.1);

% rail down in
\draw[thick] (7,-0.3) -- (6,-0.3);

%branch S up
\draw[thick] (7,-0.3) to[out=0, in=180] (8,0.5);

%branch S down
\draw[thick] (7,-0.3) to[out=0, in=180] (8,-1.1);

\node[rectangle, draw, minimum size=6mm,fill=cyan!30] at (7,-0.3){$\hat{S}_M$};

% rail aux
\draw[thick] (8,0.5) -- (8.5,0.5);

%branch S_dag down
\draw[thick] (8.5,0.5) to[out=0, in=180] (9.5,1.3);

%branch S_dag up
\draw[thick] (8.5,2.1) to[out=0, in=180] (9.5,1.3);

% new rail down
\draw[thick] (8,-1.1) -- (13.5,-1.1);

% new rail up
    \draw[thick] (9.5,1.3) -- (13.5,1.3);

\node[rectangle, draw, minimum size=6mm,fill=cyan!30] at (9.5,1.3){$\hat{S}^{\dagger}_M$};

\node[rectangle, draw, minimum size=6mm,fill=green!30] at (11,1.3){$\hat{X}_c$};

%%%%%%%%  LABEL DELLE COSE %%%%%%%
   % Label memorie
    \node at (1.5,-1.6) {{\scalebox{0.8}{$M_{3}^{(\boldsymbol{a})}$}}};

    \node at (2.95,-1.6) {{\scalebox{0.8}{$M_{2}^{(\boldsymbol{a})}$}}};
    
    \node at (4.4,-1.6) {{\scalebox{0.8}{$M_{1}^{(\boldsymbol{a})}$}}};

    % Label pallini rail down

    \node at (1.45,0) {{\scalebox{0.8}{$D_3$}}};

    \node at (2.9,0) {{\scalebox{0.8}{$D_2$}}};

    \node at (4.35,0) {{\scalebox{0.8}{$D_1$}}};

    \node at (5.8,0) {{\scalebox{0.8}{$D_0$}}};

%label gate

\node at (5.5,-1.6) {{\scalebox{0.8}{$\hat{U}_{\text{copy}}$}}};

%label rails
\node at (0.3,-0.6) {$r=\downarrow$};

\node at (0.3,1.8) {$r=\uparrow$};

\node at (8,0.7) {$r=aux$};

\node at (9,2.5) {Output circuit};

\node at (0,3.5) {{\scalebox{1.5}{$\boldsymbol{(a)}$}}};
\end{tikzpicture}
}
\end{subfigure}\hfill\vrule\hfill
\begin{subfigure}[c]{0.48\linewidth}
        \centering
        \resizebox{\linewidth}{!}{
\begin{tikzpicture}
 \def\w{8}
  \def\h{1.5}
  \def\dr{3pt}
  \def\dist{0.3}

\draw[dashed](14.75,-6.5)--(14.75,2.5);

\draw[dashed](6.5,-2)--(22.5,-2);
  
%%%%%%%% PRIMO SCOPE %%%%%%%
\begin{scope}[shift={(1,0)}]
   % Rettangolo output circuit
   \node[draw, dashed, rounded corners, fill=pink!20, minimum width=55mm, minimum height=40mm, opacity=0.9] at (9.25,0.6) {};

% rail up in
\draw[thick] (5,2.1) -- (8.5,2.1);

% rail down in
\draw[thick] (7,-0.3) -- (5,-0.3);

%branch S up
\draw[thick] (7,-0.3) to[out=0, in=180] (8,0.5);

%branch S down
\draw[thick] (7,-0.3) to[out=0, in=180] (8,-1.1);

\node[rectangle, draw, minimum size=6mm,fill=cyan!30] at (7,-0.3){$\hat{S}_M$};

% rail aux
\draw[thick] (8,0.5) -- (8.5,0.5);

%branch S_dag down
\draw[thick] (8.5,0.5) to[out=0, in=180] (9.5,1.3);

%branch S_dag up
\draw[thick] (8.5,2.1) to[out=0, in=180] (9.5,1.3);

% new rail down
\draw[thick] (8,-1.1) -- (13.5,-1.1);

% new rail up
    \draw[thick] (9.5,1.3) -- (13.5,1.3);

\node[rectangle, draw, minimum size=6mm,fill=cyan!30] at (9.5,1.3){$\hat{S}^{\dagger}_M$};

\node[rectangle, draw, minimum size=6mm,fill=green!30] at (11.5,1.3){$\hat{X}_c$};

\draw[thick,fill=red] (6.25,-0.3) circle (\dr);
\draw[thick,fill=blue] (5.95,-0.3) circle (\dr);
\draw[thick,fill=red] (5.65,-0.3) circle (\dr);
\draw[thick,fill=blue] (5.35,-0.3) circle (\dr);

 \node at (5.35,-0.6) {{\scalebox{0.8}{$D_3$}}};
 \node at (5.65,0) {{\scalebox{0.8}{$D_2$}}};
\node at (5.95,-0.6) {{\scalebox{0.8}{$D_1$}}};
\node at (6.25,0) {{\scalebox{0.8}{$D_0$}}};

\end{scope}

%%%%%% SECONDO SCOPE %%%%%%%
\begin{scope}[shift={(10,0)}]
   % Rettangolo output circuit
      \node[draw, dashed, rounded corners, fill=pink!20, minimum width=55mm, minimum height=40mm, opacity=0.9] at (9.25,0.6) {};

% rail up in
\draw[thick] (5,2.1) -- (8.5,2.1);

% rail down in
\draw[thick] (7,-0.3) -- (5,-0.3);

%branch S up
\draw[thick] (7,-0.3) to[out=0, in=180] (8,0.5);

%branch S down
\draw[thick] (7,-0.3) to[out=0, in=180] (8,-1.1);

\node[rectangle, draw, minimum size=6mm,fill=cyan!30] at (7,-0.3){$\hat{S}_M$};

% rail aux
\draw[thick] (8,0.5) -- (8.5,0.5);

%branch S_dag down
\draw[thick] (8.5,0.5) to[out=0, in=180] (9.5,1.3);

%branch S_dag up
\draw[thick] (8.5,2.1) to[out=0, in=180] (9.5,1.3);

% new rail down
\draw[thick] (8,-1.1) -- (13.5,-1.1);

% new rail up
    \draw[thick] (9.5,1.3) -- (13.5,1.3);

\node[rectangle, draw, minimum size=6mm,fill=cyan!30] at (9.5,1.3){$\hat{S}^{\dagger}_M$};

\node[rectangle, draw, minimum size=6mm,fill=green!30] at (11.5,1.3){$\hat{X}_c$};

\draw[thick,fill=red] (8.8,-1.1) circle (\dr);
\draw[thick,fill=red] (8.5,0.5) circle (\dr);
\draw[thick,fill=red] (8.2,-1.1) circle (\dr);
\draw[thick,fill=red] (7.9,0.5) circle (\dr);

\node at (7.9,0.2) {{\scalebox{0.8}{$D_3$}}};
\node at (8.2,-0.8) {{\scalebox{0.8}{$D_2$}}};
\node at (8.5,0.2) {{\scalebox{0.8}{$D_1$}}};
\node at (8.8,-0.8) {{\scalebox{0.8}{$D_0$}}};

\end{scope}
%%%% TERZO SCOPE %%%%%%%%%
\begin{scope}[shift={(1,-5)}]
   % Rettangolo output circuit
      \node[draw, dashed, rounded corners, fill=pink!20, minimum width=55mm, minimum height=40mm, opacity=0.9] at (9.25,0.6) {};

% rail up in
\draw[thick] (5,2.1) -- (8.5,2.1);

% rail down in
\draw[thick] (7,-0.3) -- (5,-0.3);

%branch S up
\draw[thick] (7,-0.3) to[out=0, in=180] (8,0.5);

%branch S down
\draw[thick] (7,-0.3) to[out=0, in=180] (8,-1.1);

\node[rectangle, draw, minimum size=6mm,fill=cyan!30] at (7,-0.3){$\hat{S}_M$};

% rail aux
\draw[thick] (8,0.5) -- (8.5,0.5);

%branch S_dag down
\draw[thick] (8.5,0.5) to[out=0, in=180] (9.5,1.3);

%branch S_dag up
\draw[thick] (8.5,2.1) to[out=0, in=180] (9.5,1.3);

% new rail down
\draw[thick] (8,-1.1) -- (13.5,-1.1);

% new rail up
    \draw[thick] (9.5,1.3) -- (13.5,1.3);

\node[rectangle, draw, minimum size=6mm,fill=cyan!30] at (9.5,1.3){$\hat{S}^{\dagger}_M$};

\node[rectangle, draw, minimum size=6mm,fill=green!30] at (11.5,1.3){$\hat{X}_c$}; 

\draw[thick,fill=red] (11,-1.1) circle (\dr);
\draw[thick,fill=blue] (10.7,1.3) circle (\dr);
\draw[thick,fill=red] (10.4,-1.1) circle (\dr);
\draw[thick,fill=blue] (10.1,1.3) circle (\dr);

\node at (10.1,1) {{\scalebox{0.8}{$D_3$}}};
\node at (10.4,-0.8) {{\scalebox{0.8}{$D_2$}}};
\node at (10.7,1) {{\scalebox{0.8}{$D_1$}}};
\node at (11,-0.8) {{\scalebox{0.8}{$D_0$}}};

\end{scope}

%%%%%%%% QUARTO SCOPE %%%%%%%%
\begin{scope}[shift={(10,-5)}]
   % Rettangolo output circuit
      \node[draw, dashed, rounded corners, fill=pink!20, minimum width=55mm, minimum height=40mm, opacity=0.9] at (9.25,0.6) {};
      
% rail up in
\draw[thick] (5,2.1) -- (8.5,2.1);

% rail down in
\draw[thick] (7,-0.3) -- (5,-0.3);

%branch S up
\draw[thick] (7,-0.3) to[out=0, in=180] (8,0.5);

%branch S down
\draw[thick] (7,-0.3) to[out=0, in=180] (8,-1.1);

\node[rectangle, draw, minimum size=6mm,fill=cyan!30] at (7,-0.3){$\hat{S}_M$};

% rail aux
\draw[thick] (8,0.5) -- (8.5,0.5);

%branch S_dag down
\draw[thick] (8.5,0.5) to[out=0, in=180] (9.5,1.3);

%branch S_dag up
\draw[thick] (8.5,2.1) to[out=0, in=180] (9.5,1.3);

% new rail down
\draw[thick] (8,-1.1) -- (13.5,-1.1);

% new rail up
    \draw[thick] (9.5,1.3) -- (13.5,1.3);

\node[rectangle, draw, minimum size=6mm,fill=cyan!30] at (9.5,1.3){$\hat{S}^{\dagger}_M$};

\node[rectangle, draw, minimum size=6mm,fill=green!30] at (11.5,1.3){$\hat{X}_c$};

\draw[thick,fill=red] (13.2,-1.1) circle (\dr);
\draw[thick,fill=red] (12.9,1.3) circle (\dr);
\draw[thick,fill=red] (12.6,-1.1) circle (\dr);
\draw[thick,fill=red] (12.3,1.3) circle (\dr);

\node at (12.3,1) {{\scalebox{0.8}{$D_3$}}};
\node at (12.6,-0.8) {{\scalebox{0.8}{$D_2$}}};
\node at (12.9,1) {{\scalebox{0.8}{$D_1$}}};
\node at (13.2,-0.8) {{\scalebox{0.8}{$D_0$}}};
\end{scope}

\node at (7,3.5) {{\scalebox{1.5}{$\boldsymbol{(b)}$}}};
    \end{tikzpicture}
    }
\end{subfigure}
\caption{Schematic of the memory cell operation in the dual-rail protocol. (a) A global copy operation $\hat{U}_{copy}$ transfers information from the memory walkers to the data walkers color states, according to Eq.~(\ref{dual_rail_copy_mapping}). (b) The output circuit redirects data walkers based on their color: a scattering gate $\hat{S}_M$ routes particles encoding a $1$ to the lower rail, while $\hat{S}^{\dagger}_M$ routes the remaining particles to the upper rail. Finally, a local gate $\hat{X}_C$ restores the correct color encoding.}
    \label{copy_dualrail}
\end{figure*}
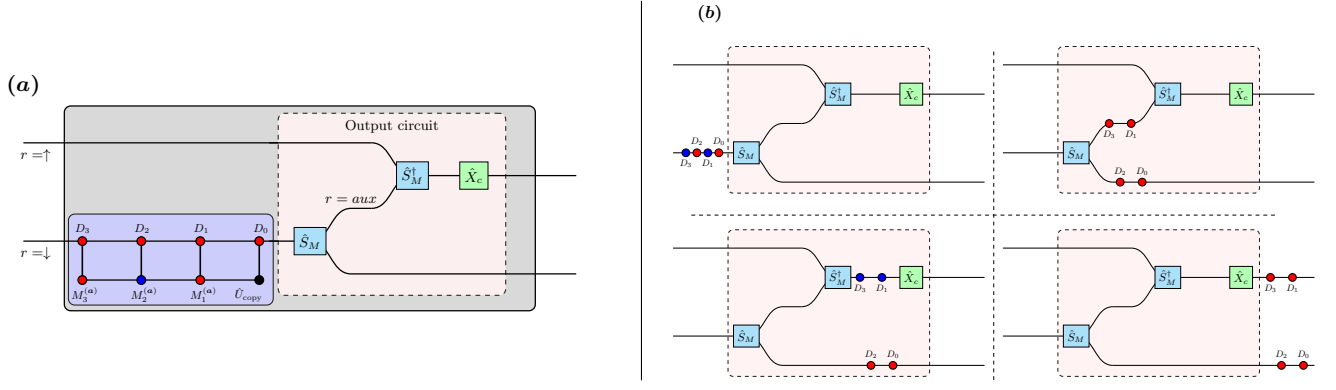

\subsubsection{Memory operations}\label{dual_rail_mem}

In the dual-rail implementation, memory operations rely on logical primitives similar to those introduced for the {single tree}
protocol. The main difference lies in how the copied message is encoded spatially across the rails after the copy stage. At a given leaf of the tree indexed by address $\boldsymbol{a}$, we have a memory cell $\boldsymbol{M}^{(\boldsymbol{a})}$ storing a classical $m$-bit message $\boldsymbol{b}^{(\boldsymbol{a})}\in\{0,1\}^m$. This message is physically encoded in the internal color degree of freedom of a fixed, $m$-\gdr{bit} memory register. 

By design, incoming data walkers always arrive at the memory cells via the lower rail $r=\downarrow$. 
%Thus, we can assume that the geometry of the cell is such that the two routing rails $(\uparrow,\downarrow)$ lie on two parallel longitudinal planes, while the memory register is placed along the lower rail, in a way such that it can directly interact with the walkers incoming through this $r=\downarrow$ rail through short-range operations. 
After the interaction with the memory register, a crucial step is to route the data walkers onto the correct rail according to the extracted bit, ensuring they all exit the memory cell in the fiduciary color state $\ket{R}$, ready for inverse routing. Specifically, we encode the memory cell information as it follows: a bit equal to logical $1$ is mapped into a data walker travelling in the lower rail, namely $r=\downarrow$, while, a bit equal to logical $0$ to the upper rail, namely $r=\uparrow$.

%{\color{orange} To engineer this movement, one must take into account that a fundamental constraint of the dual-rail architecture is ensuring that the address walkers arrive at the memory cells in the color state $\ket{R}$ and pass through the cells without shifting rails. To guarantee this, the passive scattering gates at the leaves must be engineered such that the $\ket{R}$ state always moves in a straight trajectory. Consequently, any rail-switching of the data walkers must be actively triggered by flipping their color to $\ket{B}$ only when a rail change is required.}
Since data walkers arrive at the memory cells from the lower rail $r=\downarrow$ in state $\ket{R}$ and we want a logical $1$ to remain on $r=\downarrow$. The copy operation must leave the data walker's color unchanged when reading a memory bit equal to $1$, and flip it to $\ket{B}$ when reading a memory bit equal to $0$, triggering the rail switch to $r=\uparrow$ for the target data walker where the information has to be copied. As we can see in Fig.~\ref{copy_dualrail}, this is the first step of the memory operation needed to route data walkers accordingly to the adopted dual-rail encoding of the information copied from the memory cell to the data register.

Assuming that the logical $0$ of the memory is encoded as $\ket{R}_{M}$ and the logical $1$ as $\ket{B}_{M}$, the local copy unitary $\hat{U}_{copy}^{loc}$ must act as a controlled not activated by color $\ket{R}$. The target mapping is:
\begin{equation}\label{dual_rail_copy_mapping}
\begin{split}
&\ket{R}_{D_j}\ket{m_j=B}_{M_j^{(\boldsymbol{a})}}\xrightarrow{\mathrm{copy}}\ket{R}_{D_j}\ket{m_j=B}_{M_j^{(\boldsymbol{a})}}\\&
\ket{R}_{D_j}\ket{m_j=R}_{M_j^{(\boldsymbol{a})}}\xrightarrow{\mathrm{copy}}\ket{B}_{D_j}\ket{m_j=R}_{M_j^{(\boldsymbol{a})}}
\end{split}
\end{equation}
where we have omitted the rail degree of freedom, which at this stage is always equal to $r=\downarrow$.
As discussed in Sec.~\ref{SecMemoStandard}, we can ensure that the copy operations take place on the correct targeted memory cells, selected by the input address, either through the use of a global copy gate $\hat{U}^{(\boldsymbol{a})}_{copy}$ triggered by an auxiliary flag walker $D_0$, or via an internal switch degree of freedom. Its explicit form is given by:
\begin{equation}
\label{localmemorydualrail}
\hat{U}_{copy}^{loc}(D_i\vert M_i^{(\boldsymbol{a})}):=\ket{R}_{M_i^{(\boldsymbol{a})}}\bra{R}\otimes \hat{X}_{D_i}+\ket{B}_{M_i^{(\boldsymbol{a})}}\bra{B}\otimes \hat{I}_{D_i}
\end{equation}
where again, we omit the rail degree of freedom for clarity. The operation acts strictly on the color degree of freedom of the walkers, flipping the data walker to $\ket{B}_{D_i}$ only if the corresponding memory qubit encodes a logical 0. Notice that the form of $\hat{U}_{copy}^{loc}$ is explicitly unitary and preserves the memory registers.
Finally, since these local operations act on disjoint pairs of walkers, they can be applied simultaneously, yielding an $\mathcal{O}(1)$ circuit depth for the copy phase.

Finally, we detail the sub-circuit responsible for the spatial routing, comprised of two scattering gates $\hat{S}_M$, an auxiliary rail $r=aux$, and a local clocked color-flip gate $\hat{X}_c$, as depicted in Fig.\ref{copy_dualrail}. The final goal of this sequence of gates is to route red walkers $\ket{R}$ coming out from the memory cell copy operations to the lower binary tree and blue walkers to the upper tree.
A scattering gate $\hat S_M$ is applied to the lower rail $r=\downarrow$, routing the data particles with the newly encoded message in the following way: data particles $D_i$ in state $\ket{R}_{D_i}$ are routed on the lower rail; conversely walkers in state $\ket{B}_{D_i}$ are routed onto the auxiliary rail $r=aux$, and their color is reset to $R$. The walkers diverted onto the auxiliary rail $r=aux$ must now be routed toward the upper rail. This operation can be executed with a second scattering gate $\hat{S}^{\dagger}_M$ having both $r=\uparrow$ and $r=aux$ as input rails. A walker coming from $r=\uparrow$ with color $\ket{R}$ is routed into the output rail retaining its color $R$, while a walker with color $\ket{R}$ coming from the auxiliary rail $r=aux$ is directed to the output rail, with its color changed to $\ket{B}$. The output node is the $r=\uparrow$ rail belonging to the inverted routing phase. 

More formally, the action of these gates are defined as 
\begin{equation*}
    \begin{split}
&\hat{S}_M\ket{\downarrow,R}=\ket{\downarrow,R},\quad \hat{S}_M\ket{\downarrow,B}=\ket{aux,R};\\&
\hat{S}_M^{\dagger}\ket{\uparrow,R}=\ket{\uparrow,R},\quad\hat{S}_M^{\dagger}\ket{aux,R}=\ket{\uparrow,B}.
\end{split}
\end{equation*}

This sequence of gates guarantees the target encoding of the memory cell information, which dictates the specific rail where each data walker travels. As a final step, in order to restore the state $\ket{R}$ for the data walkers joining the upper rail a local color $\hat X_c$ gate is applied on the upper rail $r=\uparrow$. This gate is time-dependent: it is triggered by an external clock exclusively during the time slots designated for the passage of these data walkers.

\subsubsection{Resource scaling}
In this section, we evaluate the hardware resources required for the dual-rail implementation of our protocol, focusing on walker count, gate complexity, spatial extent, and circuit depth. 
The dual-rail variant operates with a fixed number of walkers. Specifically, it employs an $n$-walker address register $\boldsymbol{A}$ and an $(m+1)$-walker data register $\boldsymbol{D}$, resulting in a total of $n+m+1 \sim \mathcal{O}(n+m)$ walkers. 

The routing architecture consists of two parallel binary trees of depth $n$, where
each bifurcation is associated with a unitary block comprising the same $\hat{S}$
and $\hat{U}^{(d)}$ gates. Since the structure effectively duplicates the
single-tree architecture, the total gate count is doubled, yielding
$4(2^n-1)\sim\mathcal{O}(2^n)$ stationary routing gates.
Similar to the \gdr{basic} protocol, the reliance on long-range interactions dictates both the spatial resources and the execution time. 

Regarding the spatial extent, the dual-rail architecture employs a dual-layer geometry with two parallel trees. While this physically doubles the global hardware, the maximum number of walkers traversing a specific branch remains unchanged. Consequently, the required path extent is identical to the \gdr{basic} protocol, yielding $n_{sites}^{tot}= (n^2+2nm-n)/2\sim\mathcal{O}(n^2+nm)$.

Furthermore, we evaluate the total circuit depth $T_{serial}$. The routing process requires the same number of steps $T_{rout}$ as the \gdr{basic} protocol, to which we must add the sequential overhead $T_{prep}$ for the initial state preparation via the input scattering gate. This yields a total exact depth of $T_{serial}=T_{prep}+T_{rout}=%[n(n+m)-\frac{n(n+1)}{2}] + (n+m)
(n^2+2nm+n+2m)/2
\sim\mathcal{O}(n^2+nm)$. Notice that this linear overhead is absorbed by the overall quadratic depth of the protocol.

Finally, the number of dynamically active elements per query is only modified by a constant factor (due to the presence of the parallel trees), remaining $\mathcal{O}(n)$ for a classical or sparse query, and scaling up to $\mathcal{O}(2^n)$ for a dense superposition of addresses.

\subsection{Backup Variant for dual-rail implementation}
\label{subsec:backup_dual_rail}

The {address-transmission gates of the dual-rail protocol need long-range couplings between walkers for their implementation}. {It is then natural to explore whether the backup scheme introduced in Section~\ref{sec:backup} can be adapted} to this setting to reduce noise and complexity requirements. A first naive attempt {to adapt the scheme to this variant} is to attach a single backup walker to each logical carrier and place all backup {walkers} on the same rail, for instance, the rail used by {the} data walker register $\boldsymbol{D}$.
This scheme allows for a short-range propagation of path information whenever an address carries a logical $1$, i.e. when it is placed on rail $r=\downarrow$, since only the gates on that rail act on the data train as explained in Sec.~\ref{DualRailRout}. 

However, this arrangement fails at the output stage. At that point, the copied message is encoded in the rail position of the data walkers, which must be routed toward the exit node during the output phase. Depending on the particular information retrieved from the memory cells, data walkers are routed
onto different rails, as shown in Fig.~\ref{copy_dualrail}. Hence, gaps would inevitably appear in the sequence of backup walkers. Since the protocol strictly relies on the presence of backups between successive applications of the block gate $\hat{U}_B$, this configuration would fail to coherently recollect all data walkers at the output node.

A simple and robust fix is to provide two backups per logical carrier, one initialized on each rail. With this choice, each logical carrier is accompanied by a backup on rail $r=\uparrow$ and a backup on rail $r=\downarrow$. In this way, whichever rail the logical walker occupies, there is always a corresponding backup that can be used to propagate the path information locally. That allows local applications of the two gates $\hat{U}_{in}$ and $\hat{U}_B$
separately on each rail, while also preserving the ability to recombine and recollect walkers at the output, keeping all interactions short-range. The cost is a multiplicative overhead in resources that do not change the scaling with respect to the \gdr{basic} protocol, while being the natural analogue of the backup strategy in the dual-rail geometry.

\begin{figure*}[t]
\centering
        \resizebox{\linewidth}{!}{
     \begin{tikzpicture}
         \def\w{8}
  \def\h{1.5}
  \def\dr{3pt}
  \def\dist{0.3}
  
%%%%%%%% PRIMO SCOPE %%%%%%%
\begin{scope}[shift={(1,0)}]

% Parametri per coerenza
    \def\dotRadius{0.5pt}
% rail aux
\draw[thick] (7,-0.3) -- (3.3,-0.3);
%\draw[thick] (2,-0.3) -- (2.7,-0.3);

    \foreach \x in {2.8, 3, 3.2} {
        \fill ( \x, -0.3) circle (\dotRadius);
    }

\node[rectangle, draw, minimum width=21mm,minimum height=25mm,rounded corners, dashed, fill=orange!20] at (8.75,-0.1){};

%branch S up
\draw[thick] (7,-0.3) to[out=0, in=180] (8,0.5);

%branch S down
\draw[thick] (7,-0.3) to[out=0, in=180] (8,-1.1);

\node[rectangle, draw, minimum size=6mm,fill=cyan!30] at (7,-0.3){$\hat{S}^{(in)}$};

% rail up
\draw[thick] (8,0.5) -- (11,0.5);
% new rail down
\draw[thick] (8,-1.1) -- (11,-1.1);

\draw[-stealth,thick] (4,-0.8)--(6,-0.8);

% walkers
\draw[thick,fill=red] (6.3,-0.3) circle (\dr);
\draw[thick,fill=red] (5.8,-0.3) circle (\dr);
\draw[thick,fill=blue] (5.3,-0.3) circle (\dr);
\draw[thick,fill=blue] (4.8,-0.3) circle (\dr);
\draw[thick,fill=red] (4.3,-0.3) circle (\dr);
\draw[thick,fill=blue] (3.8,-0.3) circle (\dr);

%labels
\node at (3.8,0) {{\scalebox{0.7}{$\tilde A_2^{\downarrow}$}}};
\node at (4.3,0) {{\scalebox{0.7}{$\tilde A_2^{\uparrow}$}}};
\node at (4.8,0) {{\scalebox{0.7}{$A_2$}}};
\node at (5.3,0) {{\scalebox{0.7}{$\tilde A_1^{\downarrow}$}}};
\node at (5.8,0) {{\scalebox{0.7}{$\tilde A_1^{\uparrow}$}}};
\node at (6.3,0) {{\scalebox{0.7}{$A_1$}}};

\node at (3,0) {{\scalebox{0.8}{$r=in$}}};
\node at (10.6,0.8) {{\scalebox{0.8}{$r=\uparrow$}}};
\node at (10.6,-0.8) {{\scalebox{0.8}{$r=\downarrow$}}};
\node at (8.7,0.9) {{\scalebox{0.8}{Buffer area}}};

\draw[decorate, decoration={brace, amplitude=5pt}, thick] (5.2,0.2) -- (6.4,0.2)node[midway, above=6pt, font=\scriptsize] {$BRa_1$};
\draw[decorate, decoration={brace, amplitude=5pt}, thick] (3.7,0.2) -- (4.9,0.2)node[midway, above=6pt, font=\scriptsize] {$BRa_2$};

\node at (3,1.3) {{\scalebox{1}{$\boldsymbol{(a)}$}}};
\end{scope}

%%%%%%%% SECONDO SCOPE %%%%%%%
\begin{scope}[shift={(10,0)}]

% Parametri per coerenza
    \def\dotRadius{0.5pt}
% rail aux
\draw[thick] (7,-0.3) -- (3.3,-0.3);
%\draw[thick] (2,-0.3) -- (2.7,-0.3);

    \foreach \x in {2.8, 3, 3.2} {
        \fill ( \x, -0.3) circle (\dotRadius);
    }

\node[rectangle, draw, minimum width=21mm,minimum height=25mm,rounded corners, dashed, fill=orange!20] at (8.75,-0.1){};

%branch S up
\draw[thick] (7,-0.3) to[out=0, in=180] (8,0.5);

%branch S down
\draw[thick] (7,-0.3) to[out=0, in=180] (8,-1.1);

\node[rectangle, draw, minimum size=6mm,fill=cyan!30] at (7,-0.3){$\hat{S}^{(in)}$};

% rail up
\draw[thick] (8,0.5) -- (11,0.5);
% new rail down
\draw[thick] (8,-1.1) -- (11,-1.1);

% walkers
\draw[thick,fill=red] (9,0.5) circle (\dr);
\draw[thick,fill=red] (8.5,0.5) circle (\dr);
\draw[thick,fill=red] (8,-1.1) circle (\dr);
\draw[thick,fill=blue] (6.3,-0.3) circle (\dr);
\draw[thick,fill=red] (5.8,-0.3) circle (\dr);
\draw[thick,fill=blue] (5.3,-0.3) circle (\dr);

%labels
\node at (5.3,0) {{\scalebox{0.7}{$\tilde A_2^{\downarrow}$}}};
\node at (5.8,0) {{\scalebox{0.7}{$\tilde A_2^{\uparrow}$}}};
\node at (6.3,0) {{\scalebox{0.7}{$A_2$}}};
\node at (8,-0.8) {{\scalebox{0.7}{$\tilde A_1^{\downarrow}$}}};
\node at (8.5,0.2) {{\scalebox{0.7}{$\tilde A_1^{\uparrow}$}}};
\node at (9,0.2) {{\scalebox{0.7}{$A_1$}}};

\node at (3,0) {{\scalebox{0.8}{$r=in$}}};
\node at (10.6,0.8) {{\scalebox{0.8}{$r=\uparrow$}}};
\node at (10.6,-0.8) {{\scalebox{0.8}{$r=\downarrow$}}};
\node at (8.7,0.9) {{\scalebox{0.8}{Buffer area}}};

\draw[thick, red, densely dashdotted](9.2,0.7)--(9.2,-1.3);

\node at (3,1.3) {{\scalebox{1}{$\boldsymbol{(b)}$}}};
\end{scope}

%%%% Terzo scope
\begin{scope}[shift={(1,-4)}]

% Parametri per coerenza
    \def\dotRadius{0.5pt}
% rail aux
\draw[thick] (7,-0.3) -- (3.3,-0.3);
%\draw[thick] (2,-0.3) -- (2.7,-0.3);

    \foreach \x in {2.8, 3, 3.2} {
        \fill ( \x, -0.3) circle (\dotRadius);
    }

\node[rectangle, draw, minimum width=21mm,minimum height=25mm,rounded corners, dashed, fill=orange!20] at (8.75,-0.1){};

%branch S up
\draw[thick] (7,-0.3) to[out=0, in=180] (8,0.5);

%branch S down
\draw[thick] (7,-0.3) to[out=0, in=180] (8,-1.1);

\node[rectangle, draw, minimum size=6mm,fill=cyan!30] at (7,-0.3){$\hat{S}^{(in)}$};

\node[rectangle, minimum width=8mm,minimum height=19mm,rounded corners, draw=black,densely dotted, thick] at (8.75,-0.3){};

% rail up
\draw[thick] (8,0.5) -- (11,0.5);
% new rail down
\draw[thick] (8,-1.1) -- (11,-1.1);

% walkers
\draw[thick,fill=red] (9,0.5) circle (\dr);
\draw[thick,fill=red] (8.5,0.5) circle (\dr);
\draw[thick,fill=red] (8.5,-1.1) circle (\dr);
\draw[thick,fill=red] (8,-1.1) circle (\dr);
\draw[thick,fill=red] (6.3,-0.3) circle (\dr);
\draw[thick,fill=blue] (5.8,-0.3) circle (\dr);

%labels
\node at (5.8,0) {{\scalebox{0.7}{$\tilde A_2^{\downarrow}$}}};
\node at (6.3,0) {{\scalebox{0.7}{$\tilde A_2^{\uparrow}$}}};
\node at (8,-0.8) {{\scalebox{0.7}{$A_2$}}};
\node at (8.5,-0.8) {{\scalebox{0.7}{$\tilde A_1^{\downarrow}$}}};
\node at (8.5,0.2) {{\scalebox{0.7}{$\tilde A_1^{\uparrow}$}}};
\node at (9,0.2) {{\scalebox{0.7}{$A_1$}}};

\node at (3,0) {{\scalebox{0.8}{$r=in$}}};
\node at (10.6,0.8) {{\scalebox{0.8}{$r=\uparrow$}}};
\node at (10.6,-0.8) {{\scalebox{0.8}{$r=\downarrow$}}};
\node at (8.7,0.9) {{\scalebox{0.8}{Buffer area}}};

\draw[thick, red, densely dashdotted](9.2,0.7)--(9.2,-1.3);

\node at (3,1.3) {{\scalebox{1}{$\boldsymbol{(c)}$}}};

\end{scope}

%%% Quarto scope %%%
\begin{scope}[shift={(10,-4)}]

% Parametri per coerenza
    \def\dotRadius{0.5pt}
% rail aux
\draw[thick] (7,-0.3) -- (3.3,-0.3);
%\draw[thick] (2,-0.3) -- (2.7,-0.3);

    \foreach \x in {2.8, 3, 3.2} {
        \fill ( \x, -0.3) circle (\dotRadius);
    }

\node[rectangle, draw, minimum width=21mm,minimum height=25mm,rounded corners, dashed, fill=orange!20] at (8.75,-0.1){};

%branch S up
\draw[thick] (7,-0.3) to[out=0, in=180] (8,0.5);

%branch S down
\draw[thick] (7,-0.3) to[out=0, in=180] (8,-1.1);

\node[rectangle, draw, minimum size=6mm,fill=cyan!30] at (7,-0.3){$\hat{S}^{(in)}$};

\node[rectangle, minimum width=8mm,minimum height=19mm,rounded corners, draw=black,densely dotted, thick] at (9.25,-0.3){};
\node[rectangle, minimum width=8mm,minimum height=19mm,rounded corners, draw=black,densely dotted, thick] at (8.25,-0.3){};

% rail up
\draw[thick] (8,0.5) -- (11,0.5);
% new rail down
\draw[thick] (8,-1.1) -- (11,-1.1);

% walkers
\draw[thick,fill=red] (9.5,0.5) circle (\dr);
\draw[thick,fill=red] (9,0.5) circle (\dr);
\draw[thick,fill=red] (9,-1.1) circle (\dr);
\draw[thick,fill=red] (8.5,-1.1) circle (\dr);
\draw[thick,fill=red] (8,0.5) circle (\dr);
\draw[thick,fill=red] (8,-1.1) circle (\dr);

%labels
\node at (8,-0.8) {{\scalebox{0.7}{$\tilde A_2^{\downarrow}$}}};
\node at (8,0.2) {{\scalebox{0.7}{$\tilde A_2^{\uparrow}$}}};
\node at (8.5,-0.8) {{\scalebox{0.7}{$A_2$}}};
\node at (9,-0.8) {{\scalebox{0.7}{$\tilde A_1^{\downarrow}$}}};
\node at (9,0.2) {{\scalebox{0.7}{$\tilde A_1^{\uparrow}$}}};
\node at (9.5,0.2) {{\scalebox{0.7}{$A_1$}}};

\node at (3,0) {{\scalebox{0.8}{$r=in$}}};
\node at (10.6,0.8) {{\scalebox{0.8}{$r=\uparrow$}}};
\node at (10.6,-0.8) {{\scalebox{0.8}{$r=\downarrow$}}};
\node at (8.7,0.9) {{\scalebox{0.8}{Buffer area}}};

\node at (3,1.3) {{\scalebox{1}{$\boldsymbol{(d)}$}}};

\end{scope}

\draw[thick](12.5,1.5)--(12.5,-5.5);
\draw[thick](3,-2.25)--(22,-2.25);
     \end{tikzpicture}
     }
     \caption{Initialization of the backup variant for the dual-rail protocol. (a) Routing information is encoded in the color of address and data carriers, accompanied by backups prepared in red and blue states, forming three-particle sets. (b) A transversal scattering gate $\hat{S}^{(in)}$ routes particles onto the target rails; however, this introduces a spatial gap between backups of the same carrier. (c) A buffering procedure restores the correct spacing: the main carrier and its upper-rail backup are halted, while the lower-rail backup and the subsequent carrier advance by one site. (d) This process iterates until the entire train is aligned for routing.}
    \label{dual_rail_init_backup}
\end{figure*}
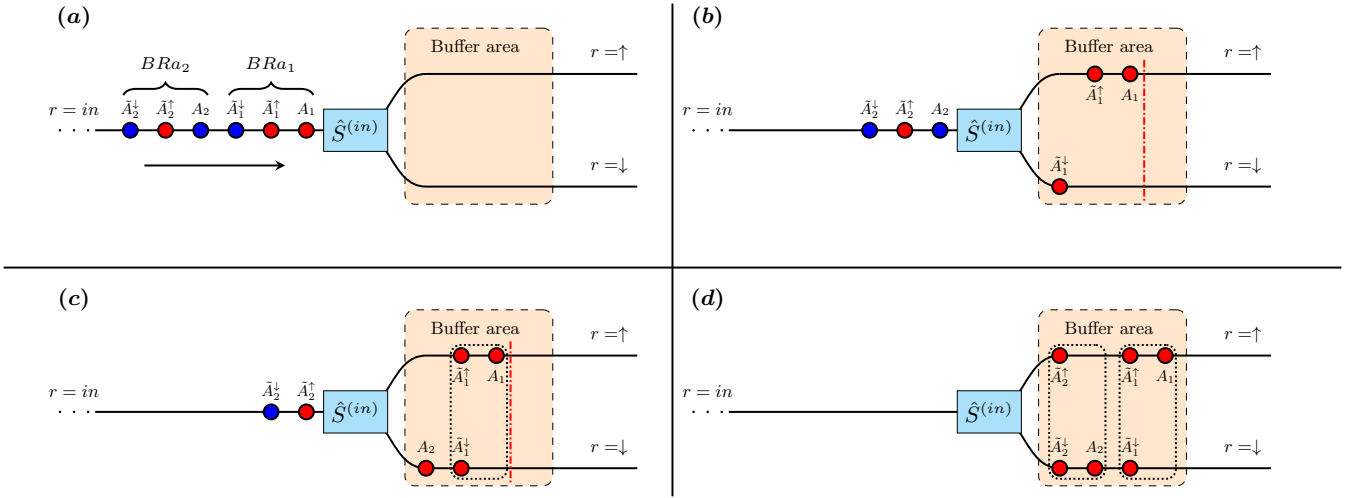

{\subsubsection{Encoding and initialization}}

In this variant, each logical information carrier is implemented together with two backup walkers, one for each rail. Consequently, encoding the full address and data registers requires a total of $3(n+m)$ walkers.
As proposed for the \gdr{basic} dual-rail implementation of the protocol, it is practically convenient to first prepare the initial state as a train of walkers placed on a single input rail, and subsequently route them onto the correct operational parallel rails. This is achieved through the same two-step procedure described in Sec.~\ref{encodindDualStandard}: encoding the target rail assignment into the walkers' color degree of freedom, and then processing the sequence through a scattering gate $\hat{S}$. 
At the input stage, namely before the routing along the parallel binary trees takes place (see Fig.~\ref{dual_rail_init_backup}a), walkers are arranged in triples: each logical carrier $W_i$ is followed by its two corresponding backups $\tilde{W}^{\uparrow}_i$ and $\tilde{W}^{\downarrow}_i$, where the arrows indicate the target rails onto which the backups will be routed after the initialization procedure.
First, the initial state information is encoded in the color of the walkers placed on the input rail $r=in$, according to the mapping in Eq.~(\ref{inputrailencoding}). This implies that the particles are prepared with a precise spatial order and color sequence, taking the form
\begin{equation*}
\underbrace{\tilde{D}_m^{\downarrow}\tilde{D}_m^{\uparrow}D_m}_{B R B} \dots \underbrace{\tilde{D}_1^{\downarrow}\tilde{D}_1^{\uparrow}D_1}_{B R B} \, \underbrace{\tilde{A}_n^{\downarrow}\tilde{A}_n^{\uparrow}A_n}_{B R a_n} \dots \underbrace{\tilde{A}_1^{\downarrow}\tilde{A}_1^{\uparrow}A_1}_{B R a_1}.
\end{equation*}
Second, this train travels through the $\hat{S}^{(in)}$ gate introduced in Eq.~(\ref{ScatterInputMapping}), which routes red particles to the upper rail $r=\uparrow$ and blue particles to the lower rail $r=\downarrow$.

Unlike the \gdr{basic} protocol, however, this basic routing procedure is insufficient to fully prepare the target dual-rail initial state~(\ref{dualrailbackprerout}). Although the particles are successfully sent to the correct rails, their sequential injection introduces a relative spatial displacement along the propagation direction (see Fig.~\ref{dual_rail_init_backup}b). This mismatch violates the geometric requirements of the protocol, since we want backup walkers corresponding to the same address or data walker to be perfectly aligned at parallel positions on their respective rails. This issue can be resolved using a buffering procedure right after the placement of the gate $\hat{S}$. 

More precisely, we require the buffering procedure to fulfill the following two conditions: i) for each triple $(\tilde W_k^{\downarrow}, \tilde W_k^{\uparrow}, W_k)$ with $k=1,\dots,n+m+1$, the backup walkers must occupy strictly parallel positions on their respective rails, and ii) two consecutive triples must have the same spatial separation of a given $\Delta x$.
This can be achieved by allocating a synchronized buffering area at the beginning of the tree, which realigns the particles as they emerge from the first scattering gate (see Fig.~\ref{dual_rail_init_backup}c). 

This staging area temporarily collects the incoming walkers and, every three time steps, that is, once a full triple is structurally aligned, it shifts the assembled configuration forward by exactly one spatial site.
By repeating this process for all the triples, we successfully prepare the desired initial configuration with no empty spatial gaps between adjacent particles (see Fig.~\ref{dual_rail_init_backup}d). The main routing phase is triggered only after the entire train of walkers has been assembled. Since each triple requires three time steps to enter but shifts forward only by one spatial site $\Delta x$, the total time overhead introduced by this procedure is $T_{\text{prep}}=3(n+m)$ extra steps. Consequently, this linear delay preserves the overall linear time scaling of the protocol. 

At the output of this staging and buffering phase, the target dual-rail initial state is successfully prepared. Notice that all walkers, regardless of the register they belong to, are in the red color state $\ket{R}$.  
While their internal states are now uniform, their rail placement reflects the corresponding encoded information.
The overall initial state of the protocol can be expressed as:
\begin{multline}\label{dualrailbackprerout}
\ket{\psi_{in}}=\bigotimes_{j=0}^m
\left(\ket{\downarrow,R}_{\tilde D_j^{\downarrow}}\otimes\ket{\uparrow,R}_{\tilde D_j^{\uparrow}}\otimes \ket{\downarrow,R}_{D_j}\right)\otimes \\\bigotimes_{i=1}^n
\left(\ket{\downarrow,R}_{\tilde A_i^{\downarrow}}\otimes\ket{\uparrow,R}_{\tilde A_i^{\uparrow}}\otimes \ket{a_i,R}_{A_i}\right).
\end{multline}
We remind that the redundancy of walkers on both rails ensures that all operations can be decomposed according to the backup logic introduced in Section \ref{sec:backup}, both for routing and output phases.

\subsubsection{Address-transmission gate decomposition}

As with the \gdr{basic} protocol, the main advantage of the backup variant lies in the possibility of decomposing the long-range address-transmission gate $\hat{U}^{(d)}$ into a sequence of at most range-2 interactions. Specifically, this decomposition consists of an initialization short-range operation $\hat{U}_{in}$ acting on two walkers, followed by the repeated application of a range-2 unitary operation $\hat{U}_B$ acting on groups of three walkers. As before, we can choose $\hat{U}_{in}$ to be a $CX$ gate activated when the control walker is in the state $\ket{R}$, and $\hat{U}_{B}$ to be a $CXX$ gate activated when the control is in the state $\ket{B}$.
The crucial difference from the \gdr{basic} protocol stems from the dual-rail encoding: these gates must now be applied independently on both rails, acting exclusively on the particles propagating within them. As a consequence, these decomposed operations are subspace-preserving. Particles encoding a logical $0$, whose states belong to $\mathcal{H}_W^{(\uparrow)}$, retain that logical state throughout the transmission, and the same holds for walkers encoding a logical $1$, which are confined to $\mathcal{H}_W^{(\downarrow)}$.

The initialization gate acts on the address walker $A_d$ and its corresponding backup $\tilde A_d^{(a_d)}$, where $a_d\in\{\uparrow,\downarrow\}$ indicates the rail occupied by $A_d$. The operation can be explicitly written as
\begin{equation}
    \hat{U}_{in}(\tilde A_d\vert A_d):=\sum_{r,c}\hat{V}_{c}^{(\tilde A_d)}\otimes\hat{\Pi}_{r,c}^{(A_d)}
\end{equation}
where $\{\Pi_{r,c}^{(A_d)}\}$ is the set of projectors onto the subspaces of the address walker $A_d$, and the operator $\hat{V}_c^{(\tilde A_d)}$ in defined as
\begin{equation*}
    \hat{V}^{(\tilde A_d)}_{c=R}\equiv \hat{X}_{\tilde A_d},\quad \hat{V}_{c=B}^{(\tilde A_d)}\equiv \hat{I}_{\tilde A_d}.
\end{equation*}
As established in the \gdr{basic} variant, we can split this gate into its components applied to the different rails, expressing it as a direct sum of two subspace-dependent operations:
\begin{equation*}  \hat{U}_{in}=\hat{U}_{in}^{(\uparrow)}\oplus \hat{U}_{in}^{(\downarrow)}.
\end{equation*}

Once the address information is transferred to its corresponding backup, it is sequentially propagated to all other particles in the train through the block operation $\hat{U}_B$. This gate takes the backup walker $\tilde A_d^{(r)}$ as the control and acts on the two particles following it. More generally, it acts on triplets of the form $(\tilde W^{(r)}_{i+1}, W_{i+1}, \tilde W_i^{(r)})$ located on the same rail $r$. The operation is constructed as:
\begin{equation}
    \hat{U}_B(\tilde W^{(r)}_{i+1},W_{i+1}\vert \tilde W_i^{(r)}):=\sum_{r,c}\hat{V}_c^{(\tilde W^{(r)}_{i+1},W_{i+1})}\otimes \hat{\Pi}_{r,c}^{(\tilde W_i^{(r)})}
\end{equation}
where $\{\hat \Pi_{r,c}^{(\tilde W_i^{(r)})}\}$ is the set of projectors onto the states of the $i$-th backup $\tilde W_i^{(r)}$ placed on rail $r$, and the unitary $\hat V_c^{(\tilde W^{(r)}_{i+1},W_{i+1})}$
is defined as 
\begin{equation*}
\begin{split}
   & \hat V_{c=B}^{(\tilde W^{(r)}_{i+1},W_{i+1})}\equiv \hat{X}^{(r)}_{\tilde W_{i+1}^{(r)}}\otimes \hat{X}_{W_{i+1}}^{(r)},\\&
   \hat V_{c=R}^{(\tilde W^{(r)}_{i+1},W_{i+1})}\equiv \hat{I}_{\tilde W_{i+1}^{(r)}}\otimes \hat{I}_{W_{i+1}}.
\end{split}
\end{equation*}
As for the initialization operation the gate can be expressed as a direct sum of gates acting on different logical subspaces
\begin{equation*}
    \hat{U}_{B}=\hat{U}_{B}^{(\uparrow)}\oplus \hat{U}_{B}^{(\downarrow)}.
\end{equation*}

A notable physical peculiarity of this protocol is that, while the backup walkers are permanently present on both rails, the primary address or data walkers $W_{i+1}$ may be absent on a given rail depending on the logical bit being processed. 
In such cases, the corresponding spatial mode is occupied by the vacuum state $\ket{\varnothing}$, and the local operations act trivially on this empty subsystem. Conversely, whenever a primary walker is absent on one rail, it is always present on the opposite one. This behavior perfectly summarizes the complementary nature of the dual-rail encoding, along with the idea that the two parallel protocols operate as exact ``photographic negatives" of one another.

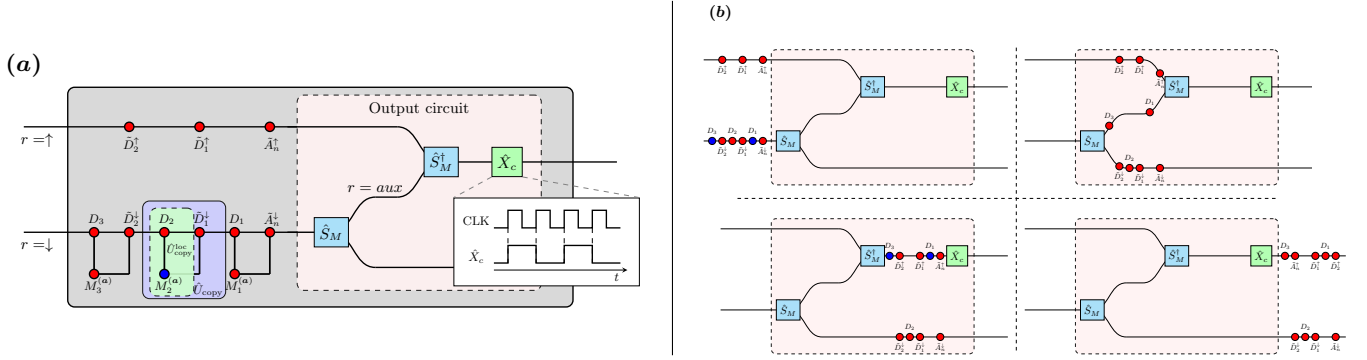
\begin{figure*}[htb!]
\begin{subfigure}[c]{0.48\linewidth}
        \centering
     \resizebox{\linewidth}{!}{
    \begin{tikzpicture}[x=1cm,y=1cm, scale=0.9]
  \def\w{8}
  \def\h{1.5}
  \def\dr{3pt}
  \def\dist{0.3}
    % Rettangolo
    \draw[rounded corners, thick, fill=gray!30] (1,-2) rectangle ({\w+4.5},\h+1.5);

% Rettangolo gate U_copy
   \node[draw, rounded corners, fill=blue!20, minimum width=17mm, minimum height=20mm, opacity=0.9] at (3.65,-0.7) {};

      % Linee verticali backup 
    \foreach \i in {1,...,3} {
      \pgfmathsetmacro{\x}{\i/5*\w}
      \draw[line width=1pt] ({\x+0.8}, -\h/2-0.5) -- ({\x+0.8}, -\dist);
    }

       % Linee spesse pallini inferiori
    \foreach \j in {1,3,5} {
      \pgfmathsetmacro{\xStart}{(\j+1)/5*(\w/2)}
      \pgfmathsetmacro{\xEnd}{(\j+2)/5*(\w/2)}
      \draw[line width=1pt] (\xStart, {-\h/2-0.5}) -- (\xEnd, {-\h/2-0.5});
    }

% Rettangolo gate U_copy^loc
   \node[draw, dashed, rounded corners, fill=green!20, minimum width=9mm, minimum height=18mm, opacity=0.9] at (3.37,-0.75) {};

 % Linee verticali dai pallini sopra a quelli sotto
    \foreach \i in {1,...,3} {
      \pgfmathsetmacro{\x}{\i/5*\w}
      \draw[line width=1pt] (\x, -\h/2-0.5) -- (\x, -\dist);
    }

    % Pallini memoria
    \foreach[count=\k from 1] \col in {red, blue, red} {
        \pgfmathsetmacro{\x}{\k/5*\w}  % \n non definito, quindi uso 6 fisso coerente con sopra
        \filldraw[fill=\col, draw=black, thick] (\x, {-\h/2-0.5}) circle [radius=\dr];
      }

    % rail down
    \draw[thick] (0,-\dist) -- (\w-0.7,-\dist);

 % rail up
    \draw[thick] (0,2.1) -- (8.5,2.1);

   % Pallini rail down
    \foreach[count=\k from 1] \col in { red, red,red,red,red, red} {
        \pgfmathsetmacro{\x}{(\k+1)/5*(\w/2)}
        \pgfmathsetmacro{\y}{-\dist}
        \filldraw[fill=\col, draw=black, thick] (\x, \y) circle [radius=\dr];
      }

% Pallini rail up
    \foreach[count=\k from 1] \col in {red, red, red} {
        \pgfmathsetmacro{\x}{\k/5*\w}  % \n non definito, quindi uso 6 fisso coerente con sopra
        \filldraw[fill=\col, draw=black, thick] (\x+0.8, {2.1}) circle [radius=\dr];
      }

%%% CIRCUITO DI OUTPUT %%%%

% Rettangolo output circuit
   \node[draw, dashed, rounded corners, fill=pink!20, minimum width=50mm, minimum height=40mm, opacity=0.9] at (9,0.6) {};

% rail up in
\draw[thick] (6,2.1) -- (8.5,2.1);

% rail down in
\draw[thick] (7,-0.3) -- (6,-0.3);

%branch S up
\draw[thick] (7,-0.3) to[out=0, in=180] (8,0.5);

%branch S down
\draw[thick] (7,-0.3) to[out=0, in=180] (8,-1.1);

\node[rectangle, draw, minimum size=6mm,fill=cyan!30] at (7,-0.3){$\hat{S}_M$};

% rail aux
\draw[thick] (8,0.5) -- (8.5,0.5);

%branch S_dag down
\draw[thick] (8.5,0.5) to[out=0, in=180] (9.5,1.3);

%branch S_dag up
\draw[thick] (8.5,2.1) to[out=0, in=180] (9.5,1.3);

% new rail down
\draw[thick] (8,-1.1) -- (13.5,-1.1);

% new rail up
    \draw[thick] (9.5,1.3) -- (13.5,1.3);

\node[rectangle, draw, minimum size=6mm,fill=cyan!30] at (9.5,1.3){$\hat{S}^{\dagger}_M$};

\node[rectangle, draw, minimum size=6mm,fill=green!30](colorgate) at (11,1.3){$\hat{X}_c$};

%%%%%%%%  LABEL DELLE COSE %%%%%%%
   % Label memorie
    \node at (1.7,-1.5) {{\scalebox{0.8}{$M_{3}^{(\boldsymbol{a})}$}}};

    \node at (3.3,-1.5) {{\scalebox{0.8}{$M_{2}^{(\boldsymbol{a})}$}}};
    
    \node at (4.9,-1.5) {{\scalebox{0.8}{$M_{1}^{(\boldsymbol{a})}$}}};

    % Label pallini rail down

    \node at (1.65,0) {{\scalebox{0.8}{$D_3$}}};

    \node at (2.45,0) {{\scalebox{0.8}{$\tilde{D}^{\downarrow}_2$}}};

    \node at (3.25,0) {{\scalebox{0.8}{$D_2$}}};

    \node at (4.05,0) {{\scalebox{0.8}{$\tilde{D}^{\downarrow}_1$}}};

    \node at (4.85,0) {{\scalebox{0.8}{$D_1$}}};

    \node at (5.65,0) {{\scalebox{0.8}{$\tilde{A}^{\downarrow}_n$}}};

    % Label pallini rail up

    \node at (2.45,1.7) {{\scalebox{0.8}{$\tilde{D}_2^{\uparrow}$}}};

    \node at (4.05,1.7) {{\scalebox{0.8}{$\tilde{D}_1^{\uparrow}$}}};

    \node at (5.65,1.7) {{\scalebox{0.8}{$\tilde{A}^{\uparrow}_n$}}};
    
%label gate

\node at (4.2,-1.6) {{\scalebox{0.8}{$\hat{U}_{\text{copy}}$}}};

\node at (3.55,-0.7) {{\scalebox{0.7}{$\hat{U}_{\text{copy}}^{\text{loc}}$}}};

%label rails
\node at (0.3,-0.6) {$r=\downarrow$};

\node at (0.3,1.8) {$r=\uparrow$};

\node at (8,0.7) {$r=aux$};

\node at (9,2.5) {Output circuit};

\begin{scope}[shift={(3.5,-1)}, scale=0.8]
\node[rectangle, draw, minimum width=38mm, minimum height=18mm,fill=white!30](clock) at (10.5,0.6){};
\draw[thick] (9, 1) -- ++(0.4,0) -- ++(0,0.5) -- ++(0.4,0) -- ++(0,-0.5) -- ++(0.4,0)-- ++(0,0.5)--++(0.4,0)--++(0,-0.5)--++(0.4,0)--++(0,0.5)--++(0.4,0)--++(0,-0.5)--++(0.4,0)--++(0,0.5)--++(0.4,0)--++(0,-0.5)--++(0.4,0);

\draw[ thick] (9, 0) -- ++(0.4,0) -- ++(0,0.5) -- ++(0.8,0) -- ++(0,-0.5) -- ++(0.8,0)-- ++(0,0.5)--++(0.8,0)--++(0,-0.5)--++(0.8,0);

\draw [dashed](9.4,1)--(9.4,0);
\draw [dashed](10.2,1)--(10.2,0);
\draw [dashed](11,1)--(11,0);
\draw [dashed](11.8,1)--(11.8,0);

\draw[-stealth](9,-0.2)--(12.8,-0.2);
\node at (12.5,-0.4) {{\scalebox{0.8}{$t$}}};

\node at (8.5,1.2) {{\scalebox{0.8}{CLK}}};

\node at (8.5,0.2) {{\scalebox{0.8}{$\hat{X}_c$}}};
\end{scope}

\draw[gray, thin, dashed] (colorgate.south west) -- (clock.north west);
\draw[gray, thin, dashed] (colorgate.south east) -- (clock.north east);

\node at (0,3.5) {{\scalebox{1.5}{$\boldsymbol{(a)}$}}};

\end{tikzpicture}
}
\end{subfigure}\hfill\vrule\hfill
\begin{subfigure}[c]{0.48\linewidth}
        \centering
        \resizebox{\linewidth}{!}{
\begin{tikzpicture}
 \def\w{8}
  \def\h{1.5}
  \def\dr{3pt}
  \def\dist{0.3}

\draw[dashed](14.75,-6.5)--(14.75,2.5);

\draw[dashed](6.5,-2)--(22.5,-2);
  
%%%%%%%% PRIMO SCOPE %%%%%%%
\begin{scope}[shift={(1,0)}]
   % Rettangolo output circuit
   \node[draw, dashed, rounded corners, fill=pink!20, minimum width=60mm, minimum height=40mm, opacity=0.9] at (9.5,0.4) {};

% rail up in
\draw[thick] (4.5,2.1) -- (8.5,2.1);

% rail down in
\draw[thick] (7,-0.3) -- (4.5,-0.3);

%branch S up
\draw[thick] (7,-0.3) to[out=0, in=180] (8,0.5);

%branch S down
\draw[thick] (7,-0.3) to[out=0, in=180] (8,-1.1);

\node[rectangle, draw, minimum size=6mm,fill=cyan!30] at (7,-0.3){$\hat{S}_M$};

% rail aux
\draw[thick] (8,0.5) -- (8.5,0.5);

%branch S_dag down
\draw[thick] (8.5,0.5) to[out=0, in=180] (9.5,1.3);

%branch S_dag up
\draw[thick] (8.5,2.1) to[out=0, in=180] (9.5,1.3);

% new rail down
\draw[thick] (8,-1.1) -- (13.5,-1.1);

% new rail up
    \draw[thick] (9.5,1.3) -- (13.5,1.3);

\node[rectangle, draw, minimum size=6mm,fill=cyan!30] at (9.5,1.3){$\hat{S}^{\dagger}_M$};

\node[rectangle, draw, minimum size=6mm,fill=green!30] at (12,1.3){$\hat{X}_c$}; 

%%% Pallini rail sopra
\draw[thick,fill=red] (6.25,2.1) circle (\dr);
\draw[thick,fill=red] (5.65,2.1) circle (\dr);
\draw[thick,fill=red] (5.05,2.1) circle (\dr);

\node at (5.05,1.8) {{\scalebox{0.7}{$\tilde D_2^{\uparrow}$}}};
\node at (5.65,1.8) {{\scalebox{0.7}{$\tilde D_1^{\uparrow}$}}};
 \node at (6.25,1.8) {{\scalebox{0.7}{$\tilde A_n^{\uparrow}$}}};

%%% Pallini rail sotto

\draw[thick,fill=red] (6.25,-0.3) circle (\dr);
\draw[thick,fill=blue] (5.95,-0.3) circle (\dr);
\draw[thick,fill=red] (5.65,-0.3) circle (\dr);
\draw[thick,fill=red] (5.35,-0.3) circle (\dr);
\draw[thick,fill=red] (5.05,-0.3) circle (\dr);
\draw[thick,fill=blue] (4.75,-0.3) circle (\dr);

\node at (4.75,0) {{\scalebox{0.7}{$D_3$}}};
\node at (5.05,-0.6) {{\scalebox{0.7}{$\tilde D_2^{\downarrow}$}}};
 \node at (5.35,0) {{\scalebox{0.7}{$D_2$}}};
 \node at (5.65,-0.6) {{\scalebox{0.7}{$\tilde D_1^{\downarrow}$}}};
\node at (5.95,0) {{\scalebox{0.7}{$D_1$}}};
\node at (6.25,-0.6) {{\scalebox{0.7}{$\tilde A_n^{\downarrow}$}}};

\end{scope}

%%%%%% SECONDO SCOPE %%%%%%%
\begin{scope}[shift={(10,0)}]
   % Rettangolo output circuit
      \node[draw, dashed, rounded corners, fill=pink!20, minimum width=60mm, minimum height=40mm, opacity=0.9] at (9.5,0.4) {};

% rail up in
\draw[thick] (5,2.1) -- (8.5,2.1);

% rail down in
\draw[thick] (7,-0.3) -- (5,-0.3);

%branch S up
\draw[thick] (7,-0.3) to[out=0, in=180] (8,0.5);

%branch S down
\draw[thick] (7,-0.3) to[out=0, in=180] (8,-1.1);

\node[rectangle, draw, minimum size=6mm,fill=cyan!30] at (7,-0.3){$\hat{S}_M$};

% rail aux
\draw[thick] (8,0.5) -- (8.5,0.5);

%branch S_dag down
\draw[thick] (8.5,0.5) to[out=0, in=180] (9.5,1.3);

%branch S_dag up
\draw[thick] (8.5,2.1) to[out=0, in=180] (9.5,1.3);

% new rail down
\draw[thick] (8,-1.1) -- (13.5,-1.1);

% new rail up
    \draw[thick] (9.5,1.3) -- (13.5,1.3);

\node[rectangle, draw, minimum size=6mm,fill=cyan!30] at (9.5,1.3){$\hat{S}^{\dagger}_M$};

\node[rectangle, draw, minimum size=6mm,fill=green!30] at (12,1.3){$\hat{X}_c$};

%%% Pallini rail sopra
\draw[thick,fill=red] (9,1.7) circle (\dr);
\draw[thick,fill=red] (8.4,2.1) circle (\dr);
\draw[thick,fill=red] (7.8,2.1) circle (\dr);

\node at (7.8,1.8) {{\scalebox{0.7}{$\tilde D_2^{\uparrow}$}}};
\node at (8.4,1.8) {{\scalebox{0.7}{$\tilde D_1^{\uparrow}$}}};
 \node at (9,1.4) {{\scalebox{0.7}{$\tilde A_n^{\uparrow}$}}};

%%% Pallini rail sotto
\draw[thick,fill=red] (9,-1.1) circle (\dr);
\draw[thick,fill=red] (8.7,0.55) circle (\dr);
\draw[thick,fill=red] (8.4,-1.1) circle (\dr);
\draw[thick,fill=red] (8.1,-1.1) circle (\dr);
\draw[thick,fill=red] (7.8,-1.05) circle (\dr);
\draw[thick,fill=red] (7.5,0.15) circle (\dr);

\node at (7.5,0.45) {{\scalebox{0.7}{$D_3$}}};
\node at (7.8,-1.35) {{\scalebox{0.7}{$\tilde D_2^{\downarrow}$}}};
\node at (8.1,-0.8) {{\scalebox{0.7}{$D_2$}}};
\node at (8.4,-1.4) {{\scalebox{0.7}{$\tilde D_1^{\downarrow}$}}};
\node at (8.7,0.85) {{\scalebox{0.7}{$D_1$}}};
\node at (9,-1.4) {{\scalebox{0.7}{$\tilde A_n^{\downarrow}$}}};

\end{scope}
%%%% TERZO SCOPE %%%%%%%%%
\begin{scope}[shift={(1,-5)}]
   % Rettangolo output circuit
      \node[draw, dashed, rounded corners, fill=pink!20, minimum width=60mm, minimum height=40mm, opacity=0.9] at (9.5,0.4) {};

% rail up in
\draw[thick] (5,2.1) -- (8.5,2.1);

% rail down in
\draw[thick] (7,-0.3) -- (5,-0.3);

%branch S up
\draw[thick] (7,-0.3) to[out=0, in=180] (8,0.5);

%branch S down
\draw[thick] (7,-0.3) to[out=0, in=180] (8,-1.1);

\node[rectangle, draw, minimum size=6mm,fill=cyan!30] at (7,-0.3){$\hat{S}_M$};

% rail aux
\draw[thick] (8,0.5) -- (8.5,0.5);

%branch S_dag down
\draw[thick] (8.5,0.5) to[out=0, in=180] (9.5,1.3);

%branch S_dag up
\draw[thick] (8.5,2.1) to[out=0, in=180] (9.5,1.3);

% new rail down
\draw[thick] (8,-1.1) -- (13.5,-1.1);

% new rail up
    \draw[thick] (9.5,1.3) -- (13.5,1.3);

\node[rectangle, draw, minimum size=6mm,fill=cyan!30] at (9.5,1.3){$\hat{S}^{\dagger}_M$};

\node[rectangle, draw, minimum size=6mm,fill=green!30] at (12,1.3){$\hat{X}_c$}; 

%%% Pallini rail sopra
\draw[thick,fill=red] (11.5,1.3) circle (\dr);
\draw[thick,fill=red] (10.9,1.3) circle (\dr);
\draw[thick,fill=red] (10.3,1.3) circle (\dr);

\node at (10.3,1) {{\scalebox{0.7}{$\tilde D_2^{\uparrow}$}}};
\node at (10.9,1) {{\scalebox{0.7}{$\tilde D_1^{\uparrow}$}}};
 \node at (11.5,1) {{\scalebox{0.7}{$\tilde A_n^{\uparrow}$}}};

%%% Pallini rail sotto
\draw[thick,fill=red] (11.5,-1.1) circle (\dr);
\draw[thick,fill=blue] (11.2,1.3) circle (\dr);
\draw[thick,fill=red] (10.9,-1.1) circle (\dr);
\draw[thick,fill=red] (10.6,-1.1) circle (\dr);
\draw[thick,fill=red] (10.3,-1.1) circle (\dr);
\draw[thick,fill=blue] (10,1.3) circle (\dr);

\node at (10,1.6) {{\scalebox{0.7}{$D_3$}}};
\node at (10.3,-1.4) {{\scalebox{0.7}{$\tilde D_2^{\downarrow}$}}};
\node at (10.6,-0.8) {{\scalebox{0.7}{$D_2$}}};
\node at (10.9,-1.4) {{\scalebox{0.7}{$\tilde D_1^{\downarrow}$}}};
\node at (11.2,1.6) {{\scalebox{0.7}{$D_1$}}};
\node at (11.5,-1.4) {{\scalebox{0.7}{$\tilde A_n^{\downarrow}$}}};

\end{scope}

%%%%%%%% QUARTO SCOPE %%%%%%%%
\begin{scope}[shift={(10,-5)}]
   % Rettangolo output circuit
      \node[draw, dashed, rounded corners, fill=pink!20, minimum width=60mm, minimum height=40mm, opacity=0.9] at (9.5,0.4) {};
      
% rail up in
\draw[thick] (5,2.1) -- (8.5,2.1);

% rail down in
\draw[thick] (7,-0.3) -- (5,-0.3);

%branch S up
\draw[thick] (7,-0.3) to[out=0, in=180] (8,0.5);

%branch S down
\draw[thick] (7,-0.3) to[out=0, in=180] (8,-1.1);

\node[rectangle, draw, minimum size=6mm,fill=cyan!30] at (7,-0.3){$\hat{S}_M$};

% rail aux
\draw[thick] (8,0.5) -- (8.5,0.5);

%branch S_dag down
\draw[thick] (8.5,0.5) to[out=0, in=180] (9.5,1.3);

%branch S_dag up
\draw[thick] (8.5,2.1) to[out=0, in=180] (9.5,1.3);

% new rail down
\draw[thick] (8,-1.1) -- (14.5,-1.1);

% new rail up
    \draw[thick] (9.5,1.3) -- (14.5,1.3);

\node[rectangle, draw, minimum size=6mm,fill=cyan!30] at (9.5,1.3){$\hat{S}^{\dagger}_M$};

\node[rectangle, draw, minimum size=6mm,fill=green!30] at (12,1.3){$\hat{X}_c$};

%%% Pallini rail sopra
\draw[thick,fill=red] (14.2,1.3) circle (\dr);
\draw[thick,fill=red] (13.6,1.3) circle (\dr);
\draw[thick,fill=red] (13,1.3) circle (\dr);

\node at (14.2,1) {{\scalebox{0.7}{$\tilde D_2^{\uparrow}$}}};
\node at (13.6,1) {{\scalebox{0.7}{$\tilde D_1^{\uparrow}$}}};
 \node at (13,1) {{\scalebox{0.7}{$\tilde A_n^{\uparrow}$}}};

%%% Pallini rail sotto

\draw[thick,fill=red] (14.2,-1.1) circle (\dr);
\draw[thick,fill=red] (13.9,1.3) circle (\dr);
\draw[thick,fill=red] (13.6,-1.1) circle (\dr);
\draw[thick,fill=red] (13.3,-1.1) circle (\dr);
\draw[thick,fill=red] (13,-1.1) circle (\dr);
\draw[thick,fill=red] (12.7,1.3) circle (\dr);

\node at (12.7,1.6) {{\scalebox{0.7}{$D_3$}}};
\node at (13,-1.4) {{\scalebox{0.7}{$\tilde D_2^{\downarrow}$}}};
\node at (13.3,-0.8) {{\scalebox{0.7}{$D_2$}}};
\node at (13.6,-1.4) {{\scalebox{0.7}{$\tilde D_1^{\downarrow}$}}};
\node at (13.9,1.6) {{\scalebox{0.7}{$D_1$}}};
\node at (14.2,-1.4) {{\scalebox{0.7}{$\tilde A_n^{\downarrow}$}}};

\end{scope}

\node at (6,3.5) {{\scalebox{1.5}{$\boldsymbol{(b)}$}}};
    \end{tikzpicture}
    }
\end{subfigure}
    \caption{Schematic of the memory cell operation in the backup variant of the dual-rail protocol. (a) The backup particle on the lower rail triggers a local copy operation, $\hat{U}_{copy}^{loc}$, which transfers information from the memory walkers to the data walkers' color states according to Eq.~(\ref{dual_rail_copy_mapping}). (b) The output circuit redirects the data walkers based on their color: a scattering gate $\hat{S}_M$ routes particles encoding a $1$ to the lower rail, while $\hat{S}_M^{\dagger}$ routes the remaining particles to the upper rail. Finally, a local time-dependent gate $\hat{X}_C$ restores the correct color encoding of the data walkers.}
    \label{copy_dualrail_backup}
    
\end{figure*}

\subsubsection{Memory operations}

Similarly to the backup variant of the \gdr{basic} protocol, we exploit the redundancy introduced by the backup walkers to reduce the complexity of the memory operations and make them strictly short-range. Specifically, the copy procedure is executed via an operation involving the data walker $D_i$, its immediately preceding backup on the lower rail $\tilde{D}_{i-1}^{\downarrow}$, and the corresponding message walker in the memory cell $M_i^{(\boldsymbol{a})}$.

We use the presence of the lower-rail backup walker as a control for a local unitary operation $\hat{U}_{copy}^{loc}(D_i\vert M_i^{(\boldsymbol{a})})$ having as targets the data walker $D_i$ and the memory walker $M_i^{(\boldsymbol{a})}$, which contains the message bit $b_i^{(\boldsymbol{a})}$. In other words, the message is copied onto the data walker if and only if the valid backup walker $\tilde{D}_{i-1}^{\downarrow}$ is detected on the same rail. Formally, this controlled-copy operation is defined as
\begin{equation}
\begin{split}
\hat{U}_{copy}&(D_i,M_i^{(\boldsymbol{a})}\vert \tilde D_{i-1}^{\downarrow}) := \\
&\ket{\downarrow,R}_{\tilde D_{i-1}^{\downarrow}}\bra{\downarrow,R} \otimes \hat{U}_{copy}^{loc}(D_i\vert M_i^{(\boldsymbol{a})})\\
&+\left(\hat{I}_{\tilde D_{i-1}^{\downarrow}} - \ket{\downarrow,R}_{\tilde D_{i-1}^{\downarrow}}\bra{\downarrow,R} \right) \otimes \hat{I}_{D_i, M_i^{(\boldsymbol{a})}}
\end{split}
\end{equation}
where the operation $\hat{U}_{copy}^{loc}(D_i\vert M_i^{(\boldsymbol{a})})$ is the local unitary defined in Eq.~(\ref{localmemorydualrail}). 

All the properties of the \gdr{basic} backup memory operations are preserved in this dual-rail architecture; consequently, the backup of the final address walker always arrives at the correct memory cell, serving as the control for the copy operation onto the first data walker.
Once the entire message has been copied onto the corresponding data walkers, the particles are routed based on their color state to the corresponding rail using the output circuit introduced in Sec.\ref{dual_rail_mem}. Specifically, the circuit construction ensures that red particles propagate straight along their current rail, whereas blue particles are diverted from the lower to the upper rail. 

Since the strict spatial separation between walkers is preserved throughout the memory operations, this synchronized movement guarantees that no collisions occur between the data walkers coming from the lower rail and the backup walkers already propagating on the upper rail.
Furthermore, the deterministic evolution of the protocol provides exact knowledge of the arrival times at any given site. This allows us to apply the time-dependent color gate $\hat{X}_c$, placed immediately after the scattering gate $\hat{S}_M^{\dagger}$, only to the data walkers, thus changing their color. Ultimately, this procedure outputs a train with both the precise spatial arrangement and the exact color encoding required to execute the inverse routing phase. The copy procedure and output circuit functioning are presented in Fig.\ref{copy_dualrail_backup}.

\subsubsection{Resource scalings}
We now detail the resources required to implement the backup variant in the dual-rail setting, evaluating the architecture in terms of walker count, gate complexity, spatial extent, and circuit depth. To implement this variant, each logical carrier requires two backup walkers. \gdr{Consequently, the address register $\boldsymbol{A}$ consists of $3n$ walkers, while the data register $\boldsymbol{D}$ requires $3m$ walkers. The total walker count is therefore linear, $3(n+m)\sim\mathcal{O}(n+m)$. As for the single tree backup variant, the final data walker does not need to propagate its information to any subsequent carrier, thus the two
backups of the last data walker are not needed for the execution of the protocol and are
kept only for uniformity of notation.}

The routing architecture comprises two parallel binary trees of depth $n$, where each bifurcation is associated with a unitary block containing three primary operations: the scattering gate $\hat{S}$, alongside the decomposition of the address-transmission gate into $\hat{U}_{in}$ and $\hat{U}_B$. This structure results in a total gate count of $6(2^n-1)\sim\mathcal{O}(2^n)$. 

Similar to the backup variant of the \gdr{basic} protocol, we can optimize the spatial extent and circuit depth, in analogy to the procedure detailed in Sec.~\ref{DepthReduction}. The extent required for each tree edge can be made equal to $n^d_{\text{sites}} = 4 = \mathcal{O}(1)$ and the total path extent per tree is then equal to $n^{\text{tot}}_{\text{sites}} = 4n$, yielding a linear scaling of $\mathcal{O}(n)$. As a consequence, the circuit depth is reduced from quadratic to linear, achieving an execution time for the routing phase of $T_{\text{rout}} = 10n+4m \sim \mathcal{O}(n+m)$, as for the backup variant of the single tree protocol.

To determine the overall depth, we must account for the initialization overhead. Since the routing phase is triggered only after the entire sequence of triples has been assembled in the buffering zone, this preparation introduces a strictly sequential delay of $T_{\text{prep}} = 3(n+m)$. Therefore, the total circuit depth evaluates to $T_{\text{parallel}} = T_{\text{prep}} + T_{\text{rout}} = 13n+7m \sim \mathcal{O}(n+m)$. Finally, the number of active routing elements depends on the nature of the query, scaling linearly as $\mathcal{O}(n)$ for classical or sparse queries, and exponentially as $\mathcal{O}(2^n)$ when querying a dense superposition of addresses.

\section{Qudit variant}
\label{sec:qudit}

Another possible generalization is through the use of four-level qudits as information carriers. The four states of each qudit are separated into two logical subspaces: one used for encoding a logical $0$ and the other for encoding a logical $1$. Each logical subspace has, in turn, two levels corresponding to a state used for the routing along the binary tree. Again, we make use of two main unitary gates employed for routing information carriers; however, these operations are now defined differently with respect to the
previously explained protocols since they act on a larger Hilbert space. The underlying logic, however, is similar to that used for defining the dual-rail scheme: scattering operations are employed to route carriers according to their internal state, while conditional multi-qudit gates implement the address-transmission and copying steps.

In this section, we describe the information carriers used, how the information is encoded in them, and how they are initialized. We then present explicit definitions of the gates and their action on the basis states of our qudit system. Finally, we discuss memory access procedures and how to extend the protocol to implement the corresponding backup variant.

\subsubsection{Encoding and initialization}

In analogy to the other protocols, in this variant we use two registers: an $n$-walker address register $\boldsymbol{A}$ and an $m$-walker data register $\boldsymbol{D}$, where each walker in the registers possesses now a four-level internal degree of freedom. The total Hilbert space of a single walker factorizes as
$\mathcal{H}_W=\mathcal{H}_S\otimes\mathcal{H}_P$, where $\mathcal{H}_S$ is the local four-dimensional state space of a single qudit and $\mathcal{H}_P$ is the space of the positions along the binary tree.
We can define the space for a single walker $W$ as
\begin{multline*}  \mathcal{H}_W=\mathrm{span}\left\{\ket{s^{l,d}}_W:s\in\{1,2,3,4\},\right.\\\left.d\in\{0,\dots,n\},l\in\{1,\dots, 2^{d}\}\right\}
\end{multline*}
where $\ket{s^{l,d}}_W\equiv\ket{s}_S\otimes\ket{l,d}_P$.
Concretely, we label the four basis states of the qudit as
\begin{equation*}
    \{\ket{1}_S,\ket{2}_S,\ket{3}_S,\ket{4}_S\}\leftrightarrow \{\ket{\redcircleopen}_S,\ket{\bluecircleopen}_S,\ket{\redcirclefill}_S,\ket{\bluecirclefill}_S\}.
\end{equation*}

The local state space of each walker is naturally decomposed into two logical subspaces, respectively representing a logical $0$ and a logical $1$.
Each subspace contains two physical levels that implement the routing degree of freedom. 
More precisely, let $\circleopen\equiv \{\redcircleopen,\bluecircleopen\}$ denote a $0$, and $\circlefill\equiv \{\redcirclefill,\bluecirclefill\}$ denote a $1$, where the colors $R/B$ denote the routing degree of freedom in that subspace. Accordingly, the local Hilbert space of a single qudit can be decomposed as 
\begin{equation*}  \mathcal{H}_S=\mathcal{H}^{(0)}_S\oplus\mathcal{H}^{(1)}_S\equiv\mathcal{H}_S^{(\circleopen)}\oplus\mathcal{H}_S^{(\circlefill)}
\end{equation*}
where $\mathcal{H}^{(\circleopen)}_S=\mathrm{span}\{\ket{\redcircleopen}_S,\ket{\bluecircleopen}_S\}$ is the subspace encoding a logical~$0$ and $\mathcal{H}_S^{(\circlefill)}=\mathrm{span}\{\ket{\redcirclefill}_S,\ket{\bluecirclefill}_S\}$ encoding logical~$1$. This decomposition induces a partition for the overall space of the walker,
\begin{equation*}
    \mathcal{H}_W=(\mathcal{H}_S^{(\circleopen)}\otimes\mathcal{H}_P)\oplus(\mathcal{H}_S^{(\circlefill)}\otimes \mathcal{H}_P).
\end{equation*}

Notice that this encoding is in one-to-one correspondence with the dual-rail representation: the logical empty/full state plays the same role as the rail index, while the routing $R/B$ state plays the role of the color. Explicitly,
\begin{equation*}
\begin{split}
&\ket{\redcircleopen}\leftrightarrow\ket{r=\uparrow,c=R},\quad \ket{\bluecircleopen}\leftrightarrow \ket{r=\uparrow,c=B},\\
&\ket{\redcirclefill}\leftrightarrow \ket{r=\downarrow,c=R},\quad \ket{\bluecirclefill}\leftrightarrow \ket{r=\downarrow,c=B}.
\end{split}
\end{equation*}

During the initialization phase, the address information is encoded by setting each walker as
\begin{equation*}
    a_i=0\mapsto \ket{\redcircleopen}_{A_i},\quad a_i=1\mapsto \ket{\redcirclefill}_{A_i}
\end{equation*}
so that logical 0/1 are represented by empty/full states in the red routing mode. All data walkers $D_i$ are instead initialized in the logical $1$ red state $\ket{\redcirclefill}_{D_i}$. This is equivalent to the initialization scheme adopted in the dual-rail variant, in which we encode information in the rail degree of freedom while fixing the color to red.

\subsubsection{Long-range routing Operations}\label{subsec:qudit_long}

As for the other variants, the routing mechanism requires two core operations: an address-transmission gate $\hat{U}^{(d)}$ and a scattering gate $\hat{S}$. In this qudit variant, these operations are defined over a larger local space, and their actions are strictly dependent on the walker's logical subspace, $\mathcal{H}^{(\circleopen)}_W$ corresponding to the logical $0$, and $\mathcal{H}^{(\circlefill)}_W$ corresponding to the logical $1$. Since we are operating within a larger local space, the qudit variant gives us significantly more freedom in designing both these operations compared to the other protocols. 
To maintain consistency with the standard presentation used in the previous sections, we first present a straightforward generalization of those operations.

The address-transmission gate {$\hat U^{(d)}$} is a controlled multi-target unitary operation, taking as control particle the walker carrying the path information corresponding to level $d$ and as targets all the walkers following it. As for the scattering gate $\hat{S}$, the behavior of $\hat{U}^{(d)}$ is different depending on which logical subspace the control walker state is in. If the state of walker $A_d$ is in the subspace $\mathcal{H}_W^{(\circleopen)}$ corresponding to a logical $0=\circleopen$, the unitary will have a nontrivial action only on the particles following $A_d$ whose state is also in $\mathcal{H}_W^{(\circleopen)}$. Conversely, if the state of $A_d$ is in the logical subspace $\mathcal{H}_W^{(\circlefill)}$ corresponding to a logical $1=\circlefill$, the unitary $\hat{U}^{(d)}$ will have a non-trivial action only on walkers whose state is in the same logical subspace. More precisely, the target walker can be either in the state $\ket{\redcircleopen}_{A_d}$ or $\ket{\redcirclefill}_{A_d}$, since the color is reset to red after each operation gadget iteration. 

{Let $T_d=\boldsymbol{D}\cup\{A_i:i>d\}$} be the set of the target walkers following $A_d$. %This set can be naturally partitioned as the ordered union of the state-dependent sets
%\begin{equation*}
%    T_d=T_d^{(\circleopen)}\cup T_d^{(\circlefill)}
%\end{equation*}
%where 
%\begin{equation*}
%   T_d^{(\circleopen)}:=\{D_j:\ket{d_j}_{D_j}\in\mathcal{H}_W^{(\circleopen)}\}\cup\{A_i:i>d,\ket{a_i}_{A_i}\in\mathcal{H}_{W}^{(\circleopen)}\}
%\end{equation*}
%is the set of walkers encoding a logical $0$ and 
%\begin{equation*}
%   T_d^{(\circlefill)}:=\{D_j:\ket{d_j}_{D_j}\in\mathcal{H}_W^{(\circlefill)}\}\cup\{A_i:i>d,\ket{a_i}_{A_i}\in\mathcal{H}_{W}^{(\circlefill)}\}
%\end{equation*}
%is the set of walkers encoding a logical $1$. 
We define the set of projectors $\{\hat{\Pi}_{a_d}^{(A_d)}:=\ket{a_d}_{A_d}\bra{a_d}, a_d\in\{\redcircleopen,\redcirclefill,\bluecircleopen,\bluecirclefill\}\}$ onto the possible basis states of the qudit $A_d$. 
Since the action of $\hat{U}^{(d)}$ preserves the logical subspace of both control and target walkers, this operator  can be expressed as a direct sum of two operators
\begin{equation*}    \hat{U}^{(d)}=\hat{U}^{(d)}_{\circleopen}\oplus\hat{U}^{(d)}_{\circlefill}
\end{equation*}
where $\hat{U}^{(d)}_{\circleopen}$ acts non trivially 
only on the walkers belonging to the logical $0$ subspace $\mathcal{H}^{(\circleopen)}_W$, while $\hat{U}^{(d)}_{\circlefill}$ only on those belonging to the logical $1$ subspace $\mathcal{H}_W^{(\circlefill)}$.
We then define the operation $\hat{U}^{(d)}_{\circleopen}$ as
\begin{equation}
\hat{U}^{(d)}_{\circleopen}=\sum_{a_d\in\{\redcircleopen,\bluecircleopen\}} \hat{\Pi}^{(A_d)}_{a_d}\otimes \hat{V}^{(T_d)}_{a_d}
\end{equation}
where $\hat{V}^{(T_d)}$ is a multi-target unitary acting on the walkers in $T_d$. \gdr{The operator is defined as
\begin{equation*}   \hat{V}^{(T_d)}_{a_d=\redcircleopen}=\bigotimes_{j\in T_d}\hat{X}^{C,0}_j,\quad \hat{V}^{(T_d)}_{a_d=\bluecircleopen}=\bigotimes_{j\in T_d}\hat{I}_j 
\end{equation*}
where $\hat{X}^{C,0}$ is a NOT operation acting non-trivially on the routing color degree of freedom only if the carried logical state is zero:
\begin{equation*}  
\begin{split}
&\hat{X}^{C,0}\ket{\redcircleopen}=\ket{\bluecircleopen},\quad \hat{X}^{C,0}\ket{\redcirclefill}=\ket{\redcirclefill},\\&
\hat{X}^{C,0}\ket{\bluecircleopen}=\ket{\redcircleopen},\quad \hat{X}^{C,0}\ket{\bluecirclefill}=\ket{\bluecirclefill}.
\end{split}
\end{equation*}}
Analogously, the operation $\hat{U}^{(d)}_{\circlefill}$ is defined as
\begin{equation*}
\hat{U}^{(d)}_{\circlefill}=\sum_{a_d\in\{\redcirclefill,\bluecirclefill\} }\hat{\Pi}^{(A_d)}_{a_d}\otimes \hat{V}^{(T_d)}_{a_d}
\end{equation*}
with 
\gdr{\begin{equation*}   \hat{V}^{(T_d)}_{a_d=\redcirclefill}=\bigotimes_{j\in T_d}\hat{X}^{C,1}_j,\quad \hat{V}^{(T_d)}_{a_d=\bluecirclefill}=\bigotimes_{j\in T_d}\hat{I}_j 
\end{equation*}
where $\hat{X}^{C,1}$ is a NOT operation acting non-trivially on the routing color degree of freedom only for walkers carrying a logical one state:
\begin{equation*}
    \begin{split} &\hat{X}^{C,1}\ket{\redcircleopen}=\ket{\redcircleopen},\quad \hat{X}^{C,1}\ket{\redcirclefill}=\ket{\bluecirclefill},\\&
\hat{X}^{C,1}\ket{\bluecircleopen}=\ket{\bluecircleopen},\quad \hat{X}^{C,1}\ket{\bluecirclefill}=\ket{\redcirclefill}.
    \end{split}
\end{equation*}
Notice that $\hat X^{C,0}$ and $\hat X^{C,1}$ act as the identity on the walkers
encoding a logical $1$ and a logical $0$, respectively. The tensor products above
can therefore be taken over the whole target set $T_d$ with no need to restrict
them to the walkers carrying the corresponding logical value.}
By construction, the two blocks act on orthogonal logical subspaces of the control walker, and each block is unitary on its respective target subspace; therefore, the full address-transmission gate $\hat{U}^{(d)}$ is unitary.

The scattering gate $\hat{S}$ acts locally on each walker and routes it according to its color state, following different rules inside the two logical subspaces. Concretely, when the qudit encodes a logical $0=\circleopen$, a $\redcircleopen$ state is routed to the left with its color unchanged, while a $\bluecircleopen$ state is routed to the right and its color is reset to red. Symmetrically, when the walker encodes a logical $1=\circlefill$, a $\redcirclefill$ state is routed to the right with its color unchanged, whereas a $\bluecirclefill$ state is moved to the left and its color is reset to red. More formally, the action of $\hat{S}$ on the four basis states of the qudit can be written as 
\begin{equation}
    \begin{split}
    &\hat{S}\ket{\redcircleopen^{l,d}}=\ket{\redcircleopen^{2l-1,d+1}},
    \hat{S}\ket{\bluecircleopen^{l,d}}=\ket{\redcircleopen^{2l,d+1}},\\&
    \hat{S}\ket{\redcirclefill^{l,d}}=\ket{\redcirclefill^{2l,d+1}},
    \hat{S}\ket{\bluecirclefill^{l,d}}=\ket{\redcirclefill^{2l-1,d+1}}.
    \end{split}
\end{equation}

Notice that, by construction, this gate preserves the logical subspace over which it acts: the states encoding a logical $0$ are mapped to other states encoding a logical $0$, while states encoding a logical $1$ are mapped in other states encoding a logical $1$. This allows us to rewrite the operator as a direct sum of operators acting on the different logical subspaces:
\begin{equation*}    \hat{S}=\hat{S}^{(\circleopen)}\oplus\hat{S}^{(\circlefill)}
\end{equation*}
where $\hat{S}^{(\circleopen)}$ acts only on $\mathcal{H}^{(\circleopen)}_W$ while $\hat{S}^{(\circlefill)}$ only on $\mathcal{H}^{(\circlefill)}_W$, each of them vanishing on the complementary subspace.
%More explicitly, the two contributions can be written as 
%\begin{equation}
%    \begin{split}
    %&\hat{S}^{(\circleopen)}=\ket{\redcircleopen^{2l-1,d+1}}\bra{\redcircleopen^{l,d}}+\ket{\redcircleopen^{2l,d+1}}\bra{\bluecircleopen^{l,d}};
    %\\&\hat{S}^{(\circlefill)}=\ket{\redcirclefill^{2l,d+1}}\bra{\redcirclefill^{l,d}}+\ket{\redcirclefill^{2l-1,d+1}}\bra{\bluecirclefill^{l,d}}.
    %\end{split}
%\end{equation}
%Notice that both operators are permutations of the basis states of the subspaces over which they act, and thus are explicitly unitary.

{This routing approach is essentially equivalent to the long-range dual-rail implementation of the protocol, with the main difference being that here all the routing takes place onto a single binary tree. Notice that, because of this compression, we operate with a constant number of walkers, meaning that the absence of a particle is no longer used to encode information or perform routing. Instead, the rail differences are now encoded in both the orthogonal subspaces into which the overall Hilbert space of the qudit is divided and in the definition of the routing operations. 
More precisely, we can directly map the subspace $\mathcal{H}_W^{(\circleopen)}$ to the state of a walker moving on the upper rail $r=\uparrow$, and, symmetrically, we can identify the subspace $\mathcal{H}_W^{(\circlefill)}$ as the one corresponding to a walker moving on the lower rail $r=\downarrow$. Moreover, the operations that act differently on the two rails in the dual-rail variant are now unified into single gates, whose action is conditionally dependent on the logical subspace to which the walker's state belongs.}

\begin{figure*}[ht]
    \centering
\begin{tikzpicture}[scale=0.75, transform shape,
  level 1/.style={level distance=1.5cm, sibling distance=3cm},
  level 2/.style={level distance=1cm, sibling distance=1cm},
  grow=east, % Cresce verso destra
  edge from parent/.style={draw, thick}, % Stile dei rami
  every node/.style={rectangle, draw, minimum size=6mm,fill=cyan!30}, % Nodi interni
  leaf/.style={rectangle, draw, fill=gray!30, minimum size=6mm}, % Foglie grigie
  block/.style={rectangle, draw=none,dashed,fill=none, minimum width=1.8cm, minimum height=5mm, rotate around={0:(0,0)}}, % Blocco blu chiaro
  gateX/.style={rectangle, draw, fill=green!20, minimum size=5mm, rotate around={0:(0,0)}} % Blocco quadrato rosso
]

\begin{scope}[shift={(0,0)}]

% Nodo radice
\node (root1) {$\hat{S}$}
  child {
    % Primo nodo intermedio (verso l'alto)
    node {$\hat{S}$}
    child { 
      %node[leaf] {$\boldsymbol{x}_3$} % Foglia sinistra-sinistra (grigia)
      edge from parent {}
    }
    child { 
      %node[leaf] {$\boldsymbol{x}_2$} % Foglia sinistra-destra (grigia)
      edge from parent{
%node[midway,xshift=-1cm,gateX, yshift=-0.5cm,rotate=27] {$\hat{X}$} % Blocco rosso ruotato
      }
    }
    edge from parent {
      node[midway,xshift=0cm, yshift=-0cm,  block, rotate=-45] {} % Blocco blu ruotato
    }
  }
  child {
    % Secondo nodo intermedio (verso l'alto)
    node {$\hat{S}$}
    child { 
      %node[leaf] {$\boldsymbol{x}_1$} % Foglia destra-sinistra (grigia)
      %edge from parent {}
    }
    child { 
      %node[leaf] {$\boldsymbol{x}_0$} % Foglia destra-destra (grigia)
      edge from parent{ 
      %node[midway,xshift=-1cm, yshift=-0.5cm,gateX, rotate=27] {$\hat{X}$} % Blocco rosso ruotato
      }
    }
    edge from parent {
      %node[midway, xshift=-1.3cm,yshift=-1.3cm,gateX, rotate=45] {$\hat{X}$} % Blocco rosso ruotato
      node[midway, xshift=0cm,yshift=0cm, block, rotate=45] {} % Blocco blu ruotato
    }
  };

% Ramo in ingresso al nodo radice
\draw (-6, 0) -- (root1) node(Ugateroot)[xshift=-1.37cm, block, rotate=0] {};

%nodes before root node

\draw[fill=black] (-0.7,0) circle (2pt)node[above=3pt, fill = none, draw=none]{};
\draw[fill=black] (-1.4,0) circle (2pt)node[above=3pt, fill = none, draw=none]{};

\draw[thick, fill=red] (-2.1,0) circle (3pt) node[below=3pt, fill = none, draw=none]{\scriptsize$A_1$};
\draw[thick, draw=red, fill=white] (-2.8,0) circle (3pt) node[below=3pt, fill = none, draw=none]{\scriptsize$A_2$};
\draw[thick, fill=red] (-3.5,0) circle (3pt)node[below=3pt, fill = none, draw=none]{\scriptsize$A_{3}$};
\draw[thick, fill=red] (-4.2,0) circle (3pt)node[below=3pt, fill = none, draw=none]{\scriptsize$A_{4}$};
\draw[thick, fill=white, draw=red] (-4.9,0) circle (3pt)node[below=3pt, fill = none, draw=none]{\scriptsize$A_{5}$};
\draw[thick, fill=red] (-5.6,0) circle (3pt)node[below=3pt, fill = none, draw=none]{\scriptsize$A_{6}$};

%nodes first level upper branch

\draw[fill=black] (1,1) circle (2pt);
\draw[fill=black] (0.5,0.5) circle (2pt);

%nodes first level lower branch

\draw[fill=black] (1,-1) circle (2pt);
\draw[fill=black] (0.5,-0.5) circle (2pt);

\node[draw=none, fill=none] at (-1.1,-0.8){\scalebox{1.2}{$(1,1)$}};

\node[draw=none, fill=none, rotate=-45] at (0.5,-1.5){\scalebox{1.2}{$(1,2)$}};

\node[draw=none, fill=none, rotate=45] at (0.5,1.5){\scalebox{1.2}{$(2,2)$}};

%in-out arrows

\draw[->, >={Stealth[length=6pt]}, thick] (-4.5,0.5) -- (-3,0.5)node[fill=none,draw=none,midway,above] {};

\node[draw=none, fill=none, align=left] at (-4.5,-2)
{\large Step 1: initialization of the walkers\\ \large at the root node };

\node[draw=none, fill=none] at (-7,2){\Large $(a)$};

\end{scope}

%%%%%% SECONDO DISEGNO%%%%%%%%%%%
\begin{scope}[shift={(0,-6)}]

% Nodo radice
\node (root1) {$\hat{S}$}
  child {
    % Primo nodo intermedio (verso l'alto)
    node {$\hat{S}$}
    child { 
      %node[leaf] {$\boldsymbol{x}_3$} % Foglia sinistra-sinistra (grigia)
      edge from parent {}
    }
    child { 
      %node[leaf] {$\boldsymbol{x}_2$} % Foglia sinistra-destra (grigia)
      edge from parent{
%node[midway,xshift=-1cm,gateX, yshift=-0.5cm,rotate=27] {$\hat{X}$} % Blocco rosso ruotato
      }
    }
    edge from parent {
      node[midway,xshift=0cm, yshift=-0cm,  block, rotate=-45] {} % Blocco blu ruotato
    }
  }
  child {
    % Secondo nodo intermedio (verso l'alto)
    node {$\hat{S}$}
    child { 
      %node[leaf] {$\boldsymbol{x}_1$} % Foglia destra-sinistra (grigia)
      %edge from parent {}
    }
    child { 
      %node[leaf] {$\boldsymbol{x}_0$} % Foglia destra-destra (grigia)
      edge from parent{ 
      %node[midway,xshift=-1cm, yshift=-0.5cm,gateX, rotate=27] {$\hat{X}$} % Blocco rosso ruotato
      }
    }
    edge from parent {
      %node[midway, xshift=-1.3cm,yshift=-1.3cm,gateX, rotate=45] {$\hat{X}$} % Blocco rosso ruotato
      node[midway, xshift=0cm,yshift=0cm, block, rotate=45] {} % Blocco blu ruotato
    }
  };

% Ramo in ingresso al nodo radice
\draw (-6, 0) -- (root1) node(Ugateroot)[xshift=-1.37cm, block, rotate=0] {};

%gates

\node[block,draw, thick,solid, minimum width = 30pt, fill=orange!20,label={[yshift=1mm]above:$\hat{U}_{in}$}] at (-1.05,0){};

%nodes before root node

\draw[fill=red,thick] (-0.7,0) circle (3pt)node[below=4pt, fill = none, draw=none]{\scriptsize$A_1$};
\draw[fill=white,draw=blue,thick] (-1.4,0) circle (3pt)node[below=4pt, fill = none, draw=none]{\scriptsize$A_2$};
\draw[thick, fill=red] (-2.1,0) circle (3pt);
\draw[thick, fill=red] (-2.8,0) circle (3pt);
\draw[thick, fill=white,draw=red] (-3.5,0) circle (3pt);
\draw[thick, fill=red] (-4.2,0) circle (3pt);

%nodes first level upper branch

\draw[fill=black] (1,1) circle (2pt);
\draw[fill=black] (0.5,0.5) circle (2pt);

%nodes first level lower branch

\draw[fill=black] (1,-1) circle (2pt);
\draw[fill=black] (0.5,-0.5) circle (2pt);

\node[draw=none, fill=none, align=left] at (-4.5,-2)
{\large Step 2: shift of one position and \\ \large application of $\hat{U}_{in}$};

\node[draw=none, fill=none] at (-7,2){\Large $(b)$};
   
\end{scope}

%%%%%%%%%%% TERZO DISEGNO %%%%%%%%%

\begin{scope}[shift={(0,-12)}]

% Nodo radice
\node (root1) {$\hat{S}$}
  child {
    % Primo nodo intermedio (verso l'alto)
    node {$\hat{S}$}
    child { 
      %node[leaf] {$\boldsymbol{x}_3$} % Foglia sinistra-sinistra (grigia)
      edge from parent {}
    }
    child { 
      %node[leaf] {$\boldsymbol{x}_2$} % Foglia sinistra-destra (grigia)
      edge from parent{
%node[midway,xshift=-1cm,gateX, yshift=-0.5cm,rotate=27] {$\hat{X}$} % Blocco rosso ruotato
      }
    }
    edge from parent {
      node[midway,xshift=0cm, yshift=-0cm,  block, rotate=-45] {} % Blocco blu ruotato
    }
  }
  child {
    % Secondo nodo intermedio (verso l'alto)
    node {$\hat{S}$}
    child { 
      %node[leaf] {$\boldsymbol{x}_1$} % Foglia destra-sinistra (grigia)
      %edge from parent {}
    }
    child { 
      %node[leaf] {$\boldsymbol{x}_0$} % Foglia destra-destra (grigia)
      edge from parent{ 
      %node[midway,xshift=-1cm, yshift=-0.5cm,gateX, rotate=27] {$\hat{X}$} % Blocco rosso ruotato
      }
    }
    edge from parent {
      %node[midway, xshift=-1.3cm,yshift=-1.3cm,gateX, rotate=45] {$\hat{X}$} % Blocco rosso ruotato
      node[midway, xshift=0cm,yshift=0cm, block, rotate=45] {} % Blocco blu ruotato
    }
  };

% Ramo in ingresso al nodo radice
\draw (-6, 0) -- (root1) node(Ugateroot)[xshift=-1.37cm, block, rotate=0] {};

%gates

\node[block,draw, thick,solid, minimum width = 30pt, fill=blue!20,label={[yshift=1mm]above:$\hat{U}_{B}$}] at (-1.05,0){};

%nodes before root node

\draw[fill=white,draw=blue,thick] (-0.7,0) circle (3pt)node[below=4pt, fill = none, draw=none]{\scriptsize$A_2$};
\draw[fill=blue,thick] (-1.4,0) circle (3pt)node[below=4pt, fill = none, draw=none]{\scriptsize$A_3$};
\draw[thick, fill=red] (-2.1,0) circle (3pt);
\draw[thick, fill=white, draw=red] (-2.8,0) circle (3pt);
\draw[thick, fill=red] (-3.5,0) circle (3pt);
\draw[fill=black] (-4.2,0) circle (2pt);

%nodes first level upper branch

\draw[fill=black] (1,1) circle (2pt);
\draw[fill=black] (0.5,0.5) circle (2pt);

%nodes first level lower branch

\draw[fill=black] (1,-1) circle (2pt);
\draw[fill=red, thick] (0.5,-0.5) circle (3pt);

\node[draw=none, fill=none, align=left] at (-4.5,-2)
{\large Step 3: shift of one position and \\ \large application of $\hat{U}_{B}$};

\node[draw=none, fill=none] at (-7,2){\Large $(c)$};

\end{scope}

%%%%%%%%% QUARTO DISEGNO %%%%%

\begin{scope}[shift={(12,0)}]

% Nodo radice
\node (root1) {$\hat{S}$}
  child {
    % Primo nodo intermedio (verso l'alto)
    node {$\hat{S}$}
    child { 
      %node[leaf] {$\boldsymbol{x}_3$} % Foglia sinistra-sinistra (grigia)
      edge from parent {}
    }
    child { 
      %node[leaf] {$\boldsymbol{x}_2$} % Foglia sinistra-destra (grigia)
      edge from parent{
%node[midway,xshift=-1cm,gateX, yshift=-0.5cm,rotate=27] {$\hat{X}$} % Blocco rosso ruotato
      }
    }
    edge from parent {
      node[midway,xshift=0cm, yshift=-0cm,  block, rotate=-45] {} % Blocco blu ruotato
    }
  }
  child {
    % Secondo nodo intermedio (verso l'alto)
    node {$\hat{S}$}
    child { 
      %node[leaf] {$\boldsymbol{x}_1$} % Foglia destra-sinistra (grigia)
      %edge from parent {}
    }
    child { 
      %node[leaf] {$\boldsymbol{x}_0$} % Foglia destra-destra (grigia)
      edge from parent{ 
      %node[midway,xshift=-1cm, yshift=-0.5cm,gateX, rotate=27] {$\hat{X}$} % Blocco rosso ruotato
      }
    }
    edge from parent {
      %node[midway, xshift=-1.3cm,yshift=-1.3cm,gateX, rotate=45] {$\hat{X}$} % Blocco rosso ruotato
      node[midway, xshift=0cm,yshift=0cm, block, rotate=45] {} % Blocco blu ruotato
    }
  };

% Ramo in ingresso al nodo radice
\draw (-6, 0) -- (root1) node(Ugateroot)[xshift=-1.37cm, block, rotate=0] {};

%gates

\node[block,draw, thick,solid, minimum width = 30pt, fill=blue!20,label={[yshift=1mm]above:$\hat{U}_{B}$}] at (-1.05,0){};

%nodes before root node

\draw[fill=blue,thick] (-0.7,0) circle (3pt)node[below=4pt, fill = none, draw=none]{\scriptsize$A_3$};
\draw[fill=blue,thick] (-1.4,0) circle (3pt)node[below=4pt, fill = none, draw=none]{\scriptsize$A_4$};
\draw[thick, fill=white, draw=red] (-2.1,0) circle (3pt);
\draw[thick, fill=red] (-2.8,0) circle (3pt);
\draw[fill=black] (-3.5,0) circle (2pt);
\draw[fill=black] (-4.2,0) circle (2pt);

%nodes first level upper branch

\draw[fill=black] (1,1) circle (2pt);
\draw[fill=white, draw=red, thick] (0.5,0.5) circle (3pt);

%nodes first level lower branch

\draw[fill=red, thick] (1,-1) circle (3pt);
\draw[fill=black] (0.5,-0.5) circle (2pt);

\node[draw=none, fill=none, align=left] at (-4.5,-2)
{\large Step 4: shift of one position and \\ \large application of $\hat{U}_{B}$};

\node[draw=none, fill=none] at (-7,2){\Large $(d)$};

\end{scope}

%%%%%%%%%% QUINTO DISEGNO %%%%%%%%

\begin{scope}[shift={(12,-6)}]

% Nodo radice
\node (root1) {$\hat{S}$}
  child {
    % Primo nodo intermedio (verso l'alto)
    node {$\hat{S}$}
    child { 
      %node[leaf] {$\boldsymbol{x}_3$} % Foglia sinistra-sinistra (grigia)
      edge from parent {}
    }
    child { 
      %node[leaf] {$\boldsymbol{x}_2$} % Foglia sinistra-destra (grigia)
      edge from parent{
%node[midway,xshift=-1cm,gateX, yshift=-0.5cm,rotate=27] {$\hat{X}$} % Blocco rosso ruotato
      }
    }
    edge from parent {
      node[midway,xshift=0cm, yshift=-0cm,  block, rotate=-45] {} % Blocco blu ruotato
    }
  }
  child {
    % Secondo nodo intermedio (verso l'alto)
    node {$\hat{S}$}
    child { 
      %node[leaf] {$\boldsymbol{x}_1$} % Foglia destra-sinistra (grigia)
      %edge from parent {}
    }
    child { 
      %node[leaf] {$\boldsymbol{x}_0$} % Foglia destra-destra (grigia)
      edge from parent{ 
      %node[midway,xshift=-1cm, yshift=-0.5cm,gateX, rotate=27] {$\hat{X}$} % Blocco rosso ruotato
      }
    }
    edge from parent {
      %node[midway, xshift=-1.3cm,yshift=-1.3cm,gateX, rotate=45] {$\hat{X}$} % Blocco rosso ruotato
      node[midway, xshift=0cm,yshift=0cm, block, rotate=45] {} % Blocco blu ruotato
    }
  };

% Ramo in ingresso al nodo radice
\draw (-6, 0) -- (root1) node(Ugateroot)[xshift=-1.37cm, block, rotate=0] {};

%gates

\node[block,draw, thick,solid, minimum width = 30pt, fill=blue!20,label={[yshift=1mm]above:$\hat{U}_{B}$}] at (-1.05,0){};

\node[block,draw,thick,minimum width = 30pt, solid, fill=orange!20, rotate = 45,label={[yshift=1mm]above:$\hat{U}_{in}$}] at (0.75,0.75){};

%nodes before root node

\draw[fill=blue,thick] (-0.7,0) circle (3pt)node[below=4pt, fill = none, draw=none]{\scriptsize$A_4$};
\draw[fill=white,draw=blue, thick] (-1.4,0) circle (3pt)node[below=4pt, fill = none, draw=none]{\scriptsize$A_5$};
\draw[thick, fill=red] (-2.1,0) circle (3pt);
\draw[ fill=black] (-2.8,0) circle (3pt);
\draw[fill=black] (-3.5,0) circle (2pt);
\draw[fill=black] (-4.2,0) circle (2pt);

%nodes first level upper branch

\draw[fill=white, draw=red, thick] (1,1) circle (3pt)node[below right=4pt, fill = none, draw=none]{\scriptsize$A_2$};
\draw[fill=red, thick] (0.5,0.5) circle (3pt)node[below right=4pt, fill = none, draw=none]{\scriptsize$A_3$};

%nodes first level lower branch

\draw[fill=black] (1,-1) circle (2pt);
\draw[fill=black] (0.5,-0.5) circle (2pt);

\node[draw=none, fill=none, align=left] at (-4.5,-2)
{\large Step 5:  shift of one position and \\ \large parallel application of $\hat{U}_{in}$ and $\hat{U}_B$ \\ \large at different levels {\color{white} $\hat{U}_B$}};

\node[draw=none, fill=none] at (-7,2){\Large $(e)$};

\end{scope}

%%%%%%%%%% SESTO DISEGNO %%%%%%%%

\begin{scope}[shift={(12,-12)}]

% Nodo radice
\node (root1) {$\hat{S}$}
  child {
    % Primo nodo intermedio (verso l'alto)
    node {$\hat{S}$}
    child { 
      %node[leaf] {$\boldsymbol{x}_3$} % Foglia sinistra-sinistra (grigia)
      edge from parent {}
    }
    child { 
      %node[leaf] {$\boldsymbol{x}_2$} % Foglia sinistra-destra (grigia)
      edge from parent{
%node[midway,xshift=-1cm,gateX, yshift=-0.5cm,rotate=27] {$\hat{X}$} % Blocco rosso ruotato
      }
    }
    edge from parent {
      node[midway,xshift=0cm, yshift=-0cm,  block, rotate=-45] {} % Blocco blu ruotato
    }
  }
  child {
    % Secondo nodo intermedio (verso l'alto)
    node {$\hat{S}$}
    child { 
      %node[leaf] {$\boldsymbol{x}_1$} % Foglia destra-sinistra (grigia)
      %edge from parent {}
    }
    child { 
      %node[leaf] {$\boldsymbol{x}_0$} % Foglia destra-destra (grigia)
      edge from parent{ 
      %node[midway,xshift=-1cm, yshift=-0.5cm,gateX, rotate=27] {$\hat{X}$} % Blocco rosso ruotato
      }
    }
    edge from parent {
      %node[midway, xshift=-1.3cm,yshift=-1.3cm,gateX, rotate=45] {$\hat{X}$} % Blocco rosso ruotato
      node[midway, xshift=0cm,yshift=0cm, block, rotate=45] {} % Blocco blu ruotato
    }
  };

% Ramo in ingresso al nodo radice
\draw (-6, 0) -- (root1) node(Ugateroot)[xshift=-1.37cm, block, rotate=0] {};

%gates

\node[block,draw, thick,solid, minimum width = 30pt, fill=blue!20,label={[yshift=1mm]above:$\hat{U}_{B}$}] at (-1.05,0){};

\node[block,draw,thick,minimum width = 30pt, solid, fill=blue!20, rotate = 45,label={[yshift=1mm]above:$\hat{U}_{B}$}] at (0.75,0.75){};

%nodes before root node

\draw[fill=white,draw=blue,thick] (-0.7,0) circle (3pt)node[below=4pt, fill = none, draw=none]{\scriptsize$A_5$};
\draw[fill=blue, thick] (-1.4,0) circle (3pt)node[below=4pt, fill = none, draw=none]{\scriptsize$A_6$};
\draw[ fill=black] (-2.1,0) circle (2pt);
\draw[ fill=black] (-2.8,0) circle (2pt);
\draw[fill=black] (-3.5,0) circle (2pt);
\draw[fill=black] (-4.2,0) circle (2pt);

%nodes first level upper branch

\draw[fill=red, thick] (1,1) circle (3pt)node[below right=4pt, fill = none, draw=none]{\scriptsize$A_3$};
\draw[fill=red, thick] (0.5,0.5) circle (3pt)node[below right=4pt, fill = none, draw=none]{\scriptsize$A_4$};

%nodes first level lower branch

\draw[fill=black] (1,-1) circle (2pt);
\draw[fill=black] (0.5,-0.5) circle (2pt);

\node[draw=none, fill=none, align=left] at (-4.5,-2)
{\large Step 6:  shift of one position and \\ \large parallel application of $\hat{U}_B$ \\ \large at different levels{\color{white}$\hat{U}_B$}};

\node[draw=none, fill=none] at (-7,2){\Large $(f)$};

\end{scope}

\draw[dashed] (4,3)--(4,-15);

\draw[dashed] (-8,-3)--(16,-3);

\draw[dashed] (-8,-9)--(16,-9);

\end{tikzpicture}
\caption{
Example of parallelization schedule, for qudits with short range operations, between the first two levels of the binary tree: (a) the walkers are initialized at level $d=1$ of the tree; (b) after a site shift, the initialization gate $\hat{U}_{in}$ is applied to the pair $(A_2,A_1)$; (c) following another site shift, the block gate $\hat{U}_B$ acts on the couple $(A_{3}, A_2)$; (d) after the next shift $\hat{U}_B$ is applied to the couple $(A_{4}, A_{3})$; (e) subsequent shifts allow the simultaneous application $\hat{U}_{B}$ at level $d=1$ and $\hat{U}_{in}$ at level $d=2$; (f) from this point onward the applications of gates $\hat{U}_B$ at levels $d=1$ and $d=2$ can be executed in parallel.}
\label{qudit_short}
\end{figure*}
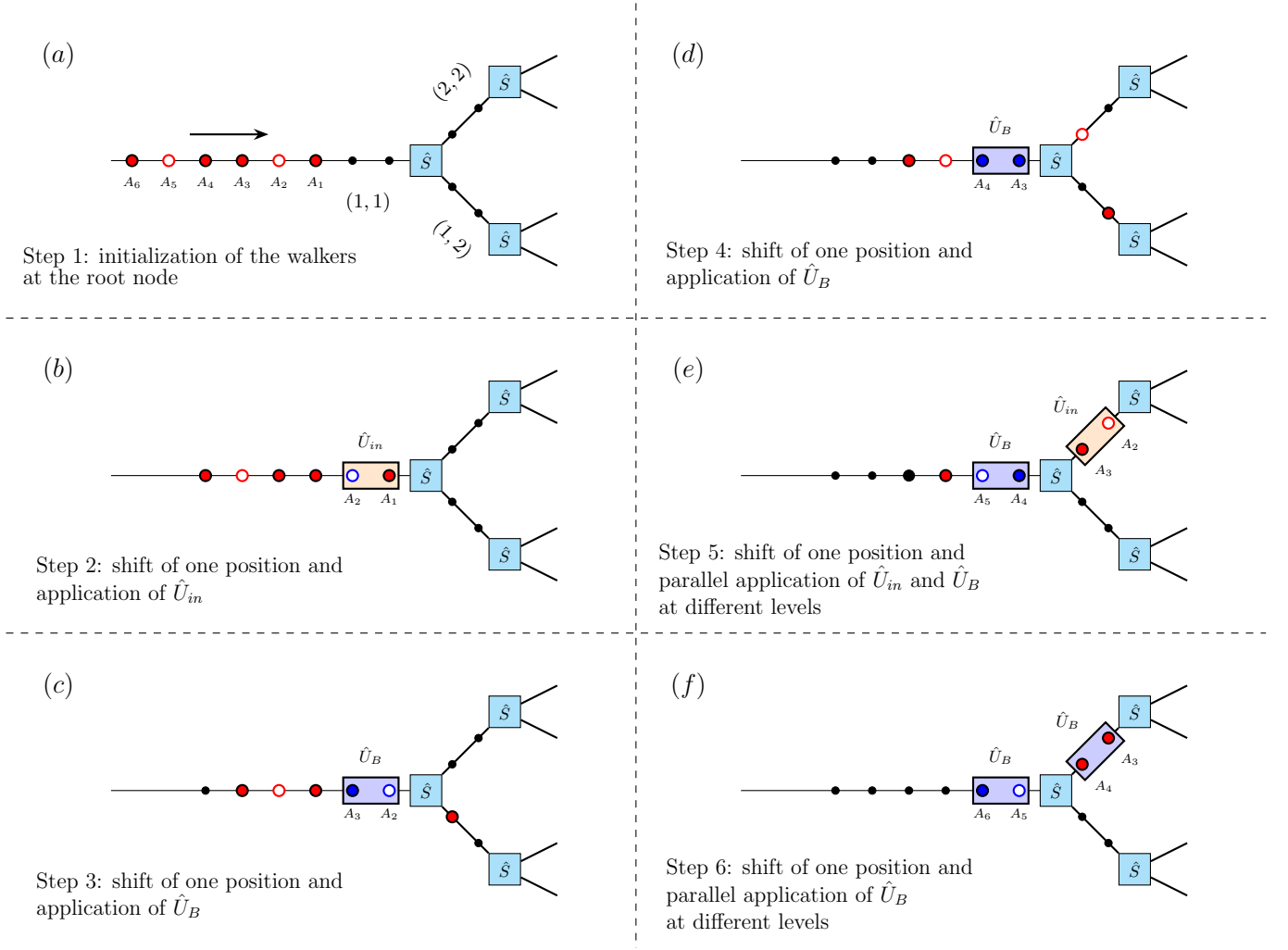

\subsubsection{Short-range routing operations}

The larger Hilbert space provided by the use of qudits as information carriers allows us to design more sophisticated operations. These operations, in turn, enable us to greatly simplify the routing structure, achieving a reduction in complexity analogous to the backup variant, without the need to introduce backup walkers at all. Indeed, we demonstrate how to design short-range routing operations that do not require additional backup walkers to operate.
As introduced earlier, we can partition the overall Hilbert space into two subspaces representing the degrees of freedom used to encode the logical information (empty/filled). Each of these subspaces is further equipped with an internal color state. While our previous protocols relied on strictly subspace-preserving operations, i.e., acting exclusively on qudits within a specific logical subspace, here we introduce operations that can act across both logical subspaces simultaneously.

We decompose the global transmission gate $\hat{U}^{(d)}$ into a sequence of short-range, nearest-neighbor interactions. This sequence consists of a single-shot initialization operator $\hat{U}_{in}$, acting on the first pair of walkers $A_d$ and $A_{d+1}$, followed by a block operator $\hat{U}_B$ applied sequentially to all subsequent adjacent pairs up to the final data qudit $D_m$. The initialization gate $\hat{U}_{in}$ is controlled exclusively by the logical state of the address $A_d$. While the protocol ensures that $A_d$ always reaches the control site in the red state $\ket{R}$, the gate is conditioned only on its logical state
and acts solely on the color degree of freedom of the target qudit $A_{d+1}$.
Explicitly, it can be formalized as:
\begin{equation}\label{Uinqudit}
\hat{U}_{in}(A_{d+1}|A_d) := \sum_{a_d \in \{\circleopen, \circlefill\}} \hat{\Pi}_{a_d}^{(A_d)} \otimes \hat{V}_{a_d}^{(A_{d+1})}
\end{equation}
where $\{\hat{\Pi}^{(A_d)}_{a_d}\}$ is the set of projectors onto the logical subspaces of the address walker $A_d$ and $\hat{V}_{a_d}^{(A_{d+1})}$ is a target unitary defined as
\begin{equation*}
\hat{V}_{a_d=\circlefill}^{(A_{d+1})} = \hat{X}_{A_{d+1}}^C, \quad \hat{V}_{a_d=\circleopen}^{(A_{d+1})} = \hat{I}_{A_{d+1}},
\end{equation*}
where $\hat{X}^C$ acts as a color NOT gate: it flips the color state of the target qudit independently of its logical subspace, not altering its logical information. \gdr{Its action on the basis states is therefore
\begin{equation*}
\begin{split}
&\hat{X}^C\ket{\redcircleopen} = \ket{\bluecircleopen}, \quad \hat{X}^C\ket{\redcirclefill} = \ket{\bluecirclefill},\\&
\hat{X}^C\ket{\bluecircleopen} = \ket{\redcircleopen},\quad \hat{X}^C\ket{\bluecirclefill} = \ket{\redcirclefill}.
\end{split}
\end{equation*}
Notice that since the operators $\hat{X}^{C,0}$ and $\hat{X}^{C,1}$ introduced in Sec.~\ref{subsec:qudit_long} act on mutually exclusive logical subspaces, their product naturally yields the full color NOT operation across the entire state space, satisfying $\hat{X}^C=\hat{X}^{C,0}\hat{X}^{C,1}$.}

The action of this first routing gate $\hat U_{in}$ transcribes the logical routing instruction of $A_d$ onto the color degree of freedom of the subsequent particle $A_{d+1}$, without altering the logical information carried by $A_{d+1}$ itself, which must be preserved for the next routing level. Given that all particles are initially prepared with color red, the routing instruction is effectively encoded by flipping the target qudit's color to blue if and only if the address particle occupies a state in the logical subspace $1$, namely $a_d=\circlefill$.

Once this information is successfully encoded into the color of $A_{d+1}$, we iteratively apply the block gate $\hat{U}_B$ across adjacent pairs of walkers $(W_{i+1},W_i)$, $i\in\{d+1,\dots,n+m\}$.
This gate propagates the color state down the train, passing the instruction from $A_{d+1}$ to all remaining walkers following it. More precisely, we define this block gate as:
\begin{equation*}
    \hat{U}_B(W_{i+1}\vert W_i):=\sum_{c\in\{R,B\}}\hat{\Pi}_{c}^{(W_i)}\otimes \hat{V}_c^{(W_{i+1})}
\end{equation*}
where now $\{\hat{\Pi}_c^{(W_i)}\}$ is the set of projectors onto the color subspaces of the $i$.th qudit and the unitary $\hat{V}_c^{(W_{i+1})}$ is defined as
\begin{equation*}  \hat{V}_{c=R}^{(W_{i+1})}=\hat{I}_{W_{i+1}},\quad \hat{V}_{c=B}^{(W_{i+1})}=\hat{X}^C_{W_{i+1}}.
\end{equation*}
The gate $\hat U_B$ acts exclusively on the color degree of freedom, leaving the logical information carried by the qudits strictly untouched. Hence, a blue control walker triggers a Pauli $\hat{X}$ operation on the target following walker. Assuming the target is initially red, this operation flips its color to blue, effectively copying the routing bit. By applying this gate iteratively, the routing information propagates along the entire train, completing the transmission after $t_B = n+m-2$ iterations. 

The scattering gate, in this short range variant, $\hat{S}_{sr}$ acts locally on each qudit. Unlike the versions introduced earlier, this $\hat{S}_{sr}$ gate ignores the logical subspace of the qudit, routing it based solely on its color state. Specifically, if a particle arriving at the node in a red state $\ket{\redcircleopen}$ or $\ket{\redcirclefill}$ it is routed to the left branch, whereas a particle in a blue state $\ket{\bluecircleopen}$ or $\ket{\bluecirclefill}$ is moved to the right branch and its color is reset to red. Formally, the action of the gate on the basis states is:
\begin{equation}
\begin{split}
&\hat{S}_{sr}\ket{\redcircleopen^{l,d}}=\ket{\redcircleopen^{2l-1,d+1}}, \quad \hat{S}_{sr}\ket{\redcirclefill^{l,d}}=\ket{\redcirclefill^{2l-1,d+1}},\\
&\hat{S}_{sr}\ket{\bluecircleopen^{l,d}}=\ket{\redcircleopen^{2l,d+1}}, \quad \hat{S}_{sr}\ket{\bluecirclefill^{l,d}}=\ket{\redcirclefill^{2l,d+1}}.
\end{split}
\end{equation}
As this version of the gate is also subspace preserving, it can again be written as
the direct sum of two routing operations acting separately on the logical subspaces
\begin{equation*}
\hat{S}_{sr}=\hat{S}_{sr}^{(\circleopen)}\oplus\hat{S}_{sr}^{(\circlefill)} .
\end{equation*}
%\begin{equation}
%\begin{split}
%    &\hat{S}^{(\circleopen)} =\ket{\redcircleopen^{2l-1,d+1}}\bra{\redcircleopen^{l,d}} + \ket{\redcircleopen^{2l,d+1}}\bra{\bluecircleopen^{l,d}};
 %   \\&
 %   \hat{S}^{(\circlefill)} = \ket{\redcirclefill^{2l-1,d+1}}\bra{\redcirclefill^{l,d}} + \ket{\redcirclefill^{2l,d+1}}\bra{\bluecirclefill^{l,d}}. 
 %\end{split}
%\end{equation}
%Both operators are unitary permutations of the basis states.
\gdr{Notice that the same routing rule now applies in both logical subspaces. This is possible because $\hat U_{in}$
and $\hat U_B$ act across both subspaces, so that every carrier receives the same
color instruction regardless of the logical value it encodes.}

\subsubsection{Memory Operations}

Even in this qudit implementation of the protocol, we can perform operations over the memory cells with the same primitives as in the \gdr{basic} one. As for the rail exchange in the dual-rail implementation, here we need operations that act only on the logical subspace and leave the color (routing) unchanged, i.e., that perform operations from $\mathcal{H}_W^{(\circleopen)}$ to $\mathcal{H}^{(\circlefill)}_W$ and vice versa.

More in details, at the memory cell $\boldsymbol{M}^{(\boldsymbol{a})}$ we have a classical $m$-bit message $\boldsymbol{b}^{(a)}\in\{0,1\}^m$ carried by a collection of $m$ qudits with fixed spatial positions. The overall message contained in the cell is defined as $\ket{\boldsymbol{b}^{(a)}}_{\boldsymbol{M}^{(\boldsymbol{a})}}\in\{\ket{\redcircleopen},\ket{\redcirclefill}\}^{\otimes m}$, that is, the classical message is encoded by $m$ qudits encoding either a logical $0$ or a logical $1$, with color state fixed as red. Once the qudits of the data register $\boldsymbol{D}$ reach the target memory cell $\boldsymbol{M}^{(\boldsymbol{a})}$ we perform a copy of the logical state of the memory qudits onto the state of the data qudits through the use of $\mathcal{O}(m)$ local operations, triggered by an additional flag qudit or by the activation of the cell degree of freedom, analogously to what is done in the previous protocols. More precisely we execute the operation
\begin{equation}
    \begin{split}
        &\ket{\redcirclefill}_{D_j}\ket{m_j=\redcircleopen}_{M_j^{(\boldsymbol{a})}}\xrightarrow{\mathrm{copy}}\ket{\redcircleopen}_{D_j}\ket{\redcircleopen}_{M_j^{\boldsymbol{(a)}}}\\
        &\ket{\redcirclefill}_{D_j}\ket{m_j=\redcirclefill}_{M_j^{(\boldsymbol{a})}}\xrightarrow{\mathrm{copy}}\ket{\redcirclefill}_{D_j}\ket{\redcirclefill}_{M_j^{\boldsymbol{(a)}}}.
    \end{split}
\end{equation}

This local operation can be implemented by introducing a local unitary gate acting on couples $(D_i, M_i^{(\boldsymbol{a})})$, whose action preserves the color of the qudit but changes its logical subspace. Explicitly, we can write the operation as
\begin{equation}\label{quditloccopy}
    \hat{U}_{copy}^{loc}(D_i \vert M_{i}^{(\boldsymbol{a})}):=\ket{\redcircleopen}_{M_{i}^{(\boldsymbol{a})}}\bra{\redcircleopen}\otimes \hat{X}^L_{D_i}+\ket{\redcirclefill}_{M_{i}^{(\boldsymbol{a})}}\bra{\redcirclefill}\otimes \hat{I}_{D_i}
\end{equation}
where now $\hat{X}^L_{D_i}$ is an operation that can be regarded as a logical bit flip that swaps the orthogonal logical subspaces %$\circleopen\leftrightarrow\circlefill$ 
$\mathcal{H}_S^{(\circleopen)}\leftrightarrow\mathcal{H}_S^{(\circlefill)}$ while preserving the color. Its action on the four qudits states is 
\begin{equation*}
\begin{split}
    &\hat{X}^L \ket{\redcircleopen}=\ket{\redcirclefill},\quad \hat{X}^L\ket{\redcirclefill}=\ket{\redcircleopen},\\
&\hat{X}^L\ket{\bluecircleopen}=\ket{\bluecirclefill},\quad \hat{X}^L\ket{\bluecirclefill}=\ket{\bluecircleopen}.
\end{split}
\end{equation*}

This qudit-based implementation of memory operations is logically consistent with the dual-rail approach while simplifying its physical realization. By including the rail degree of freedom into the state space of the qudit, the protocol allows the retrieval and copying of stored information within a single binary tree. This eliminates the need to physically move data walkers between separate hardware structures during the memory call. This approach, however, shifts the requirements of the protocol toward controlling higher-dimensional Hilbert space, necessitating enhanced precision in manipulating the internal states of the qudits through logical operations.

\subsubsection{Resource scalings}
We now evaluate the resources required for the qudit-based implementation in terms of walker count, gate complexity, path extent, and execution time. Specifically, we must account for the scalings of both the long- and short-range variants. 

Both implementations operate with a fixed number of physical walkers: an $n$-walker address register $\boldsymbol{A}$ and an $m$-walker data register $\boldsymbol{D}$, totaling $n+m \sim \mathcal{O}(n+m)$ qudits. 

The protocol is executed on a single binary tree of depth $n$, where, as usual, each bifurcation is associated with a unitary block consisting of two operations, $\hat{S}$ and $\hat{U}^{(d)}$.
In the long-range implementation, the total gate count matches that of the \gdr{basic} protocol, yielding $2(2^n-1) \sim \mathcal{O}(2^n)$. Conversely, in the short-range implementation, decomposing $\hat{U}^{(d)}$ into applications of $\hat{U}_{in}$ and $\hat{U}_B$, in analogy to the backup variant, increases the number of gates to $3(2^n-1)\sim\mathcal{O}(2^n)$.

The total spatial extent of the tree and the type of operation used bound the execution time of the protocol. In the long-range qudit implementation, the number of sites per branch scales linearly with the train size. This leads to a total path extent of $n_{\text{sites}}^{\text{path}}=n(n+m)-\frac{n(n+1)}{2}\sim\mathcal{O}(n^2+nm)$. Consequently, the total circuit depth also exhibits a quadratic scaling $T_{\text{serial}}\sim \mathcal{O}(n(n+m))$. 

Instead, the short-range protocol drastically reduces the spatial overhead. The number of sites per branch is constant $n_{\text{sites}}^d=2\sim\mathcal{O}(1)$, leading to a path extent of $n_{\text{sites}}^{\text{path}}=2n\sim\mathcal{O}(n)$.
The circuit depth can be reduced by parallelizing the operations, as shown in Fig.~\ref{qudit_short}. 
In analogy to what was done in Sec.~\ref{DepthReduction}, assuming that a single time step consists of a spatial shift and a local gate application, the leading particle requires exactly $2n$ steps to traverse the tree and arrive at the memory cells.
Once there, the walker traverses the $m$ memory sites in $m$ time steps. Finally the inverse routing requires other $2n$ steps for a total of $4n+m$ steps. Because the rest of the train follows continuously, the remaining particles follow sequentially over the next $n+m$ steps. This yields an overall depth of $T_{\text{parallel}} = 4n + m + (n+m) = 5n+2m \sim \mathcal{O}(n+m)$.

Finally, consistently with our previous variants, the number of active elements scales linearly as $\mathcal{O}(n)$ for a classical or sparse query, while the scaling becomes exponential, $\mathcal{O}(2^n)$, for a dense superposition.

\begin{table*}[htb!]
    \centering
    \renewcommand{\arraystretch}{1.3}
    \setlength{\tabcolsep}{8pt} 
    \resizebox{\textwidth}{!}{
    \begin{tabular}{l c c c c c c c}
        \toprule
        \textbf{Variant} & \textbf{Info carriers} & \textbf{Binary trees} &\textbf{Total Gates} & \textbf{Path Extent} & \textbf{Depth} & \multicolumn{2}{c}{\textbf{Active Elements}}  \\
        \cmidrule(lr){7-8}
        & & & & & & \textit{Sparse} & \textit{Full}  \\
        \midrule\midrule
       
        % --- CONFRONTI BB e ASY ---
        \textbf{BB}  & \multicolumn{1}{c}{\begin{tabular}{c}
             $2^n-1$ qutrits  \\
             $ n + m$ qubits
         \end{tabular}} & $1$ & $\mathcal{O}(2^n)$ & $\mathcal{O}(n)$ & $\mathcal{O}(n+m)$ & $\mathcal{O}(n)$ & $\mathcal{O}(2^n)$ \\
        \textbf{ASY} & $n+m$ & $2(n+m)$ & $\mathcal{O}((n+m)2^n)$ & $\mathcal{O}(n)$ & $\mathcal{O}(n\log (n+m))$ & $\mathcal{O}(n^2+nm)$ & $\mathcal{O}(2^n)$ \\
        \midrule
        
        % --- SEZIONE STANDARD ---
        \multicolumn{7}{l}{\textbf{\gdr{Basic}}} \\
        Long range  & $\leq (n + m)$ & $1$ & $2(2^n-1)$ & $n^2/2+nm+\mathcal{O}(n)$ & $n^2/2+nm+\mathcal{O}(n+m)$ & $\mathcal{O}(n)$ & $\mathcal{O}(2^n)$ \\
        Short range & $\leq 2(n+m)$   & $1$ & $3(2^n-1)$ & $4n$      & $10n+4m$ & $\mathcal{O}(n)$ & $\mathcal{O}(2^n)$ \\
        \addlinespace % Aggiunge un po' di spazio bianco elegante al posto di una linea
        
        % --- SEZIONE DUAL RAIL ---
        \multicolumn{7}{l}{\textbf{Dual-rail}} \\
        Long range  & $n+m$   & $2$ & $4(2^n-1)$ & $n^2/2+nm+\mathcal{O}(n)$      & $n^2/2+nm+\mathcal{O}(n+m)$ & $\mathcal{O}(n)$ & $\mathcal{O}(2^n)$ \\
        Short range & $3(n+m)$   & $2$ & $6(2^n-1)$ & $4n$ & $13n+7m$ & $\mathcal{O}(n)$ & $\mathcal{O}(2^n)$ \\
        \addlinespace
        
        % --- SEZIONE QUDIT ---
        \multicolumn{7}{l}{\textbf{Qudit}} \\
        Long range  & $n+m$   & $1$ & $2(2^n-1)$ & $n^2/2+nm+\mathcal{O}(n)$      & $n^2/2+nm+\mathcal{O}(n+m)$ & $\mathcal{O}(n)$ & $\mathcal{O}(2^n)$ \\
        Short range & $n+m$   & $1$ & $3(2^n-1)$ & $2n$      & $5n+2m$ & $\mathcal{O}(n)$ & $\mathcal{O}(2^n)$ \\
      
        \bottomrule
    \end{tabular}
    }
    \caption{Comparison of resource scaling for different carrier models.\gdr{The number of information carriers is reported up to $\mathcal{O}(1)$ terms, which depend on implementation details such as the flag walkers employed in the copy operation. For the BB and ASY proposals, which do not specify the spatial arrangement of their routing elements, the path extent is estimated assuming a constant number of sites per edge.}}
    \label{resource_comparison}
\end{table*}

\section{Resource scalings of our qRAM proposals}\label{sec:scaling}

In Table~\ref{resource_comparison} we summarize the asymptotic resource scalings of our proposed qRAM protocols alongside state-of-the-art architectures, such as the Bucket Brigade (BB) and the ASY models. We compare these models across the key metrics defined in the previous sections: the total number of information carriers, the number of required binary trees, the overall gate count, the spatial extent of the routing paths, namely the number of sites from the root to the memory cells, the circuit depth, and the number of active elements necessary for sparse (classical) versus full (superposition) address queries. %For the sake of clarity, we display only the asymptotic behaviors, whereas the exact dependence from $n$ and $m$ are detailed in the corresponding scaling sections presented above.

\subsection{Comparison with other architectures}

The most prominent advantage of our architecture over the Bucket Brigade (BB) scheme lies in the elimination of the exponential hardware bottleneck associated with active routing elements. In the standard BB architecture, routing information is stored within stationary qutrits located at every node of the binary tree. Consequently, preserving the coherence of a full superposition query requires actively maintaining the quantum state of $\mathcal{O}(2^n)$ physical elements for the entire duration of the protocol, a major challenge for physical implementations. We circumvent this limitation by fundamentally shifting the storage of routing information: rather than being held by qutrits at nodes, the routing instructions are encoded into the internal states of the propagating information carriers and distributed dynamically throughout the query.

\gdr{The nodes of our binary tree host only the scattering gates: these are passive elements, carrying no internal degrees of freedom whose coherence has to
be maintained, while the address-transmission gates act on the carriers
themselves along the edges. While the presence of $\mathcal{O}(2^n)$ gates is
intrinsically unavoidable for any architecture addressing $2^n$ memory cells, the
number of active quantum carriers that must be kept coherent in our model is
strictly bounded to a linear scaling of $\mathcal{O}(n+m)$ walkers.}
Furthermore, our short range configurations successfully match the optimal $\mathcal{O}(n+m)$ circuit depth of the parallelized BB protocol ~\cite{Hann_noise_resilience}%\dds{\it[Inserire citazione del PRX]}
, demonstrating that this drastic hardware reduction does not compromise the overall operational speed.

Furthermore, our protocols offer advantages in spatial resource management and operational speed when compared to previously proposed walker-based architectures. While the ASY architecture successfully reduces the required number of information carriers to a linear scaling, it trades this reduction with a considerable spatial overhead, where the routing mechanism relies on a linearly scaling number $\mathcal{O}(n+m)$ of parallel binary trees. Particularly for queries involving large target registers, $m \gg n$, this requirement becomes a prohibitive hardware bottleneck. 

In contrast, all our proposed architectures require a constant $\mathcal{O}(1)$ number of binary trees.
Regarding operational speed, the ASY architecture exhibits a sub-optimal circuit depth of $\mathcal{O}(n \log(n+m))$ under the assumption that arbitrarily long-range interactions are available at no additional cost. However, if decomposed into short-range local operations its depth reduces to a quadratic scaling of $\mathcal{O}(n^2+nm)$. Conversely, all our short range variants successfully maintain the optimal linear circuit depth of $\mathcal{O}(n+m)$, while exclusively employing short-range couplings.

\subsection{Comparison between our variants}

Each proposal introduced in this work has specific trade-offs in terms of number/type of walkers or routing gates adopted, which we compare in this section and are detailed in Table~\ref{resource_comparison}. \gdr{Transposing our basic protocol into alternative physical encodings}, such as the dual-rail and qudit implementations, inevitably introduces minor structural overheads. For instance, the dual-rail scheme necessitates two parallel binary trees instead of one. Furthermore, while the \gdr{basic} protocol implicitly encodes part of the information via the absence of particles, the particle-conserving dual-rail and qudit implementations explicitly require exactly $n+m$ physical carriers. Nevertheless, these physical trade offs affect only constant prefactors, fully preserving the asymptotic resource scalings. 

A crucial distinction within our framework lies between the long-range and short-range interaction models. 
While conceptually simpler, the long-range protocols suffer from a sub-optimal circuit depth of $\mathcal{O}(n^2+nm)$, further burdened by the intrinsic physical difficulty of implementing non-local couplings. Our short range variants successfully resolve those bottleneck, obtaining the optimal linear depth of $\mathcal{O}(n+m)$ while operating exclusively via local range-2 interactions. In the \gdr{basic} protocol, this optimization is achieved at the cost of doubling the information carriers to $2(n+m)$ to include auxiliary walkers, while the dual-rail architecture requires tripling them to $3(n+m)$. Both represent physically reasonable trade-offs that maintain the fundamental linear scaling of the carriers. Notably, the qudit short range variant emerges as the most resource-efficient model in this regard: it successfully achieves the linear depth optimization without incurring any carrier overhead. This ideal performance is made possible by leveraging the augmented local Hilbert space dimensionality of the qudit systems, which effectively absorbs the complexity of local routing operations, which makes use of solely range-1, namely nearest neighbor, operations.

\section{Discussion and conclusion}
\label{sec:conclusion}

In this work, we provided a comprehensive technical description of a novel quantum random access memory (qRAM) architecture based on quantum walkers on a binary tree, originally introduced in Ref.~\cite{ourPRL}. Building upon our preliminary proposal, we formally established the general requirements necessary to implement the core routing protocol. Crucially, we expanded our original framework by introducing several novel architectural variants and explicit memory-access mechanisms. As detailed in the following, these new additions are specifically designed to extend the applicability of our model to diverse information carriers and to adapt it for physical platforms with strictly local connectivity.

A central achievement of our proposal is the elimination of the exponential number of active routing elements required by the standard Bucket Brigade (BB) architecture. Crucially, we achieve this resource reduction while operating on a constant number of binary trees, a single tree for the \gdr{basic} and qudit variants, and two for the dual-rail one, thus avoiding the spatial overhead characteristic of the ASY model, which requires $\mathcal{O}(n+m)$ parallel binary trees. This is made possible by shifting the routing information from auxiliary particles stored at the tree nodes directly onto the internal state of the information carriers themselves. Depending on the variant, this information is stored either in the presence/absence pattern of the walkers (\gdr{basic} architecture) or in their rail/qudit degree of freedom (particle-conserving variants); in all cases, the need for the $\mathcal{O}(2^n)$ active qutrits of the BB scheme is entirely bypassed. Consequently, the nodes act as purely passive scattering gates that simply route the $\mathcal{O}(n+m)$ propagating walkers. A further distinctive feature of this dynamical routing is the behaviour of the address walkers: once a walker has transmitted its path information, in general it separates from the rest of the train and continue to propagate across different branches. The fixed ordering and the constant spatial separation $\Delta x$ of the train, combined with the mirror symmetry between the input and output trees, ensure that every address walker is deterministically recollected at the correct level during the inverse routing, so that the whole register exits coherently at the output port.

While the \gdr{basic} protocol relies on long-range interactions, we demonstrated how a minimal structural modification can replace these with local, at most range-2 operations, involving gates acting on at most three walkers at a time. By introducing a constant multiplicative overhead in the number of carriers, we successfully decomposed the main address-transmission gate into smaller, localized operations acting on subsets of walkers. This adaptation makes the architecture highly suitable for hardware platforms with native nearest-neighbor connectivities. Furthermore, this decomposition naturally enables the parallelization of operations along the tree. As a result, the overall circuit depth is drastically reduced from quadratic $\mathcal{O}(n^2+nm)$ to linear $\mathcal{O}(n+m)$, achieving the optimal theoretical depth scaling for a qRAM.

Furthermore, we extended our framework to accommodate alternative encoding structures and information carriers. In the dual-rail variant, we exchange the single-tree topology for a train of particles that are deterministically present at all times. This eliminates the need of spatial presence/absence encoding, at the cost of introducing a second binary tree to handle the dual-rail information.

To optimize the potential experimental overhead of the dual rail paradigm, we designed a novel initialization scheme useful to route particles into the two binary trees exploited. This strategy bypasses the need for a complex state preparation procedure executed simultaneously across both physical rails. Instead, it shifts the preparation complexity entirely to the internal color state of the adopted quantum walkers, localized on a single auxiliary rail. Hence, the subsequent splitting of this preliminary rail gives rise to the dual-rail routing that characterizes this model.
The price of this initialization strategy is a linear-time buffering overhead, needed to realign the walker triples before routing; this delay is nonetheless absorbed by the overall depth of the protocol. The dual-rail architecture benefits from a similar parallelization scheme through the corresponding backup variant. This ensures that the protocol retains a highly favorable linear scaling in circuit depth, $\mathcal{O}(n+m)$, albeit with larger constant prefactors compared to the \gdr{basic} single-rail counterpart.

Finally, we extended our architecture to accommodate information carriers with an enlarged internal Hilbert space, specifically focusing on a four-level qudit implementation. Remarkably, this variant circumvents the reliance on spatial presence/absence encoding without requiring the architectural overhead of a second binary tree; the entire protocol unfolds within a single, standard binary tree structure. The primary advantage of this approach lies in the richer operational capacity enabled by the qudits' higher local degrees of freedom. By accepting a more demanding control overhead over a larger state space, we can implement a strictly short-range, namely nearest-neighbor,  routing protocol without introducing any auxiliary backup particle. Crucially, this structural simplification preserves the optimal linear scaling of the circuit depth, $\mathcal{O}(n+m)$, while yielding the smallest constant prefactors among all the strategies proposed in this work. Consequently, the qudit-based architecture represents an exceptionally compelling trade-off, successfully optimizing physical footprint and operational efficiency.

The exact resource scalings for all variants introduced in this work, alongside a comparison with state-of-the-art architectures, are summarized in Table~\ref{resource_comparison}.

It is worth clarifying the physical scaling and the distinct advantages of our architecture. Naturally, to address $2^n$ memory cells, any qRAM requires $\mathcal{O}(2^n)$ nodes and a physical tree depth scaling as $\mathcal{O}(n)$. Moreover, querying a fully delocalized superposition inevitably requires the information to spread across the entire structure, activating $\mathcal{O}(2^n)$ gates coherently, which is a fundamental cost shared by all existing proposals.

However, our architecture distinguishes itself fundamentally in how it allocates and manages these quantum resources. In standard schemes like the Bucket Brigade, one must actively maintain the quantum coherence of $\mathcal{O}(2^n)$ stationary routing elements for the entire duration of the query. Alternative walker-based proposals, such as the ASY model, reduce the number of active carriers but introduce a prohibitive spatial and operational overhead of $\mathcal{O}(n+m)$ parallel binary trees requiring interaction at each tree bifurcation. Our framework successfully overcomes those bottlenecks: not only it operates on a constant number of binary trees, regardless of the address and message length, but also limits interactions between them strictly to memory operations.
Furthermore, it ensures that all $\mathcal{O}(2^n)$ routing nodes act as entirely passive scattering gates. By confining the active quantum degrees of freedom strictly to the $\mathcal{O}(n+m)$ propagating walkers, our architecture drastically relaxes the hardware coherence requirements compared to previous models. Crucially, this reduction in hardware complexity does not compromise operational speed, as our short-range variants successfully match the optimal $\mathcal{O}(n+m)$ circuit depth. Finally, the only external structural requirement is a global classical clock to synchronize the walker train movement, introducing no additional quantum overhead.

Several directions remain open for future investigation. A crucial next step consists in identifying the most suitable physical platforms capable of experimentally implementing the localized logical operations required by our protocols. Establishing such a concrete hardware mapping will provide the necessary foundation for a rigorous, quantitative noise analysis. This will allow us to characterize how platform-specific imperfections, such as gate errors, localized decoherence, and timing jitter in the classical synchronization clock, propagate through the circuit and affect the fidelity of the retrieved message as a function of the address size $n$ and message length $m$. A complementary and equally important direction concerns the study of the architecture's resilience to generic, hardware-independent noise models, analogous to the analysis carried out for the Bucket Brigade in Ref.~\cite{Hann_noise_resilience}. Finally, a more ambitious long-term goal is establishing whether our architecture admits a fully fault-tolerant implementation with an acceptable overhead, which remains a fundamental open question for assessing the ultimate practical viability of walker-based qRAM proposals.

\section{Acknowledgments}
We acknowledge financial support by MUR (Ministero dell’ Università e della Ricerca) through the PNRR MUR project PE0000023-NQSTI.

\bibliographystyle{apsrev4-2}
\bibliography{biblio_pra}

\end{document}